\documentclass[aps,onecolumn,10pt]{revtex4}
\usepackage{amsmath}
\usepackage{amssymb}
\usepackage{hyperref}
\hypersetup{colorlinks=true} 
\numberwithin{equation}{section}
\numberwithin{equation}{section}

\begin{document}
\allowdisplaybreaks
\setcounter{equation}{0}

\title{Three-dimensional and four-dimensional scalar, vector, tensor cosmological fluctuations and the cosmological decomposition theorem}

\author{Matthew G. Phelps, Asanka Amarasinghe  and Philip D. Mannheim}
\affiliation{Department of Physics, University of Connecticut, Storrs, CT 06269, USA \\
matthew.phelps@uconn.edu, asanka.amarasinghe@uconn.edu, philip.mannheim@uconn.edu\\ }

\date{December 21 2019}

\begin{abstract}

In cosmological perturbation theory it is convenient to use the scalar, vector, tensor  basis as defined according to how these components transform under three-dimensional rotations. In attempting to solve the fluctuation equations that are automatically written in terms of gauge-invariant combinations of these components, the equations are taken to break up into separate scalar, vector and tensor sectors, the decomposition theorem. Here, without needing to specify a gauge, we solve the fluctuation equations exactly for some standard cosmologies, to show that in general the various gauge-invariant combinations only separate at a higher-derivative level. To achieve separation at the level of the fluctuation equations themselves one has to assume boundary conditions for the higher-derivative equations. While asymptotic conditions suffice for fluctuations around a de Sitter background or a spatially flat Robertson-Walker background, for fluctuations around a spatially non-flat Robertson-Walker background one additionally has to require that the fluctuations be well-behaved at the origin. We show that in certain cases the gauge-invariant combinations themselves involve both scalars and vectors. For such cases there is no decomposition theorem for the individual scalar, vector and tensor components  themselves, but for the gauge-invariant combinations there still can be. Given the lack of manifest covariance in defining a basis with respect to three-dimensional rotations, we introduce an alternate scalar, vector, tensor basis whose components are defined according to how they transform under four-dimensional general coordinate transformations. With this basis the fluctuation equations greatly simplify, and while one can again break them up into separate gauge-invariant sectors at the higher-derivative level, in general we find that even with boundary conditions we do not obtain a decomposition theorem in which the fluctuations separate at the level of the fluctuation equations themselves.

\end{abstract}

\maketitle

\section{Introduction}
\label{S1}

In analyzing cosmological perturbations it is very convenient to use the scalar, vector, tensor basis for the fluctuations as developed in  \cite{Lifshitz1946}  and \cite{Bardeen1980}. In this basis the fluctuations are characterized according to how they transform under three-dimensional rotations (SVT3), and in this form the basis  has been applied extensively in cosmological perturbation theory (see e.g. \cite{Kodama1984,Bertschinger1996,Ellis2012}). For the simplest case of fluctuations around a flat Minkowski background geometry of the form $ds^2=dt^2-\delta_{ij}dx^idx^j$, the background plus fluctuation line element can be written as 
\begin{eqnarray}
ds^2 &=&(-\eta_{\mu\nu}-h_{\mu\nu})dx^{\mu}dx^{\nu}
\nonumber\\
&=&(1+2\phi) dt^2 -2(\partial_i B +B_i)dt dx^i - [(1-2\psi)\delta_{ij} +2\partial_i\partial_j E + \partial_i E_j + \partial_j E_i + 2E_{ij}]dx^i dx^j,
\label{1.1}
\end{eqnarray}
where the elements of (\ref{1.1}) are required to obey
\begin{eqnarray}
\partial_i B^i = 0,\quad \partial_i E^i = 0, \quad E_{ij}=E_{ji},\quad \partial^jE_{ij} = 0, \quad \delta^{ij}E_{ij} = 0.
\label{1.2}
\end{eqnarray}
As written, (\ref{1.1}) contains ten elements, whose transformations are defined with respect to the background spatial sector as four three-dimensional scalars ($\phi$, $B$, $\psi$, $E$) each with one degree of freedom, two transverse three-dimensional vectors ($B_i$, $E_i$) each with two independent degrees of freedom, and one symmetric three-dimensional transverse-traceless tensor ($E_{ij}$) with two degrees of freedom. The great utility of this basis is that since the cosmological fluctuation equations are gauge invariant, only gauge-invariant scalar, vector, or tensor combinations of the components of the scalar, vector, tensor basis can appear in the fluctuation equations. For fluctuations around flat spacetime for instance the combinations are the one-component $\psi$, the one-component $\phi+\dot{B}  -\ddot{E}$, the two-component $B_{i} -  \dot{E}_{i}$, and the two-component $E_{ij}$, for a total of six independent combinations, just as should be the case since one is permitted four coordinate transformations on an initial ten-component metric fluctuation $h_{\mu\nu}$. In cosmological studies that have been reported in the literature it is common to assume a specific gauge. In our work here we make no such assumption, and work with the gauge-invariant combinations as is. One of the noteworthy features of the SVT3 basis is that as a basis its various components are related to the components of the $h_{\mu\nu}$ basis through spatial integrals of the various components of $h_{\mu\nu}$ (cf. (\ref{2.7}) and (\ref{2.8}) below). The SVT3 basis is thus intrinsically non-local, and its very existence requires that spatially asymptotic boundary conditions are such that these various integrals actually exist. 

One of the key features of this present study is in exploring the role that these boundary conditions play in establishing the so-called decomposition theorem. Specifically, in attempts to solve cosmological  fluctuation equations that have been presented in the literature appeal is commonly made to the decomposition theorem in which it is assumed that the fluctuation equations are solved independently by the separate scalar, vector and tensor sectors. Thus for the schematic example in which the fluctuation equations  take the generic flat space form
\begin{eqnarray}
B_i+\partial_iB=C_i+\partial_iC,
\label{1.3}
\end{eqnarray}
where the $B$ and $B_i$ are given by (\ref{1.1}), and where the $C$ and $C_i$ are functions given by the evolution equations with $C_i$ obeying  $\partial_iC^i=0$, the decomposition theorem requires that one set
\begin{eqnarray}
B_i= C_i,\quad \partial_iB=\partial_iC.
\label{1.4}
\end{eqnarray}
However, (\ref{1.4}) does not follow from (\ref{1.3}), since on applying $\partial^i$ and $\epsilon^{ijk}\partial_j$  to (\ref{1.3}) we obtain 
\begin{eqnarray}
\partial^i\partial_i(B-C)=0,\quad \epsilon^{ijk}\partial_j(B_k-C_k)=0,
\label{1.5}
\end{eqnarray}
and from this we can only conclude that $B$ and $C$ can differ by any function $D$ that obeys $\partial^i\partial_iD=0$, while $B_i$ and $C_i$ can differ by any function $D_i$ that obeys $\epsilon^{ijk}\partial_jD_k=0$, i.e. by any $D_k$ that can be written as the gradient of a scalar. Thus in (\ref{1.5}) we have separated the various scalar and vector components that are present in (\ref{1.3}), to thus obtain a decomposition for the components. However, we have not obtained (\ref{1.4}) itself, viz. the conditions that would be required by the decomposition theorem, but have instead obtained a derivative version of it. 

To be able to proceed from (\ref{1.5}) to (\ref{1.4}) we will need some further information. For an example such as the one given in (\ref{1.3}) that information can be provided by asymptotic boundary conditions. Since three-dimensional plane waves form a complete basis for the operator $\partial_i\partial^i$, we can write a general solution for $D$ in the form
\begin{eqnarray}
D=\sum _{\bf k}a_{\bf k}e^{i\textbf{k}\cdot \textbf{x}},
\label{1.6}
\end{eqnarray}
where the $a_{\bf k}$ are constrained to obey 
\begin{eqnarray}
k^2a_{\bf k}=0.
\label{1.7}
\end{eqnarray}
However, in and of itself (\ref{1.7}) does not lead to $a_{\bf k}=0$ (and thus to $D=0$) as this is not the only allowed solution to (\ref{1.7}). Rather, since $k^2\delta(k)=0$, $k^2\delta(k)/k=0$ we can set
\begin{eqnarray}
a_{\bf k}=\alpha_k\delta(k_x)\delta(k_y)\delta(k_z) +\beta_k\left[\frac{\delta (k_x)\delta (k_y)\delta (k_z)}{k_x}+\frac{\delta (k_x)\delta (k_y)\delta (k_z)}{k_y}+\frac{\delta (k_x)\delta (k_y)\delta (k_z)}{k_z}\right],
\label{1.8}
\end{eqnarray}
where $\alpha_k$ and $\beta_k$ are constants. Inserting the $\alpha_k$ term  into (\ref{1.6}) would remove the ${\bf x}$ dependence from $D$ and provide a constant $D$ that would then not vanish at spatial infinity. Inserting the $\beta_k$ term  into (\ref{1.6}) would provide a $D$ that grows linearly in ${\bf x}$, to thus also not vanish at spatial infinity. Thus a spatially convergent  form for $D$ that would, for instance  be provided by taking  $a_{\bf k}$ to be a convergent $\exp(-k^2a^2)$ Gaussian in momentum space would be excluded, with  $\partial_i\partial^iD=0$ having no localized solutions at all. If we can exclude non-zero $D$, we can set $B=C$, and thus via (\ref{1.3})  can set $B_i=C_i$.  A spatially asymptotic boundary condition is thus needed in order to recover a decomposition theorem. Consequently, we can correlate the establishing of the decomposition theorem with the very existence of the SVT3 basis in the first place as both require asymptotic boundary conditions.

Further insight into the role of boundary conditions can be provided by studying the behavior of $\partial_i\partial^iD=0$ in coordinate space. To this end  we recall the identity
\begin{eqnarray}
A \partial_i\partial^iB-B \partial_i\partial^iA=\partial_i(A\partial^iB-B\partial^iA).
\label{1.9}
\end{eqnarray}
Taking $A$ to be a function that obeys $\partial_i\partial^iA=0$ and taking $B$ to be the propagator $D^{(3)}(\mathbf{x}-\mathbf{y})$ that obeys 
\begin{eqnarray}
\partial_i\partial^iD^{(3)}(\mathbf{x}-\mathbf{y})=\delta^3(\mathbf{x}-\mathbf{y}),
\label{1.10}
\end{eqnarray}
enables us to write $A$ as an asymptotic surface term of the form 
\begin{eqnarray}
A({\bf x}) =\int dS_y^i\left[A({\bf y})\partial^y_iD^{(3)}(\mathbf{x}-\mathbf{y})-D^{(3)}(\mathbf{x}-\mathbf{y})\partial^y_iA({\bf y})\right],
\label{1.11}
\end{eqnarray}
as integrated over a closed surface $S$. The vanishing of the asymptotic surface term then makes $A$ vanish identically. Thus the two non-trivial solutions to $\partial_i\partial^iA=0$, viz. $A$  is a constant or of the form ${\bf n}\cdot {\bf x}$ where ${\bf n}$ is a reference vector (the coordinate analogs of  (\ref{1.8})), are then excluded by an asymptotic boundary condition. Requiring that the asymptotic surface term in (\ref{1.11}) vanish then forces the only solution to $\partial_i\partial^iA=0$ to be $A=0$.

In cosmology where it is convenient to use  polar coordinates, one has to adapt (\ref{1.11}). When written  in polar coordinates with still flat metric $\gamma_{ij}$ and metric determinant $\gamma$, (\ref{1.11}) is replaced by 

\begin{eqnarray}
A(\textbf{x})=\int dS\left[A(\mathbf{y})\frac{\partial D^{(3)}(\mathbf{x},\mathbf{y})}{\partial  n} -D^{(3)}(\mathbf{x},\mathbf{y})\frac{\partial A(\mathbf{y})}{\partial n}\right],
\label{1.12a}
\end{eqnarray}
where $\partial/\partial n$ is the out-directed normal derivative on the surface S, and the propagator obeys
\begin{eqnarray}
\nabla_i\nabla^iD^{(3)}(\mathbf{x},\mathbf{y})=\gamma^{-1/2}\delta^3(\mathbf{x}-\mathbf{y}).
\label{1.13a}
\end{eqnarray}
For $D^{(3)}(\mathbf{x},\mathbf{y})=-1/4\pi|\mathbf{x}-\mathbf{y}|$, (\ref{1.12a}) takes the form
\begin{eqnarray}
A(\textbf{x})=\frac{1}{4\pi} \int dS\left[\frac{1}{|\mathbf{x}-\mathbf{y}|}\frac{\partial A(\mathbf{y})}{\partial n}-
A(\mathbf{y})\frac{\partial}{\partial  n}\frac{1}{|\mathbf{x}-\mathbf{y}|}\right].
\label{1.14a}
\end{eqnarray}
The asymptotic surface term will thus vanish if $A(\mathbf{y})$ behaves as $1/r^{n}$ where $n$ is positive.

A case that we will discuss in detail below is one where the background is a Robertson-Walker geometry with a non-vanishing spatial curvature $k$ and a non-flat $\gamma_{ij}$, viz.
\begin{align}
ds^2=dt^2-a^2(t)\left[\frac{dr^2}{1-kr^2}+r^2d\theta^2+r^2\sin^2\theta d\phi^2\right]=dt^2-a^2(t)\gamma_{ij}dx^idx^j.
\label{1.15a}
\end{align}
In this case the condition that covariantizes the flat $\delta^{ij}\partial_i\partial_jD=0$ is
\begin{align}
\gamma^{-1/2}\partial_i\gamma^{1/2}\gamma^{ij}\partial_jD=0,
\label{1.16a}
\end{align}
where $\gamma=r^4\sin^2\theta/(1-kr^2)$. For $k=-1$ first and with $r=\sinh\chi$, the solutions to (\ref{1.16a}) are of the form $D=D_{\ell}(\chi)Y^m_{\ell}(\theta,\phi)$, where
\begin{align}
\left[\frac{\partial^2}{\partial \chi^2}+2\frac{\cosh\chi}{\sinh\chi}\frac{\partial}{\partial \chi}-\frac{\ell(\ell+1)}{\sinh^2\chi}\right]D_{\ell}(\chi)=0.
\label{1.17a}
\end{align}
For the typical $\ell=0$, $\ell=1$ modes  the solutions are of the form
\begin{align}
D^{(1)}_{0}(\chi)=1,\quad D^{(2)}_{0}(\chi)=\frac{\cosh\chi}{\sinh\chi},\quad D^{(1)}_{1}(\chi)=-\frac{\cosh\chi}{\sinh\chi}+\frac{\chi}{\sinh^2\chi},\quad D^{(2)}_{1}(\chi)=\frac{1}{\sinh^2\chi}.
\label{1.18a}
\end{align}
Of these solutions $D^{(2)}_{1}(\chi)=1/\sinh^2\chi=1/r^2$ is well-behaved at $\chi=\infty$ (i.e, at $r=\infty$) and thus could not be excluded by an asymptotic boundary condition. (In general for asymptotic solutions of the form $D_{\ell}(\chi)=e^{\lambda\chi}$ the coefficient $\lambda$ obeys $\lambda^2+2\lambda=\lambda(\lambda+2)=0$, so only the solution with $\lambda=-2$ is bounded at infinity.) However $D^{(2)}_{1}(\chi)$ diverges at $\chi=0$ (i.e., at $r=0$) and could only be excluded by imposing a convergence requirement not at $r=\infty$ but at $r=0$ (though while we require vanishing at $r=\infty$ we should only require finiteness at $r=0$, so $D_{\ell}(\chi)$ only needs to be well-behaved at $\chi=0$ without needing to vanish there). In general at $\chi=0$ the solutions to (\ref{1.17a}) behave as $\chi^{\ell}$ and as $\chi^{-\ell-1}$ for arbitrary $\ell$, and thus for each $\ell$ there will always be one convergent and one divergent solution at $\chi=0$. For $\ell=0$ and $\ell=1$ the well-behaved solutions at $\chi=0$ are $D^{(1)}_{0}(\chi)$ and $D^{(1)}_{1}(\chi)$, and thus they could only be excluded by a condition at $r=\infty$. We thus need conditions at both $r=0$ and $r=\infty$ (we can impose two conditions since (\ref{1.17a}) is a second-order differential equation) if we are to have $D=0$ be the only allowed solution to (\ref{1.16a}), to then enable us to establish a decomposition theorem.

For $k=+1$ the discussion is analogous. With $r=\sin\chi$ the wave equation is of the form
\begin{align}
\left[\frac{\partial^2}{\partial \chi^2}+2\frac{\cos\chi}{\sin\chi}\frac{\partial}{\partial \chi}-\frac{\ell(\ell+1)}{\sin^2\chi}\right]D_{\ell}(\chi)=0,
\label{1.19a}
\end{align}
and the typical $\ell=0$, $\ell=1$ mode solutions are of the form
\begin{align}
D^{(1)}_{0}(\chi)=1,\quad D^{(2)}_{0}(\chi)=\frac{\cos\chi}{\sin\chi},\quad D^{(1)}_{1}(\chi)=-\frac{\cos\chi}{\sin\chi}+\frac{\chi}{\sin^2\chi},\quad D^{(2)}_{1}(\chi)=\frac{1}{\sin^2\chi}.
\label{1.20a}
\end{align}
Now a space with $k=+1$ is a closed space and the boundary is at $r=1$, i.e. at $\chi=\pi/2$. Of the solutions only $D^{(2)}_{0}(\chi)$ vanishes at $\chi=\pi/2$, and of the solutions only $D^{(1)}_{0}(\chi)$ and $D^{(1)}_{1}(\chi)$ are finite at $\chi=0$. Thus  we need conditions at both $r=0$ and $r=1$ if we are to have $D=0$ be the only allowed solution to (\ref{1.16a}), to then enable us to establish a decomposition theorem.

Now while we see that we can get a decomposition theorem if the evolution equation is of the generic form $\gamma^{-1/2}\partial_i\gamma^{1/2}\gamma^{ij}\partial_jD=0$, in the actual studies that we present below the Robertson-Walker scalar fluctuation equation (and its vector and tensor analogs) is actually of the generic form   
\begin{align}
[\gamma^{-1/2}\partial_i\gamma^{1/2}\gamma^{ij}\partial_j-A_Sk]D=0,
\label{1.21a}
\end{align}
where $A_S$ is a scalar sector constant \cite{footnote1}. The situation is now radically different than when $A_S$ is zero, with there being different outcomes dependent on the sign of $A_Sk$. To see this explicitly we consider $k=-1$, set $r=\sinh\chi$, and represent the background metric by the form
\begin{eqnarray}
ds^2= dt^2-a^2(t)[d\chi^2+\sinh^2\chi d\theta^2+\sinh^2\chi\sin^2\theta d\phi^2].
\label{1.22a}
\end{eqnarray}
For a generic scalar field, $D(\chi,\theta,\phi)$  (\ref{1.21a}) takes the form 
\begin{eqnarray}
\left[ \frac{\partial^2}{\partial\chi^2}+\frac{2\cosh\chi}{\sinh\chi}\frac{\partial}{\partial\chi}+\frac{1}{\sinh^2\chi}\frac{\partial^2}{\partial\theta^2}+\frac{\cos\theta}{\sinh^2\chi\sin\theta}\frac{\partial}{\partial\theta}+\frac{1}{\sinh^2\chi}\frac{\partial^2}{\partial\phi^2}+A_S\right]D(\chi,\theta,\phi)=0.
\label{1.23a}
\end{eqnarray}
In a configuration $D(\chi,\theta,\phi)=D_{\ell}(\chi)Y_{\ell}^m(\theta,\phi)$ (\ref{1.23a}) takes the form 
\begin{eqnarray}
[\gamma^{-1/2}\partial_i\gamma^{1/2}\gamma^{ij}\partial_j+A_S]D(\chi,\theta,\phi)=\left[ \frac{\partial^2}{\partial\chi^2}+\frac{2\cosh\chi}{\sinh\chi}\frac{\partial}{\partial\chi}-\frac{\ell(\ell+1)}{\sinh^2\chi}+A_S\right]D_{\ell}(\chi)Y_{\ell}^m(\theta,\phi)=0.
\label{1.24a}
\end{eqnarray}
We take solutions to (\ref{1.24a}) to behave as $e^{\lambda\chi}$ (times an irrelevant polynomial in $\chi$) and as $\chi^n$ in the $\chi\rightarrow \infty$, $\chi\rightarrow 0$ limits. For (\ref{1.24a}) the limits give
\begin{eqnarray}
&&\lambda^2+2\lambda+A_S=0,\quad \lambda=-1\pm(1-A_S)^{1/2},
\nonumber\\
&&n(n-1)+2n-\ell(\ell+1)=0,\quad n=\ell, -\ell-1.
\label{1.25a}
\end{eqnarray}
At $\chi=0$ (i.e., $r=0$) one of the solutions is well-behaved for all $\ell$, while the other solution diverges for all $\ell$. At $\chi=\infty$ (i.e., $r=\infty$) one of the solutions converges if $A_S<0$ while the other diverges. However if $0<A_S<1$ both solutions are well-behaved asymptotically, while if $A_S>1$ the solutions behave as $e^{-\chi}$ times an oscillating function, to thus again be convergent. Thus for $A_S>0$ both solutions are well-behaved asymptotically, and since one of them is well-behaved at $\chi=0$ for any $\ell$, there will then be a solution that vanishes at $\chi=\infty$ and is well-behaved at $\chi=0$ while not being zero. Such a solution is thus not excluded by the boundary conditions, and for it one does not have a decomposition theorem, i.e., one is not forced to $D=0$. Now it could be the case that for $A_S<0$ the one convergent solution at $\chi=\infty$ matches onto the one convergent solution at $\chi=0$. To see that this does not have to be the case we construct a few typical solutions. Some typical solutions to (\ref{1.24a}) with $\ell=1$ (viz. $n=1,-2$) are
\begin{eqnarray}
D^{(1)}_1(A_S=-3)=\sinh\chi,\quad D^{(2)}_1(A_S=-3)=\cosh\chi-\frac{\cosh^3\chi}{3\sinh^2\chi},
\nonumber\\
D^{(1)}_1(A_S=1)=\frac{\cosh \chi}{\sinh^2\chi},\quad D^{(2)}_1(A_S=1)=\frac{1}{\sinh\chi}-\frac{\chi\cosh\chi}{\sinh^2\chi}.
\label{1.26a}
\end{eqnarray}
In the limits $D^{(1)}_1(A_S=-3)$ behaves as $\lambda=1$, $n=1$, $D^{(2)}_1(A_S=-3)$ behaves as $\lambda=1$, $n=-2$, $D^{(1)}_1(A_S=1)$ behaves as $\lambda=-1$, $n=-2$, and $D^{(2)}_1(A_S=1)$ behaves as $\lambda=-1$, $n=1$. In this case we see that with $A_S$ negative there is no solution that converges in both limits. However, when $A_S$ is positive $D^{(2)}_1(A_S=1)$ is bounded at both limits, and one thus does not have to realize (\ref{1.24a}) by $D^{(2)}_1(A_S=1)=0$, with the decomposition theorem then not holding. In our studies below we will find some situations in which the analog $A_S$ is expressly positive, 
specifically the vector and tensor sector fluctuations around a Robertson-Walker background with non-vanishing $k$, and for such cases the decomposition theorem does not hold in those sectors. However, this analysis only shows that we do not get a decomposition theorem for the vector or the tensor sectors when considered separately. Nonetheless, despite this, below we will find that because of the way that these various sectors are coupled to each other in the fluctuation equations themselves, the decomposition theorem is still obtained. Thus, and in a sense quite remarkably, it is only because of their being coupled in the fluctuation equations of motion that the various sectors are able to obey the same equations that they would have obeyed had they been treated as being completely decoupled from each other in those same equations.

While these examples show that we need some form of boundary conditions in order to be able to obtain a decomposition theorem, we note that this can be avoided in certain specific circumstances. This is because in treating the fluctuation equations there are two generic types of perturbation that one needs to consider. If we start with the Einstein equations in the presence of some background energy-momentum tensor $T_{\mu\nu}$, viz. 
\begin{eqnarray}
 G_{\mu\nu}=R_{\mu\nu}-\frac{1}{2}g_{\mu\nu}R^{\alpha}_{\phantom{\alpha}\alpha}
 =-8\pi G T_{\mu\nu},
\label{1.27a}
\end{eqnarray}
the first type to consider is perturbations $\delta G_{\mu\nu}$ and $\delta T_{\mu\nu}$ to the background that obey 
\begin{eqnarray}
\delta G_{\mu\nu}=-8\pi G \delta T_{\mu\nu}
\label{1.28a}
\end{eqnarray}
in a background that obeys $G_{\mu\nu}=-8\pi G T_{\mu\nu}$. In this case we are solving $G_{\mu\nu}+\delta G_{\mu\nu}+8\pi G (T_{\mu\nu}+\delta T_{\mu\nu})=0$ to both zeroth and first order perturbations in $G_{\mu\nu}+8\pi G T_{\mu\nu}=0$.  In this first case the need for boundary conditions will in general arise as above, and in the body of this paper we will study (\ref{1.28a}) in some standard cosmologies to see the degree to which we can first of all establish a decomposition for components (analog of (\ref{1.5})), and can then proceed from there to a decomposition theorem for the evolution equations themselves (analog of (\ref{1.3})). 

The second kind of perturbation is one in which we introduce some new perturbation $\delta \bar{T}_{\mu\nu}$ to a background that obeys $G_{\mu\nu}+8\pi G T_{\mu\nu}=0$. This $\delta \bar{T}_{\mu\nu}$ will modify both the background $G_{\mu\nu}$ and the background $T_{\mu\nu}$ and will lead to a fluctuation equation of the form 
\begin{eqnarray}
\delta G_{\mu\nu}+8\pi G \delta T_{\mu\nu}=-8 \pi G \delta \bar{T}_{\mu\nu}. 
\label{1.29a}
\end{eqnarray}
Thus for our illustrative example (\ref{1.3}) will be replaced by
\begin{eqnarray}
B_i+\partial_iB-C_i-\partial_iC=\bar{C}_i+\partial_i\bar{C},
\label{1.30a}
\end{eqnarray}
where now $\partial^i\bar{C}_i=0$. Now this time we would obtain $\partial_i\partial^i(B-C-\bar{C})=0$, and initially could only say that $B-C-\bar{C}$ is equal to some function $D$ that obeys $\partial_i\partial^iD=0$. However, in this case it is the very presence of $\bar{C}$ that is causing there to be a non-vanishing $\delta G_{\mu\nu}$ and $\delta T_{\mu\nu}$ in the first place, with $B$ and $C$ needing to be zero if we were to let $\bar{C}$ go to zero. Thus we must have $D=0$ as $B-C$ must be proportional to  $\bar{C}$. Thus (\ref{1.30a}) separates into 
\begin{eqnarray}
B_i-C_i=\bar{C}_i, \quad 
\partial_iB-\partial_iC=\partial_i\bar{C},
\label{1.31a}
\end{eqnarray}
and a decomposition theorem is obtained. As we see, in this case we do not need to impose any  boundary conditions in order to obtain a decomposition theorem.

Now even though classifying the scalar, vector, tensor expansion of the fluctuations according to their behavior under three-dimensional rotations is not manifestly covariant, it is in fact covariant as it leads to fluctuation equations that are gauge invariant, something we will explicitly demonstrate below in some specific cases. Nonetheless, it would be of value to classify the scalar, vector, tensor expansion according to a behavior that is manifestly covariant, i.e. according to an expansion that is classified according to four-dimensional general coordinate scalars, vectors and tensors. We introduced such an SVT4 expansion in \cite{Amarasinghe2018} and will develop it in detail in the body of the text below. And in fact we will find that when written in the SVT4 basis the fluctuation equations are simpler than when written according to SVT3. However, for the moment we note only that for fluctuations around a four-dimensional flat Minkowski background the SVT4 expansion takes the form 
\begin{eqnarray}
h_{\mu\nu}=-2\eta_{\mu\nu}\chi+2\partial_{\mu}\partial_{\nu}F
+ \partial_{\mu}F_{\nu}+\partial_{\nu}F_{\mu}+2F_{\mu\nu},
\label{1.32a}
\end{eqnarray}
where 
\begin{eqnarray}
\partial_{\mu} F^{\mu}= 0, \quad F_{\mu\nu}=F_{\nu\mu},\quad \partial^{\nu}F_{\mu\nu} = 0, \quad \eta^{\mu\nu}F_{\mu\nu} = 0.
\label{1.33a}
\end{eqnarray}
As written, (\ref{1.32a}) contains ten elements, whose transformations are defined with respect to the background as two four-dimensional scalars ($\chi$, $F$) each with one degree of freedom, one transverse four-dimensional vector  ($F_{\mu}$) with three independent degrees of freedom, and one symmetric four-dimensional transverse-traceless tensor ($F_{\mu\nu}$) with five degrees of freedom. Since the gauge-invariant equations have to  contain a total of six gauge-invariant degrees of freedom, they must contain the five-component $F_{\mu\nu}$ and one combination of the five other components that appear in (\ref{1.33a}) (without $F_{\mu\nu}$ we cannot get to six). As we will see, for fluctuations around a flat background the gauge-invariant combinations will be $F_{\mu\nu}$ and $\chi$. For fluctuations around a curved background the gauge-invariant combinations must again include the five-component $F_{\mu\nu}$, but in general the sixth gauge-invariant combination will be a linear combination of the other five components that appear in (\ref{1.33a}) (see (\ref{6.54}) below for a specific example).

To see whether this basis can also lead to a decomposition theorem we consider a four-dimensional analog of (\ref{1.3}): 
\begin{eqnarray}
F_{\mu}+\partial_{\mu}F=C_{\mu}+\partial_{\mu}C,
\label{1.34a}
\end{eqnarray}
where the $F$ and $F_{\mu}$ are given by (\ref{1.32a}), and where the $C$ and $C_{\mu}$ are functions given by the evolution equations with $C_{\mu}$ obeying  $\partial_{\mu}C^{\mu}=0$. For the decomposition theorem to hold one has to be able to set
\begin{eqnarray}
F_{\mu}= C_{\mu},\quad \partial_{\mu}F=\partial_{\mu}C.
\label{1.35a}
\end{eqnarray}
However, (\ref{1.35a}) does not follow from (\ref{1.34a}), since on applying $\partial_{\mu}$  and $\epsilon_{\mu\nu\sigma\tau}n^{\nu}\partial^{\sigma}$ ($n^{\nu}$ is a reference vector) we obtain
\begin{eqnarray}
\partial_{\mu}\partial^{\mu}(F-C)=0,\quad \epsilon_{\mu\nu\sigma\tau}n^{\nu}\partial^{\sigma}(F^{\tau}-C^{\tau})=0.
\label{1.36a}
\end{eqnarray}
While we thus have a decomposition of components, this time we do not get a decomposition theorem in the form given in (\ref{1.35a}) since $F-C$ need not be zero as it could  be equal to an arbitrary function $D$ that is harmonic and thus unconstrained. Specifically, since the set of four-dimensional plane waves provides a complete basis for the $\partial_{\mu}\partial^{\mu}$ operator, in general we can set  
\begin{eqnarray}
D=\sum _{\bf k}a_{\bf k}e^{i\textbf{k}\cdot \textbf{x}-ikt},
\label{1.37a}
\end{eqnarray}
where $k=|{\bf k}|$. However, unlike the $a_{\bf k}$ in (\ref{1.6}) which obey $k_ik^ia_{\bf k}=0$, this time there is no constraint on the $a_{\bf k}$ at all, as the $a_{\bf k}$ obey $k_{\mu}k^{\mu}a_{\bf k}=0$ where $k_{\mu}k^{\mu}={\bf k}^2-k^2$ is identically equal to zero because of the Minkowski signature of the spacetime. Moreover, without violating $\partial_{\mu}\partial^{\mu}D=0$ we can set $a_{\bf k}=\exp(-a^2{\bf k}\cdot{\bf k})$, with the real part of $D$ thus being localized in space according to
\begin{eqnarray}
{\rm Re}[D]={\rm Re}\left[\int \frac{d^3k}{(2\pi)^3}e^{-a^2k^2+i\textbf{k}\cdot \textbf{x}-ikt}\right]=
\frac{1}{16\pi^{3/2}a^3}\left[\frac{r+t}{r}e^{-(r+t)^2/4a^2}+\frac{r-t}{r}e^{-(r-t)^2/4a^2}\right],
\label{1.38a}
\end{eqnarray}
and thus not being constrained by any spatially asymptotic boundary condition at all. (As $r\rightarrow \infty$
${\rm Re}[D]$ falls off as $\exp(-r^2)$, both for fixed $t$ and for points on the light cone where $r=\pm t$.) Thus because of the Minkowski signature there in general is no decomposition theorem. And whether there might be one in any given situation has to be explored on a case by case basis, and we will do this below in the body of the text for some characteristic cosmological models. In fact for SVT4 fluctuations around a de Sitter background for instance we will actually find that we do not in general get a decomposition theorem, though we will find that one can still get one not via boundary conditions at all but via initial conditions instead. However, the appropriate initial conditions have to be chosen extremely judiciously, and there would appear to be no rationale for making such a choice other than a desire to recover the decomposition theorem.

The status of the decomposition theorem changes completely  if we bring in an external source $\delta \bar{T}_{\mu\nu}$ as above, with (\ref{1.34a}) being replaced by 
\begin{eqnarray}
F_{\mu}-C_{\mu}+\partial_{\mu}F-\partial_{\mu}C=\bar{C}_{\mu}+\partial_{\mu}\bar{C}.
\label{1.39a}
\end{eqnarray}
Then, since now it is $\delta \bar{T}_{\mu\nu}$ that is causing the perturbation in the first place $F-C$ must be proportional to $\bar{C}$, so there now is no harmonic function ambiguity. Thus in the scalar sector we have
\begin{eqnarray}
\partial_{\mu}\partial^{\mu}(F-C-\bar{C})=0, 
\label{1.40a}
\end{eqnarray}
with solution $F-C-\bar{C}=0$. Consequently,  the decomposition theorem in the form 
\begin{eqnarray}
F_{\mu}-C_{\mu}= \bar{C}_{\mu},\quad \partial_{\mu}(F-C)=\partial_{\mu}\bar{C}
\label{1.41a}
\end{eqnarray}
then follows, doing so in fact without any need to impose any boundary or initial condition at all.

When the background is not flat we will need to generalize the SVT3 (\ref{1.1}) and SVT4 (\ref{1.32a}). One way  is to simply covariantize them with the use of covariant derivatives instead of ordinary derivatives and the use of a curved space metric instead of the flat space one, as discussed below. However, for cosmology the interesting geometries are de Sitter and Robertson-Walker, and they just happen to be conformal to flat, i.e. for them one can find coordinate systems in which the background metric is written as $ds^2=\Omega^2({\bf x},t)[dt^2-dx^2-dy^2-dz^2]$ where $\Omega ({\bf x},t)$ is an appropriate conformal factor. Thus for such geometries we can replace (\ref{1.1}) and (\ref{1.32a}) by 
\begin{eqnarray}
ds^2 &=&\Omega^2({\bf x},t)\left[(1+2\phi) dt^2 -2(\partial_i B +B_i)dt dx^i - [(1-2\psi)\delta_{ij} +2\partial_i\partial_j E + \partial_i E_j + \partial_j E_i + 2E_{ij}]dx^i dx^j\right],
\label{1.42a}
\end{eqnarray}
and 
\begin{eqnarray}
h_{\mu\nu}=\Omega^2({\bf x},t)\left[-2\eta_{\mu\nu}\chi+2\partial_{\mu}\partial_{\nu}F
+ \partial_{\mu}F_{\nu}+\partial_{\nu}F_{\mu}+2F_{\mu\nu}\right].
\label{1.43a}
\end{eqnarray}
We shall have occasion to use these fluctuation metrics in the study of specific cosmological models that we provide in this paper.

The present paper is organized as follows. In Sec. \ref{S2} we study the SVT3 expansion, gauge invariance and spatially asymptotic behavior. In Sec. \ref{S3} we study the fully covariant D-dimensional SVTD expansion. In Sec. \ref{S4} we relate the SVT4 and SVT3 bases. In Sec. \ref{S5} we solve the SVT4 fluctuation equations in a flat background. In Sec. \ref{S6} we set up and solve the SVT4 fluctuation equations in a de Sitter background. In Secs. \ref{S7} and \ref{S8} we solve the SVT3 fluctuation equations in a de Sitter background and  in a spatially flat radiation era Robertson-Walker background, and find that with asymptotic boundary conditions we can establish the validity of the decomposition theorem in both of the cases. In Sec. \ref{S9} we set up  the SVT3 fluctuation equations in a spatially non-zero Robertson-Walker background in both the radiation and matter eras, and in Sec. \ref{S10} we solve them, so that in Sec. \ref{S11} we are  able to test for and in fact establish the validity of the decomposition theorem in the spatially non-zero Robertson-Walker background case, though now we not only need boundary conditions both asymptotically and at the origin, we also need to take into consideration how the scalar, vector and tensor components interplay with each other in the fluctuation equations. In Sec. \ref{S12} we study SVT4 fluctuation equations around a general Robertson-Walker background, and find them to be far more compact than their SVT3 counterparts given in Sec. \ref{S9}. Since the formulations of cosmological SVT3 and SVT4 fluctuations are not contingent on the specific choice of evolution equations, in Sec. \ref{S13} we show how they work in an alternative to standard Einstein gravity, namely conformal gravity.  In Sec. \ref{S14} we show that in certain cases the gauge-invariant combinations involve both scalars and vectors, to thus preclude a decomposition theorem for the scalars, vectors and tensors themselves. However, instead one still can have a decomposition theorem for the gauge-invariant combinations. In Sec. \ref{S15} we present our conclusions. As noted in \cite{Amarasinghe2018}, as well as implement gauge invariance using non-local operators as per (\ref{2.7}) and (\ref{2.8}) for instance, one can also implement gauge invariance using projection operators, an equally non-local approach.   Thus in an Appendix  we show that this projection approach is equivalent to the SVT approach. The general calculations that we present in this paper are based on symbolic machine calculations for which we  had to quite extensively adapt the $xAct$ tensor calculus software package.

\section{The SVT3 expansion, gauge invariance and asymptotic behavior}
\label{S2}

For the simplest case of fluctuations around a flat Minkowski background geometry of the form $ds^2=dt^2-\delta_{ij}dx^idx^j$, the background plus fluctuation line element can be written (in a slightly more general notation that can be adapted to other geometries or to different coordinate bases such as polar) as 
\begin{eqnarray}
ds^2 &=&(-\eta_{\mu\nu}-h_{\mu\nu})dx^{\mu}dx^{\nu}
\nonumber\\
&=&(1+2\phi) dt^2 -2(\tilde{\nabla}_i B +B_i)dt dx^i - [(1-2\psi)\delta_{ij} +2\tilde{\nabla}_i\tilde{\nabla}_j E + \tilde{\nabla}_i E_j + \tilde{\nabla}_j E_i + 2E_{ij}]dx^i dx^j,
\label{2.1}
\end{eqnarray}
where $\tilde{\nabla}_i=\partial/\partial x^i$ and  $\tilde{\nabla}^i=\delta^{ij}\tilde{\nabla}_j$  (with Latin indices) are defined with respect to the background three-space metric $\delta_{ij}$ (which for flat Minkowski would be a Kronecker delta), and where the elements of (\ref{2.1}) are required to obey
\begin{eqnarray}
\delta^{ij}\tilde{\nabla}_j B_i = 0,\quad \delta^{ij}\tilde{\nabla}_j E_i = 0, \quad E_{ij}=E_{ji},\quad \delta^{jk}\tilde{\nabla}_kE_{ij} = 0, \quad \delta^{ij}E_{ij} = 0.
\label{2.2}
\end{eqnarray}

Direct calculation of the perturbed Einstein tensor in a flat Minkowski background case gives (see e.g. \cite{Amarasinghe2018})
\begin{eqnarray}
\delta G_{00}&=&- 2 \delta^{ab} \tilde{\nabla}_{b}\tilde{\nabla}_{a}\psi,
\nonumber\\
\delta G_{0i}&=&- 2 \tilde{\nabla}_{i}\dot{\psi}+ \tfrac{1}{2} \delta^{ab} \tilde{\nabla}_{b}\tilde{\nabla}_{a}(B_{i} -  \dot{E}_{i}),
\nonumber\\
\delta G_{ij}&=&- 2 \delta_{ij} \ddot{\psi} -  \delta^{ab} \delta_{ij} \tilde{\nabla}_{b}\tilde{\nabla}_{a}(\phi+\dot{B}  -\ddot{E})+ \delta^{ab} \delta_{ij} \tilde{\nabla}_{b}\tilde{\nabla}_{a}\psi 
 + \tilde{\nabla}_{j}\tilde{\nabla}_{i}(\phi+\dot{B} -  \ddot{E})  -  \tilde{\nabla}_{j}\tilde{\nabla}_{i}\psi
\nonumber\\
&+& \tfrac{1}{2} \tilde{\nabla}_{i}(\dot{B}_{j} - \ddot{E}_{j}) + \tfrac{1}{2} \tilde{\nabla}_{j}(\dot{B}_{i}  
- \ddot{E}_{i})- \ddot{E}_{ij} + \delta^{ab} \tilde{\nabla}_{b}\tilde{\nabla}_{a}E_{ij},
\nonumber\\
g^{\mu\nu}\delta G_{\mu\nu}&=&-\delta G_{00}+\delta^{ij}\delta G_{ij}=4 \delta^{ab} \tilde{\nabla}_{b}\tilde{\nabla}_{a}\psi -6\ddot{\psi}-2 \delta^{ab} \tilde{\nabla}_{b}\tilde{\nabla}_{a}(\phi+\dot{B}  -\ddot{E}),
\label{2.3}
\end{eqnarray}
where here and throughout we use the notation given in \cite{Weinberg1972}, and where the dot denotes the time derivative $\partial/\partial x^0$. While a general fluctuation $h_{\mu\nu}$ would have ten components, because of the freedom to make four gauge transformations on the coordinates, the above $\delta G_{\mu\nu}$ can only depend on six of them, with the six being given by the combinations $\psi$, $\phi+\dot{B}  -\ddot{E}$, $B_{i} -  \dot{E}_{i}$, and $E_{ij}$. Since for fluctuations around flat the perturbed $\delta G_{\mu\nu}$ is invariant under the gauge transformation $h_{\mu\nu}\rightarrow h_{\mu\nu}-\partial_{\mu}\epsilon_{\nu}-\partial_{\nu}\epsilon_{\mu}$, one ordinarily takes these six combinations to be gauge invariant. However, the gauge invariance of $\delta G_{\mu\nu}$ entails that only when taken with the various derivatives that appear in (\ref{2.3}) will these combinations be gauge invariant. Thus for instance it is $\delta G_{00}$ that is gauge invariant, and thus it is $\delta^{ab} \tilde{\nabla}_{b}\tilde{\nabla}_{a}\psi$ that is necessarily gauge invariant rather then $\psi$ itself.

To see why a gauge invariance issue might arise, it is instructive to express each of the various SVT3 components in terms  of combinations of the original components of $h_{\mu\nu}$, and to expressly do so without referencing the Einstein equations at all. We follow the derivation given in \cite{Amarasinghe2018}, and given only the definitions of various combinations of the SVT3 fluctuation components  in terms of the $h_{\mu\nu}$ as
\begin{eqnarray}
2\phi&=&-h_{00},\quad B_i+\tilde{\nabla}_iB=h_{0i},\quad h_{ij}=-2\psi\delta_{ij} +2\tilde{\nabla}_i\tilde{\nabla}_j E + \tilde{\nabla}_i E_j + \tilde{\nabla}_j E_i + 2E_{ij},
\label{2.4}
\end{eqnarray}
we obtain
\begin{eqnarray}
\delta^{ij}h_{ij}&=&-6\psi+2\tilde{\nabla}_i\tilde{\nabla}^iE,\quad
 \tilde{\nabla}^jh_{ij}=-2\tilde{\nabla}_i\psi+2\tilde{\nabla}_i\tilde{\nabla}_k\tilde{\nabla}^kE+\tilde{\nabla}_k\tilde{\nabla}^kE_{i},
 \nonumber\\
\tilde{\nabla}^i \tilde{\nabla}^jh_{ij}&=&-2\tilde{\nabla}_k\tilde{\nabla}^k\psi+2\tilde{\nabla}_k\tilde{\nabla}^k\tilde{\nabla}_{\ell}\tilde{\nabla}^{\ell}E,
\label{2.5}
\end{eqnarray}
and can thus set 
\begin{eqnarray}
\tilde{\nabla}_k\tilde{\nabla}^k\psi&=&\frac{1}{4} \left[\tilde{\nabla}^i \tilde{\nabla}^jh_{ij}-\tilde{\nabla}_k\tilde{\nabla}^k(\delta^{ij}h_{ij})\right],
\nonumber\\
\tilde{\nabla}_k\tilde{\nabla}^k\tilde{\nabla}_{\ell}\tilde{\nabla}^{\ell}E&=&\frac{3}{4} \tilde{\nabla}^i \tilde{\nabla}^jh_{ij}-\frac{1}{4}\tilde{\nabla}_k\tilde{\nabla}^k(\delta^{ij}h_{ij}),
\nonumber\\
\tilde{\nabla}_k\tilde{\nabla}^kB&=&\tilde{\nabla}^kh_{0k},
\nonumber\\
\tilde{\nabla}_k\tilde{\nabla}^kB_i&=&\tilde{\nabla}_k\tilde{\nabla}^kh_{0i}-\tilde{\nabla}_i\tilde{\nabla}^kh_{0k},
  \nonumber\\
 \tilde{\nabla}_k\tilde{\nabla}^k\tilde{\nabla}_{\ell}\tilde{\nabla}^{\ell}E_i&=&\tilde{\nabla}_k\tilde{\nabla}^k\nabla^jh_{ij}-\nabla_i\tilde{\nabla}^k\tilde{\nabla}^{\ell}h_{k\ell},
 \nonumber\\
 \tilde{\nabla}_k\tilde{\nabla}^kE_{ij}&=&\frac{1}{2}\left[\tilde{\nabla}_k\tilde{\nabla}^kh_{ij}-\tilde{\nabla}_i\tilde{\nabla}^kh_{kj}-\tilde{\nabla}_j\tilde{\nabla}^kh_{ki}+\tilde{\nabla}_i\tilde{\nabla}_j(\delta^{k\ell}h_{k\ell})\right]
 +\delta_{ij}\tilde{\nabla}_k\tilde{\nabla}^k\psi+\tilde{\nabla}_i\tilde{\nabla}_j\psi,
 \nonumber\\
 \tilde{\nabla}_{\ell}\tilde{\nabla}^{\ell}\tilde{\nabla}_k\tilde{\nabla}^kE_{ij}&=&
\frac{1}{2} \tilde{\nabla}_{\ell}\tilde{\nabla}^{\ell}\left[\tilde{\nabla}_k\tilde{\nabla}^kh_{ij}-\tilde{\nabla}_i\tilde{\nabla}^kh_{kj}-\tilde{\nabla}_j\tilde{\nabla}^kh_{ki}+\tilde{\nabla}_i\tilde{\nabla}_j(\delta^{k\ell}h_{k\ell})\right]
 \nonumber\\
&+&\frac{1}{4}\left[\delta_{ij}\tilde{\nabla}_{\ell}\tilde{\nabla}^{\ell}+\tilde{\nabla}_i\tilde{\nabla}_j \right]\left[\tilde{\nabla}^m \tilde{\nabla}^nh_{mn}-\tilde{\nabla}_k\tilde{\nabla}^k(\delta^{mn}h_{mn}) \right],
\nonumber\\
\tilde{\nabla}_{\ell}\tilde{\nabla}^{\ell} \tilde{\nabla}_k\tilde{\nabla}^k(B_i-\dot{E}_i)&=&
\tilde{\nabla}_{\ell}\tilde{\nabla}^{\ell}\tilde{\nabla}_k\tilde{\nabla}^kh_{0i}
-\tilde{\nabla}_{\ell}\tilde{\nabla}^{\ell}\tilde{\nabla}_i\tilde{\nabla}^kh_{0k}
-\partial_0\tilde{\nabla}_{\ell}\tilde{\nabla}^{\ell}\tilde{\nabla}^jh_{ij}
+\partial_0\tilde{\nabla}_{i}\tilde{\nabla}^{k}\tilde{\nabla}^{\ell}h_{k\ell},
\nonumber\\
 \tilde{\nabla}_k\tilde{\nabla}^k\tilde{\nabla}_{\ell}\tilde{\nabla}^{\ell}(\phi+\dot{B}-\ddot{E})&=&
 -\tfrac{1}{2}\tilde{\nabla}_k\tilde{\nabla}^k\tilde{\nabla}_{\ell}\tilde{\nabla}^{\ell}h_{00}
 +\tilde{\nabla}_{\ell}\tilde{\nabla}^{\ell}\partial_0\tilde{\nabla}^kh_{0k}
 -\tfrac{3}{4}\partial_0^2\tilde{\nabla}^i\tilde{\nabla}^jh_{ij}
  +\tfrac{1}{4}\partial_0^2\tilde{\nabla}_{k}\tilde{\nabla}^{k}(\delta^{ij}h_{ij}).
\label{2.6}
 \end{eqnarray}
As we see, we need to go to fairly high derivatives in order to be able to express each of the SVT3 components entirely in terms of combinations of components of the $h_{\mu\nu}$.

Given (\ref{2.6}) one can readily check that under a gauge transformation $h_{\mu\nu}\rightarrow h_{\mu\nu}-\partial_{\mu}\epsilon_{\nu}-\partial_{\nu}\epsilon_{\mu}$ the combinations  $\tilde{\nabla}_k\tilde{\nabla}^k\psi $, $\tilde{\nabla}_{\ell}\tilde{\nabla}^{\ell}\tilde{\nabla}_k\tilde{\nabla}^kE_{ij}$, $\tilde{\nabla}_{\ell}\tilde{\nabla}^{\ell}\tilde{\nabla}_k\tilde{\nabla}^k(B_i-\dot{E}_i)$ and $ \tilde{\nabla}_k\tilde{\nabla}^k\tilde{\nabla}_{\ell}\tilde{\nabla}^{\ell}(\phi+\dot{B}-\ddot{E})$ are gauge invariant. Thus, as we noted above, it is not the quantities $\psi$, $E_{ij}$, $B_i-\dot{E}_i$ and $\phi+\dot{B}-\ddot{E}$ themselves that are necessarily gauge invariant. Rather, it is their derivatives that are  gauge invariant. Comparison with (\ref{2.3}) shows that it is the quantity $\tilde{\nabla}_k\tilde{\nabla}^k\psi$ that appears in $\delta G_{00}$ and that it is the combination $ \tilde{\nabla}_k\tilde{\nabla}^kE_{ij}-\delta_{ij}\tilde{\nabla}_k\tilde{\nabla}^k\psi-\tilde{\nabla}_i\tilde{\nabla}_j\psi$ that appears in  $\delta G_{ij}$. Thus these  combinations are automatically gauge invariant.

Now we could integrate the relevant equations in (\ref{2.6}), to check gauge invariance for $\psi$, $\phi+\dot{B}-\ddot{E}$, $B_{i}-\dot{E_i}$ and $E_{ij}$ themselves, since we can set

\begin{eqnarray}
\psi&=&\frac{1}{4}\int d^3yD^{(3)}(\mathbf{x}-\mathbf{y})\left[\tilde{\nabla}_y^k \tilde{\nabla}_y^{\ell}h_{k\ell}-\tilde{\nabla}^y_m\tilde{\nabla}_y^m(\delta^{k\ell}h_{k\ell})\right],
\nonumber\\
\phi+\dot{B}-\ddot{E}&=&-\frac{1}{2} h_{00}
+\partial_0\left[\int d^3y D^{(3)}(\mathbf x - \mathbf y) \tilde\nabla^k_y h_{0k}\right]
\nonumber\\
&-&\partial_0^2\left[\int d^3y D^{(3)}(\mathbf x - \mathbf y) \int d^3z D^{(3)}(\mathbf y - \mathbf z)
\left[ \frac{3}{4} \tilde{\nabla}^i \tilde{\nabla}^jh_{ij}-\frac{1}{4}\tilde{\nabla}_k\tilde{\nabla}^k(\delta^{ij}h_{ij})
\right]\right]
\nonumber\\
&=&-\tfrac{1}{2}\tilde{\nabla}_{\ell}\tilde{\nabla}^{\ell} \tilde{\nabla}_k\tilde{\nabla}^k\int d^3y D^{(3)}(\mathbf x - \mathbf y) \int d^3z D^{(3)}(\mathbf y - \mathbf z)h_{00}
\nonumber\\
&+&\partial_0\tilde{\nabla}_{\ell}\tilde{\nabla}^{\ell}\int d^3y D^{(3)}(\mathbf x - \mathbf y) \int d^3z D^{(3)}(\mathbf y - \mathbf z)\nabla^k_z h_{0k}
\nonumber\\
&-&\partial_0^2\left[\int d^3y D^{(3)}(\mathbf x - \mathbf y) \int d^3z D^{(3)}(\mathbf y - \mathbf z)
\left[ \frac{3}{4} \tilde{\nabla}^i \tilde{\nabla}^jh_{ij}-\frac{1}{4}\tilde{\nabla}_k\tilde{\nabla}^k(\delta^{ij}h_{ij})
\right]\right],
\label{2.7}
\end{eqnarray}
and 
\begin{eqnarray}
B_i -\dot{E}_i&=& \int d^3y D^{(3)}(\mathbf x - \mathbf y)\left[ \tilde\nabla^k_y \tilde\nabla_k^y h_{0i}
- \tilde\nabla_i^y \tilde\nabla^k_y h_{0k} \right]
\nonumber\\
&-&\partial_0\left[\int d^3y D^{(3)}(\mathbf x - \mathbf y) \int d^3z D^{(3)}(\mathbf y - \mathbf z)
\left[ \tilde\nabla^k_z \tilde\nabla_k^z \tilde\nabla^j_z h_{ij}-\tilde\nabla_i^z \tilde\nabla^k_z \tilde\nabla^{\ell}_z h_{k\ell}\right]\right]
\nonumber\\
&=&\tilde{\nabla}_{\ell}\tilde{\nabla}^{\ell} \int d^3y D^{(3)}(\mathbf x - \mathbf y) \int d^3z D^{(3)}(\mathbf y - \mathbf z)\left[ \tilde\nabla^k_z \tilde\nabla_k^z h_{0i}
- \tilde\nabla_i^z \tilde\nabla^k_z h_{0k} \right]
\nonumber\\
&-&\partial_0\left[\int d^3y D^{(3)}(\mathbf x - \mathbf y) \int d^3z D^{(3)}(\mathbf y - \mathbf z)
\left[ \tilde\nabla^k_z \tilde\nabla_k^z \tilde\nabla^j_z h_{ij}-\tilde\nabla_i^z \tilde\nabla^k_z \tilde\nabla^{\ell}_z h_{k\ell}\right]\right],
\nonumber\\
E_{ij}&=&\frac{1}{2}\int d^3yD^{(3)}(\mathbf{x}-\mathbf{y})\left[\tilde{\nabla}^y_k\tilde{\nabla}_y^kh_{ij}-\tilde{\nabla}^y_i\tilde{\nabla}_y^kh_{kj}-\tilde{\nabla}^y_j\tilde{\nabla}_y^kh_{ki}+\tilde{\nabla}^y_i\tilde{\nabla}^y_j(\delta^{k\ell}h_{k\ell})\right]
\nonumber\\
&+&\frac{1}{4}\int d^3yD^{(3)}(\mathbf{x}-\mathbf{y})\left[\delta_{ij}\tilde{\nabla}^y_{\ell}\tilde{\nabla}_y^{\ell}+\tilde{\nabla}^y_i\tilde{\nabla}^y_j\right]\int d^3zD^{(3)}(\mathbf{y}-\mathbf{z})\left[\tilde{\nabla}_z^m \tilde{\nabla}_z^{n}h_{mn}-\tilde{\nabla}^z_k\tilde{\nabla}_z^k(\delta^{mn}h_{mn})\right],
\label{2.8}
\end{eqnarray}
where $D^{(3)}(\mathbf{x}-\mathbf{y})$ obeys 
\begin{eqnarray}
\delta^{ij}\tilde{\nabla}_i\tilde{\nabla}_jD^{(3)}(\mathbf{x}-\mathbf{y})=\delta^3(\mathbf{x}-\mathbf{y}),\quad
D^{(3)}(\mathbf{x}-\mathbf{y})=-\frac{1}{4\pi |\mathbf{x}-\mathbf{y}|},\quad \int d^3\mathbf{y}e^{i\mathbf{q}\cdot\mathbf{y}}D^{(3)}(\mathbf{x}-\mathbf{y})=-\frac{e^{i\mathbf{q}\cdot\mathbf{x}}}{q^2},
\label{2.9}
\end{eqnarray}
where $q^2=\delta^{ij}q_{i}q_{j}$, and where in a symbol such as $\tilde{\nabla}_y^i$ for instance the $y$ indicates that the derivative is taken with respect to the $y$ coordinate.

However, there initially is a shortcoming to (\ref{2.7}) and (\ref{2.8}), since while $\psi$ and $E_{ij}$ are manifestly gauge invariant as is (they are expressed in terms of the gauge-invariant flat 3-space $\delta R_{ij}$ and $\delta^{ij}\delta R_{ij}$), to show the gauge invariance of $\phi+\dot{B}-\ddot{E}$ and $B_i -\dot{E}_i$ we would need to be able to integrate by parts. (For $\phi+\dot{B}-\ddot{E}$  we would need to bring $\tilde{\nabla}_{\ell}\tilde{\nabla}^{\ell} \tilde{\nabla}_k\tilde{\nabla}^k$ and $\tilde{\nabla}_{\ell}\tilde{\nabla}^{\ell}$ inside the double integral, while for $B_i-\dot{E}_i$ we would need to bring $\tilde{\nabla}_{\ell}\tilde{\nabla}^{\ell}$ inside.) Similarly, to show that $B_i -\dot{E}_i$ and $E_{ij}$ are transverse we would also need to be able to integrate by parts. Thus we would either have to put constraints on how $h_{\mu\nu}$ is to behave asymptotically, or restrict to requiring in the $E_{ij}$ sector that only $\tilde{\nabla}_{\ell}\tilde{\nabla}^{\ell}\tilde{\nabla}_k\tilde{\nabla}^kE_{ij}$ be gauge invariant and that only $\tilde{\nabla}_{\ell}\tilde{\nabla}^{\ell}\tilde{\nabla}_k\tilde{\nabla}^kE_{ij}$ be transverse, or in the $E_{ij}$ plus $\psi$ sector restrict to requiring that only $\tilde{\nabla}_k\tilde{\nabla}^kE_{ij}-\delta_{ij}\tilde{\nabla}_k\tilde{\nabla}^k\psi-\tilde{\nabla}_i\tilde{\nabla}_j\psi$ be gauge invariant and that only $\tilde{\nabla}_k\tilde{\nabla}^kE_{ij}$ be transverse. 

However, if we take $h_{\mu\nu}$  to be localized in space and oscillating in time, in the integrals given in (\ref{2.7}) and (\ref{2.8}) we can integrate by parts. To be specific, suppose for each mode in a localized packet we set $h_{ij}=\epsilon_{ij}(q)e^{i\mathbf{q}\cdot\mathbf{x}-i\omega(q) t}$ where in this case $\omega(q)$ is not necessarily equal to $q$, and where $\epsilon_{ij}(q)$ is a polarization tensor, as constrained by the restriction that there be none of the $\delta(q)$ or $\delta(q)/q$ type terms that appear in (\ref{1.8}). For such a fluctuation the representative quantities $\psi$ and $E_{ij}$ as given in (\ref{2.7}) and (\ref{2.8}) then evaluate to
\begin{eqnarray}
\psi&=&e^{i\mathbf{q}\cdot\mathbf{x}-i\omega(q) t}\frac{[q^kq^{\ell}\epsilon_{k\ell}(q)-q^2\delta^{k\ell}\epsilon_{k\ell}(q)]}{4q^2},
\nonumber\\
E_{ij}&=&e^{i\mathbf{q}\cdot\mathbf{x}-i\omega(q) t}\bigg{[}\frac{[q^2\epsilon_{ij}(q)-q_iq^k\epsilon_{kj}(q)-q_jq^k\epsilon_{ki}(q)+q_iq_j\delta^{k\ell}\epsilon_{k\ell}(q)]}{2q^2}
\nonumber\\
&+&\frac{(\delta_{ij}q^2+q_iq_j)[q^kq^{\ell}\epsilon_{k\ell}(q)-q^2\delta^{k\ell}\epsilon_{k\ell}(q)]}{4q^4}\bigg{]},
\label{2.10}
\end{eqnarray}
and one can readily check that $\tilde{\nabla}^jE_{ij}=0$. Thus for a wave packet  $h_{ij}=\sum_qa_q\epsilon_{ij}(q)e^{i\mathbf{q}\cdot\mathbf{x}-i\omega(q) t}$ (which for a choice of $a_q$ could be localized), we obtain 
\begin{eqnarray}
\psi&=&\sum_qa_qe^{i\mathbf{q}\cdot\mathbf{x}-i\omega(q) t}\frac{[q^kq^{\ell}\epsilon_{k\ell}(q)-q^2\delta^{k\ell}\epsilon_{k\ell}(q)]}{4q^2},
\nonumber\\
E_{ij}&=&\sum_qa_qe^{i\mathbf{q}\cdot\mathbf{x}-i\omega(q) t}\bigg{[}\frac{[q^2\epsilon_{ij}(q)-q_iq^k\epsilon_{kj}(q)-q_jq^k\epsilon_{ki}(q)+q_iq_j\delta^{k\ell}\epsilon_{k\ell}(q)]}{2q^2}
\nonumber\\
&+&\frac{(\delta_{ij}q^2+q_iq_j)[q^kq^{\ell}\epsilon_{k\ell}(q)-q^2\delta^{k\ell}\epsilon_{k\ell}(q)]}{4q^4}\bigg{]},
\label{2.11}
\end{eqnarray}
and again $\tilde{\nabla}^jE_{ij}=0$. Now, for fluctuations around flat spacetime the set of all $e^{i\mathbf{q}\cdot\mathbf{x}-i\omega (q)t}$ plane waves is complete. Thus since any mode could be expanded as a general sum $h_{ij}=\sum_qa_q\epsilon_{ij}(q)e^{i\mathbf{q}\cdot\mathbf{x}-i\omega(q) t}$ over the plane waves, (\ref{2.11}) holds for the complete basis. Hence absent any $\delta(q)$ or $\delta (q)/q$ type terms we can  integrate by parts.

By the same token we can also integrate the other SVT3 components, and obtain 
 \begin{eqnarray}
2\phi&=&-h_{00},\quad
B=\int d^3yD^{(3)}(\mathbf{x}-\mathbf{y})\tilde{\nabla}_y^ih_{0i},\quad B_i=h_{0i}-\tilde{\nabla}_i\int d^3yD^{(3)}(\mathbf{x}-\mathbf{y})\tilde{\nabla}_y^ih_{0i},
\nonumber\\
E&=&\frac{1}{4}\int d^3yD^{(3)}(\mathbf{x}-\mathbf{y})\int d^3zD^{(3)}(\mathbf{y}-\mathbf{z})\left[3\tilde{\nabla}_z^k\tilde{\nabla}_z^{\ell}h_{k\ell}-\tilde{\nabla}^z_k\tilde{\nabla}_z^k(\delta^{k\ell}h_{k\ell})\right],
\nonumber\\
E_i&=&\int d^3yD^{(3)}(\mathbf{x}-\mathbf{y})\int d^3zD^{(3)}(\mathbf{y}-\mathbf{z})\left[\tilde{\nabla}^z_k\tilde{\nabla}_z^k\nabla_z^jh_{ij}-\nabla^z_i\tilde{\nabla}_z^k\tilde{\nabla}_z^{\ell}h_{k\ell}\right].
\label{2.12}
\end{eqnarray}
While it immediately follows that $\tilde{\nabla}^iB_i=0$,  we need to be able to integrate by parts in order to be able to show that $\nabla^iE_i=0$. However, from (\ref{2.4}) and (\ref{2.6}) we can show directly that both $\tilde{\nabla}_k\tilde{\nabla}^k\tilde{\nabla}_{\ell}\tilde{\nabla}^{\ell}(\phi+\dot{B}-\ddot{E})$ and $\tilde{\nabla}_k\tilde{\nabla}^k\tilde{\nabla}_{\ell}\tilde{\nabla}^{\ell}(B_i-\dot{E}_i)$ are gauge invariant, with the gauge invariance of $\phi+\dot{B}-\ddot{E}$ and $B_i-\dot{E}_i$ themselves then following if we define $B$, $B_i$, $E$ and $E_i$ according to (\ref{2.12}). Thus modulo issues of integrations by parts, we establish that for fluctuations around flat spacetime all of the six $\psi$, $E_{ij}$,  $\phi+\dot{B}-\ddot{E}$ and $B_i-\dot{E}_i$ quantities that appear in $\delta G_{\mu\nu}$ as given in (\ref{2.3}) are gauge invariant. (In counting components $E_{ij}$ and $B_i-\dot{E_i}$ each have two components.)

Now we had noted in Sec. \ref{S1} that we would need spatially asymptotic boundary conditions in order to be able to obtain a decomposition theorem for the SVT3 basis. We now see that we need this very same asymptotic condition in order to make transverseness and gauge invariance compatible \cite{footnote2}. In fact, we actually need such boundary conditions in order to establish the SVT3 decomposition in the first place. Specifically, suppose that we are given some general vector $A_i$ and we want to extract out its transverse and longitudinal components and set $A_i=V_i+\partial_iL$. With $\partial_iV^i=0$ it follows that
\begin{eqnarray}
\partial_i\partial^iL=\partial_iA^i.
\label{2.13}
\end{eqnarray}
Given (\ref{1.9}), the general  solution to (\ref{2.13}) is of the the form 
\begin{eqnarray}
L({\bf x})=\int d^3yD^{(3)}(\mathbf{x}-\mathbf{y})\partial^y_jA^j({\bf y})+\int dS_y^i\left[L({\bf y})\partial^y_iD^{(3)}(\mathbf{x}-\mathbf{y})-D^{(3)}(\mathbf{x}-\mathbf{y})\partial^y_iL({\bf y})\right],
\label{2.14}
\end{eqnarray}
from which it follows that 
\begin{eqnarray}
A_i({\bf x})&=&V_i({\bf x})+\partial^x_iL=V_i({\bf x})+\partial^x_i\int d^3yD^{(3)}(\mathbf{x}-\mathbf{y})\partial^y_jA^j({\bf y})
\nonumber\\
&+&\partial^x_i\int dS_y^i\left[L({\bf y})\partial^y_iD^{(3)}(\mathbf{x}-\mathbf{y})-D^{(3)}(\mathbf{x}-\mathbf{y})\partial^y_iL({\bf y})\right].
\label{2.15}
\end{eqnarray}
Now with the $\partial^x_i\int dS^i(L\partial_iD^{(3)}-D^{(3)}\partial_iL)$ term being the derivative of a scalar, initially it would appear that this term is longitudinal. However, applying $\partial_x^i$ to (\ref{2.15}) gives 
\begin{eqnarray}
&&\partial_x^i\partial^x_i\int dS_y^i\left[L({\bf y})\partial^y_iD^{(3)}(\mathbf{x}-\mathbf{y})-D^{(3)}(\mathbf{x}-\mathbf{y})\partial^y_iL({\bf y})\right]=0.
\label{2.16}
\end{eqnarray}
And thus in fact $\partial_i\int dS^i(L\partial_iD^{(3)}-D^{(3)}\partial_iL)$ is transverse. However, we had already defined $V_i$ as the transverse part of $A_i$, and thus $A_i$ could not have a second transverse piece, so that the surface term must be zero. And thus our very ability to write $A_i$ as
\begin{eqnarray}
A_i({\bf x})=V_i({\bf x})+\partial_iL=V_i({\bf x})+\partial_i\int d^3yD^{(3)}(\mathbf{x}-\mathbf{y})\partial^y_jA^j({\bf y})
\label{2.17}
\end{eqnarray}
in the first place presupposes that the surface term in (\ref{2.15}) is zero, viz.
\begin{eqnarray}
&&\int dS_y^i\left[L({\bf y})\partial^y_iD^{(3)}(\mathbf{x}-\mathbf{y})-D^{(3)}(\mathbf{x}-\mathbf{y})\partial^y_iL({\bf y})\right]
=0,
\label{2.18}
\end{eqnarray}
with $A_i$ thus needing to be well-behaved at spatial infinity  \cite{footnote3}.

Moreover, not only would  $A_i({\bf x})$ need to be asymptotically bounded so that we could uniquely decompose it into transverse and longitudinal components, as we noted for the SVT3 example given in Sec. \ref{S1} this is also a necessary condition for the decomposition theorem to be valid. To be specific, we note that if we take the theory to be just fluctuations around a flat background with no matter energy-momentum tensor, we can separate the various gauge-invariant combinations that appear in (\ref{2.3}) by taking derivatives of $\Delta G_{\mu\nu}$ ($=\delta G_{\mu\nu}$ if $\delta T_{\mu\nu}=0$), to obtain
\begin{eqnarray}
&&\delta^{ab} \tilde{\nabla}_{b}\tilde{\nabla}_{a}\psi=0,
\nonumber\\
 &&\delta^{ab} \tilde{\nabla}_{b}\tilde{\nabla}_{a} \delta^{cd} \tilde{\nabla}_{c}\tilde{\nabla}_{d}(\phi+\dot{B}  -\ddot{E})=0,
 \nonumber\\
 &&\delta^{ab} \tilde{\nabla}_{b}\tilde{\nabla}_{a} \delta^{cd} \tilde{\nabla}_{c}\tilde{\nabla}_{d}(B_i-\dot{E}_i)=0,
 \nonumber\\
 &&\delta^{ab} \tilde{\nabla}_{b}\tilde{\nabla}_{a} \delta^{cd} \tilde{\nabla}_{c}\tilde{\nabla}_{d}(-\ddot{E}_{ij}+\delta^{ef} \tilde{\nabla}_{e}\tilde{\nabla}_{f}E_{ij})=0,
\label{2.19}
\end{eqnarray}
and note that just as in (\ref{2.6}), we need to go to fourth-order derivatives. With the decomposition theorem requiring
\begin{align}
- 2 \delta^{ab} \tilde{\nabla}_{b}\tilde{\nabla}_{a}\psi&=0,
\nonumber\\
- 2 \tilde{\nabla}_{i}\dot{\psi}=0,\quad \tfrac{1}{2} \delta^{ab} \tilde{\nabla}_{b}\tilde{\nabla}_{a}(B_{i} -  \dot{E}_{i})&=0,
\nonumber\\
 -2 \delta_{ij} \ddot{\psi} -  \delta^{ab} \delta_{ij} \tilde{\nabla}_{b}\tilde{\nabla}_{a}(\phi+\dot{B}  -\ddot{E})+ \delta^{ab} \delta_{ij} \tilde{\nabla}_{b}\tilde{\nabla}_{a}\psi +\tilde{\nabla}_{j}\tilde{\nabla}_{i}(\phi+\dot{B} -  \ddot{E})  -  \tilde{\nabla}_{j}\tilde{\nabla}_{i}\psi&=0.
\nonumber\\
 \tfrac{1}{2} \tilde{\nabla}_{i}(\dot{B}_{j} - \ddot{E}_{j}) + \tfrac{1}{2} \tilde{\nabla}_{j}(\dot{B}_{i} 
- \ddot{E}_{i})&=0,
\nonumber\\
- \ddot{E}_{ij} + \delta^{ab} \tilde{\nabla}_{b}\tilde{\nabla}_{a}E_{ij}&=0,
\label{2.20}
\end{align}
we see that if for any quantity $D$ that obeys  $\delta^{ab} \tilde{\nabla}_{a}\tilde{\nabla}_{b}D=0$ (or $\delta^{ab} \tilde{\nabla}_{a}\tilde{\nabla}_{b}\delta^{cd} \tilde{\nabla}_{c}\tilde{\nabla}_{d}D=0$) we impose spatial boundary conditions on $D$ so that $D$ (or $\delta^{ab} \tilde{\nabla}_{a}\tilde{\nabla}_{b}D$) vanishes,  the decomposition theorem will then follow for the $\delta G_{\mu\nu}$ associated with fluctuations around a flat Minkowski background. In this paper we will explore the degree to which this will also be the case for SVT3 fluctuations around some cosmologically interesting backgrounds where the fluctuation equations are more complicated than in the flat background case.

As we will see immediately in Sec. \ref{S3}, we will also need an asymptotic condition in order to establish the very existence of an SVT4 decomposition for the individual components of the fluctuations. However, as we have already noted in Sec. \ref{S1} this is not in general sufficient to provide for a decomposition theorem for the fluctuation equation itself in the SVT4 case. So we turn now to analyze the SVT4 case in detail. 

\section{SVTD expansion around a D-dimensional flat spacetime}
\label{S3}

The discussion given above is not manifestly covariant as the SVT3 components are defined with respect to a three-dimensional subspace of four-dimensional spacetime. (The gauge invariance of the SVT3 formalism shows that it is covariant, just not manifestly so.) It would thus be instructive to develop a formalism that is manifestly covariant, one in which the SVT components are defined with respect to the full space rather than a subspace of it. To this end we adapt the discussion we gave in \cite{Amarasinghe2018}, and so as to be as general as possible consider the SVTD basis in a D-dimensional space.  With Greek indices that range over the full D-dimensional space we first construct a symmetric rank two tensor $F_{\mu\nu}$  that is transverse and traceless in the full D-dimensional space. (Our previously introduced $E_{ij}$ was only transverse and traceless in a 3-dimensional subspace.) The $F_{\mu\nu}$ tensor will have $D(D+1)/2-D-1=(D+1)(D-2)/2$ components, and thus for the full $h_{\mu\nu}$ we need $D+1$ additional pieces of information. For our purposes here we can provide the needed information while at the same time simplifying the discussion given in \cite{Amarasinghe2018} by introducing a D-dimensional  vector $W_{\mu}$, with the one extra needed piece of information being provided by $h=g^{\mu\nu}h_{\mu\nu}$. In terms of this $W_{\mu}$ and $h$ we have found it very convenient to define a general $h_{\mu\nu}$ fluctuation around a flat D-dimensional space to be of the form
\begin{eqnarray}
h_{\mu\nu}=2F_{\mu\nu}+\nabla_{\nu}W_{\mu}+\nabla_{\mu}W_{\nu}+\frac{2-D}{D-1}\nabla_{\mu}\nabla_{\nu}\int d^DyD^{(D)}(x-y)\nabla^{\alpha}W_{\alpha}
\nonumber\\
-\frac{g_{\mu\nu}}{D-1}(\nabla^{\alpha}W_{\alpha}-h)-\frac{\nabla_{\mu}\nabla_{\nu}}{D-1}\int d^DyD^{(D)}(x-y)h,
\label{3.1}
\end{eqnarray}
where the flat spacetime $D^{(D)}(x-y)$ obeys 
\begin{eqnarray}
g^{\mu\nu}\nabla_{\mu}\nabla_{\nu}D^{(D)}(x-y)=\delta^{(D)}(x-y).
\label{3.2}
\end{eqnarray}
As with the SVT3 case discussed in Sec. \ref{S2}, implicit in the form given for $h_{\mu\nu}$  is that the now D-dimensional integrals exist, with $\nabla^{\alpha}W_{\alpha}$ being sufficiently well-behaved at infinity.
To make the $F_{\mu\nu}$ that is defined by (\ref{3.1}) be transverse and traceless requires D+1 conditions, D to be supplied by $W_{\mu}$ and one  to be supplied by $h$. Taking the trace of (\ref{3.1}) shows that as defined $F_{\mu\nu}$ already is traceless (because of the way that $h$ has judiciously been introduced in (\ref{3.1})), while applying $\nabla^{\nu}$  to (\ref{3.1}) yields
\begin{eqnarray}
\nabla^{\nu}h_{\nu\mu}=\nabla_{\alpha}\nabla^{\alpha}W_{\mu},
\label{3.3}
\end{eqnarray}
to thus fix the D components of $W_{\mu}$. The assumed boundedness of $\nabla^{\alpha}W_{\alpha}$ thus correlates with the boundedness of $h_{\mu\nu}$, and  for any sufficiently bounded $W_{\mu}$ that obeys (\ref{3.3}) the D-dimensional rank two tensor $F_{\mu\nu}$ is transverse and traceless.

On now applying $\nabla_{\alpha}\nabla^{\alpha}$ to (\ref{3.1}) we obtain
\begin{eqnarray}
\nabla_{\alpha}\nabla^{\alpha}h_{\mu\nu}&=&2\nabla_{\alpha}\nabla^{\alpha}F_{\mu\nu}+\nabla_{\nu}\nabla^{\alpha}h_{\alpha\mu}+\nabla_{\mu}\nabla^{\alpha}h_{\alpha\nu}+\frac{2-D}{D-1}\nabla_{\mu}\nabla_{\nu}\nabla^{\alpha}W_{\alpha}
\nonumber\\
&-&\frac{g_{\mu\nu}}{D-1}(\nabla^{\alpha}\nabla^{\beta}h_{\alpha\beta}-\nabla_{\alpha}\nabla^{\alpha}h)-\frac{\nabla_{\mu}\nabla_{\nu}}{D-1}h,
\label{3.4}
\end{eqnarray}
and on rearranging we obtain
\begin{eqnarray}
&&\nabla_{\alpha}\nabla^{\alpha}h_{\mu\nu}-\nabla_{\nu}\nabla^{\alpha}h_{\alpha\mu}-\nabla_{\mu}\nabla^{\alpha}h_{\alpha\nu}+\nabla_{\mu}\nabla_{\nu}h
\nonumber\\
&=&2\nabla_{\alpha}\nabla^{\alpha}F_{\mu\nu}+\frac{2-D}{D-1}\nabla_{\mu}\nabla_{\nu}[\nabla^{\alpha}W_{\alpha}-h]-\frac{g_{\mu\nu}}{D-1}(\nabla^{\alpha}\nabla^{\beta}h_{\alpha\beta}-\nabla_{\alpha}\nabla^{\alpha}h).
\label{3.5}
\end{eqnarray}
Now $\nabla_{\alpha}\nabla^{\alpha}h_{\mu\nu}-\nabla_{\nu}\nabla^{\alpha}h_{\alpha\mu}-\nabla_{\mu}\nabla^{\alpha}h_{\alpha\nu}+\nabla_{\mu}\nabla_{\nu}h$ and $\nabla^{\alpha}\nabla^{\beta}h_{\alpha\beta}-\nabla_{\alpha}\nabla^{\alpha}h$ are both gauge invariant (the first term is equal to the D-dimensional fluctuation $2\delta R_{\mu\nu}$ around flat spacetime and the second to $-\delta R$). Now since
\begin{eqnarray}
\nabla_{\beta}\nabla^{\beta}[\nabla^{\alpha}W_{\alpha}-h]=\nabla^{\alpha}\nabla^{\beta}h_{\alpha\beta}-\nabla_{\alpha}\nabla^{\alpha}h,
\label{3.6}
\end{eqnarray}
we define
\begin{eqnarray}
\nabla^{\alpha}W_{\alpha}-h=\int d^DyD^{(D)}(x-y)[\nabla^{\alpha}\nabla^{\beta}h_{\alpha\beta}-\nabla_{\alpha}\nabla^{\alpha}h],
\label{3.7}
\end{eqnarray}
and with this solution we see that  $\nabla_{\alpha}\nabla^{\alpha}F_{\mu\nu}$ is gauge invariant. However as with $E_{ij}$ in the SVT3 case, to show that $\nabla_{\alpha}\nabla^{\alpha}F_{\mu\nu}$ is transverse requires that we can integrate by parts.

We now make the following definitions
\begin{eqnarray}
2\chi&=&\frac{1}{D-1}[\nabla^{\alpha}W_{\alpha}-h],\quad 
\quad 2F=\frac{1}{D-1}\int d^DyD^{(D)}(x-y)[D\nabla^{\alpha}W_{\alpha}-h],
\nonumber\\
F_{\mu}&=&W_{\mu}-\nabla_{\mu}\int d^DyD^{(D)}(x-y)\nabla^{\alpha}W_{\alpha}.
\label{3.8}
\end{eqnarray}
From (\ref{3.8})  it follows that  $\nabla^{\mu}F_{\mu}=0$, with, as per (\ref{3.7}),  $\chi$ being the integral of a gauge-invariant function so that $\nabla_{\alpha}\nabla^{\alpha}\chi$ is automatically gauge invariant. Given (\ref{3.8}) we can rewrite (\ref{3.1}) as 
\begin{eqnarray}
h_{\mu\nu}=-2g_{\mu\nu}\chi+2\nabla_{\mu}\nabla_{\nu}F
+ \nabla_{\mu}F_{\nu}+\nabla_{\nu}F_{\mu}+2F_{\mu\nu},
\label{3.9}
\end{eqnarray}
to thus write $h_{\mu\nu}$ in an SVTD  basis. In a general D-dimensional basis $F_{\mu\nu}$ has $(D+1)(D-2)/2$ components, the transverse $F_{\mu}$ has $D-1$ components, the two scalars $\chi$ and $F$ each have one component, and together they comprise the $D(D+1)/2$ components of a general $h_{\mu\nu}$. If we set $D=3$, we recognize (\ref{3.9}) as the spatial piece of SVT3 given in (\ref{2.1}), just as it should be.

Now in a fluctuation around flat D-dimensional spacetime $\delta G_{\mu\nu}$ can only contain $D(D+1)/2-D=D(D-1)/2$ independent gauge-invariant combinations. With $F_{\mu\nu}$ having $(D+1)(D-2)/2$ components and $\chi$ having one, viz. precisely a total of $D(D-1)/2$, and  with the derivatives of both them being gauge invariant, it follows that $\delta G_{\mu\nu}$ can only depend on $F_{\mu\nu}$ and $\chi$. And given (\ref{3.8}) and (\ref{3.9}), via (\ref{3.5}), (\ref{3.6}) and (\ref{3.7}) we obtain  the following gauge-invariant relations
\begin{eqnarray}
2\nabla_{\alpha}\nabla^{\alpha}\chi&=&\frac{1}{D-1}\left[\nabla^{\alpha}\nabla^{\beta}h_{\alpha\beta}-\nabla_{\alpha}\nabla^{\alpha}h\right],
\nonumber\\
2\nabla_{\alpha}\nabla^{\alpha}\nabla_{\beta}\nabla^{\beta}F_{\mu\nu}&=&\nabla_{\beta}\nabla^{\beta}\left[\nabla_{\alpha}\nabla^{\alpha}h_{\mu\nu}-\nabla_{\nu}\nabla^{\alpha}h_{\alpha\mu}-\nabla_{\mu}\nabla^{\alpha}h_{\alpha\nu}+\nabla_{\mu}\nabla_{\nu}h\right]
\nonumber\\
&+&\frac{1}{D-1}\left[(D-2)\nabla_{\mu}\nabla_{\nu}+g_{\mu\nu}\nabla_{\gamma}\nabla^{\gamma}\right][\nabla^{\alpha}\nabla^{\beta}h_{\alpha\beta}-\nabla_{\alpha}\nabla^{\alpha}h],
\nonumber\\
\delta R_{\mu\nu}&=&\frac{1}{2}[2\nabla_{\alpha}\nabla^{\alpha}F_{\mu\nu}+2(2-D)\nabla_{\mu}\nabla_{\nu}\chi-2g_{\mu\nu}\nabla_{\alpha}\nabla^{\alpha}\chi], \quad \delta R=2(1-D)\nabla_{\alpha}\nabla^{\alpha}\chi,
\nonumber\\
\delta G_{\mu\nu}&=&\delta R_{\mu\nu}-\frac{1}{2}g_{\mu\nu}g^{\alpha\beta}\delta R_{\alpha\beta}=\nabla_{\alpha}\nabla^{\alpha}F_{\mu\nu}+(D-2)(g_{\mu\nu}\nabla_{\alpha}\nabla^{\alpha}-\nabla_{\mu}\nabla_{\nu})\chi,
\nonumber\\
g^{\mu\nu}\delta G_{\mu\nu}&=&(D-2)(D-1)\nabla_{\alpha}\nabla^{\alpha}\chi,
\label{3.10}
\end{eqnarray}
kinematic relations that hold without the imposition of any fluctuation equation of motion. As we see, $\delta G_{\mu\nu}$ nicely depends  on just $F_{\mu\nu}$ and $\chi$, and one can readily check that $\delta G_{\mu\nu}$ automatically obeys $\nabla^{\nu}\delta G_{\mu\nu}=0$.  And with all the components of $\delta G_{\mu\nu}$ being gauge invariant for fluctuations around a D-dimensional flat spacetime,  from the expression for $g^{\mu\nu}\delta G_{\mu\nu}$  we can infer only that $\nabla_{\alpha}\nabla^{\alpha}\chi$ is gauge invariant. And on then applying $\nabla_{\alpha}\nabla^{\alpha}$ to the $\delta G_{\mu\nu}$ equation we can infer only that $\nabla_{\alpha}\nabla^{\alpha}\nabla_{\beta}\nabla^{\beta}F_{\mu\nu}$ is gauge invariant. As noted before, we can only proceed from these conditions to the gauge invariance of $\chi$ and $F_{\mu\nu}$ themselves if we can integrate by parts, and we explore this point further in the Appendix (see the discussion following (\ref{A.27a})). In the Appendix we also provide a generalization of the SVTD approach to the general D-dimensional curved spacetime background.

In $D=4$ we note that $F_{\mu\nu}$ has five components and $\chi$ has one. Since in a fluctuation around a flat four spacetime $\delta G_{\mu\nu}$ can only contain six independent gauge-invariant combinations, it can only depend on $F_{\mu\nu}$ and $\chi$. Thus using a manifestly covariant formalism we replace the six gauge-invariant combinations $\psi$, $E_{ij}$,  $\phi+\dot{B}-\ddot{E}$ and $B_i-\dot{E}_i$ used in SVT3 by the six gauge-invariant combinations $F_{\mu\nu}$ and $\chi$ used in SVT4. This is an altogether more compact set of gauge-invariant combinations, with $\delta G_{\mu\nu}$ as given in (\ref{3.10}) being altogether simpler than its form given in (\ref{2.3}). And being simpler to write down, it is also simpler to solve. However, before looking at solutions to the SVT4 fluctuation equations it is instructive to relate the SVT4 and SVT3 bases and determine which SVT4 components correspond to which SVT3 components.

\section{Relating the SVT4 and SVT3 bases}
\label{S4}

As constructed, for fluctuations around a flat four-dimensional background the SVT4 $F_{\mu\nu}$ has five independent components. When decomposed in an SVT3 basis it must contain a two-component transverse-traceless three-space rank two tensor, a two-component transverse three-space vector and a one-component three-space scalar. Moreover, assuming we can integrate by parts, all of these components must be gauge invariant. Thus the SVT3 rank two tensor associated with $F_{\mu\nu}$ must be $E_{ij}$ and the SVT3 vector must be  $B_i-\dot{E}_i$. However, this does not uniquely specify the three-space structure of the scalar component of $F_{\mu\nu}$ or of that of $\chi$, as gauge invariance alone does not provide sufficient information to enable us to determine what particular linear combinations of  $\psi$ and  $\phi+\dot{B}-\ddot{E}$ we should associate with each of them. Moreover, comparing the SVT3 and SVT4 expansions of $\delta G_{\mu\nu}$ as given in (\ref{2.3}) and (\ref{3.10}) respectively also does not enable us to uniquely specify the needed scalar combinations. We thus require another gauge-invariant gravitational fluctuation tensor, one in which the various combinations appear in a different way. We thus need to construct the fluctuation equations associated with varying a pure metric gravitational action other than the Einstein-Hilbert one, since any such fluctuation equations would still be gauge invariant. Moreover, it would be very helpful if we could find a fluctuation equation that only involved $F_{\mu\nu}$ and not $\chi$ or $F$ or $F_{\mu}$. Such a fluctuation equation is provided by the conformal gravity theory, since its underlying conformal symmetry requires that the gravitational fluctuation tensor,  labelled $\delta W_{\mu\nu}$, be traceless, to thus only depend on five rather than six gauge-invariant quantities, to thus necessarily not possess one of the SVT3 scalars.

Conformal gravity has been advanced \cite{Mannheim1989,Mannheim1994,Mannheim2006,Mannheim2012b,Mannheim2017} as a possible candidate alternative to standard Einstein gravity, and while we will study some of its implications for cosmology below, for the moment our interest is only in the fact that it provides us with a convenient gauge-invariant quantity $\delta W_{\mu\nu}$ \cite{footnote4}. As such, conformal gravity is a pure metric theory of gravity that possesses all of the general coordinate invariance and equivalence principle structure of standard gravity while augmenting it with an additional symmetry, local conformal invariance, in which  the action is left invariant under local conformal transformations on the metric of the form $g_{\mu\nu}(x)\rightarrow e^{2\alpha(x)}g_{\mu\nu}(x)$ with arbitrary local phase $\alpha(x)$. Under such a symmetry a gravitational action that is to be a polynomial function of the Riemann tensor is uniquely prescribed, and with use of the Gauss-Bonnet theorem is given by (see e.g. \cite{Mannheim2006}) 
\begin{eqnarray}
I_{\rm W}=-\alpha_g\int d^4x\, (-g)^{1/2}C_{\lambda\mu\nu\kappa}
C^{\lambda\mu\nu\kappa}
\equiv -2\alpha_g\int d^4x\, (-g)^{1/2}\left[R_{\mu\kappa}R^{\mu\kappa}-\frac{1}{3} (R^{\alpha}_{\phantom{\alpha}\alpha})^2\right].
\label{4.1}
\end{eqnarray}
Here $\alpha_g$ is a dimensionless  gravitational coupling constant, and
\begin{eqnarray}
C_{\lambda\mu\nu\kappa}= R_{\lambda\mu\nu\kappa}
-\frac{1}{2}\left(g_{\lambda\nu}R_{\mu\kappa}-
g_{\lambda\kappa}R_{\mu\nu}-
g_{\mu\nu}R_{\lambda\kappa}+
g_{\mu\kappa}R_{\lambda\nu}\right)
+\frac{1}{6}R^{\alpha}_{\phantom{\alpha}\alpha}\left(
g_{\lambda\nu}g_{\mu\kappa}-
g_{\lambda\kappa}g_{\mu\nu}\right)
\label{4.2}
\end{eqnarray}
is the conformal Weyl tensor. The conformal Weyl tensor has two features that are not possessed by the Einstein tensor, namely that it that vanishes in geometries that are conformal to flat (this precisely being the case for  the Robertson-Walker and de Sitter geometries that are of relevance to cosmology, with the background $T_{\mu\nu}$ then being zero),  and that for any metric $g_{\mu\nu}(x)$ it transforms as  $C^{\lambda}_{\phantom{\lambda}\mu\nu\kappa} \rightarrow  C^{\lambda}_{\phantom{\lambda}\mu\nu\kappa}$ under $g_{\mu\nu}(x)\rightarrow e^{2\alpha(x)}g_{\mu\nu}(x)$, with all derivatives of $\alpha(x)$ dropping out. With all of these derivatives dropping out $I_{\rm W}$ is locally conformal invariant \cite{footnote5}.

With the Weyl action $I_{\rm W}$ given in  (\ref{4.1}) being a fourth-order derivative function of the metric, functional variation with respect to the metric $g_{\mu\nu}(x)$ generates fourth-order derivative gravitational equations of motion of the form \cite{Mannheim2006} 
\begin{eqnarray}
-\frac{2}{(-g)^{1/2}}\frac{\delta I_{\rm W}}{\delta g_{\mu\nu}}=4\alpha_g W^{\mu\nu}=4\alpha_g\left[2\nabla_{\kappa}\nabla_{\lambda}C^{\mu\lambda\nu\kappa}-
R_{\kappa\lambda}C^{\mu\lambda\nu\kappa}\right]=4\alpha_g\left[W^{\mu
\nu}_{(2)}-\frac{1}{3}W^{\mu\nu}_{(1)}\right]=T^{\mu\nu},
\label{4.3}
\end{eqnarray}
where the functions $W^{\mu \nu}_{(1)}$ and $W^{\mu \nu}_{(2)}$ (respectively associated with the $(R^{\alpha}_{\phantom{\alpha}\alpha})^2$ and $R_{\mu\kappa}R^{\mu\kappa}$ terms in (\ref{4.1})) are given by
\begin{eqnarray}
W^{\mu \nu}_{(1)}&=&
2g^{\mu\nu}\nabla_{\beta}\nabla^{\beta}R^{\alpha}_{\phantom{\alpha}\alpha}                                             
-2\nabla^{\nu}\nabla^{\mu}R^{\alpha}_{\phantom{\alpha}\alpha}                          
-2 R^{\alpha}_{\phantom{\alpha}\alpha}R^{\mu\nu}                              
+\frac{1}{2}g^{\mu\nu}(R^{\alpha}_{\phantom{\alpha}\alpha})^2,
\nonumber\\
W^{\mu \nu}_{(2)}&=&
\frac{1}{2}g^{\mu\nu}\nabla_{\beta}\nabla^{\beta}R^{\alpha}_{\phantom{\alpha}\alpha}
+\nabla_{\beta}\nabla^{\beta}R^{\mu\nu}                    
 -\nabla_{\beta}\nabla^{\nu}R^{\mu\beta}                       
-\nabla_{\beta}\nabla^{\mu}R^{\nu \beta}                          
 - 2R^{\mu\beta}R^{\nu}_{\phantom{\nu}\beta}                                    
+\frac{1}{2}g^{\mu\nu}R_{\alpha\beta}R^{\alpha\beta},
\label{4.4}
\end{eqnarray}                                 
and where $T^{\mu\nu}$ is the conformal invariant, and thus traceless, energy-momentum tensor associated with a conformal matter source.  Since $W^{\mu\nu}=W^{\mu
\nu}_{(2)}-(1/3)W^{\mu\nu}_{(1)}$, known as the Bach tensor \cite{Bach1921},  is obtained from an action that is both general coordinate invariant and conformal invariant, in consequence, and without needing to impose any equation of motion or stationarity condition, $W^{\mu\nu}$ is automatically covariantly conserved and traceless and obeys $\nabla_{\nu}W^{\mu\nu}=0$, $g_{\mu\nu}W^{\mu\nu}=0$ on every variational path used for the functional variation of $I_{\rm W}$. Even though conformal gravity is a fourth-order derivative theory, we should note that as a quantum theory  it does not possess any of the negative norm ghost states that such higher-derivative theories are thought to have, to thus be unitary  \cite{Bender2008}. 

For fluctuations around a four-dimensional flat spacetime the gravitational $\delta W_{\mu\nu}$  takes the form \cite{Mannheim2006}
\begin{eqnarray}
\delta W_{\mu\nu}=\frac{1}{2}(\eta^{\rho}_{\phantom{\rho} \mu} \partial^{\alpha}\partial_{\alpha}-\partial^{\rho}\partial_{\mu})
(\eta^{\sigma}_{\phantom{\sigma} \nu} \partial^{\beta}\partial_{\beta}-
\partial^{\sigma}\partial_{\nu})K_{\rho \sigma}- 
\frac{1}{6}(\eta_{\mu \nu} \partial^{\gamma}\partial_{\gamma}-
\partial_{\mu}\partial_{\nu})(\eta^{\rho \sigma} \partial^{\delta}\partial_{\delta}-
\partial^{\rho}\partial^{\sigma})K_{\rho\sigma},
\label{4.5}
\end{eqnarray}
where $K_{\mu\nu}=h_{\mu\nu}-(1/4)g_{\mu\nu}h$. Evaluating (\ref{4.5}) in the SVT3 basis given in (\ref{2.1}) gives \cite{Amarasinghe2018}
\begin{eqnarray}
\delta W_{00}  &=& -\frac{2}{3} \delta^{mn}\delta^{\ell k}\tilde{\nabla}_m\tilde{\nabla}_n\tilde{\nabla}_{\ell}\tilde{\nabla}_k (\phi + \psi +\dot{B}-\ddot{E}),
\nonumber\\	
\delta W_{0i} &=&  -\frac{2}{3} \delta^{mn}\tilde{\nabla}_i\tilde{\nabla}_m\tilde{\nabla}_n\partial_0(\phi +\psi +\dot{B}-\ddot{E})
	+\frac{1}{2}\left[\delta^{mn}\delta^{\ell k}\tilde{\nabla}_m\tilde{\nabla}_n\tilde{\nabla}_{\ell}\tilde{\nabla}_k(B_i - \dot{E}_i) -  \delta^{\ell k}\tilde{\nabla}_{\ell}\tilde{\nabla}_k \partial_0^2(B_i - \dot{E}_i)\right],
\nonumber\\	
\delta W_{ij}  &=& \frac{1}{3}\bigg{[} \delta_{ij}\delta^{\ell k}\tilde{\nabla}_{\ell}\tilde{\nabla}_k  \partial_0^2(\phi+ \psi+\dot{B}-\ddot{E}) + \delta^{\ell k}\tilde{\nabla}_{\ell}\tilde{\nabla}_k \tilde{\nabla}_i\tilde{\nabla}_j (\phi + \psi +\dot{B}-\ddot{E}) 
\nonumber\\
&&- \delta_{ij} \delta^{mn}\delta^{\ell k}\tilde{\nabla}_m\tilde{\nabla}_n\tilde{\nabla}_{\ell}\tilde{\nabla}_k(\phi + \psi +\dot{B}-\ddot{E}) -3\tilde{\nabla}_i\tilde{\nabla}_j \partial_0^2(\phi + \psi +\dot{B}-\ddot{E})\bigg{] }
\nonumber\\
&&+\frac{1}{2}\left[ \delta^{\ell k}\tilde{\nabla}_{\ell}\tilde{\nabla}_k \tilde{\nabla}_i   \partial_0(B_j - \dot{E}_j)+ \delta^{\ell k}\tilde{\nabla}_{\ell}\tilde{\nabla}_k \tilde{\nabla}_j \partial_0(B_i - \dot{E}_i) - \tilde{\nabla}_i\partial_0^3(B_j - \dot{E}_j)-\tilde{\nabla}_j\partial_0^3(B_i - \dot{E}_i)\right]
\nonumber\\
&&+\left[\delta^{mn}\tilde{\nabla}_m\tilde{\nabla}_n-\partial_0^2\right]^2E_{ij},
\label{4.6}
\end{eqnarray}
with $\delta W_{\mu\nu}$ being gauge invariant on its own since for fluctuations around flat spacetime the background $T_{\mu\nu}$ and thus the fluctuation $\delta T_{\mu\nu}$ are both zero. Similarly,  evaluating (\ref{4.5}) in the SVT4 basis given in (\ref{3.9}) gives 
\begin{eqnarray}
\delta W_{\mu\nu}=\nabla_{\alpha}\nabla^{\alpha}\nabla_{\beta}\nabla^{\beta}F_{\mu\nu},
\label{4.7}
\end{eqnarray}
an expression that we note is structurally  simpler than its Einstein $\delta G_{\mu\nu}$ counterpart given  in (\ref{3.10}) \cite{footnote6}.

Because of its tracelessness, in both SVT3 and SVT4 $\delta W_{\mu\nu}$ only contains five gauge-invariant combinations. And from its SVT3 structure we can now unambiguously identify $\phi + \psi +\dot{B}-\ddot{E}$ as the three-dimensional scalar piece of $F_{\mu\nu}$, and thus can identify $\chi$ with the complementary combination $\phi - \psi +\dot{B}-\ddot{E}$. Thus from the three-dimensional perspective, for fluctuations around flat spacetime $F_{\mu\nu}$ contains $E_{ij}$, $B_i-\dot{E}_i$ and $\phi + \psi +\dot{B}-\ddot{E}$ \cite{footnote7}. 

\section{Solving the SVT4 fluctuation equations -- flat background}
\label{S5}

As we had noted in Sec. \ref{S1}, in treating the fluctuation equations there are two types of perturbation that one needs to consider. If we start with the Einstein equations in the presence of some general non-zero background $T_{\mu\nu}$, viz. $G_{\mu\nu}+8\pi G T_{\mu\nu}=0$, the first type is to consider perturbations $\delta G_{\mu\nu}$ and $\delta T_{\mu\nu}$ to the background and look for solutions to  
\begin{eqnarray}
\delta G_{\mu\nu}+8\pi G \delta T_{\mu\nu}=0
\label{5.1}
\end{eqnarray}
in a background that obeys $G_{\mu\nu}+8\pi G T_{\mu\nu}=0$. If the background is not flat, the fluctuation $\delta G_{\mu\nu}$ will not be gauge invariant  but it will instead be the combination $\delta G_{\mu\nu}+8\pi G \delta T_{\mu\nu}$  that will be expressible in the gauge-invariant SVT3 or SVT4 bases as appropriately generalized to a non-flat background.

The second kind of perturbation is one in which we introduce some new perturbation $\delta \bar{T}_{\mu\nu}$ to a background that obeys $G_{\mu\nu}+8\pi G T_{\mu\nu}=0$. This $\delta \bar{T}_{\mu\nu}$ will modify both the background $G_{\mu\nu}$ and the background $T_{\mu\nu}$ and will lead to a fluctuation equation of the form 
\begin{eqnarray}
\delta G_{\mu\nu}+8\pi G \delta T_{\mu\nu}=-8 \pi G \delta \bar{T}_{\mu\nu}. 
\label{5.2}
\end{eqnarray}
In (\ref{5.2}) the combination $\delta G_{\mu\nu}+8\pi G \delta T_{\mu\nu}$ will be gauge invariant since structurally it will be of the same form as it would be in the absence of $\delta \bar{T}_{\mu\nu}$, and would thus be gauge invariant since it already was in the absence of $\delta \bar{T}_{\mu\nu}$. In consequence of this any $\delta \bar{T}_{\mu\nu}$ that we could introduce would have to be gauge invariant all on its own.

If there is no background $T_{\mu\nu}$ so that the background metric is flat, the only perturbation that one could consider is $\delta \bar{T}_{\mu\nu}$, with the fluctuation equation then being of the form 
\begin{eqnarray}
\delta G_{\mu\nu}=-8 \pi G \delta \bar{T}_{\mu\nu}. 
\label{5.3}
\end{eqnarray}
With  $\delta \bar{T}_{\mu\nu}$ obeying $\nabla^{\nu}\delta \bar{T}_{\mu\nu}=0$, in analog to (\ref{3.10}) in general in the SVT4 case $\delta \bar{T}_{\mu\nu}$ must be of the form $\nabla_{\alpha}\nabla^{\alpha}\bar{F}_{\mu\nu}+2(g_{\mu\nu}\nabla_{\alpha}\nabla^{\alpha}-\nabla_{\mu}\nabla_{\nu})\bar{\chi}$, with (\ref{5.3}) taking the form 
\begin{eqnarray}
\nabla_{\alpha}\nabla^{\alpha}F_{\mu\nu}+2(g_{\mu\nu}\nabla_{\alpha}\nabla^{\alpha}-\nabla_{\mu}\nabla_{\nu})\chi=-8 \pi G[\nabla_{\alpha}\nabla^{\alpha}\bar{F}_{\mu\nu}+2(g_{\mu\nu}\nabla_{\alpha}\nabla^{\alpha}-\nabla_{\mu}\nabla_{\nu})\bar{\chi}]. 
\label{5.4}
\end{eqnarray}
The idea behind the decomposition theorem is that the tensor and scalar sectors of (\ref{5.4}) satisfy (\ref{5.4}) independently, so that one can set 
\begin{eqnarray}
\nabla_{\alpha}\nabla^{\alpha}F_{\mu\nu}=-8 \pi G\nabla_{\alpha}\nabla^{\alpha}\bar{F}_{\mu\nu},\quad
(g_{\mu\nu}\nabla_{\alpha}\nabla^{\alpha}-\nabla_{\mu}\nabla_{\nu})\chi=-8 \pi G(g_{\mu\nu}\nabla_{\alpha}\nabla^{\alpha}-\nabla_{\mu}\nabla_{\nu})\bar{\chi}. 
\label{5.5}
\end{eqnarray}
To see whether this is the case we take the trace of (\ref{5.4}), to obtain 
\begin{eqnarray}
\nabla_{\alpha}\nabla^{\alpha}\chi=-8 \pi G\nabla_{\alpha}\nabla^{\alpha}\bar{\chi}. 
\label{5.6}
\end{eqnarray}
If we now apply $\nabla_{\alpha}\nabla^{\alpha}$ to (\ref{5.4}), then given (\ref{5.6})  we obtain
\begin{eqnarray}
\nabla_{\alpha}\nabla^{\alpha}\nabla_{\beta}\nabla^{\beta}F_{\mu\nu}=-8\pi G \nabla_{\alpha}\nabla^{\alpha}\nabla_{\beta}\nabla^{\beta}\bar{F}_{\mu\nu}.
\label{5.7}
\end{eqnarray}
Now while this does give us an equation that involves $F_{\mu\nu}$ alone, this equation is not the second-order derivative equation $\nabla_{\alpha}\nabla^{\alpha}F_{\mu\nu}=-8 \pi G\nabla_{\alpha}\nabla^{\alpha}\bar{F}_{\mu\nu}$ that one is looking for. Moreover, getting to (\ref{5.6}) and (\ref{5.7}) is initially as far as we can go, since according to  (\ref{3.10}) only $\nabla_{\alpha}\nabla^{\alpha}\nabla_{\beta}\nabla^{\beta}F_{\mu\nu}$ and $\nabla_{\alpha}\nabla^{\alpha}\chi$ are automatically gauge invariant.

Now initially (\ref{5.6})  does not imply that $\chi$ is necessarily equal to $-8\pi G\bar{\chi}$, since they could differ by any function $f$ that obeys $\nabla_{\alpha}\nabla^{\alpha}f=0$, i.e., by any harmonic function of the form $f(\mathbf{q}\cdot\mathbf{x}-q t)$. However, it is the very introduction of $\bar{\chi}$ that is causing $\chi$ to be non-zero in the first place, and thus $\chi$ must be proportional to $\bar{\chi}$. Hence harmonic functions can be ignored. Then with  $\chi=-8\pi G\bar{\chi}$, it follows from (\ref{5.4}) that $\nabla_{\alpha}\nabla^{\alpha}F_{\mu\nu}=-8 \pi G\nabla_{\alpha}\nabla^{\alpha}\bar{F}_{\mu\nu}$. And again, since it is the very introduction of $\bar{F}_{\mu\nu}$ that is causing $\delta G_{\mu\nu}$ to be non-zero in the first place, it must be the case that $F_{\mu\nu}=-8 \pi G\bar{F}_{\mu\nu}$. As we see, (\ref{5.5}) does hold, and thus for an external $\delta \bar{T}_{\mu\nu}$ perturbation to a flat background we  obtain the decomposition theorem.

However, in the absence of any explicit external $\delta \bar{T}_{\mu\nu}$ the discussion is different, and is only of relevance in those cases where there is a background $T_{\mu\nu}$, as otherwise $\delta T_{\mu\nu}$ would be zero. When the background $T_{\mu\nu}$ is non-zero and accordingly the background is not flat, the fluctuation quantity $\delta G_{\mu\nu}+8\pi G \delta T_{\mu\nu}$ can still only depend on six gauge-invariant SVT4 combinations, viz. the curved space generalizations of the above $F_{\mu\nu}$ and one combination of $\chi$, $F$ and $F_{\mu}$. Thus in the following we will explore the SVT4 formulation in some non-flat backgrounds that are of cosmological interest.

\section{Solving the SVT4 fluctuation equations -- de Sitter background}
\label{S6}
\subsection{Defining the SVT4 Fluctuations Without a Conformal Factor}

Since a background de Sitter metric can be written as a comoving coordinate system metric with no conformal prefactor [viz. $ds^2=dt^2-e^{2Ht}(dx^2+dy^2+dz^2)$],  or written with a conformal prefactor as  a conformal to flat Minkowski metric [$ds^2=(1/\tau H)^2(d\tau^2-dx^2-dy^2-dz^2)$ where $\tau=e^{-Ht}/H$],  in setting up the SVT4 description of fluctuations around a de Sitter background there are then two options. One is to define the fluctuations in terms of $\chi$, $F$, $F_{\mu}$ and $F_{\mu\nu}$ with no multiplying conformal prefactor so that
\begin{eqnarray}
h_{\mu\nu}=-2g_{\mu\nu}\chi+2\nabla_{\mu}\nabla_{\nu}F
+ \nabla_{\mu}F_{\nu}+\nabla_{\nu}F_{\mu}+2F_{\mu\nu},
\label{6.1}
\end{eqnarray}
with the $\nabla_{\mu}$ derivatives being fully covariant with respect to the de Sitter background so that $\nabla^{\mu}F_{\mu}=0$, $\nabla^{\nu}F_{\mu\nu}=0$. The second is to define the fluctuations with a conformal prefactor so that the fluctuation metric is written as conformal to a flat Minkowski metric according to
\begin{eqnarray}
h_{\mu\nu}=\frac{1}{(\tau H)^2}[-2\eta_{\mu\nu}\chi+2\tilde{\nabla}_{\mu}\tilde{\nabla}_{\nu}F
+ \tilde{\nabla}_{\mu}F_{\nu}+\tilde{\nabla}_{\nu}F_{\mu}+2F_{\mu\nu}],
\label{6.2}
\end{eqnarray}
with the $\tilde{\nabla}_{\mu}$ derivatives being with respect to flat Minkowski so that $\tilde{\nabla}^{\mu}F_{\mu}=0$, $\tilde{\nabla}^{\nu}F_{\mu\nu}=0$, i.e. $-\dot{F}_0+\tilde{\nabla}^jF_j=0$, $-\dot{F}_{00}+\tilde{\nabla}^jF_{0j}=0$, $-\dot{F}_{0i}+\tilde{\nabla}^jF_{ij}=0$. We shall discuss both options below starting with (\ref{6.1}). 

However, before doing so and in order to be as general as possible we shall initially work in $D$ dimensions where the de Sitter space Riemann tensor takes the form
\begin{eqnarray}
R_{\lambda\mu\nu\kappa}=H^2(g_{\mu\nu}g_{\lambda\kappa}-g_{\lambda\nu}g_{\mu\kappa}),
\quad R_{\mu\kappa}=H^2(1-D)g_{\mu\kappa},\quad R^{\alpha}_{\phantom{\alpha}\alpha}=H^2D(1-D).
\label{6.3}
\end{eqnarray}
To construct fluctuations we have found it convenient to generalize (\ref{3.1}) to 
\begin{eqnarray}
h_{\mu\nu}&=&2F_{\mu\nu}+\nabla_{\nu}W_{\mu}+\nabla_{\mu}W_{\nu}+\frac{2-D}{D-1}\left[\nabla_{\mu}\nabla_{\nu}
+g_{\mu\nu}H^2\right]\int d^Dy(-g)^{1/2}D^{(E)}(x,y)\nabla^{\alpha}W_{\alpha}
\nonumber\\
&-&\frac{g_{\mu\nu}}{D-1}(\nabla^{\alpha}W_{\alpha}-h)-\frac{1}{D-1}\left[\nabla_{\mu}\nabla_{\nu}+g_{\mu\nu}H^2\right]\int d^Dy(-g)^{1/2}D^{(E)}(x,y)h,
\label{6.4}
\end{eqnarray}
where in the curved background the Green's function obeys
\begin{eqnarray}
\left(\nabla_{\nu}\nabla^{\nu}+H^2D\right)D^{(E)}(x,y)=(-g)^{-1/2}\delta^{(D)}(x-y).
\label{6.5}
\end{eqnarray}
With this definition $F_{\mu\nu}$ is automatically traceless. On applying $\nabla^{\nu}$ to (\ref{6.4}) and recalling that for any vector or scalar in a de Sitter space  we have
\begin{eqnarray}
(\nabla^{\nu}\nabla_{\mu}-\nabla_{\mu}\nabla^{\nu})W_{\nu}=H^2(D-1)W_{\mu},\quad 
(\nabla^{\nu}\nabla_{\mu}\nabla_{\nu}-\nabla_{\mu}\nabla^{\nu}\nabla_{\nu})V=H^2(D-1)\nabla_{\mu}V,
\label{6.6}
\end{eqnarray}
we obtain
\begin{eqnarray}
\nabla^{\nu}h_{\mu\nu}=\nabla_{\nu}\nabla^{\nu}W_{\mu}+H^2(D-1)W_{\mu},
\label{6.7}
\end{eqnarray}
with (\ref{6.7}) serving to define $W_{\mu}$. To decompose $W_{\mu}$ into transverse and longitudinal components we set $W_{\mu}=F_{\mu}+\nabla_{\mu}A$ where $\nabla^{\mu}F_{\mu}=0$, $\nabla^{\mu}W_{\mu}=\nabla^{\mu}\nabla_{\mu}A$, and thus set
\begin{eqnarray}
W_{\mu}=F_{\mu}+\nabla_{\mu}\int d^Dy(-g)^{1/2}D^{(D)}(x,y)\nabla^{\alpha}W_{\alpha},
\label{6.8}
\end{eqnarray}
where
\begin{eqnarray}
\nabla_{\nu}\nabla^{\nu}D^{(D)}(x,y)=(-g)^{-1/2}\delta^{(D)}(x-y).
\label{6.9}
\end{eqnarray}
Finally, with
\begin{eqnarray}
-2\chi&=&\frac{(2-D)H^2}{D-1}\int d^Dy(-g)^{1/2}D^{(E)}(x,y)\nabla^{\alpha}W_{\alpha}-\frac{\nabla^{\alpha}W_{\alpha}-h}{D-1}-\frac{H^2}{D-1}\int d^Dy(-g)^{1/2}D^{(E)}(x,y)h,
\nonumber\\
2F&=&\frac{2-D}{D-1}\int d^Dy(-g)^{1/2}D^{(E)}(x,y)\nabla^{\alpha}W_{\alpha}-\frac{1}{D-1}\int d^Dy(-g)^{1/2}D^{(E)}(x,y)h
\nonumber\\
&+&2\int d^Dy(-g)^{1/2}D^{(D)}(x,y)\nabla^{\alpha}W_{\alpha},
\nonumber\\
F_{\mu}&=&W_{\mu}-\nabla_{\mu}\int d^Dy(-g)^{1/2}D^{(D)}(x,y)\nabla^{\alpha}W_{\alpha},
\label{6.10}
\end{eqnarray}
we can now write $h_{\mu\nu}$ as given (\ref{6.1}), with $F_{\mu\nu}$ being given by the transverse-traceless
\begin{eqnarray}
2F_{\mu\nu}=h_{\mu\nu}+2g_{\mu\nu}\chi-2\nabla_{\mu}\nabla_{\nu}F
- \nabla_{\mu}F_{\nu}-\nabla_{\nu}F_{\mu},
\label{6.11}
\end{eqnarray}
where $W_{\mu}$ is determined from (\ref{6.7}). In this way then we can decompose $h_{\mu\nu}$ into a covariant SVTD in the de Sitter background case.

As well as the above formulation, which involves the Green's function $D^{(E)}(x,y)$, we should note that there is also an alternate formulation that does not involve it at all, one that implements the tracelessness of $F_{\mu\nu}$ using $D^{(D)}(x,y)$ alone, though it does so at the expense of leading to a more complicated expression for $W_{\mu}$. To this end we replace (\ref{6.4}) by 
\begin{eqnarray}
h_{\mu\nu}&=&2F_{\mu\nu}+\nabla_{\nu}W_{\mu}+\nabla_{\mu}W_{\nu}+\frac{2-D}{D-1}\nabla_{\mu}\nabla_{\nu}\int d^Dy(-g)^{1/2}D^{(D)}(x,y)\nabla^{\alpha}W_{\alpha}
\nonumber\\
&-&\frac{g_{\mu\nu}}{D-1}(\nabla^{\alpha}W_{\alpha}-h)-\frac{1}{D-1}\nabla_{\mu}\nabla_{\nu}\int d^Dy(-g)^{1/2}D^{(D)}(x,y)h,
\label{6.12}
\end{eqnarray}
with $F_{\mu\nu}$ automatically being traceless \cite{footnote8}. To fix $W_{\mu}$ we evaluate 
\begin{eqnarray}
\nabla^{\nu}h_{\mu\nu}&=&\nabla_{\nu}\nabla^{\nu}W_{\mu}+H^2(D-1)W_{\mu}
+H^2(2-D)\nabla_{\mu}\int d^Dy(-g)^{1/2}D^{(D)}(x,y)\nabla^{\alpha}W_{\alpha}
\nonumber\\
&-&H^2\nabla_{\mu}\int d^Dy(-g)^{1/2}D^{(D)}(x,y)h,
\nonumber\\
\nabla^{\mu}\nabla^{\nu}h_{\mu\nu}&=&\nabla^{\mu}\nabla_{\nu}\nabla^{\nu}W_{\mu}+H^2(\nabla^{\nu}W_{\nu}-h).
\label{6.13}
\end{eqnarray}
In terms of (\ref{6.12}) and (\ref{6.8}) we can set 
\begin{eqnarray}
2\chi&=&\frac{1}{D-1}[\nabla^{\alpha}W_{\alpha}-h],\quad 2F=\frac{1}{D-1}\int d^Dy(-g)^{1/2}D^{(D)}(x,y)[D\nabla^{\alpha}W_{\alpha}-h],
\nonumber\\
F_{\mu}&=&W_{\mu}-\nabla_{\mu}\int d^Dy(-g)^{1/2}D^{(D)}(x,y)\nabla^{\alpha}W_{\alpha},
\nonumber\\
2F_{\mu\nu}&=&h_{\mu\nu}+2g_{\mu\nu}\chi-2\nabla_{\mu}\nabla_{\nu}F
- \nabla_{\mu}F_{\nu}-\nabla_{\nu}F_{\mu},
\label{6.14}
\end{eqnarray}
with $\nabla^{\mu}F_{\mu}=0$ as before, and with (\ref{6.1}) following. Thus either way we are led to (\ref{6.1}) and we now apply it to fluctuations around a background de Sitter geometry.

\subsection{Application of SVT4 to de Sitter Fluctuation Equations}

We now restrict to  four dimensions where in a de Sitter geometry  the background Einstein equations are given by 
\begin{eqnarray}
G_{\mu\nu}=-8\pi G T_{\mu\nu}=3H^2g_{\mu\nu}.
\label{6.15}
\end{eqnarray}
The fluctuating Einstein tensor is given by 
\begin{eqnarray}
\delta G_{\mu\nu}=\frac{1}{2}\left[\nabla_{\alpha}\nabla^{\alpha}h_{\mu\nu}-\nabla_{\nu}\nabla^{\alpha}h_{\alpha\mu}-\nabla_{\mu}\nabla^{\alpha}h_{\alpha\nu}+\nabla_{\mu}\nabla_{\nu}h\right]
+\frac{g_{\mu\nu}}{2}\left[\nabla^{\alpha}\nabla^{\beta}h_{\alpha\beta}-\nabla_{\alpha}\nabla^{\alpha}h\right]
+\frac{H^2}{2}\left[4h_{\mu\nu}-g_{\mu\nu}h\right],
\label{6.16}
\end{eqnarray}
while the perturbation in the background $T_{\mu\nu}$ is given by $\delta T_{\mu\nu}=-3H^2h_{\mu\nu}$ (we conveniently set $8\pi G=1$). If we now reexpress these fluctuations in  the SVT4 basis given in (\ref{6.1}) we obtain
\begin{eqnarray}
\delta G_{\mu\nu}=2g_{\mu\nu}\nabla_{\alpha}\nabla^{\alpha}\chi-2\nabla_{\mu}\nabla_{\nu}\chi
+6H^2\nabla_{\mu}\nabla_{\nu}F
+3H^2 \nabla_{\mu}F_{\nu}+3H^2\nabla_{\nu}F_{\mu}+(\nabla_{\alpha}\nabla^{\alpha}+4H^2)F_{\mu\nu},
\label{6.17}
\end{eqnarray}
\begin{eqnarray}
3H^2h_{\mu\nu}=3H^2\left[-2g_{\mu\nu}\chi+2\nabla_{\mu}\nabla_{\nu}F
+ \nabla_{\mu}F_{\nu}+\nabla_{\nu}F_{\mu}+2F_{\mu\nu}\right],
\label{6.18}
\end{eqnarray}
and thus 
\begin{eqnarray}
\delta G_{\mu\nu}-3H^2h_{\mu\nu}=(\nabla_{\alpha}\nabla^{\alpha}-2H^2)F_{\mu\nu}+2(g_{\mu\nu}\nabla_{\alpha}\nabla^{\alpha}-\nabla_{\mu}\nabla_{\nu}+3H^2g_{\mu\nu})\chi.
\label{6.19}
\end{eqnarray}
As we see, $\delta G_{\mu\nu}+8\pi G\delta T_{\mu\nu}=\delta G_{\mu\nu}-3H^2h_{\mu\nu}$ only depends on $F_{\mu\nu}$ and $\chi$,  with it thus being these quantities that are  gauge invariant, with the thus non-gauge-invariant $F_{\mu}$ and $F$ dropping out \cite{footnote9}.

In the event that there is an additional source term $\delta\bar{T}_{\mu\nu}$, it must be gauge invariant on its own, and must obey $\nabla^{\nu}\delta \bar{T}_{\mu\nu}=0$ in the de Sitter background, to thus be of the form 
\begin{eqnarray}
\delta \bar{T}_{\mu\nu}=\bar{F}_{\mu\nu}+2(g_{\mu\nu}\nabla_{\alpha}\nabla^{\alpha}-\nabla_{\mu}\nabla_{\nu}+3H^2g_{\mu\nu})\bar{\chi}.
\label{6.20}
\end{eqnarray}
With this source the fluctuation equations take the form
\begin{eqnarray}
(\nabla_{\alpha}\nabla^{\alpha}-2H^2)F_{\mu\nu}+2(g_{\mu\nu}\nabla_{\alpha}\nabla^{\alpha}-\nabla_{\mu}\nabla_{\nu}+3H^2g_{\mu\nu})\chi&=&\bar{F}_{\mu\nu}+2(g_{\mu\nu}\nabla_{\alpha}\nabla^{\alpha}-\nabla_{\mu}\nabla_{\nu}+3H^2g_{\mu\nu})\bar{\chi},
\label{6.21}
\end{eqnarray}
with trace
\begin{eqnarray}
6(\nabla_{\alpha}\nabla^{\alpha}+4H^2)\chi&=&6(\nabla_{\alpha}\nabla^{\alpha}+4H^2)\bar{\chi}.
\label{6.22}
\end{eqnarray}

While the trace condition would only set $\chi=\bar{\chi}+f$ where $f$ obeys  $(\nabla_{\alpha}\nabla^{\alpha}+4H^2)f=0$, when there is a $\bar{\chi}$ source present then it is the cause of fluctuations in the background in the first place, and thus we can only have $\chi=\bar{\chi}$ with any possible $f$ being zero. Then, from  (\ref{6.21}) we obtain 
\begin{eqnarray}
(\nabla_{\alpha}\nabla^{\alpha}-2H^2)F_{\mu\nu}=\bar{F}_{\mu\nu},
\label{6.23}
\end{eqnarray}
and the decomposition theorem is achieved.

However, if there is no $\delta \bar{T}_{\mu\nu}$ source the fluctuation equations take the form 
\begin{eqnarray}
(\nabla_{\alpha}\nabla^{\alpha}-2H^2)F_{\mu\nu}+2(g_{\mu\nu}\nabla_{\alpha}\nabla^{\alpha}-\nabla_{\mu}\nabla_{\nu}+3H^2g_{\mu\nu})\chi=0,
\label{6.24}
\end{eqnarray}
with the trace condition  being given by 
\begin{eqnarray}
6(\nabla_{\alpha}\nabla^{\alpha}+4H^2)\chi=0,
\label{6.25a}
\end{eqnarray}
with spherical Bessel solution
\begin{eqnarray}
\chi=\sum_{\bf k} k^2\tau^2[a_2({\bf k})j_2(k\tau)+b_2 ({\bf k})y_2(k\tau)]e^{i{\bf k}\cdot {\bf x}}.
\label{6.26a}
\end{eqnarray}
(To obtain this solution for $\chi$ it is more straightforward to use $ds^2=(1/\tau^2 H^2)(d\tau^2-dx^2-dy^2-dz^2)$ as the background de Sitter metric, something we can do regardless of whether or not we include a conformal factor in the fluctuations.) Given the trace condition we can rewrite the evolution equation given in (\ref{6.24}) as 
\begin{eqnarray}
(\nabla_{\alpha}\nabla^{\alpha}-2H^2)F_{\mu\nu}-2(g_{\mu\nu}H^2+\nabla_{\mu}\nabla_{\nu})\chi=0.
\label{6.27a}
\end{eqnarray}

Since it is not automatic that $\chi$ would obey $(H^2g_{\mu\nu}+\nabla_{\mu}\nabla_{\nu})\chi=0$ even though it does obey  $g^{\mu\nu}(H^2g_{\mu\nu}+\nabla_{\mu}\nabla_{\nu})\chi=0$, it is thus not automatic that (\ref{6.24}) and (\ref{6.27a}) could be replaced by
\begin{eqnarray}
(\nabla_{\alpha}\nabla^{\alpha}-2H^2)F_{\mu\nu}=0,\quad (g_{\mu\nu}H^2+\nabla_{\mu}\nabla_{\nu})\chi=0,
\label{6.28a}
\end{eqnarray}
as would be required of a decomposition theorem. In fact, since $g_{\mu\nu}\chi$ and $\nabla_{\mu}\nabla_{\nu}\chi$ behave totally differently ($\nabla_{\mu}\nabla_{\nu}\chi$ is non-zero if $\mu\neq \nu$ while $g_{\mu\nu}\chi$ is not), the only way to get a decomposition theorem would be for $\chi$, and thus $a_2({\bf k})$ and $b_2({\bf k})$,  to be zero. As we now show, this can in fact be made to be the case, though it is only a particular solution to the full fluctuation equations.

To explore this possibility we need to obtain an expression that  only depends on $F_{\mu\nu}$, and we note that for any scalar in  $D=4$ de Sitter we  have \cite{Mannheim2012a}
\begin{eqnarray}
\nabla_{\alpha}\nabla^{\alpha}\nabla_{\mu}\nabla_{\nu}\chi=\nabla_{\mu}\nabla_{\nu}\nabla_{\alpha}\nabla^{\alpha}\chi
-2H^2g_{\mu\nu}\nabla_{\alpha}\nabla^{\alpha}\chi
+8H^2\nabla_{\mu}\nabla_{\nu}\chi.
\label{6.29a}
\end{eqnarray}
Given the trace condition shown in (\ref{6.25a})  we then find that
\begin{eqnarray}
(\nabla_{\alpha}\nabla^{\alpha}-4H^2)(g_{\mu\nu}H^2+\nabla_{\mu}\nabla_{\nu})\chi
=(\nabla_{\mu}\nabla_{\nu}-H^2g_{\mu\nu})(\nabla^{\alpha}\nabla^{\alpha}+4H^2)\chi=0,
\label{6.30a}
\end{eqnarray}
and from (\ref{6.27a}) we thus obtain the fourth-order derivative equation
\begin{eqnarray}
(\nabla_{\alpha}\nabla^{\alpha}-4H^2)(\nabla_{\alpha}\nabla^{\alpha}-2H^2)F_{\mu\nu}=0
\label{6.31a}
\end{eqnarray}
for $F_{\mu\nu}$, with a decomposition for the components of the fluctuations thus being found, only in the higher-derivative form given in (\ref{6.30a}) and (\ref{6.31a}) rather than in the second-derivative form given in (\ref{6.28a}). Now $(\nabla_{\alpha}\nabla^{\alpha}-2H^2)F_{\mu\nu}=0$ is a particular solution to (\ref{6.31a}), and for this particular solution it would follow that the only solution to (\ref{6.24}) would then be $\chi=0$, with both the $F_{\mu\nu}$ and $\chi$ sector equations given in  (\ref{6.28a}) then holding.

To determine the conditions under which $(\nabla_{\alpha}\nabla^{\alpha}-2H^2)F_{\mu\nu}=0$ might actually hold we need to look for the general solution to (\ref{6.31a}), and since (\ref{6.31a}) is a covariant equation we can evaluate it in any coordinate system, with conformal to flat Minkowski being the most convenient for the de Sitter background. To this end we recall that in any metric that is conformal to flat Minkowski ($ds^2=-g_{MN}dx^Mdx^N=-\Omega^2(x)\eta_{\mu\nu}dx^{\mu}dx^{\nu}$) one has the relation \cite{Mannheim2012a}
\begin{eqnarray}
g^{LR}\nabla_{L}\nabla_{R}A_{MN}&=&
\eta^{LR}\Omega^{-2}\partial_{L}\partial_{R}A_{MN}
-2\Omega^{-4}\partial_{M}\Omega\partial_{N}\Omega \eta^{TQ}A_{TQ}
-2\eta^{LR}\Omega^{-3}\partial_{L}\partial_{R}\Omega A_{MN}
\nonumber \\
&-&2\eta^{LR}\Omega^{-3}\partial_{R}\Omega \partial_{L}A_{MN}
+2\Omega^{-4}\eta_{MN}\eta^{TX}\partial_{X}\Omega\eta^{QY} \partial_{Y}\Omega A_{TQ}
\nonumber \\
&+&2\eta^{KQ}\Omega^{-3}\partial_{Q}\Omega \partial_{N}A_{KM}
+2\eta^{KQ}\Omega^{-3}\partial_{Q}\Omega \partial_{M}A_{KN}
\nonumber\\
&-&2\Omega^{-1}\partial_{N}\Omega \nabla_{L}A^{L}_{\phantom{L}M}
-2\Omega^{-1}\partial_{M}\Omega \nabla_{L}A^{L}_{\phantom{L}N},
\label{6.32a}
\end{eqnarray}
for any rank two tensor $A_{MN}$, with the $\nabla_{L}$ referring to covariant derivatives in the $g_{MN}$ geometry. For an $A_{MN}$ that is transverse and traceless, and for $\Omega=1/\tau H$ (\ref{6.32a}) reduces to  (the dot denotes $\partial/\partial\tau$)
\begin{eqnarray}
g^{LR}\nabla_{L}\nabla_{R}A_{MN}&=&
\eta^{LR}\tau^2 H^2\partial_{L}\partial_{R}A_{MN}
+4H^2A_{MN}
-2\tau H^2\dot{A}_{MN}
+2H^2\eta_{MN}A_{00}
\nonumber \\
&+&2\tau H^2\partial_{N}A_{0M}
+2\tau H^2 \partial_{M}A_{0N}.
\label{6.33a}
\end{eqnarray}
While the general components of $A_{MN}$ are coupled in (\ref{6.33a}), this is not the case for  $A_{00}$, and so we look at $A_{00}$ and obtain 
\begin{eqnarray}
\nabla_{L}\nabla^{L}A_{00}&=&
\eta^{LR}\tau^2 H^2\partial_{L}\partial_{R}A_{00}
+2H^2A_{00}
+2\tau H^2\dot{A}_{00}.
\label{6.34a}
\end{eqnarray}
Now in a de Sitter background the identity $\nabla_{P}\nabla_{K}\nabla^{K}A^{P}_{\phantom{P}M}
=[\nabla_{K}\nabla^{K}+5H^2]\nabla_{P}A^{P}_{\phantom{P}M}
-2H^2\nabla_{M}A^{P}_{\phantom{P}P}$ holds  \cite{Mannheim2012a}. Thus if any $A_{MN}$ is transverse and traceless then so is $\nabla_{L}\nabla^{L}A_{MN}$. So let us define $A_{MN}=[\nabla_{L}\nabla^{L}-2H^2]F_{MN}$, with this $A_{MN}$ being traverse and traceless since $F_{MN}$ is. For this $A_{MN}$ (\ref{6.31a}) takes the form
\begin{eqnarray}
(\nabla_{\alpha}\nabla^{\alpha}-4H^2)A_{\mu\nu}=0.
\label{6.35a}
\end{eqnarray}
Thus for $A_{00}$ we have
\begin{eqnarray}
\eta^{LR}\tau^2 H^2\partial_{L}\partial_{R}A_{00}+2\tau H^2\dot{A}_{00}
-2H^2A_{00}=0.
\label{6.36a}
\end{eqnarray}
In a plane wave mode $e^{i{\bf k}\cdot{\bf x}}$ the quantity $A_{00}$ thus obeys
\begin{eqnarray}
\ddot{A}_{00}-\frac{2}{\tau}\dot{A}_{00}+k^2A_{00}+\frac{2}{\tau^2}A_{00}=0.
\label{6.37a}
\end{eqnarray}
The general solution to (\ref{6.36a}) is thus
\begin{eqnarray}
A_{00}&=&[\nabla_{L}\nabla^{L}-2H^2]F_{00}=\sum_{\bf k} k^4\tau^2[a_{00}({\bf k})j_0(k\tau)+b_{00} ({\bf k})y_0(k\tau)]e^{i{\bf k}\cdot {\bf x}}
\nonumber\\
&=&\sum_{\bf k} k^3\tau[a_{00}({\bf k})\sin(k\tau)+b_{00} ({\bf k})\cos(k\tau)]e^{i{\bf k}\cdot {\bf x}},
\label{6.38a}
\end{eqnarray}
where $a_{00}({\bf k})$ and $b_{00}({\bf k})$ are polarization tensors. (Here and throughout we leave out the complex conjugate solution.)

To see if we can support this solution, or whether we are forced to (\ref{6.28a}), we need to see whether (\ref{6.38a}) is compatible  with  (\ref{6.27a}), and thus require that
\begin{eqnarray}
A_{00}=(\nabla_{\alpha}\nabla^{\alpha}-2H^2)F_{00}=2(g_{00}H^2+\nabla_{0}\nabla_{0})\chi=2\left[-\frac{1}{\tau^2}+\frac{\partial^2}{\partial \tau^2}-\Gamma^{\alpha}_{00}\partial_{\alpha}\right]\chi,
\label{6.39}
\end{eqnarray}
when evaluated in the solution  for $\chi$ as given in (\ref{6.26a}). On noting that $\Gamma^{\alpha}_{00}=-\delta^{\alpha}_0/\tau$ in a background de Sitter geometry, we evaluate
\begin{eqnarray}
2\left[-\frac{1}{\tau^2}+\frac{\partial^2}{\partial \tau^2}-\Gamma^{\alpha}_{00}\partial_{\alpha}\right][k^2\tau^2j_2(k\tau)]=2\left[\frac{\partial^2}{\partial \tau^2}+\frac{1}{\tau}\frac{\partial}{\partial \tau}-\frac{1}{\tau^2}\right][k^2\tau^2j_2(k\tau)]=2k^3\tau\sin(k\tau),
\label{6.40}
\end{eqnarray}
where we have utilized properties of Bessel functions in the last step. With an analogous expression holding for the $y_2(k\tau)$ term, we thus precisely do confirm (\ref{6.38a}), and on comparing (\ref{6.26a}) with (\ref{6.38a}) obtain
\begin{eqnarray}
a_{00}({\bf k})=2a_2({\bf k}), \quad b_{00}({\bf k})=2b_2({\bf k}).
\label{6.41}
\end{eqnarray}
As we see, in the general solution we are not at all forced to $\chi=0$ as would be required by the decomposition theorem.

For completeness we note that once we have determined $A_{00}$ we can use (\ref{6.33a}) and the $\nabla_{L}A^{L}_{\phantom{L}M}=0$ and $g^{MN}A_{MN}=0$ conditions to determine the other components of $A_{\mu\nu}$, and note only that they satisfy and behave as  
\begin{eqnarray}
&&
\eta^{\mu\nu}\partial_{\mu}A_{0\nu}+\frac{2}{\tau}A_{00}=0,\quad \eta^{\mu\nu}\partial_{\mu}A_{i\nu}+\frac{2}{\tau}A_{0i}=0,
\nonumber\\
&&\left[\frac{\partial^2}{\partial\tau^2}+k^2\right]A_{0i}=\frac{2}{\tau}\partial_iA_{00},\quad\left[\frac{\partial^2}{\partial\tau^2}+\frac{2}{\tau}\frac{\partial}{\partial \tau}+k^2\right]A_{ij} =\frac{2}{\tau^2}\delta_{ij}A_{00}+\frac{2}{\tau}\left(\partial_iA_{0j}+\partial_jA_{0i}\right),
\nonumber\\
&&A_{0i}=\sum_{\bf k} ik_ik[-k\tau a_{00}({\bf k})\cos(k\tau)+k\tau b_{00} ({\bf k})\sin(k\tau)
+a_{00}({\bf k})\sin(k\tau)+b_{00} ({\bf k})\cos(k\tau)]e^{i{\bf k}\cdot {\bf x}}, 
\nonumber\\
&&A_{ij}=\sum_{\bf k}k_ik_jk\tau[a_{00}({\bf k})\sin(k\tau)+b_{00} ({\bf k})\cos(k\tau)]e^{i{\bf k}\cdot {\bf x}}
\nonumber\\
&&+\sum_{\bf k}\left[\delta_{ij}k^2-3k_ik_j\right]\left[-a_{00} ({\bf k})\cos(k\tau)+b_{00}({\bf k})\sin(k\tau)+\frac{1}{k\tau}[a_{00} ({\bf k})\sin(k\tau)+b_{00}({\bf k})\cos(k\tau)
\right]e^{i{\bf k}\cdot {\bf x}}.
\label{6.42}
\end{eqnarray}
(In order to derive the solutions given in (\ref{6.42}) we needed to include terms that would vanish identically in the left-hand sides of the second-order differential equations so that the first-order $\nabla_{L}A^{L}_{\phantom{L}M}=0$ conditions would then be satisfied.) In this solution we then need to satisfy 
\begin{eqnarray}
(\nabla_{\alpha}\nabla^{\alpha}-2H^2)F_{\mu\nu}=A_{\mu\nu},
\label{6.43}
\end{eqnarray}
which for the representative $A_{00}$ and $F_{00}$ is of the form
\begin{eqnarray}
\ddot{F}_{00}-\frac{2}{\tau}\dot{F}_{00}+k^2F_{00}=-\sum_{\bf k}\frac{k^3}{H^2\tau} [a_{00}({\bf k})\sin(k\tau)+b_{00} ({\bf k})\cos(k\tau)]e^{i{\bf k}\cdot {\bf x}},
\label{6.44}
\end{eqnarray}
with solution
\begin{eqnarray}
F_{00}=\sum_{\bf k}\frac{k^2}{2H^2} [-a_{00}({\bf k})\cos(k\tau)+b_{00} ({\bf k})\sin(k\tau)]e^{i{\bf k}\cdot {\bf x}}.
\label{6.45}
\end{eqnarray}

Now in order to get a decomposition theorem in the form given in (\ref{6.28a}) we would need $\chi$ to vanish, i.e. we would need $a_2({\bf k})$ and $b_2({\bf k})$ to vanish. And that would mean that $a_{00}({\bf k})$ and $b_{00}({\bf k})$ would have to vanish as well, and thus not only would $A_{00}$ have to vanish but so would all the other components of $A_{\mu\nu}$ as well. A decomposition theorem would thus require that
\begin{eqnarray}
(\nabla_{\alpha}\nabla^{\alpha}-2H^2)F_{\mu\nu}=0,
\label{6.46}
\end{eqnarray}
for all components of $F_{\mu\nu}$. To look for a non-trivial solution to (\ref{6.46}) in order to show that the decomposition theorem does in fact have a solution, we note that in a plane wave (\ref{6.46}) reduces to 
\begin{eqnarray}
\ddot{F}_{00}-\frac{2}{\tau}\dot{F}_{00}+k^2F_{00}=0,
\label{6.47}
\end{eqnarray}
for the representative $F_{00}$ component. The non-trivial solution to (\ref{6.47}) is of the form 
\begin{eqnarray}
F_{00}=\sum_{\bf k} k^2\tau^2[c_{00}({\bf k})j_1(k\tau)+d_{00} ({\bf k})y_1(k\tau)]e^{i{\bf k}\cdot {\bf x}}.
\label{6.48}
\end{eqnarray}
The form for $F_{00}$ given in (\ref{6.48}) and its $F_{\mu\nu}$ analogs together with $\chi=0$ thus constitute a non-trivial solution that corresponds to the decomposition theorem, so in this sense the decomposition theorem can be recovered, as it is a specific solution to the full evolution equations. However, there is no compelling reason to restrict the solutions to (\ref{6.35a}) to the trivial $A_{\mu\nu}=0$, with it being (\ref{6.38a}), (\ref{6.42}), (\ref{6.45}) and (\ref{6.48}) that provide the most general solution in the $F_{00}$ sector and its analogs according to 
\begin{eqnarray}
F_{00}=\sum_{\bf k}\frac{k^2}{2H^2} [-a_{00}({\bf k})\cos(k\tau)+b_{00} ({\bf k})\sin(k\tau)]e^{i{\bf k}\cdot {\bf x}}+\sum_{\bf k} k^2\tau^2[c_{00}({\bf k})j_1(k\tau)+d_{00} ({\bf k})y_1(k\tau)]e^{i{\bf k}\cdot {\bf x}},
\label{6.49}
\end{eqnarray}
while at the same time (\ref{6.26a}) is the most general solution in the $\chi$ sector as constrained by (\ref{6.41}). Moreover, in this solution we can choose the coefficients in (\ref{6.41}) so that $F_{\mu\nu}$ and $\chi$ are localized in space. Thus no spatially asymptotic boundary coefficient could affect them. In fact suppose that we could have constrained the solutions by an asymptotic condition. We would need one that would force $A_{\mu\nu}$ to have to vanish in $(\nabla_{\alpha}\nabla^{\alpha}-4H^2)A_{\mu\nu}=0$ while not at the same time forcing $F_{\mu\nu}$ to have to vanish in $(\nabla_{\alpha}\nabla^{\alpha}-2H^2)F_{\mu\nu}=0$, something that would not obviously appear possible to achieve. Thus as we see, in this general solution the decomposition theorem does not hold. And just as we noted in Sec. \ref{S1},  in the SVT4 case asymptotic boundary conditions do not force us to the decomposition theorem, to thus provide a completely solvable cosmological model in which the decomposition theorem does not hold. However, we should point out that while we could not make $\chi$ vanish through spatial boundary conditions it would be possible to force $\chi$ to vanish at all times by judiciously choosing initial  conditions at an initial time.  However, there would not appear to be any compelling rationale for doing so, and  to nonetheless do so would appear to be quite contrived. Thus absent any compelling rationale for such a judicious choice or for any other choice at all for that matter (i.e., no compelling rationale that would force $\chi$ to vanish) the decomposition theorem would not hold for SVT4 fluctuations around a de Sitter background.

\subsection{Defining the SVT4 Fluctuations With a Conformal Factor}

In a 
\begin{align}
ds^2=\frac{1}{(\tau H)^2}(d\tau^2-dx^2-dy^2-dz^2)
\label{6.50}
\end{align}
de Sitter background with fluctuations of the form 
\begin{align}
h_{\mu\nu}=\frac{1}{(\tau H)^2}(-2g_{\mu\nu}\chi+2\tilde{\nabla}_{\mu}\tilde{\nabla}_{\nu}F
+ \tilde{\nabla}_{\mu}F_{\nu}+\tilde{\nabla}_{\nu}F_{\mu}+2F_{\mu\nu}), 
\label{6.51}
\end{align}
where $\tilde{\nabla}^{\mu}F_{\mu}=0$, $\tilde{\nabla}^{\nu}F_{\mu\nu}=0$, $g^{\mu\nu}F_{\mu\nu}=0$, and where, as per (\ref{6.2}), the $\tilde{\nabla}_{\mu}$ denote derivatives with respect to the flat Minkowski $\eta_{\mu\nu}dx^{\mu}dx^{\nu}$ metric, we write the fluctuation Einstein tensor as 
\begin{eqnarray}
\delta G_{00}&=& -6 \dot{\chi} \tau^{-1} - 2 \tau^{-1} \tilde{\nabla}^2\dot{F} - 2 \tilde{\nabla}^2\chi -2 \tau^{-1} \tilde{\nabla}^2F_{0}- \overset{..}{F}_{00} - 2 \dot{F}_{00} \tau^{-1} + \tilde{\nabla}^2F_{00},
\nonumber\\ 
\delta G_{0i}&=& -2 \tau^{-1} \tilde{\nabla}_{i}\overset{..}{F} + 6 \tau^{-2} \tilde{\nabla}_{i}\dot{F} - 2 \tilde{\nabla}_{i}\dot{\chi} - 2 \tau^{-1} \tilde{\nabla}_{i}\chi +3 \dot{F}_{i} \tau^{-2} - 2 \tau^{-1} \tilde{\nabla}_{i}\dot{F}_{0} + 3 \tau^{-2} \tilde{\nabla}_{i}F_{0}- \overset{..}{F}_{0i}
 \nonumber \\ 
&& + 6 F_{0i} \tau^{-2} +  \tilde{\nabla}^2F_{0i} - 2 \tau^{-1} \tilde{\nabla}_{i}F_{00},
\nonumber\\ 
\delta G_{ij}&=& -2 \overset{..}{\chi}\delta_{ij} + 6 \overset{..}{F}\delta_{ij} \tau^{-2} - 2 \overset{...}{F}\delta_{ij} \tau^{-1} + 2 \dot{\chi}\delta_{ij} \tau^{-1} + 2\delta_{ij} \tau^{-1} \tilde{\nabla}^2\dot{F} + 2\delta_{ij} \tilde{\nabla}^2\chi - 2 \tau^{-1} \tilde{\nabla}_{j}\tilde{\nabla}_{i}\dot{F}
\nonumber \\ 
&& + 6 \tau^{-2} \tilde{\nabla}_{j}\tilde{\nabla}_{i}F - 2 \tilde{\nabla}_{j}\tilde{\nabla}_{i}\chi +6 \dot{F}_{0}\delta_{ij} \tau^{-2} - 2 \overset{..}{F}_{0}\delta_{ij} \tau^{-1} + 2\delta_{ij} \tau^{-1} \tilde{\nabla}^2F_{0} + 3 \tau^{-2} \tilde{\nabla}_{i}F_{j} 
\nonumber \\ 
&& + 3 \tau^{-2} \tilde{\nabla}_{j}F_{i} - 2 \tau^{-1} \tilde{\nabla}_{j}\tilde{\nabla}_{i}F_{0}- \overset{..}{F}_{ij} + 6 F_{ij} \tau^{-2} + 6 F_{00}\delta_{ij} \tau^{-2} +2 \dot{F}_{ij} \tau^{-1} +\tilde{\nabla}^2F_{ij}
 \nonumber \\ 
&& - 2 \tau^{-1} \tilde{\nabla}_{i}F_{0j} - 2 \tau^{-1} \tilde{\nabla}_{j}F_{0i},
\nonumber\\
g^{\mu\nu}\delta G_{\mu\nu} &=& 18 H^2 \overset{..}{F} - 6 H^2 \overset{...}{F} \tau + 12 H^2 \dot{\chi} \tau - 6 H^2 \overset{..}{\chi} \tau^2 + 6 H^2 \tau \tilde{\nabla}^2\dot{F} + 6 H^2 \tilde{\nabla}^2F 
\nonumber \\ 
&& + 6 H^2 \tau^2 \tilde{\nabla}^2\chi +24 H^2 \dot{F}_{0} - 6 H^2 \overset{..}{F}_{0} \tau + 6 H^2 \tau \tilde{\nabla}_{k}^2F_{0}+24 H^2 F_{00}.
\label{6.52}
\end{eqnarray}
Here the dot denotes the derivative with respect to the conformal time $\tau$ and $\tilde{\nabla}^2=\delta^{ij}\tilde{\nabla}_i\tilde{\nabla}_j$. With a $3H^2h_{\mu\nu}$ perturbation  the fluctuation equations take the form
\begin{eqnarray}
\Delta_{00}&=& -6 \overset{..}{F} \tau^{-2} - 6 \dot{\chi} \tau^{-1} - 6 \tau^{-2} \chi - 2 \tau^{-1} \tilde{\nabla}^2\dot{F} - 2 \tilde{\nabla}^2\chi -6 \dot{F}_{0} \tau^{-2} - 2 \tau^{-1} \tilde{\nabla}^2F_{0}- \overset{..}{F}_{00} 
\nonumber \\ 
&& - 6 F_{00} \tau^{-2} - 2 \dot{F}_{00} \tau^{-1} + \tilde{\nabla}^2F_{00}=0,
\nonumber\\ 
\Delta_{0i}&=& -2 \tau^{-1} \tilde{\nabla}_{i}\overset{..}{F} - 2 \tilde{\nabla}_{i}\dot{\chi} - 2 \tau^{-1} \tilde{\nabla}_{i}\chi -2 \tau^{-1} \tilde{\nabla}_{i}\dot{F}_{0}- \overset{..}{F}_{0i} + \tilde{\nabla}^2F_{0i} - 2 \tau^{-1} \tilde{\nabla}_{i}F_{00}=0,
\nonumber\\ 
\Delta_{ij}&=& -2 \overset{..}{\chi}\delta_{ij} + 6 \overset{..}{F}\delta_{ij} \tau^{-2} - 2 \overset{...}{F}\delta_{ij} \tau^{-1} + 2 \dot{\chi}\delta_{ij} \tau^{-1} + 6\delta_{ij} \tau^{-2} \chi + 2\delta_{ij} \tau^{-1} \tilde{\nabla}^2\dot{F} + 2\delta_{ij} \tilde{\nabla}^2\chi 
\nonumber \\ 
&& - 2 \tau^{-1} \tilde{\nabla}_{j}\tilde{\nabla}_{i}\dot{F} - 2 \tilde{\nabla}_{j}\tilde{\nabla}_{i}\chi +6 \dot{F}_{0}\delta_{ij} \tau^{-2} - 2 \overset{..}{F}_{0}\delta_{ij} \tau^{-1} + 2\delta_{ij} \tau^{-1} \tilde{\nabla}^2F_{0} \nonumber \\ 
&& - 2 \tau^{-1} \tilde{\nabla}_{j}\tilde{\nabla}_{i}F_{0}- \overset{..}{F}_{ij} + 6 F_{00}\delta_{ij} \tau^{-2} + 2 \dot{F}_{ij} \tau^{-1} + \tilde{\nabla}^2F_{ij} -2 \tau^{-1} \tilde{\nabla}_{i}F_{0j} - 2 \tau^{-1} \tilde{\nabla}_{j}F_{0i}=0,
\nonumber\\
g^{\mu\nu}\Delta_{\mu\nu} &=& 24 H^2 \overset{..}{F} - 6 H^2 \overset{...}{F} \tau + 12 H^2 \dot{\chi} \tau - 6 H^2 \overset{..}{\chi} \tau^2 + 24 H^2 \chi + 6 H^2 \tau \tilde{\nabla}^2\dot{F} 
\nonumber \\ 
&& + 6 H^2 \tau^2 \tilde{\nabla}^2\chi +24 H^2 \dot{F}_{0} - 6 H^2 \overset{..}{F}_{0} \tau + 6 H^2 \tau \tilde{\nabla}^2F_{0}+24 H^2 F_{00}=0,
\label{6.53}
\end{eqnarray}
where $\Delta_{\mu\nu}=\delta G_{\mu\nu}+8\pi G \delta T_{\mu\nu}$.
On introducing $\alpha=\dot{F}+\tau\chi+F_0$ the perturbative equations simplify to 
\begin{eqnarray}
\Delta_{00}&=& -6 \dot{\alpha} \tau^{-2} - 2 \tau^{-1} \tilde{\nabla}^2\alpha - \overset{..}{F}_{00}  - 6 F_{00} \tau^{-2} - 2 \dot{F}_{00} \tau^{-1} + \tilde{\nabla}^2F_{00}=0,
\nonumber\\ 
\Delta_{0i}&=& -2 \tau^{-1} \tilde{\nabla}_{i}\dot{\alpha}- \overset{..}{F}_{0i} +  \tilde{\nabla}^2F_{0i} - 2 \tau^{-1} \tilde{\nabla}_{i}F_{00}=0,
\nonumber\\ 
\Delta_{ij}&=&\delta_{ij} \left[- 2 \ddot{\alpha}\tau^{-1}+  6 \dot{\alpha} \tau^{-2}  + 2\tau^{-1} \tilde{\nabla}^2\alpha + 6 F_{00} \tau^{-2}\right]-2\tau^{-1} \tilde{\nabla}_{i}\tilde{\nabla}_{j}\alpha
\nonumber\\
&& - \overset{..}{F}_{ij}  + 2 \dot{F}_{ij} \tau^{-1} + \tilde{\nabla}^2F_{ij} -2 \tau^{-1} \tilde{\nabla}_{i}F_{0j} - 2 \tau^{-1} \tilde{\nabla}_{j}F_{0i}=0,
\nonumber\\
H^{-2}g^{\mu\nu}\Delta_{\mu\nu} &=& 24\dot{\alpha} - 6  \overset{..}{\alpha} \tau + 6  \tau \tilde{\nabla}^2\alpha +24 F_{00}=0.
\label{6.54}
\end{eqnarray}
We thus see that $\alpha$ and $F_{\mu\nu}$ are gauge invariant for a total of six (one plus five) gauge-invariant components, just as needed. 

While we have written $\Delta_{\mu\nu}$ in the non-manifestly covariant form given (\ref{6.54}) as this will be convenient for actually solving $\Delta_{\mu\nu}=0$ below, since the SVT4 approach is covariant we are able to write the rank two  tensor $\Delta_{\mu\nu}$ in a manifestly covariant form. To do so we introduce a unit  timelike four-vector $U^{\mu}$ whose only non-zero component is $U^{0}$. In terms of this $U^{\mu}$ the gauge-invariant $\alpha$ is now given by the manifestly general coordinate scalar $\alpha=U^{\mu}\partial_{\mu}F+\chi/H\Omega+U^{\mu}F_{\mu}$, while the $F_{00}$ term in $g^{\mu\nu}\Delta_{\mu\nu}$ can be written as $U^{\mu}U^{\nu}F_{\mu\nu}$.

If there is to be a decomposition theorem then (\ref{6.54}) would have to break up into 
\begin{align}
 -6 \dot{\alpha} \tau^{-2} - 2 \tau^{-1} \tilde{\nabla}^2\alpha=0, \quad - \overset{..}{F}_{00}  - 6 F_{00} \tau^{-2} - 2 \dot{F}_{00} \tau^{-1} + \tilde{\nabla}^2F_{00}&=0,
\nonumber\\ 
 -2 \tau^{-1} \tilde{\nabla}_{i}\dot{\alpha}=0,\quad - \overset{..}{F}_{0i} +  \tilde{\nabla}^2F_{0i} - 2 \tau^{-1} \tilde{\nabla}_{i}F_{00}&=0,
\nonumber\\ 
\delta_{ij} \left[- 2 \ddot{\alpha}\tau^{-1}+  6 \dot{\alpha} \tau^{-2}  + 2\tau^{-1} \tilde{\nabla}^2\alpha \right]-2\tau^{-1} \tilde{\nabla}_{i}\tilde{\nabla}_{j}\alpha&=0,
\nonumber\\
6 \delta_{ij}F_{00} \tau^{-2} - \overset{..}{F}_{ij}  + 2 \dot{F}_{ij} \tau^{-1} + \tilde{\nabla}^2F_{ij} - 2 \tau^{-1} \tilde{\nabla}_{i}F_{0j} - 2 \tau^{-1} \tilde{\nabla}_{j}F_{0i}&=0,
\nonumber\\
24\dot{\alpha} - 6  \overset{..}{\alpha} \tau + 6  \tau \tilde{\nabla}^2\alpha=0,\quad 24 F_{00}&=0,
\label{6.55}
\end{align}
to then yield
\begin{eqnarray}
&& \dot{\alpha}=0,\quad \tilde{\nabla}_{i}\tilde{\nabla}_{j}\alpha=0,\quad \tilde{\nabla}^2\alpha=0,\quad F_{00}=0,\quad  - \ddot{F}_{0i} +  \tilde{\nabla}^2F_{0i} =0,
\nonumber\\ 
&&  - \overset{..}{F}_{ij}  + 2 \dot{F}_{ij} \tau^{-1} + \tilde{\nabla}^2F_{ij} - 2 \tau^{-1} \tilde{\nabla}_{i}F_{0j} - 2 \tau^{-1} \tilde{\nabla}_{j}F_{0i}=0,
\label{6.56}
\end{eqnarray}
with the $\epsilon^{ijk}\tilde{\nabla}_j\Delta_{0k}=0$ condition not being needed as it is satisfied identically. The solution to (\ref{6.56}) is the form 
\begin{eqnarray}
\nonumber\\ 
&& \alpha=0,\quad F_{00}=0,\quad F_{0i}=\sum _{\bf k}f_{0i}({\bf k})e^{i{\bf k}\cdot {\bf x}-ik\tau},\quad  ik^jf_{0j}({\bf k})=0,\quad F_{ij}=\sum _{\bf k}[f_{ij}({\bf k})+\tau\hat{f}_{ij}({\bf k})]e^{i{\bf k}\cdot {\bf x}-ik\tau},
\nonumber\\
&&-ikf_{ij}({\bf k})+\hat{f}_{ij}({\bf k})=ik_jf_{0i}({\bf k})+ik_if_{0j}({\bf k}),
\nonumber\\
&&\delta^{ij}f_{ij}({\bf k})=0,\quad
\delta^{ij}\hat{f}_{ij}({\bf k})=0,\quad ik^jf_{ij}({\bf k})=-ikf_{0i}({\bf k}),\quad ik^j\hat{f}_{ij}({\bf k})=0,
\label{6.57}
\end{eqnarray}
and while the most general solution for  $\alpha$ would be  a constant,  we have imposed an asymptotic spatial boundary condition, which sets the constant to zero.

We now solve the full (\ref{6.54}) exactly  to determine whether and under what conditions (\ref{6.57}) might hold. Eliminating $F_{00}$ between the $\Delta_{00}=0$ and $g^{\mu\nu}\Delta_{\mu\nu}=0$ equations in (\ref{6.54}) yields
\begin{eqnarray}
-\frac{\tau}{4}\left(\frac{\partial ^2}{\partial \tau^2}-\tilde{\nabla}^2\right)\left(\frac{\partial ^2}{\partial \tau^2}-\tilde{\nabla}^2\right)\alpha=0,
\label{6.58}
\end{eqnarray}
with general solution
\begin{eqnarray}
\alpha=\sum_{\bf k}\left(a_{\bf k}+\tau b_{\bf k}\right)e^{i{\bf k}\cdot {\bf x}-ik\tau},
\label{6.59}
\end{eqnarray}
where $a_{\bf k}$ and $b_{\bf k}$ are independent of ${\bf x}$ and $\tau$. Given $\alpha$, $F_{00}$ then evaluates to 
\begin{eqnarray}
F_{00}=\sum_{\bf k}\left[a_{00}({\bf k})+\tau b_{00}({\bf k})\right]e^{i{\bf k}\cdot {\bf x}-ik\tau},
\label{6.60}
\end{eqnarray}
where
\begin{eqnarray}
a_{00}({\bf k})=ika_{\bf k}-b_{\bf k},\quad b_{00}({\bf k})=\frac{ik}{2}b_{\bf k}.
\label{6.61}
\end{eqnarray}
Inserting these solutions for $\alpha$ and $F_{00}$ into $\Delta_{0i}=0$ then yields
\begin{eqnarray}
\ddot{F}_{0i}-\tilde{\nabla}^2F_{0i}=-\sum_{\bf k}kk_ib_{\bf k}e^{i{\bf k}\cdot {\bf x}-ik\tau},
\label{6.62}
\end{eqnarray}
with solution 
\begin{eqnarray}
F_{0i}=\sum_{\bf k}\left[a_{0i}({\bf k})+\tau b_{0i}({\bf k})\right]e^{i{\bf k}\cdot {\bf x}-ik\tau},
\label{6.63}
\end{eqnarray}
where
\begin{eqnarray}
b_{0i}({\bf k})=-\frac{ik_i}{2}b_{\bf k}.
\label{6.64}
\end{eqnarray}
With $F_{0i}$ obeying the transverse condition $\partial^iF_{0i}-\dot{F}_{00}=0$, we obtain 
\begin{eqnarray}
ik^ia_{0i}({\bf k})=k^2a_{\bf k}+\frac{3ik}{2}b_{\bf k}.
\label{6.65}
\end{eqnarray}
Finally, from $\Delta_{ij}=0$ we obtain 
\begin{eqnarray}
 \ddot{F}_{ij}  - \frac{2}{\tau} \dot{F}_{ij} - \tilde{\nabla}^2F_{ij}=
 \sum_{\bf k}\left[\delta_{ij}ikb_{\bf k}+2k_ik_ja_{\bf k}-2ik_ia_{0j}-2ik_ja_{0i}\right]\frac{1}{\tau}e^{i{\bf k}\cdot {\bf x}-ik\tau}.
\label{6.66}
\end{eqnarray}
We can thus set 
\begin{eqnarray}
F_{ij}=\sum_{\bf k}\left[a_{ij}({\bf k})+\tau b_{ij}({\bf k})\right]e^{i{\bf k}\cdot {\bf x}-ik\tau},
\label{6.67}
\end{eqnarray}
where 
\begin{eqnarray}
2ika_{ij}({\bf k})-2b_{ij}({\bf k})=\delta_{ij}ikb_{\bf k}+2k_ik_ja_{\bf k}-2ik_ia_{0j}-2ik_ja_{0i}.
\label{6.68}
\end{eqnarray}
With $F_{ij}$ obeying the transverse and traceless conditions $\partial^{j}F_{ij}=\dot{F}_{0i}$, $\delta^{ij}F_{ij}-F_{00}=0$, we obtain 
\begin{eqnarray}
ik^ja_{ij}({\bf k})=-ika_{0i}({\bf k})-\frac{ik_i}{2}b_{\bf k},\quad ik^jb_{ij}({\bf k})=-\frac{kk_i}{2}b_{\bf k}, \quad
\delta^{ij}a_{ij}({\bf k})=ika_{\bf k}-b_{\bf k},
\quad \delta^{ij}b_{ij}({\bf k})=\frac{ik}{2}b_{\bf k}.
\label{6.69}
\end{eqnarray}
Equations (\ref{6.58}) to (\ref{6.69}) provide us with the most general solution to (\ref{6.54}).

Having now obtained the exact solution, we see that  we do not get the decomposition theorem solution given in (\ref{6.57}). If we want to get the exact solution to reduce to the decomposition theorem solution we would need to set $\alpha$ and $F_{00}$ to zero, and this would be a particular solution to the fluctuation equations. However, there is no reason to set them to zero, and certainly no spatial asymptotic condition that could do so. And even if there were to be one, then such an asymptotic condition would have to suppress $\alpha$ and $F_{00}$ while at the same time not suppressing $F_{0i}$ and $F_{ij}$, even though though all of the fluctuation components have precisely the same asymptotic spatial behavior. We could possibly set $\alpha$ and $F_{00}$ to zero at all times via judiciously chosen initial conditions, but there would not appear to be any compelling reason for doing that either. As we had seen in our study of SVT4 without a conformal factor we would only be able to recover the decomposition theorem solution if we were to set $\chi$ to zero, and just as with wanting to set $\alpha$ and $F_{00}$ to zero, for $\chi$ there is also no reason  to do so. Thus in parallel with our analysis of SVT4 with no conformal factor, we find that similarly for SVT4 with a conformal factor no decomposition theorem is obtained in the de Sitter background case. However, we had noted above that when an external $\delta\bar{T}_{\mu\nu}$ source is present, we then do obtain a decomposition theorem since it is the external source that is causing the perturbation to the background geometry in the first place. So in this case the decomposition theorem is recovered.

\subsection{Some General Comments}

While we have discussed SVT4 fluctuations around a de Sitter background as this is a rich enough system to show that one does not in general get a decomposition theorem, this discussion is not the one that is relevant to the early universe  inflationary model since that model is not described by an explicit cosmological constant but by a scalar field instead. Specifically, if we have a scalar field $S(x)$ with a Lagrangian density $L(S)=K(S)-V(S)$, then at the $S=S_0$ minimum of the $V(S)$ potential with constant $S_0$ the potential acts as a cosmological constant $V(S_0)$ and one has a background de Sitter geometry. If we now perturb the background the potential will change to $V(\delta S)$ even though $V(S_0)$ will not change at all. With there also being a change $K(\delta S)$ in  the scalar field kinetic energy, all of the terms in the background $T_{\mu\nu}=\partial_{\mu}S\partial_{\nu}S-g_{\mu\nu}L(S)$ will be perturbed and $\delta T_{\mu\nu}$ will not be of the form $\delta T_{\mu\nu}=\delta g_{\mu\nu}V(S_0)$ that we studied above. Nonetheless, it would not appear that there would obviously be an SVT4 decomposition theorem in  this more general scalar field case.

To conclude this section we note that in the above study of Einstein gravity SVT4 fluctuations around a de Sitter background we found in the no conformal prefactor case that the tensor fluctuations obeyed (\ref{6.31a}), viz.  
\begin{eqnarray}
(\nabla_{\alpha}\nabla^{\alpha}-4H^2)(\nabla_{\alpha}\nabla^{\alpha}-2H^2)F_{\mu\nu}=0.
\label{6.70}
\end{eqnarray}
Even though (\ref{6.70}) was obtained in Einstein gravity, this very same structure for $F_{\mu\nu}$ also appears in conformal gravity. In \cite{Mannheim2012a}  the perturbative conformal gravity Bach tensor $\delta W_{\mu\nu}$ was calculated for fluctuations around a de Sitter background of the form $h_{\mu\nu}=K_{\mu\nu}+g_{\mu\nu}g^{\alpha\beta}h_{\alpha\beta}/4$ (i.e. a traceless but not necessarily transverse $K_{\mu\nu}$), and was found to take the  form
\begin{align}
\delta W_{\mu\nu}&=\frac{1}{2}[\nabla_{\alpha}\nabla^{\alpha}-4H^2][\nabla_{\beta}\nabla^{\beta}-2H^2]K_{\mu\nu}
-\frac{1}{2}[\nabla_{\beta}\nabla^{\beta}-4H^2][
\nabla_{\mu}\nabla_{\lambda}K^{\lambda}_{\phantom{\lambda}\nu}
+\nabla_{\nu}\nabla_{\lambda}K^{\lambda}_{\phantom{\lambda}\mu}]
\nonumber\\
&+\frac{1}{6}[g_{\mu\nu}\nabla_{\alpha}\nabla^{\alpha}+2\nabla_{\mu}\nabla_{\nu}
-6H^2g_{\mu\nu}]\nabla_{\kappa}\nabla_{\lambda}K^{\kappa\lambda}.
\label{6.71}
\end{align}
Evaluating $\delta W_{\mu\nu}$ for the fluctuation $h_{\mu\nu}$ given in (\ref{6.1}) in the same de Sitter background is found to yield 
\begin{eqnarray}
\delta W_{\mu\nu}= (\nabla_{\alpha}\nabla^{\alpha}-4H^2)(\nabla_{\alpha}\nabla^{\alpha}-2H^2)F_{\mu\nu},
\label{6.72}
\end{eqnarray}
i.e. the same structure that we would have obtained from (\ref{6.71}) had we made $K_{\mu\nu}$ transverse and replaced it by $2F_{\mu\nu}$ (even though the relation of $F_{\mu\nu}$ to  $h_{\mu\nu}$ is not the same as that of $K_{\mu\nu}$ to $h_{\mu\nu}$). We recognize the structure of the conformal gravity (\ref{6.72}) as being none other than that of the standard gravity (\ref{6.70}). The transverse-traceless sector of standard gravity (viz. gravity waves) thus has a conformal structure.

Now in a geometry that is conformal to flat such as de Sitter, the background $W_{\mu\nu}$ vanishes identically. Thus from the conformal gravity equation of motion (\ref{4.3}) for the Bach tensor it follows that the background $T_{\mu\nu}$ also vanishes identically. In the absence of a new source $\delta \bar{T}_{\mu\nu}$, for conformal gravity fluctuations around de Sitter  we can thus set 
\begin{eqnarray}
4\alpha_g\delta W_{\mu\nu}-\delta T_{\mu\nu}=0,\quad 4\alpha_g\delta W_{\mu\nu}=0,
\label{6.73}
\end{eqnarray}
since $\delta T_{\mu\nu}=0$. And since $\delta T_{\mu\nu}$ is zero,  it follows that  $\delta W_{\mu\nu}$ is gauge invariant all on its own in a background that is conformal to flat. And with it being traceless, the five degree of freedom $\delta W_{\mu\nu}$ can only depend on the five degree of freedom $F_{\mu\nu}$, just as we see in (\ref{6.72}). 

Now from  (\ref{6.19}) we can identify $(\nabla_{\alpha}\nabla^{\alpha}-2H^2)F_{\mu\nu}$ as the transverse-traceless piece of $\delta G_{\mu\nu}-3H^2h_{\mu\nu}$ in a de Sitter background, and thus can set 
\begin{eqnarray}
(\nabla_{\alpha}\nabla^{\alpha}-4H^2)(\delta G_{\mu\nu}+\delta T_{\mu\nu})^{T\theta}=(\nabla_{\alpha}\nabla^{\alpha}-4H^2)(\nabla_{\alpha}\nabla^{\alpha}-2H^2)F_{\mu\nu}
\label{6.74}
\end{eqnarray}
for the transverse ($T$) traceless ($\theta$) sector of $\delta G_{\mu\nu}+\delta T_{\mu\nu}$. Thus given (\ref{6.72}) we can set
\begin{eqnarray}
\delta W_{\mu\nu}=(\nabla_{\alpha}\nabla^{\alpha}-4H^2)(\delta G_{\mu\nu}+\delta T_{\mu\nu})^{T\theta}.
\label{6.75}
\end{eqnarray}
We thus generalize the flat space fluctuation relation $\delta W_{\mu\nu}=\nabla_{\alpha}\nabla^{\alpha}\delta G_{\mu\nu}^{T\theta}$ given in \cite{footnote6} to the de Sitter case. Finally, we note that if the  $\delta\bar{T}_{\mu\nu}$ source is present, then its tracelessness in the conformal case restricts its form in (\ref{6.20}) to $\delta \bar{T}_{\mu\nu}=\bar{F}_{\mu\nu}$, with the conformal gravity fluctuation equation in a de Sitter background then taking the form 
\begin{eqnarray}
4\alpha_g(\nabla_{\alpha}\nabla^{\alpha}-4H^2)(\nabla_{\alpha}\nabla^{\alpha}-2H^2)F_{\mu\nu}=\bar{F}_{\mu\nu}.
\label{6.76}
\end{eqnarray}
Thus with or without $\delta \bar{T}_{\mu\nu}$, in the conformal gravity SVT4 de Sitter case $\delta W_{\mu\nu}$ depends on $F_{\mu\nu}$ alone, and with there being no dependence on $\chi$ the decomposition theorem is automatic. 

\section{Solving the SVT3 fluctuation equations -- de Sitter background}
\label{S7}
In the SVT3 background de Sitter case one can write the background and fluctuation metric in the conformal to flat form
\begin{eqnarray}
ds^2 =\frac{1}{H^2\tau^2}\bigg{[}(1+2\phi) d\tau^2 -2(\tilde{\nabla}_i B +B_i)d\tau dx^i - [(1-2\psi)\delta_{ij} +2\tilde{\nabla}_i\tilde{\nabla}_j E + \tilde{\nabla}_i E_j + \tilde{\nabla}_j E_i + 2E_{ij}]dx^i dx^j\bigg{]},
\label{7.1}
\end{eqnarray}
where the $\tilde{\nabla}_{i}$ denote derivatives with respect to the flat 3-space $\delta_{ij}dx^idx^j$ metric.
In terms of the SVT3 form for the fluctuations the components of the perturbed $\delta G_{\mu\nu}$ are given by (see e.g. \cite{Amarasinghe2018})
\begin{eqnarray}
\delta G_{00}&=&-\frac{6}{\tau}\dot{\psi}-\frac{2}{\tau}\tilde{\nabla}^2(\tau \psi +B-\dot{E}),
\nonumber\\
\delta G_{0i}&=&\frac{1}{2}\tilde{\nabla}^2(B_i-\dot{E}_i)+\frac{1}{\tau^2}\tilde{\nabla}_i(3B-2\tau^2\dot{\psi}+2\tau \phi)+\frac{3}{\tau^2}B_i,
\nonumber\\
\delta G_{ij}&=&\frac{\delta_{ij}}{\tau^2}\left[-2\tau^2\ddot{\psi}+2\tau\dot{\phi}+4\tau\dot{\psi}-6\phi-6\psi+\tilde{\nabla}^2\left(2\tau B-\tau^2\dot{B}+\tau^2\ddot{E}-2\tau\dot{E}-\tau^2\phi+\tau^2\psi\right)\right]
\nonumber\\
&+&\frac{1}{\tau^2}\tilde{\nabla}_i\tilde{\nabla}_j\left[-2\tau B +\tau^2\dot{B}-\tau^2\ddot{E}+2\tau\dot{E}+6E+\tau^2\phi-\tau^2\psi\right]
\nonumber\\
&+&\frac{1}{2\tau^2}\tilde{\nabla}_i\left[-2\tau B_j+2\tau\dot{E}_j+\tau^2\dot{B}_j-\tau^2\ddot{E}_j+6E_j\right]
\nonumber\\
&+&\frac{1}{2\tau^2}\tilde{\nabla}_j\left[-2\tau B_i+2\tau\dot{E}_i+\tau^2\dot{B}_i-\tau^2\ddot{E}_i+6E_i\right]
\nonumber\\
&-&\ddot{E}_{ij}+\frac{6}{\tau^2}E_{ij}+\frac{2}{\tau}\dot{E}_{ij}+\tilde{\nabla}^2E_{ij},
\label{7.2}
\end{eqnarray}
where the dot denotes the derivative with respect to the conformal time $\tau$ and $\tilde{\nabla}^2=\delta^{ij}\tilde{\nabla}_i\tilde{\nabla}_j$. 
For the de Sitter SVT3 metric the gauge-invariant metric combinations are (see e.g. \cite{Amarasinghe2018})
\begin{eqnarray}
\alpha=\phi+\psi+\dot{B}-\ddot{E} ,\quad \beta=\tau\psi+B-\dot{E}, \quad B_i-\dot{E}_i,\quad E_{ij}.
\label{7.3}
\end{eqnarray}
(For a generic SVT3 metric with a general conformal factor $\Omega(\tau)$ the quantity $-(\Omega/\dot{\Omega})\psi+B-\dot{E}$ is gauge invariant, to thus become $\beta$ when $\Omega(\tau)=1/H\tau$, with the other gauge invariants being independent of $\Omega(\tau)$.)
In terms of the gauge-invariant combinations the fluctuation equations $\Delta_{\mu\nu}=\delta G_{\mu\nu}+\delta T_{\mu\nu}=0$ take the form
\begin{eqnarray}
 \Delta_{00}&=&-\frac{6}{\tau^2}(\dot{\beta}-\alpha)-\frac{2}{\tau}\tilde{\nabla}^2\beta=0,
\label{7.4}
\end{eqnarray}
\begin{eqnarray}
 \Delta_{0i}&=&\frac{1}{2}\tilde{\nabla}^2(B_i-\dot{E}_i)-\frac{2}{\tau}\tilde{\nabla}_i(\dot{\beta}-\alpha)=0,
\label{7.5}
\end{eqnarray}
\begin{eqnarray}
 \Delta_{ij}&=&\frac{\delta_{ij}}{\tau^2}\left[-2\tau(\ddot{\beta}-\dot{\alpha})+6(\dot{\beta}-\alpha)+\tau \tilde{\nabla}^2(2\beta-\tau \alpha)\right]
+\frac{1}{\tau}\tilde{\nabla}_i\tilde{\nabla}_j(-2 \beta +\tau\alpha)
\nonumber\\
&+&\frac{1}{2\tau}\tilde{\nabla}_i[-2(B_j-\dot{E}_j)+\tau(\dot{B}_j-\ddot{E}_j)]
+\frac{1}{2\tau}\tilde{\nabla}_j[-2(B_i-\dot{E}_i)+\tau(\dot{B}_i-\ddot{E}_i)]
\nonumber\\
&-&\ddot{E}_{ij}+\frac{2}{\tau}\dot{E}_{ij}+\tilde{\nabla}^2E_{ij}=0,
\label{7.6}
\end{eqnarray}
\begin{eqnarray}
g^{\mu\nu}\Delta_{\mu\nu}&=&H^2[-6\tau(\ddot{\beta}-\dot{\alpha})+24(\dot{\beta}-\alpha)
+6\tau \tilde{\nabla}^2\beta-2\tau^2\tilde{\nabla}^2\alpha]=0,
\label{7.7}
\end{eqnarray}
to thus be manifestly gauge invariant.

If there is to be a decomposition theorem the S, V and T components of $\Delta_{\mu\nu}$ will satisfy $\Delta_{\mu\nu}=0$ independently, to thus be required to obey
\begin{align}
-\frac{6}{\tau^2}(\dot{\beta}-\alpha)-\frac{2}{\tau}\tilde{\nabla}^2\beta=0,\quad \frac{1}{2}\tilde{\nabla}^2(B_i-\dot{E}_i)=0, \quad \frac{2}{\tau}\tilde{\nabla}_i(\dot{\beta}-\alpha)&=0,
\nonumber\\
\frac{\delta_{ij}}{\tau^2}\left[-2\tau(\ddot{\beta}-\dot{\alpha})+6(\dot{\beta}-\alpha)+\tau \tilde{\nabla}^2(2\beta-\tau\alpha)\right]+ \frac{1}{\tau^2}\tilde{\nabla}_i\tilde{\nabla}_j(-2\tau \beta +\tau^2\alpha)&=0,
\nonumber\\
\frac{1}{2\tau^2}\tilde{\nabla}_i[-2\tau (B_j-\dot{E}_j)+\tau^2(\dot{B}_j-\ddot{E}_j)]
+\frac{1}{2\tau^2}\tilde{\nabla}_j[-2\tau (B_i-\dot{E}_i)+\tau^2(\dot{B}_i-\ddot{E}_i)]&=0,
\nonumber\\
-\ddot{E}_{ij}+\frac{2}{\tau}\dot{E}_{ij}+\tilde{\nabla}^2E_{ij}&=0.
\label{7.8}
\end{align}

To determine whether or not these conditions might hold we need to solve the fluctuation equations $\Delta_{\mu\nu}=0$ directly, to see what the structure of the solutions might look like.  To this end we first apply $\tau\partial_{\tau}-1$ to $-\tau^2\Delta_{00}/2$,  to obtain
\begin{eqnarray}
\tau^2\tilde{\nabla}^2\dot{\beta}+3\tau(\ddot{\beta}-\dot{\alpha})-3(\dot{\beta}-\alpha)=0,
\label{7.9}
\end{eqnarray}
and then add $3\tau^2\Delta_{00}$ to  $g^{\mu\nu}\Delta_{\mu\nu}/H^2$ to obtain
\begin{eqnarray}
\tau^2\tilde{\nabla}^2\alpha+3\tau(\ddot{\beta}-\dot{\alpha})-3(\dot{\beta}-\alpha)=0.
\label{7.10}
\end{eqnarray}
Combining these equations and using $\Delta_{00}=0$ we thus obtain
\begin{eqnarray}
\tilde{\nabla}^2(\alpha-\dot{\beta})=0,\quad \tilde{\nabla}^2\beta=0,
\label{7.11}
\end{eqnarray}
and 
\begin{eqnarray}
\tau^2\tilde{\nabla}^2(\alpha+\dot{\beta})+6\tau(\ddot{\beta}-\dot{\alpha})-6(\dot{\beta}-\alpha)=0.
\label{7.12}
\end{eqnarray}
Applying $\tilde{\nabla}^2$ then gives
\begin{eqnarray}
\tilde{\nabla}^4(\alpha+\dot{\beta})=0,\quad \tilde{\nabla}^4(\alpha-\dot{\beta})=0.
\label{7.13}
\end{eqnarray}
Applying $\tilde{\nabla}^2$ to $\Delta_{0i}$ in turn then gives
\begin{eqnarray}
 \tilde{\nabla}^4(B_i-\dot{E}_i)=0,
\label{7.14}
\end{eqnarray}
while applying $\epsilon^{ijk}\tilde{\nabla}_j$ to $\Delta_{0k}$ gives
\begin{eqnarray}
\frac{1}{2}\epsilon^{ijk}\tilde{\nabla}_j\tilde{\nabla}^2(B_k-\dot{E}_k)=0.
\label{7.15}
\end{eqnarray}
Finally, to obtain an equation that only involves $E_{ij}$ we apply $\tilde{\nabla}^4$ to $\Delta_{ij}$, to obtain
\begin{eqnarray}
 \tilde{\nabla}^4\left(-\ddot{E}_{ij}+\frac{2}{\tau}\dot{E}_{ij}+\tilde{\nabla}^2E_{ij}\right)=0.
\label{7.16}
\end{eqnarray}
As we see, we can isolate all the individual S, V and T gauge-invariant combinations, to thus give decomposition for the individual SVT3 components. However, the relations we obtain look nothing like the relations that a decomposition theorem would require, and thus without some further input we do not obtain a decomposition theorem.

To provide some further input we impose some asymptotic boundary conditions. To this end  we recall from Sec. \ref{S1} that for any spatially asymptotically bounded function $A$ that obeys $\tilde{\nabla}^2A=0$, the only solution is $A=0$. If $A$ obeys $\tilde{\nabla}^4A=0$, we must first set $\tilde{\nabla}^2A=C$, so that $\tilde{\nabla}^2C=0$. Imposing boundary conditions for $C$ enables us to set $C$=0. In such a case we can then set $\tilde{\nabla}^2A=0$, and with sufficient asymptotic convergence can then set $A=0$. Now a function could obey $\tilde{\nabla}^2A=0$ trivially by being independent of the spatial coordinates altogether, and only depend on $\tau$. However, it then would not vanish at spatial infinity, and we can thus exclude this possibility. With such spatial convergence for all of the S, V and T components we can then set 
\begin{eqnarray}
\alpha=0,\quad \dot{\beta}=0,\quad \beta =0,\quad B_i-\dot{E}_i=0,\quad -\tau\ddot{E}_{ij}+2\dot{E}_{ij}+\tau\tilde{\nabla}^2E_{ij}=0.
\label{7.17}
\end{eqnarray}
Since this solution coincides with the solution that would be obtained to (\ref{7.8}) under the same boundary conditions, we see that under these asymptotic boundary conditions we have a decomposition theorem.

In this solution all components of the SVT3 decomposition vanish identically except the rank two tensor $E_{ij}$. Taking $E_{ij}$ to behave as $\epsilon_{ij}\tau^2f(\tau)g({\bf x})$ where $\epsilon_{ij}$ is a polarization tensor, we find that the solution obeys
\begin{eqnarray}
\frac{\tau^2 \ddot{f}+2\tau \dot{f}-2f}{\tau^2f}=\frac{\tilde{\nabla}^2g}{g}=-k^2,
\label{7.18}
\end{eqnarray}
where $k^2$ is a separation constant. Consequently $E_{ij}$ is given as
\begin{eqnarray}
E_{ij}=\epsilon_{ij}({\bf k})\tau^2[a_1({\bf k})j_1(k\tau)+b_1({\bf k})y_1(k\tau)]e^{i{\bf k}\cdot{\bf x}},
\label{7.19}
\end{eqnarray}
where ${\bf k}\cdot{\bf k}=k^2$, $j_1$ and $y_1$ are spherical Bessel functions, and $a_1({\bf k})$ and $b_1({\bf k})$ are spacetime independent constants. For $E_{ij}$ to obey the transverse and traceless conditions $\delta^{ij}E_{ij}=0$, $\tilde{\nabla}^jE_{ij}=0$ the polarization tensor $\epsilon_{ij}({\bf k})$ must obey $\delta^{ij}\epsilon_{ij}=0$, ${\bf k}^{j}\epsilon_{ij}({\bf k})=0$.
Then, by taking a family of separation constants we can form a transverse-traceless wave packet
\begin{eqnarray}
E_{ij}&=&\sum_{\bf k}\epsilon_{ij}({\bf k})\tau^2[a_1({\bf k})j_1(k\tau)+b_1({\bf k})y_1(k\tau)]e^{i{\bf k}\cdot{\bf x}}\nonumber\\
&=&
\sum_{\bf k}\epsilon_{ij}({\bf k})\left[a_1({\bf k})\left(\frac{\sin(k\tau)}{k^2}-\frac{\tau\cos(k\tau)}{k}\right)+b_1({\bf k})\left(\frac{\cos(k\tau)}{k^2}+\frac{\tau\sin(k\tau)}{k}\right)\right],
\label{7.20}
\end{eqnarray}
and can choose the $a_1({\bf k})$ and $b_1({\bf k})$ coefficients to make the packet be as well-behaved at spatial infinity as desired. Finally, since according to (\ref{7.1}) the full fluctuation is given not by $E_{ij}$ but by $2E_{ij}/H^2\tau^2$, then with $\tau=e^{-Ht}/H$, through the $\cos(k\tau)/k^2$ term we find that at large comoving time  $E_{ij}/\tau^2$ behaves as $e^{2Ht}$, viz. the standard de Sitter fluctuation exponential growth.

\section{Solving the SVT3 Fluctuation Equations -- Radiation Era Spatially Flat Robertson-Walker Background}
\label{S8}
In comoving coordinates a spatially flat Robertson-Walker background metric takes the form $ds^2=dt^2-a^2(t)\delta_{ij}dx^idx^j$. In the radiation era where a perfect fluid pressure $p$ and energy density $\rho$ are related by $\rho=3p$, the background energy-momentum tensor is given by the traceless
\begin{eqnarray}
T_{\mu\nu}=p(4U_{\mu}U_{\nu}+g_{\mu\nu}),
\label{8.1}
\end{eqnarray}
where $g^{\mu\nu}U_{\mu}U_{\nu}=-1$, $U^{0}=1$, $U_0=-1$, $U^{i}=0$, $U_i=0$. With this source the background Einstein equations $G_{\mu\nu}=-T_{\mu\nu}$ with $8\pi G=1$ fix $a(t)$ to be $a(t)=t^{1/2}$. In conformal to flat coordinates we set $\tau=\int dt/t^{1/2}=2t^{1/2}$, with the conformal factor being given by $\Omega(\tau)=\tau/2$. In conformal to flat coordinates the background pressure is of the form $p=4/\tau^4$ while $U^{0}=2/\tau$, $U_0=-\tau/2$. In this coordinate system the SVT3 fluctuation metric as written with an explicit conformal factor is of the form 
\begin{eqnarray}
ds^2 &=&\frac{\tau^2}{4}\bigg{[}(1+2\phi) d\tau^2 -2(\tilde{\nabla}_i B +B_i)d\tau dx^i - [(1-2\psi)\delta_{ij} +2\tilde{\nabla}_i\tilde{\nabla}_j E + \tilde{\nabla}_i E_j + \tilde{\nabla}_j E_i + 2E_{ij}]dx^i dx^j\bigg{]},
\label{8.2}
\end{eqnarray}
and the fluctuation energy-momentum tensor is of the form
\begin{eqnarray}
\delta T_{\mu\nu}=\delta p(4U_{\mu}U_{\nu}+g_{\mu\nu})+p(4\delta U_{\mu}U_{\nu}+4U_{\mu}\delta U_{\nu}+h_{\mu\nu}).
\label{8.3}
\end{eqnarray}
As written, we might initially expect there to be five fluctuation variables in the fluctuation energy-momentum tensor: $p$ and the four components of $\delta U_{\mu}$. However, varying  $g^{\mu\nu}U_{\mu}U_{\nu}=-1$ gives 
\begin{eqnarray}
 \delta g^{00}U_{0}U_{0}+2g^{00}U_{0}\delta U_{0}=0,
\label{8.4}
\end{eqnarray}
i.e. 
\begin{eqnarray}
\delta U_{0}=-\frac{1}{2}(g^{00})^{-1}(-g^{00}g^{00}\delta g_{00})U_{0}=-\frac{\tau\phi}{2}.
\label{8.5}
\end{eqnarray}
Thus $\delta U_{0}$ is not an independent of the metric fluctuations, and we need the fluctuation equations to only fix six (viz. ten minus four) independent gauge-invariant metric fluctuations and the  four $\delta p$ and $\delta U_i$ (counting all components). With ten $\Delta_{\mu\nu}=0$ equations, we can nicely determine all of them.

To this end we evaluate $\delta G_{\mu\nu}$, to obtain 
\begin{eqnarray}
\delta G_{00}&=&\frac{6}{\tau}\dot{\psi}+\frac{2}{\tau}\tilde{\nabla}^2(-\tau \psi +B-\dot{E}),
\nonumber\\
\delta G_{0i}&=&\frac{1}{2}\tilde{\nabla}^2(B_i-\dot{E}_i)+\frac{1}{\tau^2}\tilde{\nabla}_i(-B-2\tau^2\dot{\psi}-2\tau \phi)-\frac{1}{\tau^2}B_i,
\nonumber\\
\delta G_{ij}&=&\frac{\delta_{ij}}{\tau^2}\left[-2\tau^2\ddot{\psi}-2\tau\dot{\phi}-4\tau\dot{\psi}+2\phi+2\psi+\tilde{\nabla}^2\left(-2\tau B-\tau^2\dot{B}+\tau^2\ddot{E}+2\tau\dot{E}-\tau^2\phi+\tau^2\psi\right)\right]
\nonumber\\
&+&\frac{1}{\tau^2}\tilde{\nabla}_i\tilde{\nabla}_j\left[2\tau B +\tau^2\dot{B}-\tau^2\ddot{E}-2\tau\dot{E}-2E+\tau^2\phi-\tau^2\psi\right]
\nonumber\\
&+&\frac{1}{2\tau^2}\tilde{\nabla}_i\left[2\tau B_j-2\tau\dot{E}_j+\tau^2\dot{B}_j-\tau^2\ddot{E}_j-2E_j\right]
\nonumber\\
&+&\frac{1}{2\tau^2}\tilde{\nabla}_j\left[2\tau B_i-2\tau\dot{E}_i+\tau^2\dot{B}_i-\tau^2\ddot{E}_i-2E_i\right]
\nonumber\\
&-&\ddot{E}_{ij}-\frac{2}{\tau^2}E_{ij}-\frac{2}{\tau}\dot{E}_{ij}+\tilde{\nabla}^2E_{ij},
\label{8.6}
\end{eqnarray}
where the dot denotes $\partial_\tau$.
For a spatially flat Robertson-Walker  metric the gauge-invariant metric combinations are (see e.g. \cite{Amarasinghe2018})
\begin{eqnarray}
\alpha=\phi+\psi+\dot{B}-\ddot{E} ,\quad \gamma=-\tau\psi+B-\dot{E}, \quad B_i-\dot{E}_i,\quad E_{ij}.
\label{8.7}
\end{eqnarray}
(For a generic (\ref{8.2}) with conformal factor $\Omega(\tau)$, the quantity $-(\Omega/\dot{\Omega})\psi+B-\dot{E}$ becomes $\gamma$ when $\Omega(\tau)=\tau/2$, with the other gauge invariants being independent of $\Omega(\tau)$.)
In terms of the gauge-invariant combinations the fluctuation equations $\Delta_{\mu\nu}=\delta G_{\mu\nu}+\delta T_{\mu\nu}=0$ take the form
\begin{eqnarray}
 \Delta_{00}&=&-\frac{16}{\tau^3}\delta U_{0}-\frac{8}{\tau^2}\phi+\frac{3\tau^2}{4}\left(\delta p -\frac{16}{\tau^4}\psi\right)
 +\frac{6}{\tau^2}(\alpha-\dot{\gamma})+\frac{2}{\tau}\tilde{\nabla}^2\gamma=0,
\label{8.8}
\end{eqnarray}
\begin{eqnarray}
 \Delta_{0i}&=&-\frac{8}{\tau^3}\delta U_{i}+\frac{4}{\tau}\tilde{\nabla}_i\psi
 +\frac{1}{2}\tilde{\nabla}^2(B_i-\dot{E}_i)-\frac{2}{\tau}\tilde{\nabla}_i(\alpha-\dot{\gamma})=0,
\label{8.9}
\end{eqnarray}
\begin{eqnarray}
 \Delta_{ij}&=&\frac{\delta_{ij}}{4\tau^2}\left[\tau^4\delta p-16\psi-8\tau(\dot{\alpha}-\ddot{\gamma})+8(\alpha-\dot{\gamma})-4\tau \tilde{\nabla}^2(\tau\alpha+2\gamma)\right]
+\frac{1}{\tau}\tilde{\nabla}_i\tilde{\nabla}_j(\tau\alpha+2\gamma)
\nonumber\\
&+&\frac{1}{2\tau}\tilde{\nabla}_i[2(B_j-\dot{E}_j)+\tau(\dot{B}_j-\ddot{E}_j)]
+\frac{1}{2\tau}\tilde{\nabla}_j[2(B_i-\dot{E}_i)+\tau(\dot{B}_i-\ddot{E}_i)]
\nonumber\\
&-&\ddot{E}_{ij}-\frac{2}{\tau}\dot{E}_{ij}+\tilde{\nabla}^2E_{ij}=0,
\label{8.10}
\end{eqnarray}
\begin{eqnarray}
g^{\mu\nu}\Delta_{\mu\nu}&=&\frac{64}{\tau^5}\delta U_0+\frac{32}{\tau^4}\phi -\frac{24}{\tau^3}(\dot{\alpha}-\ddot{\gamma})-\frac{8}{\tau^2}\tilde{\nabla}^2\alpha-\frac{24}{\tau^3}\tilde{\nabla}^2\gamma=0.
\label{8.11}
\end{eqnarray}
Since $\Delta_{\mu\nu}$ is gauge invariant, we see that it is not $\delta U_0$, $\delta U_i$ and $\delta p$ themselves that are gauge invariant. Rather it is the combinations $\delta U_0+\tau\phi/2$, $\delta p-16\psi/\tau^4$, and $\delta U_i-\tau^2\tilde{\nabla}_i\psi/2$ that are gauge invariant instead. Since we have shown above that $\delta U_0+\tau\phi/2$ is actually zero, confirming that it is equal to a gauge-invariant quantity provides a nice check on our calculation. However, since $\delta U_0+\tau\phi/2$ is zero, we can replace the $\Delta_{00}$ and $g^{\mu\nu}\Delta_{\mu\nu}$ equations by 
\begin{eqnarray}
 \Delta_{00}&=&\frac{3\tau^2}{4}\left(\delta p -\frac{16}{\tau^4}\psi\right)
 +\frac{6}{\tau^2}(\alpha-\dot{\gamma})+\frac{2}{\tau}\tilde{\nabla}^2\gamma=0,
\label{8.12}
\end{eqnarray}
\begin{eqnarray}
g^{\mu\nu}\Delta_{\mu\nu}&=& -\frac{24}{\tau^3}(\dot{\alpha}-\ddot{\gamma})-\frac{8}{\tau^2}\tilde{\nabla}^2\alpha-\frac{24}{\tau^3}\tilde{\nabla}^2\gamma=0.
\label{8.13}
\end{eqnarray}

To put $\delta U_i$ in a more convenient form we decompose it into transverse and longitudinal components  as $\delta U_i=V_i+\tilde{\nabla}_iV$ where 
\begin{eqnarray}
\tilde{\nabla}^iV_i=0,\quad \tilde{\nabla}^2V=\tilde{\nabla}^i\delta U_i,\quad V({\bf x},\tau)=\int d^3{\bf y}D^{(3)}({\bf x}-{\bf y})\tilde{\nabla}^i_y\delta U_i({\bf y}, \tau).
\label{8.14}
\end{eqnarray}
In terms of these components the $\Delta_{0i}=0$ equation takes the form
\begin{eqnarray}
\Delta_{0i}&=&-\frac{8}{\tau^3}V_{i}+\frac{1}{2}\tilde{\nabla}^2(B_i-\dot{E}_i)-\frac{8}{\tau^3}\tilde{\nabla}_iV+\frac{4}{\tau}\tilde{\nabla}_i\psi
 -\frac{2}{\tau}\tilde{\nabla}_i(\alpha-\dot{\gamma})=0.
 \label{8.15}
\end{eqnarray}
Applying $\tilde{\nabla}^i$ to $\Delta_{0i}$ and then $\tilde{\nabla}^2$ in turn yields
\begin{eqnarray}
\tilde{\nabla}^i\Delta_{0i}&=&\tilde{\nabla}^2\left[-\frac{8}{\tau^3}V+\frac{4}{\tau}\psi
 -\frac{2}{\tau}(\alpha-\dot{\gamma})\right]=0,\quad \tilde{\nabla}^2\left[-\frac{8}{\tau^3}V_{i}+\frac{1}{2}\tilde{\nabla}^2(B_i-\dot{E}_i)\right]=0,
 \label{8.16}
 \end{eqnarray}
 while applying $\epsilon^{ijk}\tilde{\nabla}_j$ to $\Delta_{0k}$ yields
 \begin{eqnarray}
 \epsilon^{ijk}\tilde{\nabla}_j\Delta_{0k}&=&\epsilon^{ijk}\tilde{\nabla}_j\left[-\frac{8}{\tau^3}V_{k}+\frac{1}{2}\tilde{\nabla}^2(B_k-\dot{E}_k)\right]=0.
 \label{8.17}
\end{eqnarray}
We thus identify $V-\tau^2\psi/2$ and $V_i$ as being gauge invariant.

To determine more of the structure of the solutions to the fluctuation equations we apply $\tilde{\nabla}^i\tilde{\nabla}^j$ to the $\Delta_{ij}=0$ equation and then use the $\Delta_{00}=0$ equation to eliminate $\delta p -16\psi/\tau^4$ from $\Delta_{ij}$, to obtain
\begin{eqnarray}
&&\tilde{\nabla}^2\bigg{[}\frac{\tau^2}{4}\delta p-\frac{4}{\tau^2}\psi-\frac{2}{\tau}(\dot{\alpha}-\ddot{\gamma})+\frac{2}{\tau^2}(\alpha-\dot{\gamma})\bigg{]}
=-\frac{2}{3\tau}\left[
\tilde{\nabla}^4\gamma+3\tilde{\nabla}^2(\dot{\alpha}-\ddot{\gamma})\right]=0.
\label{8.18}
\end{eqnarray}
With the application of $\tilde{\nabla}^2$ to the $g^{\mu\nu}\Delta_{\mu\nu}=0$ equation yielding
\begin{eqnarray}
3\tilde{\nabla}^2(\dot{\alpha}-\ddot{\gamma})+\tau \tilde{\nabla}^4\alpha+3\tilde{\nabla}^4\gamma=0,
\label{8.19}
\end{eqnarray}
we obtain
\begin{eqnarray}
\tilde{\nabla}^4(\tau \alpha+2\gamma)=0.
\label{8.20}
\end{eqnarray}
On applying $\partial_{\tau}$ to (\ref{8.20}) and using (\ref{8.20}) and (\ref{8.18}) we obtain
\begin{eqnarray}
3\tau\tilde{\nabla}^4\dot{\alpha}=-3\tilde{\nabla}^4\alpha-6\tilde{\nabla}^4\dot{\gamma}
=\frac{6}{\tau}\tilde{\nabla}^4\gamma-6\tilde{\nabla}^4\dot{\gamma}
=3\tau\tilde{\nabla}^4\ddot{\gamma}-\tau\tilde{\nabla}^6\gamma.
\label{8.21}
\end{eqnarray}
The parameter $\gamma$ thus obeys
\begin{eqnarray}
\tilde{\nabla}^4\left(\tilde{\nabla}^2\gamma-3\ddot{\gamma}-\frac{6}{\tau}\dot{\gamma}+\frac{6}{\tau^2}\gamma\right)=0.
\label{8.22}
\end{eqnarray}
We can treat (\ref{8.22})  as a second-order differential equation for $\tilde{\nabla}^4\gamma$.
On setting $\tilde{\nabla}^4\gamma=f_k(\tau)g_k({\bf x})$ for a single mode, on separating  with a separation constant $-k^2$ according to
\begin{eqnarray}
\frac{\tilde{\nabla}^2g_k}{g_k}=\frac{3\ddot{f}_k+6\dot{f}_k/\tau-6f_k/\tau^2}{f_k}=-k^2,
\label{8.23}
\end{eqnarray}
we find that the $\tau$ dependence of $f_k(\tau)$ is given by $j_1(k\tau/\surd{3})$ and $y_1(k\tau/\surd{3})$, while the spatial dependence is given by plane waves. Since the set of plane waves is complete, the general solution to (\ref{8.22}) can be written as
\begin{eqnarray}
\gamma&=&\sum_{\bf k}[a_1({\bf k})j_1(k\tau/\surd{3})+b_1({\bf k})y_1(k\tau/\surd{3})]e^{i{\bf k}\cdot{\bf x}} +{\rm delta~function~terms},
\label{8.24}
\end{eqnarray}
where the delta function terms are solutions to $\tilde{\nabla}^4\gamma=0$, solutions that, in analog to  (\ref{1.8}),  are generically of the form $\delta(k)$, $\delta(k)/k$, $\delta(k)/k^2$, $\delta(k)/k^3$.

Proceeding the same way for $\alpha$  we obtain 
\begin{eqnarray}
3\tilde{\nabla}^4\dot{\alpha}=3\tilde{\nabla}^4\ddot{\gamma}-\tilde{\nabla}^6\gamma
=-\frac{3}{2}\tilde{\nabla}^4(\tau\ddot{\alpha}+2\dot{\alpha})+\frac{\tau}{2}\tilde{\nabla}^6\alpha.
\label{8.25}
\end{eqnarray}
The parameter $\alpha$ thus obeys
\begin{eqnarray}
\tilde{\nabla}^4\left(\tilde{\nabla}^2\alpha-3\ddot{\alpha}-\frac{12}{\tau}\dot{\alpha}\right)=0.
\label{8.26}
\end{eqnarray}
On setting $\tilde{\nabla}^4\alpha=d_k(\tau)e_k({\bf x})$ for a single mode, we can separate with a separation constant $-k^2$ according to
\begin{eqnarray}
\frac{\tilde{\nabla}^2e_k}{e}=\frac{3\ddot{d}_k+12\dot{d}_k/\tau}{d_k}=-k^2.
\label{8.27}
\end{eqnarray}
The $\tau$ dependence of $d_k(\tau)$ is thus given by $j_1(k\tau/\surd{3})/\tau$ and  $y_1(k\tau/\surd{3})/\tau$, while the spatial dependence is given by plane waves. The general solution to (\ref{8.26}) is thus given by 
\begin{eqnarray}
\alpha&=&\frac{1}{\tau}\sum_{\bf k}[m_1({\bf k})j_1(k\tau/\surd{3})+n_1({\bf k})y_1(k\tau/\surd{3})]e^{i{\bf k}\cdot{\bf x}} +{\rm delta~function~terms},
\label{8.28}
\end{eqnarray}
where the delta function terms are solutions to $\tilde{\nabla}^4\alpha=0$. Finally, we recall that $\alpha$ and $\gamma$ are related through $\tilde{\nabla}^4(\tau\alpha+2\gamma)=0$,  with the coefficients  thus obeying
\begin{eqnarray}
 m_1({\bf k})+2a_1({\bf k})=0,\quad n_1({\bf k})+2b_1({\bf k})=0.
\label{8.29}
\end{eqnarray}

Having determined $\alpha$ and $\gamma$, we can now determine $\delta p-16\psi/\tau^4$ from the $\Delta_{00}=0$ equation, and obtain
\begin{eqnarray}
\nonumber\\
&&\delta p -\frac{16}{\tau^4}\psi=-\frac{8}{\tau^4}(\alpha-\dot{\gamma})-\frac{8}{3\tau^3}\tilde{\nabla}^2\gamma
\nonumber\\
&=&\sum_{\bf k}\left[\frac{8}{\tau^4}\left(\frac{2}{\tau}+\frac{\partial}{\partial \tau}\right)
+\frac{8k^2}{3\tau^3}\right]\left[a_1({\bf k})j_1(k\tau/\surd{3})+b_1({\bf k})y_1(k\tau/\surd{3})\right]e^{i{\bf k}\cdot{\bf x}} +{\rm delta~function~terms}
\nonumber\\
&=&\sum_{\bf k}a_1({\bf k})\left[\frac{8k}{\tau^4\surd{3}}j_0(k\tau/\surd{3})+\frac{8k^2}{3\tau^3}j_1(k\tau/\surd{3})\right]+\sum_{\bf k}b_1({\bf k})\left[\frac{8k}{\tau^4\surd{3}}y_0(k\tau/\surd{3})+\frac{8k^2}{3\tau^3}y_1(k\tau/\surd{3})\right]
\nonumber\\
&&+{\rm delta~function~terms}.
\label{8.30}
\end{eqnarray}
To determine $B_i-\dot{E}_i$ we apply $\tilde{\nabla}^j$ to $\Delta_{ij}=0$, to obtain
\begin{eqnarray}
\tilde{\nabla}^2\left[\frac{1}{\tau}(B_i-\dot{E}_i)+\frac{1}{2}(\dot{B}_i-\ddot{E}_i)\right]=\tilde{\nabla}_i\left[\frac{2}{\tau}(\dot{\alpha}-\ddot{\gamma})+\frac{2}{3\tau}\tilde{\nabla}^2\gamma\right],
\label{8.31}
\end{eqnarray}
from which it follows that
\begin{eqnarray}
\tilde{\nabla}^i\tilde{\nabla}^2\left[\frac{1}{\tau}(B_i-\dot{E}_i)+\frac{1}{2}(\dot{B}_i-\ddot{E}_i)\right]=\tilde{\nabla}^2\left[\frac{2}{\tau}(\dot{\alpha}-\ddot{\gamma})+\frac{2}{3\tau}\tilde{\nabla}^2\gamma\right]=0.
\label{8.32}
\end{eqnarray}
On now applying $\tilde{\nabla}^2$ to (\ref{8.31}) we  obtain
\begin{eqnarray}
\tilde{\nabla}^4\left[\frac{1}{\tau}(B_i-\dot{E}_i)+\frac{1}{2}(\dot{B}_i-\ddot{E}_i)\right]=0,
\label{8.33}
\end{eqnarray}
just as required for consistency with (\ref{8.18}).
Equation (\ref{8.33}) can be satisfied through a $1/\tau^2$ conformal time dependence, and while it could also be satisfied via a spatial dependence that satisfies $\tilde{\nabla}^4(B_i-\dot{E}_i)=0$, viz. the above delta function terms. With plane waves being complete we can thus set
\begin{eqnarray}
B_i-\dot{E}_i=\frac{1}{\tau^2}\sum_{\bf k}a_i({\bf k})e^{i{\bf k}\cdot{\bf x}}
+F(\tau)\times {\rm delta~function~terms},
\label{8.34}
\end{eqnarray}
where the $a_i({\bf k})$ are transverse vectors that obeys $k^ia_i({\bf k})=0$, and where $F(\tau)$ is an arbitrary function of $\tau$.

After solving (\ref{8.31}) and (\ref{8.32}), from the second equation in (\ref{8.16}) we can then determine $V_i$ as it obeys 
\begin{eqnarray}
\frac{8}{\tau^3}\tilde{\nabla}^2V_{i}=\frac{1}{2}\tilde{\nabla}^4(B_i-\dot{E}_i)=\frac{1}{2\tau^2}\sum_{\bf k}k^4a_i({\bf k})e^{i{\bf k}\cdot{\bf x}}.
 \label{8.35}
 \end{eqnarray}
From (\ref{8.17}) we can infer that 
\begin{eqnarray}
-\frac{8}{\tau^3}V_{i}+\frac{1}{2}\tilde{\nabla}^2(B_i-\dot{E}_i)=\tilde{\nabla}_iA,
\label{8.36}
\end{eqnarray}
where $A$ is a scalar function that obeys $\tilde{\nabla}^2A=0$, with $\tilde{\nabla}_iA$ being curl free. We recognize $A$ as an integration constant for the integration of (\ref{8.35}). From the first equation in (\ref{8.16}) we additionally obtain
\begin{eqnarray}
&&\tilde{\nabla}^2(-\frac{8}{\tau^3}V+\frac{4}{\tau}\psi)
 =\frac{2}{\tau}\tilde{\nabla}^2(\alpha-\dot{\gamma})
 \nonumber\\
 &=&\frac{2}{\tau\surd{3}}\sum_{\bf k}k^3[a_1({\bf k})j_0(k\tau/\surd{3})+b_1({\bf k})y_0(k\tau/\surd{3})]e^{i{\bf k}\cdot{\bf x}} + {\rm delta~function~terms.}
 \label{8.37}
 \end{eqnarray}

To determine an equation for $E_{ij}$ we note that the $\delta_{ij}$ term in $\Delta _{ij}$ can be written as 
$(\delta_{ij}/4\tau^2)[-8\tau(\dot{\alpha}-\ddot{\gamma})-4\tau^2\tilde{\nabla}^2\alpha-(32/3)\tau\tilde{\nabla}^2\gamma]$. Through use of (\ref{8.18}), (\ref{8.19}) and (\ref{8.20}), we can show that $\tilde{\nabla}^2$ applied to this term gives zero. Then given (\ref{8.33}) and (\ref{8.20}), from (\ref{8.10}) it follows that $E_{ij}$ obeys 
\begin{eqnarray}
 \tilde{\nabla}^4\left(-\ddot{E}_{ij}-\frac{2}{\tau}\dot{E}_{ij}+\tilde{\nabla}^2E_{ij}\right)=0.
\label{8.38}
\end{eqnarray}
Setting $\tilde{\nabla}^4E_{ij}=\epsilon_{ij}({\bf k})f_k(\tau)g_k({\bf x})$ for a momentum mode, the $\tau$ dependence  is given as $j_0(k\tau)$ and $y_0(k\tau)$, with the general solution being of the form
\begin{eqnarray}
E_{ij}&=&\sum_{\bf k}[a^0_{ij}({\bf k})j_0(k\tau)+b^0_{ij}({\bf k})y_0(k\tau)]e^{i{\bf k}\cdot{\bf x}}
+ {\rm delta~function~terms}.
\label{8.39}
\end{eqnarray}
Since according to (\ref{8.2}) the full tensor fluctuation is given not by $E_{ij}$ but by $\tau^2E_{ij}/2$, then with $\tau=2t^{1/2}$, we find that at large comoving time  $\tau^2 E_{ij}$ behaves as $t^{1/2}$. Thus to summarize,  we have constructed the exact and general solution to the SVT3 $k=0$ radiation era Robertson-Walker fluctuation equations for all of the dynamical degrees of freedom $\alpha$, $\beta$, $B_i-\dot{E}_i$, $E_{ij}$, $\delta p-16\psi/\tau^4$, $V-\tau^2\psi/2$ and $V_i$. 

For a decomposition theorem to hold the condition $\Delta_{\mu\nu}=0$ would need to decompose into 

\begin{eqnarray}
 &&\frac{3\tau^2}{4}\left(\delta p -\frac{16}{\tau^4}\psi\right)
 +\frac{6}{\tau^2}(\alpha-\dot{\gamma})+\frac{2}{\tau}\tilde{\nabla}^2\gamma=0,
 \label{8.40}
\end{eqnarray}
\begin{eqnarray}
&&-\frac{8}{\tau^3}V_{i}+\frac{1}{2}\tilde{\nabla}^2(B_i-\dot{E}_i)=0,
 \label{8.41}
\end{eqnarray}
\begin{eqnarray}
&&-\frac{8}{\tau^3}\tilde{\nabla}_iV+\frac{4}{\tau}\tilde{\nabla}_i\psi
 -\frac{2}{\tau}\tilde{\nabla}_i(\alpha-\dot{\gamma})=0,
 \label{8.42}
\end{eqnarray}
\begin{eqnarray}
&&\frac{\delta_{ij}}{4\tau^2}\left[\tau^4\delta p-16\psi-8\tau(\dot{\alpha}-\ddot{\gamma})+8(\alpha-\dot{\gamma})-4\tau \tilde{\nabla}^2(\tau\alpha+2\gamma)\right]
+\frac{1}{\tau}\tilde{\nabla}_i\tilde{\nabla}_j(\tau\alpha+2\gamma)=0,
 \label{8.43}
\end{eqnarray}
\begin{eqnarray}
&&\frac{1}{2\tau}\tilde{\nabla}_i[2(B_j-\dot{E}_j)+\tau(\dot{B}_j-\ddot{E}_j)]
+\frac{1}{2\tau}\tilde{\nabla}_j[2(B_i-\dot{E}_i)+\tau(\dot{B}_i-\ddot{E}_i)]=0,
 \label{8.44}
\end{eqnarray}
\begin{eqnarray}
&&-\ddot{E}_{ij}-\frac{2}{\tau}\dot{E}_{ij}+\tilde{\nabla}^2E_{ij}=0,
 \label{8.45}
\end{eqnarray}
\begin{eqnarray}
&& -\frac{24}{\tau^3}(\dot{\alpha}-\ddot{\gamma})-\frac{8}{\tau^2}\tilde{\nabla}^2\alpha-\frac{24}{\tau^3}\tilde{\nabla}^2\gamma=0.
\label{8.46}
\end{eqnarray}

To determine whether these conditions might hold, we note that in the $\alpha$, $\gamma$ sector the (\ref{8.40}) and (\ref{8.46})  equations are the same as in the general $\Delta_{\mu\nu}=0$ case (viz. (\ref{8.12}) and (\ref{8.13})), but (\ref{8.43}) is different. If we use (\ref{8.12}) to substitute for $\delta p$ in (\ref{8.43}) we obtain 
\begin{eqnarray}
 &&\frac{\delta_{ij}}{4\tau^2}\left[-8\tau(\dot{\alpha}-\ddot{\gamma})-
 4\tau^2\tilde{\nabla}^2\alpha-\frac{32\tau}{3} \tilde{\nabla}^2\gamma\right]+\frac{1}{\tau}\tilde{\nabla}_i\tilde{\nabla}_j(\tau\alpha+2\gamma)=0.
\label{8.47}
\end{eqnarray}
Given the differing behaviors of $\delta_{ij}$ and $\tilde{\nabla}_i\tilde{\nabla}_j$ it follows that the terms that they multiply  in (\ref{8.47}) must each vanish \cite{footnote10}, and thus we can set
\begin{eqnarray}
 &&-8\tau(\dot{\alpha}-\ddot{\gamma})-
 4\tau^2\tilde{\nabla}^2\alpha-\frac{32\tau}{3} \tilde{\nabla}^2\gamma=0,
 \label{8.48}
 \end{eqnarray}
 \begin{eqnarray}
 \tau\alpha+2\gamma=0.
\label{8.49}
\end{eqnarray}
Combining these equations then gives
\begin{eqnarray}
 &&3(\dot{\alpha}-\ddot{\gamma})+ \tilde{\nabla}^2\gamma=0.
\label{8.50}
\end{eqnarray}
We recognize (\ref{8.18}) as the $\tilde{\nabla}^2$ derivative of (\ref{8.50}) and recognize (\ref{8.20}) as the $\tilde{\nabla}^4$ derivative of (\ref{8.49}).

Similarly, in the $V$,$V_i$ sector we recognize the two equations that appear in (\ref{8.16}) as the $\nabla^2$ derivative of (\ref{8.41}) and the $\tilde{\nabla}^i$ derivative of (\ref{8.42}), with (\ref{8.17}) being the curl of (\ref{8.41}). In the $B_i-\dot{E}_i$ sector we recognize (\ref{8.33}) as the $\tilde{\nabla}^j\tilde{\nabla}^2$ derivative of (\ref{8.44}), and in the $E_{ij}$ sector we recognize (\ref{8.38}) as the $\tilde{\nabla}^4$ derivative of  (\ref{8.45}). Consequently we see that if spatially asymptotic boundary conditions are such that the only solutions to $\Delta_{\mu\nu}=0$ are also solutions to (\ref{8.40}) to (\ref{8.46}) (i.e. vanishing of all delta function terms and integration constants that would lead to non-vanishing asymptotics), then the decomposition theorem follows. Otherwise it does not. Finally, we should note that, as constructed, in the matter sector we have found solutions for the gauge-invariant quantities $\delta p-16\psi/\tau^4$, and $V-\tau^2\psi/2$. However since $\psi$ is not gauge invariant, by choosing a gauge in which $\psi=0$, we would then have solutions for $\delta p$ and $V$ alone.

\section{Solving the SVT3 Fluctuation Equations --  General Robertson-Walker Backgrounds}
\label{S9}
\subsection{Setting up the Equations}

Having seen how things work in a particular background Robertson-Walker case (radiation era with $k=0$), we now present a general analysis that can be applied to any background Robertson-Walker geometry with any background perfect fluid equation of state. To characterize a general Robertson-Walker background there are two straightforward options. One is to write the background metric in a conformal to flat form $ds^2=\Omega^2(\tau,x^i)(d\tau^2-\delta_{ij}dx^idx^j)$ with $\Omega(\tau,x^i)$ depending on both the conformal time $\tau=\int dt/a(t)$ and the spatial coordinates. The other is to write the background geometry as conformal to a static Robertson-Walker geometry: 
\begin{eqnarray}
ds^2=\Omega^2(\tau)[d\tau^2-\tilde{\gamma}_{ij}dx^idx^j]=\Omega^2(\tau)\left[ d\tau^2-\frac{dr^2}{1-kr^2}-r^2d\theta^2-r^2\sin\theta^2d\phi^2\right],
\label{9.1}
\end{eqnarray}
with $\Omega(\tau)$ depending only on $\tau$, and with $\tilde{\gamma}_{ij}dx^idx^j$ denoting the spatial sector of the metric. These two formulations of the background metric are coordinate equivalent as one can transform one into the other by a general coordinate transformation without any need to make a conformal transformation on the background metric (see e.g. Sec. \ref{S13} below). For our purposes in this section we shall take (\ref{9.1}) to be the background metric, and shall take the fluctuation metric to be of the form
\begin{eqnarray}
ds^2=\Omega^2(\tau)\left[2\phi d\tau^2 -2(\tilde{\nabla}_i B +B_i)d\tau dx^i - [-2\psi\tilde{\gamma}_{ij} +2\tilde{\nabla}_i\tilde{\nabla}_j E + \tilde{\nabla}_i E_j + \tilde{\nabla}_j E_i + 2E_{ij}]dx^i dx^j\right].
\label{9.2}
\end{eqnarray}
In (\ref{9.2})  $\tilde{\nabla}_i=\partial/\partial x^i$ and  $\tilde{\nabla}^i=\tilde{\gamma}^{ij}\tilde{\nabla}_j$  (with Latin indices) are defined with respect to the background three-space metric $\tilde{\gamma}_{ij}$. And with
\begin{eqnarray}
\tilde{\gamma}^{ij}\tilde{\nabla}_j V_i=\tilde{\gamma}^{ij}[\partial_j V_i-\tilde{\Gamma}^{k}_{ij}V_k]
\label{9.3}
\end{eqnarray}
for any 3-vector $V_i$ in a 3-space with 3-space connection $\tilde{\Gamma}^{k}_{ij}$, the elements of (\ref{9.2}) are required to obey
\begin{eqnarray}
\tilde{\gamma}^{ij}\tilde{\nabla}_j B_i = 0,\quad \tilde{\gamma}^{ij}\tilde{\nabla}_j E_i = 0, \quad E_{ij}=E_{ji},\quad \tilde{\gamma}^{jk}\tilde{\nabla}_kE_{ij} = 0, \quad\tilde{\gamma}^{ij}E_{ij} = 0.
\label{9.4}
\end{eqnarray}
With the  3-space sector of the background geometry being maximally 3-symmetric, it is described by a Riemann tensor of the form
\begin{eqnarray}
\tilde{R}_{ijk\ell}=k[\tilde{\gamma}_{jk}\tilde{\gamma}_{i\ell}-\tilde{\gamma}_{ik}\tilde{\gamma}_{j\ell}].
\label{9.5}
\end{eqnarray}
In \cite{Amarasinghe2018} (and as discussed in (\ref{9.43a}) to (\ref{9.47a}) below), it was shown that for the fluctuation metric given in (\ref{9.2}) with $\Omega(\tau)$ being arbitrary function of $\tau$, the gauge-invariant metric combinations are $\phi + \psi + \dot B - \ddot E$, $ - \dot\Omega^{-1}\Omega \psi + B - \dot E$, $B_i-\dot{E}_i$, and $E_{ij}$. As we shall see, the fluctuation equations will explicitly depend on these specific combinations.

We take the background $T_{\mu\nu}$ to be of the perfect fluid form
\begin{eqnarray}
T_{\mu\nu}=(\rho+p)U_{\mu}U_{\nu}+pg_{\mu\nu},
\label{9.6}
\end{eqnarray}
with fluctuation
\begin{eqnarray}
\delta T_{\mu\nu}=(\delta\rho+\delta p)U_{\mu}U_{\nu}+\delta pg_{\mu\nu}+(\rho+p)(\delta U_{\mu}U_{\nu}+U_{\mu}\delta U_{\nu})+ph_{\mu\nu}.
\label{9.7}
\end{eqnarray}
Here  $g^{\mu\nu}U_{\mu}U_{\nu}=-1$, $U^{0}=\Omega^{-1}(\tau)$, $U_0=-\Omega(\tau)$, $U^{i}=0$, $U_i=0$ for the background, while for the fluctuation we have 
\begin{eqnarray}
 \delta g^{00}U_{0}U_{0}+2g^{00}U_{0}\delta U_{0}=0,
\label{9.8}
\end{eqnarray}
i.e. 
\begin{eqnarray}
\delta U_{0}=-\frac{1}{2}(g^{00})^{-1}(-g^{00}g^{00}\delta g_{00})U_{0}=-\Omega(\tau)\phi.
\label{9.9}
\end{eqnarray}
Thus just as in Sec. \ref{S8} we see that $\delta U_0$ is not an independent degree of freedom. As in Sec. \ref{S8} we shall set $\delta U_i=V_i+\tilde{\nabla}_iV$, where now $\tilde{\gamma}^{ij}\tilde{\nabla}_j V_i=\tilde{\gamma}^{ij}[\partial_j V_i-\tilde{\Gamma}^{k}_{ij}V_k]=0$. As constructed, in general we have 11 fluctuation variables, the six from the metric together with the three $\delta U_i$, and $\delta\rho$ and $\delta p$. But we only have ten fluctuation equations. Thus to solve the theory when there is both a $\delta \rho$ and a $\delta p$ we will need some constraint between $\delta p$ and $\delta \rho$, a point we return to below.

For the background Einstein equations we have
\begin{eqnarray}
G_{00}&=& -3k - 3 \dot{\Omega}^2\Omega^{-2},\quad G_{0i} =0,
\quad G_{ij} = \tilde{\gamma}_{ij}\left[k - \dot\Omega^2\Omega^{-2}+ 2\ddot\Omega \Omega^{-1}\right],
 \nonumber\\
G_{00}+8\pi G T_{00} &=& -3k - 3 \dot\Omega^2\Omega^{-2} + \Omega^2 \rho=0,\quad
G_{ij}+8\pi G T_{ij}= \tilde{\gamma}_{ij}\left[k - \dot\Omega^2\Omega^{-2} + 2\ddot\Omega \Omega^{-1}  + \Omega^2 p\right]=0,
\nonumber\\
\rho &=& 3k\Omega^{-2}+3\dot\Omega^2 \Omega^{-4},\quad p = -k\Omega^{-2} + \dot\Omega^2\Omega^{-4} -2\ddot\Omega \Omega^{-3},\quad p = -\rho -\frac{1}{3} \frac{\Omega}{\dot\Omega}\dot\rho,
\label{9.10}
\end{eqnarray}
(after setting $8\pi G=1$), with the last relation following from $\nabla_{\nu}T^{\mu\nu}=0$, viz. conservation of the energy-momentum tensor in the full 4-space. To solve these equations we would need an equation of state that would relate $\rho$ and $p$. However we do not need to impose one just yet, as we shall generate the fluctuation equations as subject to (\ref{9.10}) but without needing to specify the form of $\Omega(\tau)$ or a relation  between $\rho$ and $p$.

For $\delta G_{\mu\nu}$ we have
\begin{eqnarray}
\delta G_{00}&=& -6 k \phi - 6 k \psi + 6 \dot{\psi} \dot{\Omega} \Omega^{-1} + 2 \dot{\Omega} \Omega^{-1} \tilde{\nabla}_{a}\tilde{\nabla}^{a}B - 2 \dot{\Omega} \Omega^{-1} \tilde{\nabla}_{a}\tilde{\nabla}^{a}\dot{E} - 2 \tilde{\nabla}_{a}\tilde{\nabla}^{a}\psi, 
 \nonumber\\ 
\delta G_{0i}&=& 3 k \tilde{\nabla}_{i}B -  \dot{\Omega}^2 \Omega^{-2} \tilde{\nabla}_{i}B + 2 \overset{..}{\Omega} \Omega^{-1} \tilde{\nabla}_{i}B - 2 k \tilde{\nabla}_{i}\dot{E} - 2 \tilde{\nabla}_{i}\dot{\psi} - 2 \dot{\Omega} \Omega^{-1} \tilde{\nabla}_{i}\phi +2 k B_{i} -  k \dot{E}_{i} \nonumber \\ 
&& -  B_{i} \dot{\Omega}^2 \Omega^{-2} + 2 B_{i} \overset{..}{\Omega} \Omega^{-1} + \tfrac{1}{2} \tilde{\nabla}_{a}\tilde{\nabla}^{a}B_{i} -  \tfrac{1}{2} \tilde{\nabla}_{a}\tilde{\nabla}^{a}\dot{E}_{i},
 \nonumber\\ 
\delta G_{ij}&=& -2 \overset{..}{\psi}\tilde{\gamma}_{ij} + 2 \dot{\Omega}^2\tilde{\gamma}_{ij} \phi \Omega^{-2} + 2 \dot{\Omega}^2\tilde{\gamma}_{ij} \psi \Omega^{-2} - 2 \dot{\phi} \dot{\Omega}\tilde{\gamma}_{ij} \Omega^{-1} - 4 \dot{\psi} \dot{\Omega}\tilde{\gamma}_{ij} \Omega^{-1} - 4 \overset{..}{\Omega}\tilde{\gamma}_{ij} \phi \Omega^{-1} \nonumber \\ 
&& - 4 \overset{..}{\Omega}\tilde{\gamma}_{ij} \psi \Omega^{-1} - 2 \dot{\Omega}\tilde{\gamma}_{ij} \Omega^{-1} \tilde{\nabla}_{a}\tilde{\nabla}^{a}B - \tilde{\gamma}_{ij} \tilde{\nabla}_{a}\tilde{\nabla}^{a}\dot{B} +\tilde{\gamma}_{ij} \tilde{\nabla}_{a}\tilde{\nabla}^{a}\overset{..}{E} + 2 \dot{\Omega}\tilde{\gamma}_{ij} \Omega^{-1} \tilde{\nabla}_{a}\tilde{\nabla}^{a}\dot{E} 
\nonumber \\ 
&& - \tilde{\gamma}_{ij} \tilde{\nabla}_{a}\tilde{\nabla}^{a}\phi +\tilde{\gamma}_{ij} \tilde{\nabla}_{a}\tilde{\nabla}^{a}\psi + 2 \dot{\Omega} \Omega^{-1} \tilde{\nabla}_{j}\tilde{\nabla}_{i}B + \tilde{\nabla}_{j}\tilde{\nabla}_{i}\dot{B} -  \tilde{\nabla}_{j}\tilde{\nabla}_{i}\overset{..}{E} - 2 \dot{\Omega} \Omega^{-1} \tilde{\nabla}_{j}\tilde{\nabla}_{i}\dot{E} \nonumber \\ 
&& + 2 k \tilde{\nabla}_{j}\tilde{\nabla}_{i}E - 2 \dot{\Omega}^2 \Omega^{-2} \tilde{\nabla}_{j}\tilde{\nabla}_{i}E + 4 \overset{..}{\Omega} \Omega^{-1} \tilde{\nabla}_{j}\tilde{\nabla}_{i}E + \tilde{\nabla}_{j}\tilde{\nabla}_{i}\phi -  \tilde{\nabla}_{j}\tilde{\nabla}_{i}\psi +\dot{\Omega} \Omega^{-1} \tilde{\nabla}_{i}B_{j} + \tfrac{1}{2} \tilde{\nabla}_{i}\dot{B}_{j}
\nonumber \\ 
&& -  \tfrac{1}{2} \tilde{\nabla}_{i}\overset{..}{E}_{j} -  \dot{\Omega} \Omega^{-1} \tilde{\nabla}_{i}\dot{E}_{j} + k \tilde{\nabla}_{i}E_{j} -  \dot{\Omega}^2 \Omega^{-2} \tilde{\nabla}_{i}E_{j} + 2 \overset{..}{\Omega} \Omega^{-1} \tilde{\nabla}_{i}E_{j} + \dot{\Omega} \Omega^{-1} \tilde{\nabla}_{j}B_{i} + \tfrac{1}{2} \tilde{\nabla}_{j}\dot{B}_{i} \nonumber \\ 
&& -  \tfrac{1}{2} \tilde{\nabla}_{j}\overset{..}{E}_{i} -  \dot{\Omega} \Omega^{-1} \tilde{\nabla}_{j}\dot{E}_{i} + k \tilde{\nabla}_{j}E_{i} -  \dot{\Omega}^2 \Omega^{-2} \tilde{\nabla}_{j}E_{i} + 2 \overset{..}{\Omega} \Omega^{-1} \tilde{\nabla}_{j}E_{i}- \overset{..}{E}_{ij} - 2 \dot{\Omega}^2 E_{ij} \Omega^{-2} \nonumber \\ 
&& - 2 \dot{E}_{ij} \dot{\Omega} \Omega^{-1} + 4 \overset{..}{\Omega} E_{ij} \Omega^{-1} + \tilde{\nabla}_{a}\tilde{\nabla}^{a}E_{ij},
 \nonumber\\
g^{\mu\nu}\delta G_{\mu\nu} &=& 6 \dot{\Omega}^2 \phi \Omega^{-4} + 6 \dot{\Omega}^2 \psi \Omega^{-4} - 6 \dot{\phi} \dot{\Omega} \Omega^{-3} - 18 \dot{\psi} \dot{\Omega} \Omega^{-3} - 12 \overset{..}{\Omega} \phi \Omega^{-3} - 12 \overset{..}{\Omega} \psi \Omega^{-3} - 6 \overset{..}{\psi} \Omega^{-2} + 6 k \phi \Omega^{-2} \nonumber \\ 
&& + 6 k \psi \Omega^{-2} - 6 \dot{\Omega} \Omega^{-3} \tilde{\nabla}_{a}\tilde{\nabla}^{a}B - 2 \Omega^{-2} \tilde{\nabla}_{a}\tilde{\nabla}^{a}\dot{B} + 2 \Omega^{-2} \tilde{\nabla}_{a}\tilde{\nabla}^{a}\overset{..}{E} + 6 \dot{\Omega} \Omega^{-3} \tilde{\nabla}_{a}\tilde{\nabla}^{a}\dot{E} \nonumber \\ 
&& - 2 \dot{\Omega}^2 \Omega^{-4} \tilde{\nabla}_{a}\tilde{\nabla}^{a}E + 4 \overset{..}{\Omega} \Omega^{-3} \tilde{\nabla}_{a}\tilde{\nabla}^{a}E + 2 k \Omega^{-2} \tilde{\nabla}_{a}\tilde{\nabla}^{a}E - 2 \Omega^{-2} \tilde{\nabla}_{a}\tilde{\nabla}^{a}\phi + 4 \Omega^{-2} \tilde{\nabla}_{a}\tilde{\nabla}^{a}\psi. 
\label{9.11}
\end{eqnarray}
We introduce
\begin{eqnarray}
\alpha  &=& \phi + \psi + \dot B - \ddot E,\quad \gamma = - \dot\Omega^{-1}\Omega \psi + B - \dot E,\quad \hat{V} = V-\Omega^2 \dot \Omega^{-1}\psi,
 \nonumber\\
\delta \hat{\rho}&=&\delta \rho - 12 \dot{\Omega}^2 \psi \Omega^{-4} + 6 \overset{..}{\Omega} \psi \Omega^{-3} - 6 k \psi \Omega^{-2}=\delta\rho +\frac{\Omega}{\dot{\Omega}}\dot{\rho}\psi=\delta \rho-3(\rho+p)\psi,
\nonumber\\
\delta \hat{p}&=&\delta p - 4 \dot{\Omega}^2 \psi \Omega^{-4} + 8 \overset{..}{\Omega} \psi \Omega^{-3} + 2 k \psi \Omega^{-2} - 2 \overset{...}{\Omega} \dot{\Omega}^{-1} \psi \Omega^{-2}=\delta p +\frac{\Omega}{\dot{\Omega}}\dot{p}\psi,
\label{9.12}
\end{eqnarray}
(in (\ref{9.12}) we used (\ref{9.10})), where, as we show below, the functions $\delta \hat{\rho}$, $\delta \hat{p}$ and $\hat{V} $ are gauge invariant. (The $\gamma$ introduced in (\ref{8.7}) is a special case of the $\gamma$ introduced in (\ref{9.12}).) Given (\ref{9.12}) we can express the components of $\Delta_{\mu\nu}=\delta G_{\mu\nu}+8\pi G\delta T_{\mu\nu}$ quite compactly. Specifically,  on using (\ref{9.10}) for the background but without imposing any relation between the background $\rho$ and $p$, we obtain evolution equations of the form 
\begin{eqnarray}
\Delta_{00}&=& 6 \dot{\Omega}^2 \Omega^{-2}(\alpha-\dot\gamma) + \delta \hat{\rho} \Omega^2 + 2 \dot{\Omega} \Omega^{-1} \tilde{\nabla}_{a}\tilde{\nabla}^{a}\gamma=0, 
\label{9.13}
\end{eqnarray}
\begin{eqnarray}
\Delta_{0i}&=& -2 \dot{\Omega} \Omega^{-1} \tilde{\nabla}_{i}(\alpha - \dot\gamma) + 2 k \tilde{\nabla}_{i}\gamma 
+(-4 \dot{\Omega}^2 \Omega^{-3}  + 2 \overset{..}{\Omega} \Omega^{-2}  - 2 k \Omega^{-1}) \tilde{\nabla}_{i}\hat{V}
\nonumber\\
&& +k(B_i-\dot E_i)+ \tfrac{1}{2} \tilde{\nabla}_{a}\tilde{\nabla}^{a}(B_{i} - \dot{E}_{i})
+ (-4 \dot{\Omega}^2 \Omega^{-3} + 2 \overset{..}{\Omega} \Omega^{-2} - 2 k \Omega^{-1})V_{i}=0,
\label{9.14}
\end{eqnarray}
\begin{eqnarray}
\Delta_{ij}&=& \tilde{\gamma}_{ij}\big[ 2 \dot{\Omega}^2 \Omega^{-2}(\alpha-\dot\gamma)
-2  \dot{\Omega} \Omega^{-1}(\dot\alpha -\ddot\gamma)-4\ddot\Omega\Omega^{-1}(\alpha-\dot\gamma)+ \Omega^2 \delta \hat{p}-\tilde\nabla_a\tilde\nabla^a( \alpha + 2\dot\Omega \Omega^{-1}\gamma) \big] 
\nonumber\\
&&+\tilde\nabla_i\tilde\nabla_j( \alpha + 2\dot\Omega \Omega^{-1}\gamma)
+\dot{\Omega} \Omega^{-1} \tilde{\nabla}_{i}(B_{j}-\dot E_j)+\tfrac{1}{2} \tilde{\nabla}_{i}(\dot{B}_{j}-\ddot{E}_j)
+\dot{\Omega} \Omega^{-1} \tilde{\nabla}_{j}(B_{i}-\dot E_i)+\tfrac{1}{2} \tilde{\nabla}_{j}(\dot{B}_{i}-\ddot{E}_i)
\nonumber\\
&&- \overset{..}{E}_{ij} - 2 k E_{ij} - 2 \dot{E}_{ij} \dot{\Omega} \Omega^{-1} + \tilde{\nabla}_{a}\tilde{\nabla}^{a}E_{ij}=0,
\label{9.15}
\end{eqnarray}
\begin{eqnarray}
\tilde{\gamma}^{ij}\Delta_{ij} &=&  6 \dot{\Omega}^2 \Omega^{-2}(\alpha-\dot\gamma)
-6  \dot{\Omega} \Omega^{-1}(\dot\alpha -\ddot\gamma)-12\ddot\Omega\Omega^{-1}(\alpha-\dot\gamma)+ 3\Omega^2 \delta \hat{p}-2\tilde\nabla_a\tilde\nabla^a( \alpha + 2\dot\Omega \Omega^{-1}\gamma)=0,
\label{9.16}
\end{eqnarray}
\begin{eqnarray}
g^{\mu\nu}\Delta_{\mu\nu}&=& 3 \delta \hat{p} -  \delta \hat{\rho}
-12 \overset{..}{\Omega}  \Omega^{-3}(\alpha - \dot\gamma) -6 \dot{\Omega} \Omega^{-3}(\dot{\alpha} -\ddot\gamma)
-2 \Omega^{-2} \tilde{\nabla}_{a}\tilde{\nabla}^{a}(\alpha +3\dot\Omega\Omega^{-1}\gamma)=0.
\label{9.17}
\end{eqnarray}

Starting from the general identities
\begin{eqnarray}
\nabla_{k}\nabla_{n}T_{\ell m}-\nabla_{n}\nabla_{k}T_{\ell m}=T^{s}_{\phantom{s}m}R_{\ell s n k}+T_{\ell}^{\phantom{\ell}s}R_{ms n k},\quad
\nabla_{k}\nabla_{n}A_{m}-\nabla_{n}\nabla_{k}A_{m}=A^{s}R_{ms n k}
\label{9.18}
\end{eqnarray}
that hold for any rank two tensor or vector in any geometry, for the 3-space Robertson-Walker geometry where $\tilde{R}_{msnk}=k(\tilde{\gamma}_{sn}\tilde{\gamma}_{mk}-\tilde{\gamma}_{mn}\tilde{\gamma}_{sk})$ 
we obtain
\begin{eqnarray}
&&\tilde\nabla_i\tilde\nabla_a\tilde\nabla^aA_j-\tilde\nabla_a\tilde\nabla^a\tilde\nabla_iA_j 
= 2k\tilde{\gamma}_{ij}\tilde{\nabla}_aA^a-2k(\tilde\nabla_i A_j + \tilde\nabla_j A_i),
\nonumber\\
&&\tilde\nabla^j\tilde\nabla_a\tilde\nabla^aA_j=
(\tilde\nabla_a\tilde\nabla^a+2k)\tilde\nabla^j A_j,\quad \tilde{\nabla}^j\tilde{\nabla}_iA_j=\tilde{\nabla}_i\tilde{\nabla}^jA_j+2kA_i
\label{9.19}
\end{eqnarray}
for any 3-vector $A_i$ in a maximally symmetric 3-geometry with 3-curvature $k$. Similarly, noting that for any scalar $S$ in any geometry we have
\begin{eqnarray}
&&\nabla_a\nabla_b\nabla_iS=\nabla_a\nabla_i\nabla_bS=\nabla_i\nabla_a\nabla_bS+\nabla^sSR_{bsia},
\nonumber\\
&&\nabla_{\ell}\nabla_k\nabla_n\nabla_{m}S=\nabla_n\nabla_{m}\nabla_{\ell}\nabla_kS
+\nabla_{n}[\nabla^sSR_{ksm\ell}]
+\nabla^s\nabla_kSR_{msn\ell}
+\nabla_m\nabla^sSR_{ksn\ell}
+\nabla_{\ell}[\nabla^sSR_{msnk}],
\label{9.20}
\end{eqnarray}
in a Robertson-Walker 3-geometry background we obtain 
\begin{eqnarray}
\tilde{\nabla}_a\tilde{\nabla}^a\tilde{\nabla}_iS&=&\tilde{\nabla}_i\tilde{\nabla}_a\tilde{\nabla}^aS+2k\tilde{\nabla}_iS,\quad \tilde{\nabla}_a\tilde{\nabla}^a\tilde{\nabla}_i\tilde{\nabla}_jS=\tilde{\nabla}_i\tilde{\nabla}_j\tilde{\nabla}_a\tilde{\nabla}^aS+6k(\tilde{\nabla}_i\tilde{\nabla}_j-\tfrac{1}{3}\tilde{\gamma}_{ij}\tilde{\nabla}_a\tilde{\nabla}^a)S,
\nonumber\\
\tilde{\nabla}_a\tilde{\nabla}^a\tilde{\nabla}_i\tilde{\nabla}_{j}S&=&\tilde{\nabla}_i\tilde{\nabla}_{j}\tilde{\nabla}_a\tilde{\nabla}^aS
+6k\tilde{\nabla}_i\tilde{\nabla}_{j}S-2k\tilde{\gamma}_{ij}\tilde{\nabla}_a\tilde{\nabla}^aS.
\label{9.21}
\end{eqnarray}
Thus we find that 
\begin{eqnarray}
\tilde\nabla^i \Delta_{0i} &=& 
 \tilde\nabla_a\tilde\nabla^a\big[ -2 \dot{\Omega} \Omega^{-1} (\alpha - \dot\gamma) + 2 k \gamma 
+(-4 \dot{\Omega}^2 \Omega^{-3}  + 2 \overset{..}{\Omega} \Omega^{-2}  - 2 k \Omega^{-1}) \hat{V}\big]=0,
\label{9.22}
\end{eqnarray}
and thus
\begin{eqnarray}
(\tilde{\nabla}_k\tilde\nabla^k -2k)\Delta_{0i} &=& (\tilde{\nabla}_k\tilde\nabla^k-2k)\left[k(B_i-\dot E_i)+ \tfrac{1}{2} \tilde{\nabla}_{a}\tilde{\nabla}^{a}(B_{i} - \dot{E}_{i})
+ (-4 \dot{\Omega}^2 \Omega^{-3} + 2 \overset{..}{\Omega} \Omega^{-2} - 2 k \Omega^{-1})V_{i}\right]
=0.~~~~~~
\label{9.23}
\end{eqnarray}
Also we obtain
\begin{eqnarray}
\epsilon^{ij\ell}\tilde{\nabla}_j\Delta_{0i}=\epsilon^{ij\ell}\tilde{\nabla}_j\left[k(B_i-\dot E_i)+ \tfrac{1}{2} \tilde{\nabla}_{a}\tilde{\nabla}^{a}(B_{i} - \dot{E}_{i})
+ (-4 \dot{\Omega}^2 \Omega^{-3} + 2 \overset{..}{\Omega} \Omega^{-2} - 2 k \Omega^{-1})V_{i}\right]=0.
\label{9.24}
\end{eqnarray}
Now we had noted in Sec. \ref{S6} that in a de Sitter space if a tensor $A^{P}_{\phantom{P}M}$ is transverse and traceless then so is $\nabla_L\nabla^LA^{P}_{\phantom{P}M}$ Since this holds in any maximally symmetric space the quantity  $\tilde{\nabla}_a\tilde{\nabla}^aE_{ij}$ is transverse and traceless too. Thus given (\ref{9.19}) we  obtain
\begin{eqnarray}
\tilde\nabla^j\Delta_{ij}&=& \tilde{\nabla}_{i}[ 2 \dot{\Omega}^2 \Omega^{-2}(\alpha-\dot\gamma)
-2  \dot{\Omega} \Omega^{-1}(\dot\alpha -\ddot\gamma)-4\ddot\Omega\Omega^{-1}(\alpha-\dot\gamma)+ \Omega^2 \delta \hat{p}+ 2 k(\alpha + 2 \dot{\Omega}  \Omega^{-1} \gamma)]
\nonumber \\ 
&&+[ \tilde{\nabla}_{a}\tilde{\nabla}^{a}+2k][\tfrac{1}{2}(\dot{B}_i-\ddot{E}_i)+\dot{\Omega}\Omega^{-1}(B_i-\dot{E}_i)]=0,
\label{9.25}
\end{eqnarray}
\begin{eqnarray}
\tilde\nabla^i\tilde\nabla^j\Delta_{ij}&=& \tilde{\nabla}_{a}\tilde{\nabla}^{a}[2 \dot{\Omega}^2 \Omega^{-2}(\alpha-\dot\gamma)
-2  \dot{\Omega} \Omega^{-1}(\dot\alpha -\ddot\gamma)-4\ddot\Omega\Omega^{-1}(\alpha-\dot\gamma)+ \Omega^2 \delta \hat{p}
\nonumber \\ 
&& + 2 k(\alpha + 2 \dot{\Omega}  \Omega^{-1} \gamma)]=0.
\label{9.26}
\end{eqnarray}
Thus we obtain
\begin{eqnarray}
3\tilde\nabla^i\tilde\nabla^j\Delta_{ij}-\tilde\nabla_a\tilde\nabla^a(\tilde{\gamma}^{ij}\Delta_{ij})
=2\tilde{\nabla}^2[\tilde{\nabla}^2+3k](\alpha+2\dot{\Omega}\Omega^{-1}\gamma)=0,
\label{9.27}
\end{eqnarray}
\begin{eqnarray}
\tilde\nabla^i\tilde\nabla^j\Delta_{ij}+k\tilde{\gamma}^{ij}\Delta_{ij}
=[\tilde{\nabla}^2+3k][2 \dot{\Omega}^2 \Omega^{-2}(\alpha-\dot\gamma)
-2  \dot{\Omega} \Omega^{-1}(\dot\alpha -\ddot\gamma)-4\ddot\Omega\Omega^{-1}(\alpha-\dot\gamma)+ \Omega^2 \delta \hat{p}]=0.
\label{9.28}
\end{eqnarray}
We now define $A=2 \dot{\Omega}^2 \Omega^{-2}(\alpha-\dot\gamma)-2  \dot{\Omega} \Omega^{-1}(\dot\alpha -\ddot\gamma)-4\ddot\Omega\Omega^{-1}(\alpha-\dot\gamma)+ \Omega^2 \delta \hat{p}$ and $C=\alpha+2\dot{\Omega}\Omega^{-1}\gamma$. And using (\ref{9.21}) obtain
\begin{eqnarray}
(\tilde{\nabla}_a\tilde{\nabla}^a+k)\tilde{\nabla}_i(A+2kC)=\tilde{\nabla}_i(\tilde{\nabla}_a\tilde{\nabla}^a+3k)(A+2kC),
\label{9.29}
\end{eqnarray}
and thus with  (\ref{9.27}) and (\ref{9.28}) obtain
\begin{eqnarray}
(\tilde{\nabla}_a\tilde{\nabla}^a-2k)(\tilde{\nabla}_a\tilde{\nabla}^a+k)\tilde{\nabla}_i(A+2kC)=
\tilde{\nabla}_i\tilde{\nabla}_a\tilde{\nabla}^a(\tilde{\nabla}_b\tilde{\nabla}^b+3k)(A+2kC)=0.
\label{9.30}
\end{eqnarray}
Consequently, on comparing with (\ref{9.25})  we obtain
\begin{eqnarray}
(\tilde{\nabla}_a\tilde{\nabla}^a-2k)(\tilde{\nabla}_b\tilde{\nabla}^b+k)\tilde\nabla^j\Delta_{ij}=
(\tilde{\nabla}_a\tilde{\nabla}^a-2k)(\tilde{\nabla}_b\tilde{\nabla}^b+k)[\tilde{\nabla}_{c}\tilde{\nabla}^{c}+2k][\tfrac{1}{2}(\dot{B}_i-\ddot{E}_i)+\dot{\Omega}\Omega^{-1}(B_i-\dot{E}_i)]=0,
\label{9.31}
\end{eqnarray}
to give a relation that only involves $B_i-\dot{E}_{i}$.

To obtain a relation that involves $E_{ij}$ we proceed as follows. We note that sector of $\Delta_{ij}$ that contains the above $A$ and $C$ can be written as 
\begin{eqnarray}
D_{ij}=\tilde{\gamma}_{ij}(A-\tilde{\nabla}_a\tilde{\nabla}^aC)+\tilde{\nabla}_i\tilde{\nabla}_jC.
\label{9.32}
\end{eqnarray}
We thus introduce 
\begin{eqnarray}
A_{ij}&=&D_{ij}-\frac{1}{3}\tilde{\gamma}_{ij}\tilde{\gamma}^{ab}D_{ab}=(\tilde{\nabla}_i\tilde{\nabla}_j-\tfrac{1}{3} \tilde{\gamma}_{ij}\tilde{\nabla}_a\tilde{\nabla}^a)C,
\nonumber\\
B_{ij}&=&\Delta_{ij}-\frac{1}{3}\tilde{\gamma}_{ij}\tilde{\gamma}^{ab}\Delta_{ab}=(\tilde{\nabla}_i\tilde{\nabla}_j-\tfrac{1}{3} \tilde{\gamma}_{ij}\tilde{\nabla}_a\tilde{\nabla}^a)C
\nonumber\\
&+&\dot{\Omega} \Omega^{-1} \tilde{\nabla}_{i}(B_{j}-\dot E_j)+\tfrac{1}{2} \tilde{\nabla}_{i}(\dot{B}_{j}-\ddot{E}_j)
+\dot{\Omega} \Omega^{-1} \tilde{\nabla}_{j}(B_{i}-\dot E_i)+\tfrac{1}{2} \tilde{\nabla}_{j}(\dot{B}_{i}-\ddot{E}_i)
\nonumber\\
&-& \overset{..}{E}_{ij} - 2 k E_{ij} - 2 \dot{E}_{ij} \dot{\Omega} \Omega^{-1} + \tilde{\nabla}_{a}\tilde{\nabla}^{a}E_{ij}=0,
\label{9.33}
\end{eqnarray}
with (\ref{9.33}) defining $A_{ij}$ and $B_{ij}$, and with $A$ dropping out. Using (\ref{9.19}) and the third relation in (\ref{9.21}) we obtain
\begin{eqnarray}
(\tilde{\nabla}_b\tilde{\nabla}^b-3k)A_{ij}=
(\tilde{\nabla}_i\tilde{\nabla}_j-\tfrac{1}{3} \tilde{\gamma}_{ij}\tilde{\nabla}_a\tilde{\nabla}^a)(\tilde{\nabla}_b\tilde{\nabla}^b+3k)C,
\label{9.34}
\end{eqnarray}
and via (\ref{9.21}) and (\ref{9.22}) thus obtain 
\begin{eqnarray}
(\tilde{\nabla}_a\tilde{\nabla}^a-6k)(\tilde{\nabla}_b\tilde{\nabla}^b-3k)A_{ij}=
(\tilde{\nabla}_i\tilde{\nabla}_j-\tfrac{1}{3} \tilde{\gamma}_{ij}\tilde{\nabla}_a\tilde{\nabla}^a)\tilde{\nabla}_b\tilde{\nabla}^b(\tilde{\nabla}_c\tilde{\nabla}^c+3k)C=0.
\label{9.35}
\end{eqnarray}
 Comparing with the structure of $\Delta_{ij}$ and $\tilde{\gamma}^{ij}\Delta_{ij}$, we thus obtain
 \begin{eqnarray}
 &&(\tilde{\nabla}_a\tilde{\nabla}^a-6k)(\tilde{\nabla}_b\tilde{\nabla}^b-3k)[B_{ij}-A_{ij}]
 =(\tilde{\nabla}_a\tilde{\nabla}^a-6k)(\tilde{\nabla}_b\tilde{\nabla}^b-3k)
 \nonumber\\
 &&\times
 \big{[}\dot{\Omega} \Omega^{-1} \tilde{\nabla}_{i}(B_{j}-\dot E_j)+\tfrac{1}{2} \tilde{\nabla}_{i}(\dot{B}_{j}-\ddot{E}_j)
+\dot{\Omega} \Omega^{-1} \tilde{\nabla}_{j}(B_{i}-\dot E_i)+\tfrac{1}{2} \tilde{\nabla}_{j}(\dot{B}_{i}-\ddot{E}_i)
\nonumber\\
&& -\overset{..}{E}_{ij} - 2 k E_{ij} - 2  \dot{\Omega} \Omega^{-1}\dot{E}_{ij} + \tilde{\nabla}_{a}\tilde{\nabla}^{a}E_{ij}\big{]}=0.
\label{9.36}
\end{eqnarray}
We now note that for any vector $A_i$ that obeys $\tilde{\nabla}^iA_i=0$, through repeated use of the first relation in (\ref{9.19}) we obtain 
\begin{eqnarray}
&&(\tilde{\nabla}_b\tilde{\nabla}^b-3k)(\tilde{\nabla}_iA_j+\tilde{\nabla}_jA_i)=
\tilde{\nabla}_i(\tilde{\nabla}_b\tilde{\nabla}^b+k)A_j+
\tilde{\nabla}_j(\tilde{\nabla}_b\tilde{\nabla}^b+k)A_i,
\nonumber\\
&&(\tilde{\nabla}_a\tilde{\nabla}^a-6k)(\tilde{\nabla}_b\tilde{\nabla}^b-3k)(\tilde{\nabla}_iA_j+\tilde{\nabla}_jA_i)
=\tilde{\nabla}_i(\tilde{\nabla}_a\tilde{\nabla}^a-2k)(\tilde{\nabla}_b\tilde{\nabla}^b+k)A_j+
\tilde{\nabla}_j(\tilde{\nabla}_a\tilde{\nabla}^a-2k)(\tilde{\nabla}_b\tilde{\nabla}^b+k)A_i.
\nonumber\\
\label{9.37}
\end{eqnarray}
On using the first relation in (\ref{9.19}) again,  it follows that 
\begin{eqnarray}
&&(\tilde{\nabla}_c\tilde{\nabla}^c-2k)(\tilde{\nabla}_a\tilde{\nabla}^a-6k)(\tilde{\nabla}_b\tilde{\nabla}^b-3k)(\tilde{\nabla}_iA_j+\tilde{\nabla}_jA_i)
\nonumber\\
&=&\tilde{\nabla}_i(\tilde{\nabla}_c\tilde{\nabla}^c+2k)(\tilde{\nabla}_a\tilde{\nabla}^a-2k)(\tilde{\nabla}_b\tilde{\nabla}^b+k)A_j+
\tilde{\nabla}_j(\tilde{\nabla}_c\tilde{\nabla}^c+2k)(\tilde{\nabla}_a\tilde{\nabla}^a-2k)(\tilde{\nabla}_b\tilde{\nabla}^b+k)A_i.
\label{9.38}
\end{eqnarray}
On setting $A_i=\tfrac{1}{2}(\dot{B}_i-\ddot{E}_i)+\dot{\Omega}\Omega^{-1}(B_i-\dot{E}_i)$ (so that $A_i$ is such that $\tilde{\nabla}^iA_i=0$), and recalling (\ref{9.31}) we obtain
\begin{eqnarray}
&&(\tilde{\nabla}_c\tilde{\nabla}^c-2k)(\tilde{\nabla}_a\tilde{\nabla}^a-6k)(\tilde{\nabla}_b\tilde{\nabla}^b-3k)
\nonumber\\
&&\times\bigg{[}
\tilde{\nabla}_i[\tfrac{1}{2}(\dot{B}_j-\ddot{E}_j)+\dot{\Omega}\Omega^{-1}(B_j-\dot{E}_j)]+\tilde{\nabla}_j[\tfrac{1}{2}(\dot{B}_i-\ddot{E}_i)+\dot{\Omega}\Omega^{-1}(B_i-\dot{E}_i)]\bigg{]}=0.
\label{9.39}
\end{eqnarray}
Thus finally from (\ref{9.36}) we obtain
\begin{eqnarray}
&&(\tilde{\nabla}_c\tilde{\nabla}^c-2k)(\tilde{\nabla}_a\tilde{\nabla}^a-6k)(\tilde{\nabla}_b\tilde{\nabla}^b-3k)
\big{[}
- \overset{..}{E}_{ij} - 2 k E_{ij} - 2  \dot{\Omega} \Omega^{-1}\dot{E}_{ij} + \tilde{\nabla}_{a}\tilde{\nabla}^{a}E_{ij}\big{]}=0.
\label{9.40}
\end{eqnarray}
Thus with (\ref{9.27}), (\ref{9.28}), (\ref{9.31}), (\ref{9.40}) together with (\ref{9.13}), (\ref{9.22}) and (\ref{9.23}) we have succeeded in decomposing the fluctuation equations for the components, with the various components obeying derivative equations that are higher than second order.

With (\ref{9.27}) only involving $\alpha +2\dot{\Omega}\Omega^{-1}\gamma$, with (\ref{9.31}) only involving 
$B_i-\dot{E}_i$, and with (\ref{9.40}) only involving $E_{ij}$, and with all components of $\Delta_{\mu\nu}$ being gauge invariant, we recognize $C=\alpha +2\dot{\Omega}\Omega^{-1}\gamma$, $B_i-\dot{E}_i$ and $E_{ij}$ as being gauge invariant. With $B_i-\dot{E}_i$ being gauge invariant, from (\ref{9.23}) we recognize $V_i$ as being gauge invariant too. While we have identified some gauge-invariant quantities we note that by manipulating $\Delta_{\mu\nu}$ so as to obtain derivative expressions in which each of these quantities appears on its own, we cannot establish the gauge invariance of all 11 of the fluctuation variables this way since $\Delta_{\mu\nu}$ only has 10 components. However, just as with fluctuations around flat spacetime, in analog to (\ref{2.6}) below we shall obtain derivative relations between the SVT3 fluctuations and the $h_{\mu\nu}$ fluctuations by manipulating  (\ref{9.2}). As we show below, this will enable us  to establish the gauge invariance of the remaining fluctuation quantities.

\subsection{What is Needed to get a Decomposition Theorem}

To get a decomposition theorem for $\Delta_{\mu\nu}=0$ we would require 
\begin{eqnarray}
6 \dot{\Omega}^2 \Omega^{-2}(\alpha-\dot\gamma) + \delta \hat{\rho}{} \Omega^2 + 2 \dot{\Omega} \Omega^{-1} \tilde{\nabla}_{a}\tilde{\nabla}^{a}\gamma=0,&& 
\nonumber\\
 -2 \dot{\Omega} \Omega^{-1} \tilde{\nabla}_{i}(\alpha - \dot\gamma) + 2 k \tilde{\nabla}_{i}\gamma 
+(-4 \dot{\Omega}^2 \Omega^{-3}  + 2 \overset{..}{\Omega} \Omega^{-2}  - 2 k \Omega^{-1}) \tilde{\nabla}_{i}\hat{V}=0,&&
\nonumber\\
 +k(B_i-\dot E_i)+ \tfrac{1}{2} \tilde{\nabla}_{a}\tilde{\nabla}^{a}(B_{i} - \dot{E}_{i})
+ (-4 \dot{\Omega}^2 \Omega^{-3} + 2 \overset{..}{\Omega} \Omega^{-2} - 2 k \Omega^{-1})V_{i}=0,&&
\nonumber\\
 \tilde{\gamma}_{ij}\big[ 2 \dot{\Omega}^2 \Omega^{-2}(\alpha-\dot\gamma)
-2  \dot{\Omega} \Omega^{-1}(\dot\alpha -\ddot\gamma)-4\ddot\Omega\Omega^{-1}(\alpha-\dot\gamma)+ \Omega^2 \delta \hat{p}-\tilde\nabla_a\tilde\nabla^a( \alpha + 2\dot\Omega \Omega^{-1}\gamma) \big] &&
\nonumber\\
+\tilde\nabla_i\tilde\nabla_j( \alpha + 2\dot\Omega \Omega^{-1}\gamma)=0,&&
\nonumber\\
\dot{\Omega} \Omega^{-1} \tilde{\nabla}_{i}(B_{j}-\dot E_j)+\tfrac{1}{2} \tilde{\nabla}_{i}(\dot{B}_{j}-\ddot{E}_j)
+\dot{\Omega} \Omega^{-1} \tilde{\nabla}_{j}(B_{i}-\dot E_i)+\tfrac{1}{2} \tilde{\nabla}_{j}(\dot{B}_{i}-\ddot{E}_i)=0,&&
\nonumber\\
- \overset{..}{E}_{ij} - 2 k E_{ij} - 2 \dot{E}_{ij} \dot{\Omega} \Omega^{-1} + \tilde{\nabla}_{a}\tilde{\nabla}^{a}E_{ij}=0,&&
\nonumber\\
 6 \dot{\Omega}^2 \Omega^{-2}(\alpha-\dot\gamma)
-6  \dot{\Omega} \Omega^{-1}(\dot\alpha -\ddot\gamma)-12\ddot\Omega\Omega^{-1}(\alpha-\dot\gamma)+ 3\Omega^2 \delta \hat{p}-2\tilde\nabla_a\tilde\nabla^a( \alpha + 2\dot\Omega \Omega^{-1}\gamma)=0,&&
\nonumber\\
3 \delta \hat{p}{} -  \delta \hat{\rho}
-12 \overset{..}{\Omega}  \Omega^{-3}(\alpha - \dot\gamma) -6 \dot{\Omega} \Omega^{-3}(\dot{\alpha} -\ddot\gamma)
-2 \Omega^{-2} \tilde{\nabla}_{a}\tilde{\nabla}^{a}(\alpha +3\dot\Omega\Omega^{-1}\gamma)=0.&&
\label{9.41}
\end{eqnarray}
With $\tilde{\gamma}_{ij}$ and $\tilde{\nabla}_{i}\tilde{\nabla}_j$ not being equal to each other, we would immediately obtain 
\begin{eqnarray}
2 \dot{\Omega}^2 \Omega^{-2}(\alpha-\dot\gamma)
-2  \dot{\Omega} \Omega^{-1}(\dot\alpha -\ddot\gamma)-4\ddot\Omega\Omega^{-1}(\alpha-\dot\gamma)+ \Omega^2 \delta \hat{p}  =0,\quad  \alpha + 2\dot\Omega \Omega^{-1}\gamma=0.
\label{9.42}
\end{eqnarray}
We recognize the equations for the components of the fluctuations as being derivatives of the relations that are required of the decomposition theorem. We thus need to see if we can find boundary conditions that would force the solutions to the higher-derivative fluctuation equations to obey (\ref{9.41}) and (\ref{9.42}).

\subsection{Establishing Gauge Invariance}

Starting with (\ref{9.2}), setting $h_{\mu\nu}=\Omega^2(\tau)f_{\mu\nu}$,  $f=\tilde{\gamma}^{ij}f_{ij}=-6\psi+2\tilde{\nabla}_i\tilde{\nabla}^iE$ and taking appropriate derivatives, then following quite a bit of algebra we obtain
\begin{align}
&(3k+\tilde{\nabla}^b\tilde{\nabla}_b)\tilde{\nabla}^a\tilde{\nabla}_a\alpha=-\frac{1}{2}(3k+\tilde{\nabla}^b\tilde{\nabla}_b)\tilde{\nabla}^i\tilde{\nabla}_if_{00}
\nonumber\\
&+\frac{1}{4}\tilde{\nabla}^a\tilde{\nabla}_a\left(-2kf-\tilde{\nabla}^b\tilde{\nabla}_bf+\tilde{\nabla}^m\tilde{\nabla}^nf_{mn}\right)
+\partial_0(3k+\tilde{\nabla}^b\tilde{\nabla}_b)\tilde{\nabla}^if_{0i}-\frac{1}{4}\partial^2_0\left(3\tilde{\nabla}^m\tilde{\nabla}^nf_{mn}-\tilde{\nabla}^a\tilde{\nabla}_af\right),
\label{9.43a}
\end{align}
\begin{align}
&(3k+\tilde{\nabla}^b\tilde{\nabla}_b)\tilde{\nabla}^a\tilde{\nabla}_a\gamma
\nonumber\\
&=-\frac{1}{4}\Omega\dot{\Omega}^{-1}\tilde{\nabla}^a\tilde{\nabla}_a\left(-2kf-\tilde{\nabla}^b\tilde{\nabla}_bf+\tilde{\nabla}^m\tilde{\nabla}^nf_{mn}\right)
+\left[(3k+\tilde{\nabla}^b\tilde{\nabla}_b)\tilde{\nabla}^if_{0i}-\frac{1}{4}\partial_0\left(3\tilde{\nabla}^m\tilde{\nabla}^nf_{mn}-\tilde{\nabla}^a\tilde{\nabla}_af\right)\right],
\label{9.44a}
\end{align}
\begin{align}
&(\tilde{\nabla}^a\tilde{\nabla}_a-2k)(\tilde{\nabla}^i\tilde{\nabla}_i +2k)(B_j-\dot{E_j})=(\tilde{\nabla}^i\tilde{\nabla}_i +2k)(\tilde{\nabla}^a\tilde{\nabla}_af_{0j}-2kf_{0j}
-\tilde{\nabla}_j\tilde{\nabla}^af_{0a})
\nonumber
\\
&-\partial_0\tilde{\nabla}^a\tilde{\nabla}_a\tilde{\nabla}^if_{ij}
+\partial_0\tilde{\nabla}_j\tilde{\nabla}^a\tilde{\nabla}^bf_{ab}
+2k\partial_0\tilde{\nabla}^if_{ij},
\label{9.45a}
\end{align}
\begin{align}
&2(3k+\tilde{\nabla}^c\tilde{\nabla}_c)(-3k+\tilde{\nabla}^b\tilde{\nabla}_b)(\tilde{\nabla}^a\tilde{\nabla}_a-6k)(\tilde{\nabla}^d\tilde{\nabla}_d-2k)E_{ij}
\nonumber
\\
&=(3k+\tilde{\nabla}^c\tilde{\nabla}_c)(-3k+\tilde{\nabla}^b\tilde{\nabla}_b)(\tilde{\nabla}^a\tilde{\nabla}_a-6k)(\tilde{\nabla}^d\tilde{\nabla}_d-2k)f_{ij}
\nonumber
\\
&+\frac{1}{2}(-3k+\tilde{\nabla}^c\tilde{\nabla}_c)(\tilde{\nabla}^b\tilde{\nabla}_b-6k)(\tilde{\nabla}^a\tilde{\nabla}_a-2k)(-2kf-\tilde{\nabla}^d\tilde{\nabla}_df+\tilde{\nabla}^m\tilde{\nabla}^nf_{mn}) \tilde{\gamma}_{ij}
\nonumber
\\
&-2(\tilde{\nabla}^c\tilde{\nabla}_c-2k)\Bigg[\frac{1}{4}(3k+\tilde{\nabla}^b\tilde{\nabla}_b)\tilde{\nabla}_i\tilde{\nabla}_j\left(3\tilde{\nabla}^m\tilde{\nabla}^nf_{mn}-\tilde{\nabla}^a\tilde{\nabla}_af\right)
\nonumber
\\
&-k\tilde{\gamma}_{ij}\tilde{\nabla}_d\tilde{\nabla}^d((3k+\tilde{\nabla}^e\tilde{\nabla}_e)f+\frac{3}{2}(-2kf-\tilde{\nabla}^g\tilde{\nabla}_gf+\tilde{\nabla}^m\tilde{\nabla}^nf_{mn}))
\nonumber
\\
&-k\tilde{\gamma}_{ij}(-3k+\tilde{\nabla}^h\tilde{\nabla}_h)((3k+\tilde{\nabla}^m\tilde{\nabla}_m)f+\frac{3}{2}(-2kf-\tilde{\nabla}^n\tilde{\nabla}_nf+\tilde{\nabla}^m\tilde{\nabla}^nf_{mn}))\Bigg]
\nonumber
\\
&-\frac{1}{3}(-3k+\tilde{\nabla}^c\tilde{\nabla}_c)(3k+\tilde{\nabla}^b\tilde{\nabla}_b)\Bigg(\tilde{\nabla}_i\tilde{\nabla}^a\tilde{\nabla}_a(3\tilde{\nabla}^df_{dj}-\tilde{\nabla}_jf)
\nonumber
\\
&+\tilde{\nabla}_j\tilde{\nabla}^d\tilde{\nabla}_d(3\tilde{\nabla}^bf_{bi}-\tilde{\nabla}_if)-2\tilde{\nabla}_i\tilde{\nabla}_j(3\tilde{\nabla}^b\tilde{\nabla}^cf_{bc}-\tilde{\nabla}^e\tilde{\nabla}_ef)
\nonumber
\\
&-2k\tilde{\nabla}_i(3\tilde{\nabla}^af_{aj}-\tilde{\nabla}_jf)-2k\tilde{\nabla}_j(3\tilde{\nabla}^af_{ai}-\tilde{\nabla}_if)\Bigg).
\label{9.46a}
\end{align}
Despite its somewhat forbidding appearance (\ref{9.46a}) is actually a derivative of 
\begin{align}
&2(\tilde{\nabla}^a\tilde{\nabla}_a-2k)(\tilde{\nabla}^b\tilde{\nabla}_b-3k)E_{ij}
=(\tilde{\nabla}^a\tilde{\nabla}_a-2k)(\tilde{\nabla}^b\tilde{\nabla}_b-3k)f_{ij}
\nonumber\\
&+\tfrac{1}{2}\tilde{\nabla}_i\tilde{\nabla}_j\left[\tilde{\nabla}^a\tilde{\nabla}^bf_{ab}+(\tilde{\nabla}^a\tilde{\nabla}_a+4k)f\right]-(\tilde{\nabla}^a\tilde{\nabla}_a-3k)(\tilde{\nabla}_i\tilde{\nabla}^bf_{jb}+\tilde{\nabla}_j\tilde{\nabla}^bf_{ib})
\nonumber\\
&+\tfrac{1}{2}\tilde{\gamma}_{ij}\left[(\tilde{\nabla}^a\tilde{\nabla}_a-4k)\tilde{\nabla}^b\tilde{\nabla}^cf_{bc}
-(\tilde{\nabla}_a\tilde{\nabla}^a\tilde{\nabla}_b\tilde{\nabla}^b-2k\tilde{\nabla}^a\tilde{\nabla}^a+4k^2)f\right],
\label{9.47a}
\end{align}
a relation that itself can be  derived from the $D=3$ version of (\ref{6.4}) with $H^2=k$ by application  of $(\tilde{\nabla}^a\tilde{\nabla}_a-2k)(\tilde{\nabla}^a\tilde{\nabla}_a-3k)$ to (\ref{6.4}). One can check the validity of these relations by inserting 
\begin{align}
h_{0i}=\Omega^2(\tau)f_{0i}=\Omega^2(\tau)[\tilde{\nabla}_iB+B_i],\quad h_{ij}=\Omega^2(\tau)f_{ij}=\Omega^2(\tau)[-2\psi\tilde{\gamma}_{ij} +2\tilde{\nabla}_i\tilde{\nabla}_j E + \tilde{\nabla}_i E_j + \tilde{\nabla}_j E_i + 2E_{ij}]
\label{9.48a}
\end{align}

into them. And one can check their gauge invariance by inserting $h_{\mu\nu}\rightarrow h_{\mu\nu}-\nabla_{\mu}\epsilon_{\nu}-\nabla_{\mu}\epsilon_{\mu}$ into them. We thus establish that the metric fluctuations $\alpha$, $\gamma$, $B_i-\dot{E}_i$ and $E_{ij}$ are gauge invariant. And from  (\ref{9.13}), (\ref{9.22}), (\ref{9.23}) and (\ref{9.28}) can thus establish that the matter fluctuations $\delta \hat{\rho}$, $\hat{V}$, $V_i$ and $\delta \hat{p}$ are gauge invariant too. Interestingly, we see that in going from fluctuations around flat to fluctuations around Robertson-Walker with arbitrary $k$ and arbitrary dependence of $\Omega(\tau)$ on $\tau$ the gauge-invariant metric fluctuation combinations $\alpha$, $\gamma$, $B_i-\dot{E}_i$ and $E_{ij}$  remain the same, though $\gamma$ does depend generically on $\Omega(\tau)$.

\subsection{Solving the Background}

In order to actually solve the fluctuation equations we will need to determine the appropriate background $\Omega(\tau)$, and we will also need to deal with the fact that, as noted above,  the fluctuation equations contain more degrees of freedom (11) than there are evolution equations (10). For the background first we note that no matter what the value of $k$, from (\ref{9.10}) we see that if $\rho=3p$ then $\rho=3/\Omega^4$,  as written in a convenient normalization (one which differs from the one used in Sec. \ref{S8}), while if $p=0$ we have $\rho=3/\Omega^3$. Once we specify a background equation of state that relates $\rho$ and $p$ we can solve for $\Omega (\tau)$ and $t=\int \Omega(\tau)d\tau$. We thus obtain 
\begin{eqnarray}
p=\rho/3,~k=0:&&\quad \Omega=\tau,\quad p=1/\tau^4,\quad \rho=3/\tau^4,\quad t=\tau^2/2,\quad a(t)=\Omega(\tau)=(2t)^{1/2},
\nonumber\\
p=\rho/3,~k=-1:&&\quad \Omega=\sinh\tau,\quad p=1/\sinh^4\tau,\quad \rho=3/\sinh^4\tau,\quad t=\cosh\tau,\quad a(t)=\Omega(\tau)=(t^2-1)^{1/2},
\nonumber\\
p=\rho/3,~k=+1:&&\quad \Omega=\sin\tau,\quad p=1/\sin^4\tau,\quad \rho=3/\sin^4\tau,\quad t=-\cos\tau,\quad a(t)=\Omega(\tau)=(1-t^2)^{1/2},
\nonumber\\
p=0,~k=0:&&\quad \Omega=\tau^2/4,\quad p=0,\quad \rho=192/\tau^6,\quad t=\tau^{3}/12,\quad a(t)=\Omega(\tau)=(3t/2)^{2/3},
\nonumber\\
p=0,~k=-1:&&\quad \Omega=\sinh^2(\tau/2),\quad p=0,\quad \rho=3/\sinh^6(\tau/2),\quad t=\tfrac{1}{2}[\sinh\tau-\tau],\quad a(t)=\Omega(\tau),
\nonumber\\
p=0,~k=1:&&\quad \Omega=\sin^2(\tau/2),\quad p=0,\quad \rho=3/\sin^6(\tau/2),\quad t=\tfrac{1}{2}[\tau-\sin\tau],\quad a(t)=\Omega(\tau).
\label{9.49}
\end{eqnarray}
For $p=0$ and $k=\pm 1$ we cannot obtain $a(t)$ in a closed form. Consequently one ordinarily only determines $a(t)$ in parametric form. As we see, the conformal time $\tau$ can serve as the appropriate parameter.

\subsection{Relating $\delta\rho$ and $\delta p$}

To reduce the number of fluctuation variables from 11 to 10 we follow kinetic theory, and first consider a relativistic flat spacetime ideal $N$ particle classical gas of spinless particles each of mass $m$ in a volume $V$ at a temperature $T$. As discussed for instance in \cite{Mannheim2006}, for this
system one can use a basis of momentum eigenmodes, with the Helmholtz free energy $A(V,T)$ being given as 
\begin{equation}
e^{-A(V,T)/NkT}=V\int
d^3pe^{-(p^2+m^2)^{1/2}/kT}, 
\label{9.50}
\end{equation}                                 
so that the pressure takes the simple form 
\begin{equation}
p=-\left(\frac{\partial A}{ \partial
V}\right)_T=\frac{NkT}{V},
\label{9.51}
\end{equation}                                 
while the internal energy $U=\rho V$ evaluates in terms of Bessel
functions as  
\begin{equation}
U=A-T\left(\frac{\partial A}{ \partial
T}\right)_V=3NkT+Nm\frac{K_1(m/kT)}{K_2(m/kT)}.
\label{9.52}
\end{equation}                                 
In the high and low temperature limits (the radiation and matter eras)
we then find that the expression for $U$ simplifies to
\begin{eqnarray}
\rho=\frac{U}{V}\rightarrow
\frac{3NkT}{V}=3p,&&\quad \frac{m}{kT}\rightarrow 0,
\nonumber \\
\rho=\frac{U}{V} \rightarrow
\frac{Nm}{V}+\frac{3NkT}{2V}=\frac{Nm}{V}+\frac{3p}{2} \approx
\frac{Nm}{V},&&\quad \frac{m}{kT}
\rightarrow \infty.
\label{9.53}
\end{eqnarray}                                 
Consequently, while $p$ and $\rho$ are nicely proportional to each other ($p=w\rho$)
in the high temperature radiation and the low temperature matter eras
(where $w(T\rightarrow\infty)=1/3$ and $w(T\rightarrow 0)=0$), we also
see that in transition region between the two eras their relationship is
altogether more complicated. Since such a transition era occurs fairly close to recombination, it is this complicated relation that should be used there. Use of a $p=w\rho$ equation of state would at best only be valid at temperatures which are very different from
those of order $m/K$, though for massless particles it would be of
course be valid to use $p=\rho/3$ at all temperatures. 

As derived, these expressions only hold in a flat  Minkowski spacetime. However, $A(V,T)$ only involves  an integration over the spatial 3-momentum. Thus for the spatially flat $k=0$ Robertson-Walker metric $ds^2=dt^2-a^2(t)(dr^2+r^2d\theta^2+r^2\sin^2\theta d\phi^2)$, all of these kinetic theory relations will continue to hold with $T$ taken to depend on the comoving time $t$. (Typically $T\sim 1/a(t)$.) Suppose we now perturb the system and obtain a perturbed $\delta T$ that now depends on both $t$ and $r,\theta,\phi$. In the radiation era where $p=\rho/3$ we would obtain $\delta p =\delta \rho/3$ (and thus $3\delta\hat{p}=\delta\hat{\rho}$). In the matter era where $p=0$, from (\ref{9.53}) we would obtain $\delta p=2\delta \rho/3$. In the intermediate region the relation would be much more complicated. Nonetheless in all cases we would have reduced the number of independent fluctuation variables, though we note that not just in the radiation and matter eras but even in the intermediate region, it is standard in cosmological perturbation theory to use $\delta p/\delta \rho=v^2$ where $v^2$ is taken to be a spacetime-independent constant.

While we can use the above $A(V,T)$ for spatially flat cosmologies with $k=0$, for spatially curved cosmologies with non-zero $k$ we cannot use a mode basis made out of 3-momentum eigenstates at all. One has to adapt the basis to a curved 3-space by replacing
$(p^2+m^2)^{1/2}/kT$ by $(dx^{\mu}/d\tau)U^{\nu}g_{\mu\nu}/kT$ (see e.g. \cite{Mannheim2006}), while replacing
$\int d^3p$ by a sum over a complete set of basis modes associated with
the propagation of a spinless massive particle in the chosen
$g_{\mu\nu}$ background, and then follow the steps above to see what
generalization of (\ref{9.53}) might then ensue. While tractable in principle it is not straightforward to do this in practice, and we will not do it here. While one would need to do this in order to obtain a $k\neq 0$ generalization of the $k=0$  $\delta p/\delta \rho=v^2$ relation, and while such a generalization would be needed in order to solve the $k\neq 0$ fluctuation equations completely, since our purpose here is only to test for the validity of the decomposition theorem, we will not actually need to find a relation between $\delta p$ and $\delta \rho$, since as we now see, we will be able to test for the validity of the decomposition theorem without actually needing to know the specific form of such a relation at all, or even needing to specify any particular form for the background $\Omega(\tau)$ either for that matter.

\section{Implementing the Boundary Conditions for $k=-1$}
\label{S10}
\subsection{The Scalar Sector}

We have seen that the scalar sector evolution equations (\ref{9.22}), (\ref{9.26}), (\ref{9.27}) and (\ref{9.28}) involve derivatives of the form $\tilde{\nabla}^2$, $\tilde{\nabla}^2+3k$ where the coefficient of $k$ is either zero or positive, while the vector and tensor sectors equations (\ref{9.23}), (\ref{9.31}) and (\ref{9.40}) also involve derivatives such as $\tilde{\nabla}^2-2k$, $\tilde{\nabla}^2-3k$, and $\tilde{\nabla}^2-6k$ in which the coefficient of $k$ is negative. Now as we noted in Sec. \ref{S1}, the implications of boundary conditions are very sensitive to the sign of the coefficient of $k$, and we will need to monitor both positive and negative coefficient cases below. In implementing evolution equations that involve products of derivative operators such as the generic $(\tilde{\nabla}^2+\alpha)(\tilde{\nabla}^2+\beta)F=0$ ($F$ denotes scalar, vector or tensor), we can satisfy these relations by $(\tilde{\nabla}^2+\alpha)F=0$,  by $(\tilde{\nabla}^2+\beta)F=0$, or by $F=0$. The decomposition theorem will only follow if boundary conditions prevent us from satisfying $(\tilde{\nabla}^2+\alpha)F=0$ or   $(\tilde{\nabla}^2+\beta)F=0$ with non-zero $F$, leaving $F=0$ as the only remaining possibility. It is the purpose of this section to explore whether or not boundary conditions do force us to $F=0$ in any of the scalar, vector or tensor sectors. While a decomposition theorem would immediately hold if they do, as we will show in Sec. \ref{S11} the interplay of the vector and tensor sectors in the $\Delta_{ij}=0$ relation given in (\ref{9.15}) will still force us to a decomposition theorem even if they do not.

To illustrate what is involved it is sufficient to restrict $k$ to $k=-1$, and to take the metric to be of the form 
\begin{eqnarray}
ds^2=\Omega^2(\tau)\left[ d\tau^2-d\chi^2-\sinh^2\chi d\theta^2-\sinh^2\chi\sin^2\theta d\phi^2\right],
\label{10.1b}
\end{eqnarray}
where $r=\sinh \chi$. Since the analysis leading to the structure for $\Delta_{\mu\nu}$ given in (\ref{9.13}) to (\ref{9.17}) is completely covariant these equations equally hold if we represent the spatial sector of the metric as given in (\ref{10.1b}). With $k=-1$ the scalar sector evolution equations involving the $\tilde{\nabla}^2$ and $\tilde{\nabla}^2-3$ operators take the form
\begin{eqnarray}
(\tilde{\nabla}_a\tilde{\nabla}^a+A_S)S=0.
\label{10.2b}
\end{eqnarray}
(Here $S$ is to denote the full combinations of scalar sector  components that appear in  (\ref{9.22}), (\ref{9.26}), (\ref{9.27}) and (\ref{9.28}).)  In (\ref{10.2b}) we have introduced a generic scalar sector constant $A_S$, whose values in  (\ref{9.22}), (\ref{9.26}), (\ref{9.27}) and (\ref{9.28})  are $(0,-3)$. On setting $S(\chi,\theta,\phi)=S_{\ell}(\chi)Y^m_{\ell}(\theta,\phi)$ (\ref{10.2b}) reduces to 
\begin{eqnarray}
 \left[\frac{d^2}{d\chi^2}+2\frac{\cosh\chi }{\sinh\chi}\frac{d }{ d\chi}
-\frac{\ell(\ell+1)}{ \sinh^2\chi}+A_S\right]S_{\ell}=0.
\label{10.3b}
\end{eqnarray}
In the  $\chi\rightarrow \infty$ and $\chi\rightarrow 0$ limits  we take the solution to behave as $e^{\lambda \chi}$ (times an irrelevant polynomial in $\chi$), and as $\chi^n$, to thus obtain
\begin{eqnarray}
&&\lambda^2+2\lambda+A_S=0,\quad \lambda=-1\pm(1-A_S)^{1/2},
\nonumber\\
&&\lambda(A_S=0)=(-2,~0),\quad \lambda(A_S=-3)=(-3,~1),
\nonumber\\
&&n(n-1)+2n-\ell(\ell+1)=0,\quad n=\ell,-\ell-1.
\label{10.4b}
\end{eqnarray}
For each of $A_S=0$ and  $A_S=-3$ one solution converges at $\chi=\infty$ and the other diverges at $\chi=\infty$. Thus we need to see how they match up with the solutions at $\chi=0$, where one solution is well-behaved and the other is not. 

So to this end we look for exact solutions to (\ref{10.3b}). Thus,  as discussed for instance in \cite{Bander1966,Mannheim1988}, and as appropriately generalized here,  (\ref{10.3b}) 
admits of solutions (known as associated Legendre functions) of the form 
\begin{eqnarray}
S_{\ell}=\sinh^{\ell}\chi\left(\frac{1}{ \sinh\chi} \frac{d }{ d\chi}\right)^{\ell+1}f(\chi),
\label{10.5b}
\end{eqnarray}
where $f(\chi)$ obeys
\begin{eqnarray}
\left[\frac{d^3}{d\chi^3}+\nu^2\frac{d}{d\chi}\right]f(\chi)=0,\quad \nu^2=A_S-1,
\label{10.6b}
\end{eqnarray}
with $f(\chi)$ thus obeying 
\begin{eqnarray}
f(\nu^2>0)=\cos\nu\chi,~\sin\nu\chi,\quad f(\nu^2=-\mu^2<0)=\cosh\mu\chi,~\sinh\mu\chi,\quad f(\nu^2=0)=\chi,~\chi^2.
\label{10.7b}
\end{eqnarray}
For each $f(\chi)$ this would lead to solutions of the form
\begin{eqnarray}
\hat{S}_0=\frac{1}{\sinh\chi}\frac{df}{d\chi},\quad \hat{S}_1=\frac{d\hat{S}_0}{d\chi},\quad \hat{S}_2=\sinh\chi\frac{d}{d\chi}\left[\frac{\hat{S}_1}{\sinh\chi}\right],\quad \hat{S}_3=\sinh^2\chi\frac{d}{d\chi}\left[\frac{\hat{S}_2}{\sinh^2\chi}\right],....
\label{10.8b}
\end{eqnarray}
However, on evaluating these expressions it can happen that some of these solutions vanish. Thus for $A_S=0$ for instance where $f(\chi)=(\sinh\chi,\cosh\chi)$ the two solutions with $\ell=0$ are $\cosh\chi/\sinh\chi$ and $1$. However this would lead to the two solutions with $\ell=1$ being $1/\sinh^2\chi$ and $0$. To address this point we note that suppose we have obtained some non-zero solution $\hat{S}_{\ell}$. Then, a second solution of the form $\hat{f}_{\ell}(\chi)\hat{S}_{\ell}(\chi)$ may be found by inserting $\hat{f}_{\ell}(\chi)\hat{S}_{\ell}(\chi)$ into (\ref{10.3b}), to yield
\begin{eqnarray}
\hat{S}_{\ell}\frac{d^2 \hat{f}_{\ell}}{ d\chi^2}+2\hat{S}_{\ell}\frac{\cosh\chi }{ \sinh\chi}\frac{d \hat{f}_{\ell}}{ d\chi}+2\frac{d \hat{S}_{\ell}}{ d\chi}\frac{d \hat{f}_{\ell}}{ d\chi}=0,
\label{10.9b}
\end{eqnarray}
which integrates to
\begin{eqnarray}
\frac{d \hat{f}_{\ell}}{ d\chi}=\frac{1}{\sinh^2\chi\hat{S}_{\ell}^2},~~~~~\hat{f}_{\ell}\hat{S}_{\ell}=\hat{S}_{\ell}\int \frac{d\chi }{\sinh^2\chi\hat{S}_{\ell}^2}.
\label{10.10b}
\end{eqnarray}
Thus for $\ell=1$, from the non-trivial $A_S=0$ solution $\hat{S}_{1}=1/\sinh^2\chi$ we obtain a second solution of the form $\hat{f}_{\ell}\hat{S}_{\ell}=\cosh\chi/\sinh\chi-\chi/\sinh^2\chi$. However, once we have this second solution we can then return to (\ref{10.8b}) and use it to obtain the subsequent solutions associated with higher $\ell$ values, since use of the chain in (\ref{10.8b}) only requires that at any point the elements in it are solutions regardless of how they may or may not have been found.

Having the form given in (\ref{10.10b}) is  useful for another purpose, as it allows us to relate the behaviors of the solutions in the $\chi\rightarrow \infty$ and $\chi\rightarrow 0$ limits. Thus suppose that $\hat{S}_{\ell}$ behaves as $e^{\lambda\chi}$ and as $\chi^{\ell}$ in these two limits. Then $\hat{f}_{\ell}\hat{S}_{\ell}$ must behave as $e^{-(\lambda+2)\chi}$ and $\chi^{-\ell-1}$ in the two limits. Alternatively, if $\hat{S}_{\ell}$ behaves as $e^{\lambda\chi}$ and as $\chi^{-\ell-1}$ in these two limits, then $\hat{f}_{\ell}\hat{S}_{\ell}$ must behave as $e^{-(\lambda+2)\chi}$ and $\chi^{\ell}$ in the two limits. Comparing with (\ref{10.4b}), we note that if we set $\lambda=-1\pm(1-A_S)^{1/2}$ then consistently we find that $-(\lambda+2)=-1\mp(1-A_S)^{1/2}$. However, this analysis shows that we cannot directly identify which  $\chi\rightarrow \infty$ behavior is associated with which $\chi\rightarrow 0$ behavior (the insertion of either $\chi^{\ell}$ or $\chi^{-\ell -1}$ into (\ref{10.10b}) generates the other, with both behaviors thus being required in any $\hat{S}_{\ell}$, $\hat{f}_{\ell}\hat{S}_{\ell}$ pair), and to determine which is which we thus need to construct the asymptotic solutions directly.

For $A_S=0$, $\nu=i$,  the relevant $f(\nu^2)$ given in (\ref{10.7b}) are $\cosh \chi$ and $\sinh \chi$. 
Consequently, we find the first few $S^{(i)}_{\ell}$, $i=1,2$ solutions to $\tilde{\nabla}_a\tilde{\nabla}^aS=0$ to be of the form 
\begin{align}
&\hat{S}^{(1)}_{0}(A_S=0)=\frac{\cosh\chi}{\sinh\chi},\quad \hat{S}^{(2)}_{0}(A_S=0)=1,
\nonumber\\
&\hat{S}^{(1)}_{1}(A_S=0)=\frac{1}{\sinh^2\chi},\quad \hat{S}^{(2)}_{1}(A_S=0)=\frac{\cosh\chi}{\sinh\chi}-\frac{\chi}{\sinh^2\chi},
\nonumber\\
&\hat{S}^{(1)}_{2}(A_S=0)=\frac{\cosh\chi}{\sinh^3\chi},\quad \hat{S}^{(2)}_{2}(A_S=0)=1+\frac{3}{\sinh^2\chi}-\frac{3\chi\cosh\chi}{\sinh^3\chi},
\nonumber\\
&\hat{S}^{(1)}_{3}(A_S=0)=\frac{4}{\sinh^2\chi}+\frac{5}{\sinh^4\chi},\quad \hat{S}^{(2)}_{3}(A_S=0)=
\frac{2\cosh\chi}{\sinh\chi}+\frac{15\cosh\chi}{\sinh^3\chi}-\frac{12\chi}{\sinh^2\chi}-\frac{15\chi}{\sinh^4\chi}.
\label{10.11b}
\end{align}
From this pattern we see that the solutions that are bounded at $\chi=\infty$ are badly-behaved at $\chi=0$, while the solutions that are  well-behaved at $\chi=0$ are unbounded at $\chi=\infty$. Thus all of these $A_S=0$ solutions are excluded by a requirement that solutions be  bounded at $\chi=\infty$ and be well-behaved at $\chi=0$.

For $A_S=-3$, $\nu=2i$, the relevant $f(\nu^2)$ given in (\ref{10.7b}) are $\cosh 2\chi$ and $\sinh 2\chi$. Consequently, the first few solutions  to $(\tilde{\nabla}_a\tilde{\nabla}^a-3)S=0$ are of the form
\begin{eqnarray}
&&\hat{S}^{(1)}_0(A_S=-3)=\cosh\chi,\quad \hat{S}^{(2)}_0(A_S=-3)=2\sinh\chi+\frac{1}{\sinh\chi},
\nonumber\\
&&\hat{S}^{(1)}_1(A_S=-3)=\sinh\chi,\quad \hat{S}^{(2)}_1(A_S=-3)=2\cosh\chi-\frac{\cosh\chi}{\sinh^2\chi},
\nonumber\\
&&\hat{S}^{(1)}_2(A_S=-3)=2\cosh\chi-\frac{3\cosh\chi}{\sinh^2\chi}+\frac{3\chi}{\sinh^3\chi},\quad \hat{S}^{(2)}_2(A_S=-3)=\frac{1}{\sinh^3\chi},
\nonumber\\
&&\hat{S}^{(1)}_3(A_S=-3)=2\sinh\chi-\frac{5}{\sinh\chi}-\frac{15}{\sinh^3\chi}+\frac{15\chi\cosh\chi}{\sinh^4\chi},\quad \hat{S}^{(2)}_3(A_S=-3)=\frac{\cosh\chi}{\sinh^4\chi}.
\label{10.12b}
\end{eqnarray}
From this pattern we again see that the solutions that are bounded at $\chi=\infty$ are badly-behaved at $\chi=0$, while the solutions that are  well-behaved at $\chi=0$ are unbounded at $\chi=\infty$. Thus all of these $A_S=-3$ solutions are also excluded by a requirement that solutions be  bounded at $\chi=\infty$ and be well-behaved at $\chi=0$.

With all of these $A_S=0$, $A_S=-3$ solutions being excluded, as noted in Sec. \ref{S1}, we must realize (\ref{9.22}), (\ref{9.26}), (\ref{9.27}) and (\ref{9.28}) by 
\begin{eqnarray}
-2 \dot{\Omega} \Omega^{-1} (\alpha - \dot\gamma) + 2 k \gamma 
+(-4 \dot{\Omega}^2 \Omega^{-3}  + 2 \overset{..}{\Omega} \Omega^{-2}  - 2 k \Omega^{-1}) \hat{V}=0,
\label{10.13b}
\end{eqnarray}
\begin{eqnarray}
&&2 \dot{\Omega}^2 \Omega^{-2}(\alpha-\dot\gamma)
-2  \dot{\Omega} \Omega^{-1}(\dot\alpha -\ddot\gamma)-4\ddot\Omega\Omega^{-1}(\alpha-\dot\gamma)+ \Omega^2 \delta \hat{p}
 + 2 k(\alpha + 2 \dot{\Omega}  \Omega^{-1} \gamma)]=0,
\label{10.14b}
\end{eqnarray}
\begin{eqnarray}
\alpha+2\dot{\Omega}\Omega^{-1}\gamma=0,
\label{10.15b}
 \end{eqnarray}
\begin{eqnarray}
 2 \dot{\Omega}^2 \Omega^{-2}(\alpha-\dot\gamma)
-2  \dot{\Omega} \Omega^{-1}(\dot\alpha -\ddot\gamma)-4\ddot\Omega\Omega^{-1}(\alpha-\dot\gamma)+ \Omega^2 \delta \hat{p}=0.
\label{10.16b}
\end{eqnarray}
These equations are augmented by (\ref{9.13}), (\ref{9.16}) and (\ref{9.17})
\begin{eqnarray}
 6 \dot{\Omega}^2 \Omega^{-2}(\alpha-\dot\gamma) + \delta \hat{\rho} \Omega^2 + 2 \dot{\Omega} \Omega^{-1} \tilde{\nabla}_{a}\tilde{\nabla}^{a}\gamma=0, 
\label{10.17b}
\end{eqnarray}
\begin{eqnarray}
 6 \dot{\Omega}^2 \Omega^{-2}(\alpha-\dot\gamma)
-6  \dot{\Omega} \Omega^{-1}(\dot\alpha -\ddot\gamma)-12\ddot\Omega\Omega^{-1}(\alpha-\dot\gamma)+ 3\Omega^2 \delta \hat{p}-2\tilde\nabla_a\tilde\nabla^a(\alpha + 2\dot\Omega \Omega^{-1}\gamma)=0,
\label{10.18b}
\end{eqnarray}
\begin{eqnarray}
3 \delta \hat{p}-  \delta \hat{\rho}
-12 \overset{..}{\Omega}  \Omega^{-3}(\alpha - \dot\gamma) -6 \dot{\Omega} \Omega^{-3}(\dot{\alpha} -\ddot\gamma)
-2 \Omega^{-2} \tilde{\nabla}_{a}\tilde{\nabla}^{a}(\alpha +3\dot\Omega\Omega^{-1}\gamma)=0.
\label{10.19b}
\end{eqnarray}
On taking the $\tilde{\nabla}_i$ derivative of (\ref{10.13b}), we recognize (\ref{10.13b}) to (\ref{10.19b})  as precisely being the scalar sector ones given in (\ref{9.41}) and (\ref{9.42}). We thus establish the decomposition theorem in the scalar sector. 

\subsection{The Vector Sector}

To determine the structure of $k=-1$ solutions to the vector sector  (\ref{9.23}) and (\ref{9.31}), we first need to evaluate the quantity $\tilde{\nabla}_a\tilde{\nabla}^aV^i$, where $V^i$ obeys the transverse condition 
\begin{eqnarray}
\tilde\nabla_a V^a&=& \frac{V_{2} \cos\theta}{\sin\theta \sinh^2\chi} + \frac{2 V_{1} \cosh\chi}{\sinh\chi} + \partial_{1}V_{1} + \frac{\partial_{2}V_{2}}{\sinh^2\chi} + \frac{\partial_{3}V_{3}}{\sin^2\theta \sinh^2\chi}=0.
\label{10.20b}
\end{eqnarray}
On implementing this condition, the $(\chi,\theta,\phi) \equiv (1,2,3)$ components of $\tilde{\nabla}_a\tilde{\nabla}^aV^i$ take the form
\begin{eqnarray}
\tilde{\nabla}_a\tilde{\nabla}^aV^1&=&V_{1} \left(2 + \frac{2}{\sinh^2\chi}\right) + \frac{4 \cosh\chi \partial_{1}V_{1}}{\sinh\chi} + \partial_{1}\partial_{1}V_{1} + \frac{\cos\theta \partial_{2}V_{1}}{\sin\theta \sinh^2\chi} + \frac{\partial_{2}\partial_{2}V_{1}}{\sinh^2\chi} + \frac{\partial_{3}\partial_{3}V_{1}}{\sin^2\theta \sinh^2\chi},
 \nonumber\\ 
\tilde{\nabla}_a\tilde{\nabla}^aV^2&=& V_{2} \left(- \frac{2}{\sinh^4\chi} + \frac{1}{\sin^2\theta \sinh^4\chi} -  \frac{2}{\sinh^2\chi}\right) + \frac{4 V_{1} \cos\theta \cosh\chi}{\sin\theta \sinh^3\chi} + \frac{2 \cos\theta \partial_{1}V_{1}}{\sin\theta \sinh^2\chi} + \frac{\partial_{1}\partial_{1}V_{2}}{\sinh^2\chi} \nonumber \\ 
&& + \frac{2 \cosh\chi \partial_{2}V_{1}}{\sinh^3\chi} + \frac{3 \cos\theta \partial_{2}V_{2}}{\sin\theta \sinh^4\chi} + \frac{\partial_{2}\partial_{2}V_{2}}{\sinh^4\chi} + \frac{\partial_{3}\partial_{3}V_{2}}{\sin^2\theta \sinh^4\chi},
\nonumber\\ 
\tilde{\nabla}_a\tilde{\nabla}^aV^3&=& - \frac{2 V_{3}}{\sin^2\theta \sinh^2\chi} + \frac{\partial_{1}\partial_{1}V_{3}}{\sin^2\theta \sinh^2\chi} -  \frac{\cos\theta \partial_{2}V_{3}}{\sin^3\theta \sinh^4\chi} + \frac{\partial_{2}\partial_{2}V_{3}}{\sin^2\theta \sinh^4\chi} + \frac{2 \cosh\chi \partial_{3}V_{1}}{\sin^2\theta \sinh^3\chi} \nonumber \\ 
&& + \frac{2 \cos\theta \partial_{3}V_{2}}{\sin^3\theta \sinh^4\chi} + \frac{\partial_{3}\partial_{3}V_{3}}{\sin^4\theta \sinh^4\chi}.
\label{10.21b}
\end{eqnarray}
To explore the structure of the $k=-1$ vector sector we seek solutions to
\begin{eqnarray}
(\tilde{\nabla}_a\tilde{\nabla}^a+A_V)V_i=0.
\label{10.22b}
\end{eqnarray}
(Here $V_i$ is to denote the full combinations of vector components that appear in (\ref{9.23}) and (\ref{9.31}).) In (\ref{10.22b}) we have introduced a generic vector sector constant $A_V$, whose values in (\ref{9.23}) and (\ref{9.31})  are $(2,-1,-2)$.

Conveniently, we find that the equation for $V_1$ involves no mixing with $V_2$ or $V_3$, and can thus be solved directly. On setting $V_1(\chi,\theta,\phi)=g_{1,\ell}(\chi)Y_{\ell}^m(\theta,\phi)$, the equation for $V_1$ reduces to 
\begin{eqnarray}
\left[\frac{d^2}{d\chi^2}+4\frac{\cosh\chi}{ \sinh\chi}\frac{d }{d\chi}
+2+A_V+\frac{2 }{ \sinh^2\chi}-\frac{\ell(\ell+1)}{ \sinh^2\chi}\right]g_{1,\ell}=0.
\label{10.23b}
\end{eqnarray}
To check the $\chi \rightarrow \infty$ and $\chi \rightarrow 0$ limits, we take the solutions to behave as $e^{\lambda\chi}$ (times an irrelevant polynomial in $\chi$) and $\chi^n$ in these two limits. For (\ref{10.23b}) the limits give
\begin{eqnarray}
&&\lambda^2+4\lambda+2+A_V=0,\quad\lambda=-2\pm (2-A_V)^{!/2},
\nonumber\\
&&\lambda(A_V=2)=(-2,~-2),\quad \lambda(A_V=-1)=-2\pm \surd{3},\quad \lambda(A_V=-2)=(0,-4),
\nonumber\\
&&n(n-1)+4n+2-\ell(\ell+1)=0,\quad n=\ell-1, -\ell-2.
\label{10.24b}
\end{eqnarray}
Thus for $A_V=2$ and $A_V=-1$ both solutions are bounded at infinity, while for $A_V=-2$ one solution is bounded at infinity. Moreover, for each value of $A_V$ one of the solutions will  be well-behaved as $\chi\rightarrow 0$ for any $\ell\geq 1$ while the other solution will not be.  Thus for $A_V=2$ there will always be one $\ell\geq 1$ solution that is bounded at $\chi=\infty$ and well-behaved at $\chi=0$. To determine whether we can obtain a solution that is bounded in both limits for $A_V=-1$, $A_V=-2$ we need to explicitly find the solutions in closed form.

To this end we need to put (\ref{10.23b})  into the form of a differential equation whose solutions are known. We thus set $g_{1,\ell}=\alpha_{\ell}/\sinh\chi$, to find that (\ref{10.23b}) takes the form
\begin{eqnarray}
\left[\frac{d^2 }{d\chi^2}+2\frac{\cosh\chi}{ \sinh\chi}\frac{d }{d\chi}
-\frac{\ell(\ell+1) }{\sinh^2\chi}+A_V-1\right]\alpha_{\ell}=0.
\label{10.25b}
\end{eqnarray}
We recognize (\ref{10.25b}) as being in the form given in  (\ref{10.3b}), which we discussed above, with $\nu^2=A_V-2$.

Thus for $A_V=2$, viz. $\nu=0$  in (\ref{10.7b}) and $f(\nu^2=0)=\chi,~\chi^2$,  we find $V^{(i)}_{\ell}$, $i=1,2$ solutions to $(\tilde{\nabla}_a\tilde{\nabla}^a+2)V_1=0$ of the form 
\begin{eqnarray}
\hat{V}^{(1)}_0(A_V=2)&=&\frac{1}{ \sinh^2\chi},\quad \hat{V}^{(2)}_0(A_V=2)=\frac{\chi }{ \sinh^2\chi},
\nonumber\\
\hat{V}^{(1)}_1(A_V=2)&=&\frac{\cosh \chi }{ \sinh^3\chi},\quad \hat{V}^{(2)}_1(A_V=2)=\frac{1}{ \sinh^2\chi}-\frac{\chi\cosh\chi}{ \sinh^3\chi},
\nonumber\\
\hat{V}^{(1)}_2(A_V=2)&=&\frac{2}{ \sinh^2\chi}+\frac{3}{\sinh^4\chi},\quad \hat{V}^{(2)}_2(A_V=2)=\frac{3\cosh\chi}{\sinh^3\chi}-\frac{2\chi}{\sinh^2\chi}-\frac{3\chi }{\sinh^4\chi},
\nonumber\\
\hat{V}^{(1)}_3(A_V=2)&=&\frac{2\cosh\chi}{\sinh^3\chi}+\frac{5\cosh\chi}{\sinh^5\chi},\quad \hat{V}^{(2)}_3(A_V=2)=\frac{11}{\sinh^2\chi}+\frac{15}{\sinh^4\chi}-\frac{6\chi\cosh\chi}{\sinh^3\chi}-\frac{15\chi\cosh\chi }{\sinh^5\chi}.~~~
\label{10.26b}
\end{eqnarray}
The just as required by (\ref{10.24b}), the $\hat{V}^{(2)}_{\ell}(A_V=2)$ solutions with $\ell \geq1$ are bounded at  $\chi=\infty$ and well-behaved at $\chi=0$. Since they can thus not be excluded by boundary conditions at $\chi=\infty$ and $\chi=0$ (though boundary conditions do exclude modes with $\ell=0$), solutions to (\ref{9.23}) and (\ref{9.31}) do not become the vector sector solutions associated with (\ref{9.41}). Thus if we implement (\ref{9.31}) by $(\tilde{\nabla}_a\tilde{\nabla}^a+2)V_i=0$, the  decomposition theorem will fail in the vector sector for modes with $\ell \geq 1$. Thus an equation such as (\ref{9.31}) 
will be solved by 
\begin{eqnarray}
(\tilde{\nabla}_a\tilde{\nabla}^a-1)(\tilde{\nabla}_b\tilde{\nabla}^b-2)\left[\tfrac{1}{2}(\dot{B}_i-\ddot{E}_i)+\dot{\Omega}\Omega^{-1}(B_i-\dot{E}_i)\right]=V_i,
\label{10.27b}
\end{eqnarray}
and not by
\begin{eqnarray}
\tfrac{1}{2}(\dot{B}_i-\ddot{E}_i)+\dot{\Omega}\Omega^{-1}(B_i-\dot{E}_i)=0.
\label{10.28b}
\end{eqnarray}
Thus (\ref{9.31}) is solved by the $\chi$ dependence of $B_i-\dot{E}_i$ and not by its $\tau$ dependence, i.e., not by the $B_i-\dot{E_i}=1/\Omega^2$ dependence on $\tau$ that one would have obtained from the decomposition-theorem-required (\ref{10.28b}). This then raises the question of what does fix the $\tau$ dependence in the vector sector. We will address this issue below.

For $A_V=-2$ we see that $\nu^2=-4$ and that $f(\nu^2)=\cosh 2\chi,\sinh 2\chi$. However in the scalar case discussed above where $\nu^2=A_S-1$, $\nu^2$ would also obey $\nu^2=-4$ if $A_S=-3$. Thus for $A_V=-2$ we can obtain the solutions to $(\tilde{\nabla}_a\tilde{\nabla}^a-2)V_1=0$ directly from (\ref{10.12b}), and after implementing $g_{1,\ell}=\alpha_{\ell}/\sinh\chi$ we  obtain 
\begin{eqnarray}
&&\hat{V}^{(1)}_0(A_V=-2)=\frac{\cosh\chi}{\sinh\chi},\quad \hat{V}^{(2)}_0(A_V=-2)=2+\frac{1}{\sinh^2\chi},
\nonumber\\
&&\hat{V}^{(1)}_1(A_V=-2)=1,\quad \hat{V}^{(2)}_1(A_V=-2)=2\frac{\cosh\chi}{\sinh\chi}-\frac{\cosh\chi}{\sinh^3\chi},
\nonumber\\
&&\hat{V}^{(1)}_2(A_V=-2)=2\frac{\cosh\chi}{\sinh\chi}-\frac{3\cosh\chi}{\sinh^3\chi}+\frac{3\chi}{\sinh^4\chi},\quad \hat{V}^{(2)}_2(A_V=-2)=\frac{1}{\sinh^4\chi},
\nonumber\\
&&\hat{V}^{(1)}_3(A_V=-2)=2-\frac{5}{\sinh^2\chi}-\frac{15}{\sinh^4\chi}+\frac{15\chi\cosh\chi}{\sinh^5\chi},\quad \hat{V}^{(2)}_3(A_V=-2)=\frac{\cosh\chi}{\sinh^5\chi}.
\label{10.29b}
\end{eqnarray}
As required by (\ref{10.24b}), the $\hat{V}^{(2)}_2(A_V=-2)$ and $\hat{V}^{(2)}_3(A_V=-2)$ solutions  are bounded at  $\chi=\infty$. However, they are not well-behaved at $\chi=0$. Since they thus can  be excluded by boundary conditions at $\chi=\infty$ and $\chi=0$, if we implement (\ref{9.31}) by $(\tilde{\nabla}_a\tilde{\nabla}^a-2)V_i=0$,  the only allowed solution will be $V_i=0$, and the decomposition theorem will then follow.

Finally, for $A_V=-1$, viz. $\nu=i\surd{3}$, $f(\nu^2)=e^{\chi\surd{3}},e^{-\chi\surd{3}}$, the solutions to $(\tilde{\nabla}_a\tilde{\nabla}^a-1)V_1=0$ are of the form
\begin{eqnarray}
&&\hat{V}^{(1)}_0(A_V=-1)=\frac{e^{\chi\surd{3}}}{\sinh^2\chi},\quad \hat{V}^{(2)}_0(A_V=-1)=\frac{e^{-\chi\surd{3}}}{\sinh^2\chi},
\nonumber\\
&&\hat{V}^{(1)}_1(A_V=-1)=\frac{e^{\chi\surd{3}}}{\sinh^3\chi}\left[\surd{3}\sinh\chi-\cosh\chi\right],\quad \hat{V}^{(2)}_1(A_V=-1)=\frac{e^{-\chi\surd{3}}}{\sinh^3\chi}\left[-\surd{3}\sinh\chi-\cosh\chi\right],
\nonumber\\
&&\hat{V}^{(1)}_2(A_V=-1)=\frac{e^{\chi\surd{3}}}{\sinh^4\chi}\left[3-3\surd{3}\cosh\chi\sinh\chi+5\sinh^2\chi
\right],
\nonumber\\
 &&\hat{V}^{(2)}_2(A_V=-1)=\frac{e^{-\chi\surd{3}}}{\sinh^4\chi}\left[3+3\surd{3}\cosh\chi\sinh\chi+5\sinh^2\chi\right],
\nonumber\\
&&\hat{V}^{(1)}_3(A_V=-1)=\frac{e^{\chi\surd{3}}}{\sinh^5\chi}\left[15\surd{3}\sinh\chi+14\surd{3}\sinh^3\chi
-15\cosh\chi-24\cosh\chi\sinh^2\chi\right],
\nonumber\\
&&\hat{V}^{(2)}_3(A_V=-1)=\frac{e^{-\chi\surd{3}}}{\sinh^5\chi}\left[-15\surd{3}\sinh\chi-14\surd{3}\sinh^3\chi
-15\cosh\chi-24\cosh\chi\sinh^2\chi\right].
\label{10.30b}
\end{eqnarray}
All of these solutions are bounded at $\chi=\infty$ and all $\hat{V}^{(1)}_{\ell}(A_V=-1)-\hat{V}^{(2)}_{\ell}(A_V=-1)$ with $\ell\geq 1$ are well-behaved at $\chi=0$.  Thus if implement (\ref{9.31}) by $(\tilde{\nabla}_a\tilde{\nabla}^a-1)V_i=0$,  we are not forced to $V_i=0$, with the decomposition theorem not then following in this sector.

\subsection{The Tensor Sector}

For $k=-1$ the transverse-traceless tensor sector modes need to satisfy 
\begin{eqnarray}
 \tilde{\gamma}^{ab}T_{ab}&=& T_{11} + \frac{T_{22}}{\sinh^2\chi} + \frac{T_{33}}{\sin^2\theta \sinh^2\chi} =0,
\nonumber\\
\tilde\nabla_a T^{a 1}&=& - \frac{\cosh\chi T_{22}}{\sinh^3\chi} -  \frac{\cosh\chi T_{33}}{\sin^2\theta \sinh^3\chi} + \frac{\cos\theta T_{12}}{\sin\theta \sinh^2\chi} + \frac{2 \cosh\chi T_{11}}{\sinh\chi} + \partial_{1}T_{11} + \frac{\partial_{2}T_{12}}{\sinh^2\chi} \nonumber \\ 
&& + \frac{\partial_{3}T_{13}}{\sin^2\theta \sinh^2\chi}=0, \nonumber\\
\tilde\nabla_a T^{a 2}&=& - \frac{\cos\theta T_{33}}{\sin^3\theta \sinh^4\chi} + \frac{\cos\theta T_{22}}{\sin\theta \sinh^4\chi} + \frac{2 \cosh\chi T_{12}}{\sinh^3\chi} + \frac{\partial_{1}T_{12}}{\sinh^2\chi} + \frac{\partial_{2}T_{22}}{\sinh^4\chi} + \frac{\partial_{3}T_{23}}{\sin^2\theta \sinh^4\chi}=0,
\nonumber\\
\tilde\nabla_a T^{a 3}&=& \frac{\cos\theta T_{23}}{\sin^3\theta \sinh^4\chi} + \frac{2 \cosh\chi T_{13}}{\sin^2\theta \sinh^3\chi} + \frac{\partial_{1}T_{13}}{\sin^2\theta \sinh^2\chi} + \frac{\partial_{2}T_{23}}{\sin^2\theta \sinh^4\chi} + \frac{\partial_{3}T_{33}}{\sin^4\theta \sinh^4\chi}=0.
\label{10.31b}
\end{eqnarray}
Under these conditions the components of $\tilde{\nabla}_a\tilde{\nabla}^aT^{ij}$ evaluate to
\begin{eqnarray}
\tilde{\nabla}_a\tilde{\nabla}^aT^{11}&=& T_{11} \left(6 + \frac{6}{\sinh^2\chi}\right) + \frac{6 \cosh\chi \partial_{1}T_{11}}{\sinh\chi} + \partial_{1}\partial_{1}T_{11} + \frac{\cos\theta \partial_{2}T_{11}}{\sin\theta \sinh^2\chi} + \frac{\partial_{2}\partial_{2}T_{11}}{\sinh^2\chi} + \frac{\partial_{3}\partial_{3}T_{11}}{\sin^2\theta \sinh^2\chi},
 \nonumber\\ 
\tilde{\nabla}_a\tilde{\nabla}^aT^{22}&=& \frac{4 T_{22}}{\sinh^6\chi} -  \frac{4 T_{22}}{\sin^2\theta \sinh^6\chi} + \frac{4 T_{11}}{\sinh^4\chi} -  \frac{2 T_{22}}{\sinh^4\chi} -  \frac{2 T_{11}}{\sin^2\theta \sinh^4\chi} + \frac{2 T_{11}}{\sinh^2\chi} -  \frac{2 \cosh\chi \partial_{1}T_{22}}{\sinh^5\chi} \nonumber \\ 
&& + \frac{\partial_{1}\partial_{1}T_{22}}{\sinh^4\chi} + \frac{4 \cosh\chi \partial_{2}T_{12}}{\sinh^5\chi} + \frac{\cos\theta \partial_{2}T_{22}}{\sin\theta \sinh^6\chi} + \frac{\partial_{2}\partial_{2}T_{22}}{\sinh^6\chi} -  \frac{4 \cos\theta \partial_{3}T_{23}}{\sin^3\theta \sinh^6\chi} + \frac{\partial_{3}\partial_{3}T_{22}}{\sin^2\theta \sinh^6\chi},
 \nonumber\\ 
\tilde{\nabla}_a\tilde{\nabla}^aT^{33}&=& \frac{2T_{33}} {\sin^4\theta\sinh^6\chi}\left(1-{\sinh^2\chi}\right) + T_{11} \left(\frac{2}{\sin^4\theta \sinh^4\chi} + \frac{2}{\sin^2\theta \sinh^2\chi}\right) -  \frac{4 \cos\theta \cosh\chi T_{12}}{\sin^3\theta \sinh^5\chi} 
\nonumber \\ 
&& -  \frac{4 \cos\theta \partial_{1}T_{12}}{\sin^3\theta \sinh^4\chi} -  \frac{2 \cosh\chi \partial_{1}T_{33}}{\sin^4\theta \sinh^5\chi} + \frac{\partial_{1}\partial_{1}T_{33}}{\sin^4\theta \sinh^4\chi} + \frac{4 \cos\theta \partial_{2}T_{11}}{\sin^3\theta \sinh^4\chi} + \frac{\cos\theta \partial_{2}T_{33}}{\sin^5\theta \sinh^6\chi} \nonumber \\ 
&& + \frac{\partial_{2}\partial_{2}T_{33}}{\sin^4\theta \sinh^6\chi} + \frac{4 \cosh\chi \partial_{3}T_{13}}{\sin^4\theta \sinh^5\chi} + \frac{\partial_{3}\partial_{3}T_{33}}{\sin^6\theta \sinh^6\chi},
\nonumber\\ 
\tilde{\nabla}_a\tilde{\nabla}^aT^{12}&=& T_{12} \left(- \frac{1}{\sin^2\theta \sinh^4\chi} -  \frac{2}{\sinh^2\chi}\right) + \frac{2 \cosh\chi \partial_{1}T_{12}}{\sinh^3\chi} + \frac{\partial_{1}\partial_{1}T_{12}}{\sinh^2\chi} + \frac{2 \cosh\chi \partial_{2}T_{11}}{\sinh^3\chi} \nonumber \\ 
&& + \frac{\cos\theta \partial_{2}T_{12}}{\sin\theta \sinh^4\chi} + \frac{\partial_{2}\partial_{2}T_{12}}{\sinh^4\chi} -  \frac{2 \cos\theta \partial_{3}T_{13}}{\sin^3\theta \sinh^4\chi} + \frac{\partial_{3}\partial_{3}T_{12}}{\sin^2\theta \sinh^4\chi},
 \nonumber\\ 
\tilde{\nabla}_a\tilde{\nabla}^aT^{13}&=& - \frac{2 T_{13}}{\sin^2\theta \sinh^2\chi} + \frac{2 \cosh\chi \partial_{1}T_{13}}{\sin^2\theta \sinh^3\chi} + \frac{\partial_{1}\partial_{1}T_{13}}{\sin^2\theta \sinh^2\chi} -  \frac{\cos\theta \partial_{2}T_{13}}{\sin^3\theta \sinh^4\chi} + \frac{\partial_{2}\partial_{2}T_{13}}{\sin^2\theta \sinh^4\chi} \nonumber \\ 
&& + \frac{2 \cosh\chi \partial_{3}T_{11}}{\sin^2\theta \sinh^3\chi} + \frac{2 \cos\theta \partial_{3}T_{12}}{\sin^3\theta \sinh^4\chi} + \frac{\partial_{3}\partial_{3}T_{13}}{\sin^4\theta \sinh^4\chi},
 \nonumber\\ 
\tilde{\nabla}_a\tilde{\nabla}^aT^{23}&=& T_{23} \left(\frac{2(1-\sinh^2\chi)}{\sin^2\theta\sinh^6\chi} -  \frac{1}{\sin^4\theta \sinh^6\chi}\right) + \frac{2 \cos\theta \partial_{1}T_{13}}{\sin^3\theta \sinh^4\chi} -  \frac{2 \cosh\chi \partial_{1}T_{23}}{\sin^2\theta \sinh^5\chi} + \frac{\partial_{1}\partial_{1}T_{23}}{\sin^2\theta \sinh^4\chi}\nonumber \\ 
&& + \frac{2 \cosh\chi \partial_{2}T_{13}}{\sin^2\theta \sinh^5\chi} + \frac{\cos\theta \partial_{2}T_{23}}{\sin^3\theta \sinh^6\chi} + \frac{\partial_{2}\partial_{2}T_{23}}{\sin^2\theta \sinh^6\chi} + \frac{2 \cosh\chi \partial_{3}T_{12}}{\sin^2\theta \sinh^5\chi} + \frac{2 \cos\theta \partial_{3}T_{22}}{\sin^3\theta \sinh^6\chi} 
\nonumber \\ 
&& + \frac{\partial_{3}\partial_{3}T_{23}}{\sin^4\theta \sinh^6\chi}.
\label{10.32b}
\end{eqnarray}

Following our analysis of the vector sector, in the $k=-1$ tensor sector we seek solutions to
\begin{eqnarray}
(\tilde{\nabla}_a\tilde{\nabla}^a+A_T)T_{ij}=0.
\label{10.33b}
\end{eqnarray}
(Here $T_{ij}$ is to denote the full combination of  tensor components that appears in (\ref{9.40}).)  In (\ref{10.33b}) we have introduced a generic tensor sector constant $A_T$, whose values in (\ref{9.40})  are $(2,3,6)$.
Conveniently, we find that the equation for $T_{11}$ involves no mixing with any other components of $T_{ij}$, and can thus be solved directly. On setting $T_{11}(\chi,\theta,\phi)=h_{11,\ell}(\chi)Y_{\ell}^m(\theta,\phi)$, the equation for $T_{11}$ reduces to 
\begin{eqnarray}
\left[\frac{d^2}{d\chi^2}+6\frac{\cosh\chi}{ \sinh\chi}\frac{d }{d\chi}
+6+\frac{6 }{ \sinh^2\chi}-\frac{\ell(\ell+1)}{ \sinh^2\chi}+A_T\right]h_{11,\ell}=0.
\label{10.34b}
\end{eqnarray}

To determine the $\chi \rightarrow \infty$ and $\chi \rightarrow 0$ limits, we take the solutions to behave as $e^{\lambda\chi}$ (times an irrelevant polynomial in $\chi$) and $\chi^n$ in these two limits. For (\ref{10.34b}) the limits give
\begin{eqnarray}
&&\lambda^2+6\lambda+6+A_T,\quad \lambda=-3\pm(3-A_T)^{1/2},
\nonumber\\
&&\lambda(A_T=2)=(-4,~-2),\quad \lambda(A_T=3)=(-3,~-3),\quad \lambda(A_T=6)=-3\pm i\surd{3},
\nonumber\\
&&n(n-1)+6n+6-\ell(\ell+1)=0,\quad n=\ell-2, -\ell-3.
\label{10.35b}
\end{eqnarray}
Thus for any allowed $A_T$, every solution to (\ref{10.34b}) is bounded at $\chi=\infty$, while for each $A_T$ one of the solutions will be well-behaved as $\chi\rightarrow 0$ for any $\ell\geq 2$.  Thus for $\ell=2,3,4,..$  there will always be one solution for any allowed $A_T$ that is bounded at $\chi=\infty$ and well-behaved at $\chi=0$, with all solutions with $\ell=0$ or $\ell=1$ being excluded.

To solve (\ref{10.34b}) we set $h_{11,\ell}=\gamma_{\ell}/\sinh^2\chi$ to obtain:
\begin{eqnarray}
 \left[\frac{d^2}{d\chi^2}+2\frac{\cosh\chi}{\sinh\chi}\frac{d}{d\chi}
-\frac{\ell(\ell+1) }{ \sinh^2\chi}-2+A_T\right]\gamma_{\ell}=0.
\label{10.36b}
\end{eqnarray}
We recognize (\ref{10.36b}) as being (\ref{10.3b}), and can set $\nu^2=A_T-3$ in (\ref{10.7b}), viz. $\nu^2=(-1,0,3)$ for $A_T=2,3,6$. For $A_T=2$ we see that $\nu^2=-1$. However in the scalar case discussed above where $\nu^2=A_S-1$, $\nu^2$ would also obey $\nu^2=-1$ if $A_S=0$. Thus for $A_T=2$ we can obtain the solutions to $(\tilde{\nabla}_a\tilde{\nabla}^a+2)T_{11}=0$ directly from (\ref{10.11b}), and after implementing $h_{11,\ell}=\gamma_{\ell}/\sinh^2\chi$ we obtain $T^{(1)}_{\ell}$, $T^{(2)}_{\ell}$ solutions to (\ref{10.34b}) of the form 
\begin{align}
&\hat{T}^{(1)}_{0}(A_T=2)=\frac{\cosh\chi}{\sinh^3\chi},\quad \hat{T}^{(2)}_{0}(A_T=2)=\frac{1}{\sinh^2\chi},
\nonumber\\
&\hat{T}^{(1)}_{1}(A_T=2)=\frac{1}{\sinh^4\chi},\quad \hat{T}^{(2)}_{1}(A_T=2)=\frac{\cosh\chi}{\sinh^3\chi}-\frac{\chi}{\sinh^4\chi},
\nonumber\\
&\hat{T}^{(1)}_{2}(A_T=2)=\frac{\cosh\chi}{\sinh^5\chi},\quad \hat{T}^{(2)}_{2}(A_T=2)=\frac{1}{\sinh^2\chi}+\frac{3}{\sinh^4\chi}-\frac{3\chi\cosh\chi}{\sinh^5\chi},
\nonumber\\
&\hat{T}^{(1)}_{3}(A_T=2)=\frac{4}{\sinh^4\chi}+\frac{5}{\sinh^6\chi},\quad \hat{T}^{(2)}_{3}(A_T=2)=
\frac{2\cosh\chi}{\sinh^3\chi}+\frac{15\cosh\chi}{\sinh^5\chi}-\frac{12\chi}{\sinh^4\chi}-\frac{15\chi}{\sinh^6\chi}.
\label{10.37b}
\end{align}
All of these solutions are bounded at $\chi=\infty$ and all $\hat{T}^{(2)}_{\ell}(A_T=2)$ with $\ell\geq 2$ are well-behaved at $\chi=0$.  Thus if implement (\ref{9.40}) by $(\tilde{\nabla}_a\tilde{\nabla}^a+2)T_{ij}=0$,  we are not forced to $T_{ij}=0$, with the decomposition theorem not then following in the tensor sector.

For $A_T=3$ we see that $\nu^2=0$. However in the vector case discussed above where $\nu^2=A_V-2$, $\nu^2$ would also obey $\nu^2=0$ if $A_V=2$. Thus for $A_T=3$ we can obtain the solutions to $(\tilde{\nabla}_a\tilde{\nabla}^a+3)T_{11}=0$ directly from (\ref{10.26b}), and after implementing $h^{11}_{\ell}=\alpha_{\ell}/\sinh\chi$ we obtain 
\begin{eqnarray}
\hat{T}^{(1)}_0(A_T=3)&=&\frac{1}{ \sinh^3\chi},\quad \hat{T}^{(2)}_0(A_T=3)=\frac{\chi }{\sinh^3\chi},
\nonumber\\
\hat{T}^{(1)}_1(A_T=3)&=&\frac{\cosh \chi }{ \sinh^4\chi},\quad \hat{T}^{(2)}_1(A_T=3)=\frac{1}{ \sinh^3\chi}-\frac{\chi\cosh\chi}{\sinh^4\chi},
\nonumber\\
\hat{T}^{(1)}_2(A_T=3)&=&\frac{2}{ \sinh^3\chi}+\frac{3}{\sinh^5\chi},\quad \hat{T}^{(2)}_2(A_T=3)=\frac{3\cosh\chi}{\sinh^4\chi}-\frac{2\chi}{\sinh^3\chi}-\frac{3\chi }{\sinh^5\chi},
\nonumber\\
\hat{T}^{(1)}_3(A_T=3)&=&\frac{2\cosh\chi}{\sinh^4\chi}+\frac{5\cosh\chi}{\sinh^6\chi},\quad \hat{T}^{(2)}_3(A_T=3)=\frac{11}{\sinh^3\chi}+\frac{15}{\sinh^5\chi}-\frac{6\chi\cosh\chi}{\sinh^4\chi}-\frac{15\chi\cosh\chi }{\sinh^6\chi}.~~~
\label{10.38b}
\end{eqnarray}
All of these solutions are bounded at $\chi=\infty$ and all $\hat{T}^{(2)}_{\ell}(A_T=3)$ with $\ell\geq 2$ are well-behaved at $\chi=0$.  Thus if implement (\ref{9.40}) by $(\tilde{\nabla}_a\tilde{\nabla}^a+3)T_{ij}=0$,  we are not forced to $T_{ij}=0$, with the decomposition theorem not then following.

A similar outcome occurs for $A_T=6$, and even though we do not evaluate the $A_T=6$ solutions explicitly, according to (\ref{10.35b}) all solutions to $(\tilde{\nabla}_a\tilde{\nabla}^a+6)T_{11}=0$ with $A_T=6$ are bounded at $\chi=\infty$ (behaving as $e^{-3\chi}\cos(\surd{3}\chi)$ and $e^{-3\chi}\sin(\surd{3}\chi)$), with one set of these solutions being well-behaved at $\chi=0$ for all $\ell \geq 2$.  Thus if implement (\ref{9.40}) by $(\tilde{\nabla}_a\tilde{\nabla}^a+6)T_{ij}=0$,  we are not forced to $T_{ij}=0$, with the decomposition theorem again not following in the tensor sector.

\section{Recovering the Decomposition Theorem}
\label{S11}

In Sec. \ref{S10} we have seen that there are realizations of the evolution equations in the scalar, vector, and tensor sectors that would not lead to a decomposition theorem in those sectors. However, equally there are other realizations that given the boundary conditions would lead to a decomposition theorem. Thus we need to determine which realizations are the relevant ones. To this end we look not at the individual higher-derivative equations obeyed by the separate scalar, vector, and tensor sectors, but at how these various sectors interface with each other in the original second-order $\Delta_{\mu\nu}=0$ equations themselves. Any successful such interface would require that all the terms in $\Delta_{\mu\nu}=0$ would have to have the same $\chi$ behavior. Noting that the scalar modes appear with two $\tilde{\nabla}$ derivatives in $\Delta_{ij}=0$, the vector sector appears with one $\tilde{\nabla}$ derivative and the tensor appears with none, we need to compare derivatives of scalars with vectors and derivatives of vectors with tensors. 

To see how to obtain such a needed common $\chi$ behavior we differentiate the scalar field (\ref{10.3b}) with respect to $\chi$, to obtain
\begin{eqnarray}
 \left[\frac{d^2}{d\chi^2}+4\frac{\cosh\chi}{\sinh\chi}\frac{d }{ d\chi}
+\frac{2}{\sinh^2\chi}-\frac{\ell(\ell+1)}{\sinh^2\chi}+4+A_S\right]\frac{d S_{\ell}}{d \chi}
+2A_S\frac{\cosh\chi}{\sinh\chi}S_{\ell}=0.
\label{11.1}
\end{eqnarray}
Comparing with the vector (\ref{10.23b}) we see that up to an overall normalization we can identify $d S_{\ell}/d\chi$ with the vector $g_{1,\ell}$ for modes that obey $A_S=0$ and $A_V=2$, so that these particular scalar and vector modes can interface. As a check, with the vector sector needing $\ell \geq 1$ we differentiate $\hat{S}^{(2)}_1(A_S=0)$ to obtain
\begin{eqnarray}
 \frac{d}{d \chi}\hat{S}^{(2)}_1(A_S=0) =\frac{d}{d \chi}\left[\frac{\cosh\chi}{\sinh\chi}-\frac{\chi}{\sinh^2\chi}\right]
 =-\frac{2}{ \sinh^2\chi}+\frac{2\chi\cosh\chi}{ \sinh^3\chi}=-2\hat{V}^{(2)}_1(A_V=2).
\label{11.2}
\end{eqnarray}

Similarly, if we differentiate the vector field (\ref{10.23b}) with respect to $\chi$ we  obtain
\begin{eqnarray}
\left[\frac{d^2}{d\chi^2}+6\frac{\cosh\chi}{ \sinh\chi}\frac{d }{d\chi}
+10+A_V+\frac{6 }{ \sinh^2\chi}-\frac{\ell(\ell+1)}{ \sinh^2\chi}\right]\frac{d g_{1,\ell}}{d \chi}
+2(2+A_V)\frac{\cosh\chi}{\sinh\chi}g_{1,\ell}=0.
\label{11.3}
\end{eqnarray}
Comparing with the tensor (\ref{10.34b}) we see that up to an overall normalization we can identify $d g_{1,\ell}/d\chi$ with the tensor $h_{11,\ell}$ for modes that obey $A_V=-2$ and $A_T=2$, so that these particular vector and tensor modes can interface. As a check, with the tensor sector needing $\ell \geq 2$ we differentiate $\hat{V}^{(1)}_2(A_V=-2)$ to obtain
\begin{eqnarray}
 \frac{d}{d \chi}\hat{V}^{(1)}_2(A_V=-2) = \frac{d}{d \chi}\left[\frac{2\cosh\chi}{\sinh\chi}-\frac{3\cosh\chi }{\sinh^3\chi}+\frac{3\chi}{\sinh^4\chi}\right]
 =\frac{4}{\sinh^2\chi}+\frac{12}{\sinh^4\chi}-\frac{12\chi\cosh\chi}{\sinh^5\chi}=4\hat{T}^{(2)}_2(A_T=2).~~
\label{11.4}
\end{eqnarray}

Thus while we can interface $A_S=0$ and $A_V=2$, we cannot interface $A_V=2$ with any of the tensor modes. Rather, we must interface the $A_V=-2$ vector modes with the  $A_T=2$ tensor modes. With none of the scalar sector modes meeting the boundary conditions at both $\chi=\infty$ and $\chi=0$ anyway, the scalar sector must satisfy  $\Delta_{\mu\nu}=0$ by itself, with the scalar term contribution to $\Delta_{\mu\nu}=0$ then having to vanish, just as required of the decomposition theorem. However, in the vector and tensor sectors we can achieve a common $\chi$ behavior if we set $B_1-\dot{E}_1=p_1(\tau)\hat{V}^{(1)}_2(A_V=-2)$, $E_{11}=q_{11}(\tau)\hat{T}^{(2)}_2(A_T=2)$, since then the $\Delta_{11}=0$ equation reduces to
\begin{eqnarray}
\Delta_{11}=\bigg{[}\frac{1}{\sinh^2\chi}+\frac{3}{\sinh^4\chi}-\frac{3\chi\cosh\chi}{\sinh^5\chi}\bigg{]}\bigg{[} 8\dot{\Omega} \Omega^{-1}p_1(\tau)+4\dot{p}_1(\tau)- \ddot{q}_{11}(\tau) +2 q_{11}(\tau)  -2\dot{\Omega} \Omega^{-1}\dot{q}_{11}(\tau) -2q_{11}(\tau)\bigg{]}=0.
\nonumber\\
\label{11.5}
\end{eqnarray}
This relation has a non-trivial solution of the form
\begin{eqnarray}
4p_1(\tau)-\dot{q}_{11}(\tau)=\frac{1}{\Omega^2(\tau)},
\label{11.6}
\end{eqnarray}
to thereby relate the $\tau$ dependencies of the vector and tensor sectors. With the other components of $V_i$ and $T_{ij}$ being constructed in a similar manner, as such we have provided an exact interface solution in the vector and tensor sectors. However, it only falls short in one regard. Both of $\hat{V}^{(1)}_2(A_V=-2)$ and $\hat{T}^{(2)}_2(A_T=2)$ are well-behaved at $\chi=0$ and $\hat{T}^{(2)}_2(A_T=2)$ vanishes at $\chi=\infty$. However, $\hat{V}^{(1)}_2(A_V=-2)$ does not vanish at $\chi=\infty$, as it limits to a constant value. Imposing a boundary condition that the vector and tensor modes have to vanish at $\chi=\infty$ then excludes this solution, with the decomposition theorem then being recovered according to 
\begin{eqnarray}
\tfrac{1}{2}(\dot{B}_i-\ddot{E}_i)+\dot{\Omega}\Omega^{-1}(B_i-\dot{E}_i)=0,\quad 
- \overset{..}{E}_{ij} +2 E_{ij} - 2 \dot{E}_{ij} \dot{\Omega} \Omega^{-1} + \tilde{\nabla}_{a}\tilde{\nabla}^{a}E_{ij}=0,
\label{11.7}
\end{eqnarray}
with these being the equations that then serve to fix the $\tau$ dependencies in the vector and tensor sectors.  Consequently, we establish that the decomposition theorem does in fact hold for Robertson-Walker cosmologies with non-vanishing spatial 3-curvature after all. 

\section{The SVT4 Fluctuation Equations --  General Robertson-Walker Backgrounds}
\label{S12}
\subsection{The Background}

As well as discuss SVT3 fluctuations around general Robertson-Walker backgrounds in Einstein gravity,  it is of interest to discuss SVT4 fluctuations  as well. To this end we take the background metric and the 3-space Ricci tensor to be of the form 
\begin{eqnarray}
ds^2 &=&-g_{\mu\nu}dx^{\mu}dx^{\nu}=\Omega^2(\tau)\left(d\tau^2 -\tilde{\gamma}_{ij} dx^i dx^j\right),\quad \tilde{R}_{ij} = -2k \tilde{\gamma}_{ij}.
\label{12.1}
\end{eqnarray}
Given the symmetry of the 4-geometry, the 4-space Ricci tensor and the 4-space Einstein tensor can be written as 
\begin{eqnarray}
R_{\mu\nu} &=& (A+B)U_\mu U_\nu + g_{\mu\nu}B,\quad G_{\mu\nu}= \tfrac{1}{2} A g_{\mu \nu } -  \tfrac{1}{2} B g_{\mu \nu } + A U_{\mu } U_{\nu } + B U_{\mu } U_{\nu },
\label{12.2}
\end{eqnarray}
where $A$ and $B$ are functions of $\tau$ alone and $U^{\mu}$ is a unit 4-vector that obeys $g_{\mu\nu}U^{\mu}U^{\nu}=-1$. With a background perfect fluid radiation era or matter era source of the form
\begin{eqnarray} 
T_{\mu\nu} &=& (\rho+p)U_\mu U_\nu +  p g_{\mu\nu},
\label{12.3}
\end{eqnarray}
where $\rho$ and $p$ are functions of $\tau$, the background Einstein equations are of the form
\begin{eqnarray}
\Delta_{\mu\nu}^{(0)}&=&\tfrac{1}{2} A g_{\mu \nu } -  \tfrac{1}{2} B g_{\mu \nu } + g_{\mu \nu } p + A U_{\mu } U_{\nu } + B U_{\mu } U_{\nu } + p U_{\mu } U_{\nu } + U_{\mu } U_{\nu } \rho=0,
\label{12.4}
\end{eqnarray}
with solution 
\begin{eqnarray}
A &=& -\tfrac{1}{2} (3p+\rho)= -3 \dot{\Omega}^2 \Omega^{-4} + 3 \overset{..}{\Omega} \Omega^{-3}, \quad B= \tfrac{1}{2}(p-\rho)=- \dot{\Omega}^2 \Omega^{-4} -  \overset{..}{\Omega} \Omega^{-3} - 2 k \Omega^{-2}, 
\nonumber\\
\rho &=& \tfrac{1}{2} (- A - 3 B)= 3 \dot{\Omega}^2 \Omega^{-4} + 3 k \Omega^{-2},\quad p = \tfrac{1}{2} (- A + B)
= \dot{\Omega}^2 \Omega^{-4} - 2 \overset{..}{\Omega} \Omega^{-3} -  k \Omega^{-2}.
\label{12.5}
\end{eqnarray}

\subsection{The SVT4 Fluctuations}

While we have incorporated a prefactor of $\Omega^2(\tau)$ in the background metric, we have found it more convenient to not include such a prefactor in the fluctuations. We thus take the background plus fluctuation metric to be of the form
\begin{eqnarray}
ds^2 &=&-[g_{\mu\nu}+h_{\mu\nu}]dx^{\mu}dx^{\nu},\quad h_{\mu\nu}= -2 g_{\mu\nu}\chi + 2\nabla_\mu \nabla_\nu F +\nabla_\mu F_\nu +\nabla_\nu F_\mu+ 2F_{\mu\nu},
\label{12.6}
\end{eqnarray}
where the $\nabla_{\mu}$ derivatives are with respect to the full background $g_{\mu\nu}$, with respect to which $\nabla^{\mu}F_{\mu}=0$, $\nabla^{\mu}F_{\mu\nu}=0$. In analog to our discussion of SVT3 Robertson-Walker fluctuations given above, we set
\begin{eqnarray}
\delta U_{\mu} &=& (V_\mu + \nabla_\mu V) + U_\mu U^\alpha(V_\alpha + \nabla_\alpha V)-U_\mu\left(\tfrac{1}{2} U^\alpha U^\beta h_{\alpha\beta}\right), \quad Q_\mu = F_\mu + \nabla_\mu F, \quad 
\hat{V}= V-U^\alpha Q_\alpha,
\nonumber\\
\delta \hat{\rho}{} &=& \delta \rho-(\rho+p)( Q^{\alpha } U_{\alpha } \nabla_{\beta }U^{\beta }-Q^{\alpha } U^{\beta } \nabla_{\alpha }U_{\beta }),\quad 
\delta \hat{p}{} = \delta p - \tfrac{1}{3} Q^{\alpha } \nabla_{\alpha }(3p+\rho) +  \tfrac{1}{3} (\rho+p) Q^{\alpha } U_{\alpha } \nabla_{\beta }U^{\beta }.
\label{12.7}
\end{eqnarray}
With these definitions and quite a bit of algebra we find that we can write the fluctuation equation $\Delta_{\mu\nu}=0$ as
\begin{eqnarray}
\Delta_{\mu\nu}&=& (g_{\mu \nu } + U_{\mu } U_{\nu }) \delta \hat{p}{} + U_{\mu } U_{\nu } \delta \hat{\rho}{} + \bigl((A -  B) g_{\mu \nu } + 2 (A + B) U_{\mu } U_{\nu }\bigr) \chi \nonumber \\ 
&& - 2 (A + B) U_{\mu } U_{\nu } U^{\alpha } \nabla_{\alpha }\hat{V}{} + 2 g_{\mu \nu } \nabla_{\alpha }\nabla^{\alpha }\chi -  (A + B) U_{\nu } \nabla_{\mu }\hat{V}{} -  (A + B) U_{\mu } \nabla_{\nu }\hat{V}{} \nonumber \\ 
&& - 2 \nabla_{\nu }\nabla_{\mu }\chi -2 (A + B) U_{\mu } U_{\nu } U^{\alpha } V_{\alpha } -  (A + B) U_{\nu } V_{\mu } \nonumber \\ 
&& -  (A + B) U_{\mu } V_{\nu }+2 (A + B) U_{\mu } U_{\nu } U^{\alpha } U^{\beta } F_{\alpha \beta } + 2 (A + B) U_{\nu } U^{\alpha } F_{\mu \alpha } + (\tfrac{1}{3} A + B) F_{\mu \nu } \nonumber \\ 
&& + 2 (A + B) U_{\mu } U^{\alpha } F_{\nu \alpha } + \nabla_{\alpha }\nabla^{\alpha }F_{\mu \nu }=0,
\nonumber\\ 
g^{\mu\nu}\Delta_{\mu\nu}&=& 3 \delta \hat{p}{} -  \delta \hat{\rho}{} + 2 (A - 3 B) \chi + 6 \nabla_{\alpha }\nabla^{\alpha }\chi +2 (A + B) U^{\alpha } U^{\beta } F_{\alpha \beta }=0,
\label{12.8}
\end{eqnarray}
or as
\begin{eqnarray}
\Delta_{\mu\nu}&=& (g_{\mu \nu } + U_{\mu } U_{\nu }) \delta \hat{p}{} + U_{\mu } U_{\nu } \delta \hat{\rho}{} + \bigl(-2 p g_{\mu \nu } - 2 (p + \rho) U_{\mu } U_{\nu }\bigr) \chi + 2 (p + \rho) U_{\mu } U_{\nu } U^{\alpha } \nabla_{\alpha }\hat{V}{} 
\nonumber \\ 
&& + 2 g_{\mu \nu } \nabla_{\alpha }\nabla^{\alpha }\chi + (p + \rho) U_{\nu } \nabla_{\mu }\hat{V}{} + (p + \rho) U_{\mu } \nabla_{\nu }\hat{V}{} - 2 \nabla_{\nu }\nabla_{\mu }\chi +2 (p + \rho) U_{\mu } U_{\nu } U^{\alpha } V_{\alpha } \nonumber \\ 
&& + (p + \rho) U_{\nu } V_{\mu } + (p + \rho) U_{\mu } V_{\nu }-2 (p + \rho) U_{\mu } U_{\nu } U^{\alpha } U^{\beta } F_{\alpha \beta } - 2 (p + \rho) U_{\nu } U^{\alpha } F_{\mu \alpha } -  \tfrac{2}{3} \rho F_{\mu \nu } \nonumber \\ 
&& - 2 (p + \rho) U_{\mu } U^{\alpha } F_{\nu \alpha } + \nabla_{\alpha }\nabla^{\alpha }F_{\mu \nu }=0,
\nonumber\\ 
g^{\mu\nu}\Delta_{\mu\nu}&=& 3 \delta \hat{p}{} -  \delta \hat{\rho}{} + (-6 p + 2 \rho) \chi + 6 \nabla_{\alpha }\nabla^{\alpha }\chi -2 (p + \rho) U^{\alpha } U^{\beta } F_{\alpha \beta }=0.
\label{12.9}
\end{eqnarray}
As written, $\Delta_{\mu\nu}$ only depends on the metric fluctuations $F_{\mu\nu}$ and $\chi$  and  the source fluctuations $\delta \hat{\rho}$,  $\delta \hat{p}$, $\hat{V}$ and $V_i$. Comparing with the SVT3 (\ref{9.13}) to (\ref{9.17})  where there are $\alpha$, $\gamma$, $B_i-\dot{E}_i$ and $E_{ij}$ metric fluctuations and the same set of source fluctuations, we find, just as in the de Sitter background case, that  in a general Robertson-Walker background the SVT4 formalism is far more compact than the SVT3 formalism. 

As a check on our result we note that in a background de Sitter geometry with $\rho=-p=3H^2$, $k=0$, $\Omega=1/\tau H$, $\delta \hat{\rho}=\delta \rho=0$,  $\delta \hat{p}=\delta p=0$, (\ref{12.9}) reduces to 
\begin{eqnarray}
\Delta_{\mu\nu}&=& 6 H^2 g_{\mu \nu } \chi + 2 g_{\mu \nu } \nabla_{\alpha }\nabla^{\alpha }\chi - 2 \nabla_{\nu }\nabla_{\mu }\chi -2 H^2 F_{\mu \nu } + \nabla_{\alpha }\nabla^{\alpha }F_{\mu \nu }=0,
 \nonumber\\ 
g^{\mu\nu}\Delta_{\mu\nu}&=& 24 H^2 \chi + 6 \nabla_{\alpha }\nabla^{\alpha }\chi=0. 
\label{12.10}
\end{eqnarray}
We recognize (\ref{12.10}) as (\ref{6.24}) and (\ref{6.25a}), just as required.

Finally, since the SVT4 fluctuation equations involve the $\nabla_{\alpha }\nabla^{\alpha }$ operator with its curved space harmonic basis functions, as before there will again be no decomposition theorem unless we choose some judicious initial conditions.

\section{Cosmological Perturbations in Conformal Gravity}
\label{S13}

\subsection{SVT3 Conformal Gravity Cosmological Fluctuations}

Since the SVT3 and SVT4 formulations are not contingent on the choice of evolution equations, we complete our study of cosmological fluctuations by discussing how things work in an alternative to standard Einstein gravity, namely conformal gravity.  For SVT3 fluctuations around a Robertson-Walker background in the conformal gravity case  we have found it more convenient not to use the metric given in (\ref{9.1}), viz.
\begin{eqnarray}
ds^2&=&a^2(\tau)\left[d\tau^2-\frac{dr^2}{1-kr^2}-r^2d\theta^2-r^2\sin^2\theta d\phi^2\right],
\label{13.1}
\end{eqnarray}
but to instead take advantage of the fact that via a general coordinate transformation a non-zero $k$ Robertson-Walker metric can be brought into a form in which it is conformal to flat. (With $k=0$ the metric already is conformal to flat.) The needed transformations for $k<0$ and $k>0$ may for instance be found in \cite{Amarasinghe2018}.  
For the illustrative $k<0$ case for instance, it is convenient to set $k=-1/L^2$, and introduce ${\rm sinh} \chi=r/L$ and $p=\tau/L$, with the  metric given in (\ref{13.1}) then taking the form
\begin{eqnarray}
ds^2=L^2a^2(p)\left[dp^2-d\chi^2 -{\rm sinh}^2\chi d\theta^2-{\rm sinh}^2\chi \sin^2\theta d\phi^2\right].
\label{13.2}
\end{eqnarray}
Next we introduce
\begin{eqnarray}
p^{\prime}+r^{\prime}=\tanh[(p+\chi)/2],\quad p^{\prime}-r^{\prime}=\tanh[(p-\chi)/2],\quad p^{\prime}=\frac{\sinh p}{\cosh p+\cosh \chi},\quad r^{\prime}=\frac{\sinh \chi}{\cosh p+\cosh \chi},
\label{13.3}
\end{eqnarray}
so that
\begin{eqnarray}
dp^{\prime 2}-dr^{\prime 2}&=&\frac{1}{4}[dp^2-d\chi^2]{\rm sech}^2[(p+\chi)/2]{\rm sech}^2[(p-\chi)/2],
\nonumber\\
\frac{1}{4}(\cosh p+\cosh \chi)^2&=&{\rm \cosh}^2[(p+\chi)/2]{\rm \cosh}^2[(p-\chi)/2]=\frac{1}{[1-(p^{\prime}+r^{\prime})^2][1-(p^{\prime}-r^{\prime})^2]}.
\label{13.4}
\end{eqnarray}
With these transformations the line element takes the conformal to flat form
\begin{eqnarray}
ds^2=\frac{4L^2a^2(p)}{[1-(p^{\prime}+r^{\prime})^2][1-(p^{\prime}-r^{\prime})^2]}\left[dp^{\prime 2}-dr^{\prime 2} -r^{\prime 2}d\theta^2-r^{\prime 2} \sin^2\theta d\phi^2\right].
\label{13.5}
\end{eqnarray}
The spatial sector can then be written in Cartesian form
\begin{eqnarray}
ds^2=L^2a^2(p)(\cosh p+\cosh \chi)^2\left[dp^{\prime 2}-dx^{\prime 2} -dy^{\prime 2} -dz^{\prime 2}\right],
\label{13.6}
\end{eqnarray}
where $r^{\prime}=(x^{\prime 2}+ y^{\prime 2}+z^{\prime 2})^{1/2}$.  

We note that while our interest in this section is in discussing fluctuations in conformal gravity, a theory that actually has an underlying conformal symmetry, in transforming from (\ref{13.1})  to (\ref{13.6}) we have only made coordinate transformations and have not made any conformal transformation. However, since the $W_{\mu\nu}$ gravitational Bach tensor introduced in (\ref{4.3}) and (\ref{4.4}) above is associated with a conformal theory, under a conformal transformation of the form $g_{\mu\nu}\rightarrow \Omega^{2}(x)g_{\mu\nu}$, $W_{\mu\nu}$ and $\delta W_{\mu\nu}$ respectively transform into $\Omega^{-2}(x)W_{\mu\nu}$ and $\Omega^{-2}(x)\delta W_{\mu\nu}$. Moreover, since this is the case for any background metric that is conformal to flat we need not even restrict to Robertson-Walker or de Sitter, and can consider fluctuations around any background metric of the form 
\begin{eqnarray}
ds^2=\Omega^2(x)[dt^2-\delta_{ij}dx^idx^j],
\label{13.7}
\end{eqnarray}
where $\delta_{ij}$ is the Kronecker delta function and $\Omega(x)$ is a completely arbitrary function of the four $x^{\mu}$ coordinates.

In this background we take the SVT3  background plus fluctuation line element to be of the form
\begin{eqnarray}
ds^2 = \Omega^2(x) \left[ (1+2\phi) dt^2 -2(\tilde{\nabla}_i B +B_i)dt dx^i - [(1-2\psi)\delta_{ij} +2\tilde{\nabla}_i\tilde{\nabla}_j E + \tilde{\nabla}_i E_j + \tilde{\nabla}_j E_i + 2E_{ij}]dx^i dx^j\right],
\label{13.8}
\end{eqnarray}
where $\Omega(x)$ is an arbitrary function of the coordinates, where $\tilde{\nabla}_i=\partial/\partial x^i$ (with Latin index) and  $\tilde{\nabla}^i=\delta^{ij}\tilde{\nabla}_j$ (i.e. not $\Omega^{-2}\delta^{ij}\tilde{\nabla}_j$) are defined with respect to the background 3-space metric $\delta_{ij}$, and where the elements of (\ref{13.8}) obey
\begin{eqnarray}
\delta^{ij}\tilde{\nabla}_j B_i = 0,\quad \delta^{ij}\tilde{\nabla}_j E_i = 0, \quad E_{ij}=E_{ji},\quad \delta^{jk}\tilde{\nabla}_kE_{ij} = 0, \quad \delta^{ij}E_{ij} = 0.
\label{13.9}
\end{eqnarray}
For these fluctuations $\delta W_{\mu\nu}$ is readily calculated, and it is found to have the form \cite{Amarasinghe2018} 
\begin{eqnarray}
\delta W_{00}  &=& -\frac{2}{3\Omega^2} \delta^{mn}\delta^{\ell k}\tilde{\nabla}_m\tilde{\nabla}_n\tilde{\nabla}_{\ell}\tilde{\nabla}_k \alpha,
\nonumber\\	
\delta W_{0i} &=&  -\frac{2}{3\Omega^2} \delta^{mn}\tilde{\nabla}_i\tilde{\nabla}_m\tilde{\nabla}_n\partial_0\alpha
	+\frac{1}{2\Omega^2}\left[\delta^{\ell k}\tilde{\nabla}_{\ell}\tilde{\nabla}_k(\delta^{mn}\tilde{\nabla}_m\tilde{\nabla}_n-\partial_0^2)(B_i - \dot{E}_i)\right],
\nonumber\\	
\delta W_{ij}  &=& \frac{1}{3\Omega^2}\bigg{[} \delta_{ij}\delta^{\ell k}\tilde{\nabla}_{\ell}\tilde{\nabla}_k (\partial_0^2 - \delta^{mn}\tilde{\nabla}_m\tilde{\nabla}_n) 
+(\delta^{\ell k}\tilde{\nabla}_{\ell}\tilde{\nabla}_k -3\partial_0^2)\tilde{\nabla}_i\tilde{\nabla}_j  
\bigg{] }\alpha
\nonumber\\
&&+\frac{1}{2\Omega^2}\left[ \left[\delta^{\ell k}\tilde{\nabla}_{\ell}\tilde{\nabla}_k -\partial_0^2\right]\left[\tilde{\nabla}_i   \partial_0(B_j - \dot{E}_j)+ \tilde{\nabla}_j \partial_0(B_i - \dot{E}_i)\right] \right]
+\frac{1}{\Omega^2}\left[\delta^{mn}\tilde{\nabla}_m\tilde{\nabla}_n-\partial_0^2\right]^2E_{ij}.
\label{13.10}
\end{eqnarray}
where as before $\alpha=\phi + \psi +\dot{B}-\ddot{E}$. We note that the derivatives that appear in (\ref{13.10}) are conveniently with respect to the flat Minkowski metric and not with respect to the full background $ds^2=\Omega^2(x)[dt^2-\delta_{ij}dx^idx^j]$ metric, with the $\Omega(x)$ dependence only appearing as an overall factor. This must be the case since the $\delta W_{\mu\nu}$ given in (\ref{13.10}) is related to the $\delta W_{\mu\nu}$ given in (\ref{4.6}) by an $\Omega^{-2}(x)$  conformal transformation, and the $\delta W_{\mu\nu}$ given in (\ref{4.6}) is associated with fluctuations around flat spacetime.

Unlike the standard gravity case, in a background geometry that is conformal to flat, namely in a background geometry in which the Weyl tensor vanishes,  then according to (\ref{4.3}) the background $W_{\mu\nu}$ will vanish as well. The background $T_{\mu\nu}$ thus vanishes also. Fluctuations are thus described by $\delta T_{\mu\nu}=0$, and thus by $\delta W_{\mu\nu}=0$, with $\delta W_{\mu\nu}$ as given in (\ref{13.10}), and thus $\alpha$, $B_i-\dot{E}_i$ and $E_{ij}$,  thus being gauge invariant  \cite{footnote11}. To check whether a decomposition theorem might hold we thus need to solve the equation $\delta W_{\mu\nu}=0$. To this end we note that since there are derivatives with respect to the purely spatial $\delta^{\ell k}\tilde{\nabla}_{\ell}\tilde{\nabla}_k$, on imposing spatial boundary conditions the relation $\delta W_{00}=0$ immediately sets $\alpha=0$.  With $\alpha=0$, applying the spatial boundary conditions to the relation $W_{0i}=0$ immediately sets $(\delta^{mn}\tilde{\nabla}_m\tilde{\nabla}_n-\partial_0^2)(B_i - \dot{E}_i)=0$, with $\delta W_{ij}=0$ then realizing $\left[\delta^{mn}\tilde{\nabla}_m\tilde{\nabla}_n-\partial_0^2\right]^2E_{ij}=0$. Thus with asymptotic boundary conditions the solution to $\delta W_{\mu\nu}=0$ is
\begin{eqnarray}
\alpha=0,\quad (\delta^{mn}\tilde{\nabla}_m\tilde{\nabla}_n-\partial_0^2)(B_i - \dot{E}_i)=0,\quad \left[\delta^{mn}\tilde{\nabla}_m\tilde{\nabla}_n-\partial_0^2\right]^2E_{ij}=0.
\label{13.11}
\end{eqnarray}
Since decomposition would require
\begin{align}
\delta^{mn}\delta^{\ell k}\tilde{\nabla}_m\tilde{\nabla}_n\tilde{\nabla}_{\ell}\tilde{\nabla}_k \alpha=0, \quad  \delta^{mn}\tilde{\nabla}_i\tilde{\nabla}_m\tilde{\nabla}_n\partial_0\alpha&=0,
\nonumber\\
\delta^{\ell k}\tilde{\nabla}_{\ell}\tilde{\nabla}_k(\delta^{mn}\tilde{\nabla}_m\tilde{\nabla}_n-\partial_0^2)(B_i - \dot{E}_i)=0,\quad
\bigg{[} \delta_{ij}\delta^{\ell k}\tilde{\nabla}_{\ell}\tilde{\nabla}_k (\partial_0^2 - \delta^{mn}\tilde{\nabla}_m\tilde{\nabla}_n) 
+(\delta^{\ell k}\tilde{\nabla}_{\ell}\tilde{\nabla}_k -3\partial_0^2)\tilde{\nabla}_i\tilde{\nabla}_j  
\bigg{] }\alpha&=0,
\nonumber\\
\left[\delta^{\ell k}\tilde{\nabla}_{\ell}\tilde{\nabla}_k -\partial_0^2\right]\left[\tilde{\nabla}_i   \partial_0(B_j - \dot{E}_j)+ \tilde{\nabla}_j \partial_0(B_i - \dot{E}_i)\right]=0,\quad \left[\delta^{mn}\tilde{\nabla}_m\tilde{\nabla}_n-\partial_0^2\right]^2E_{ij}&=0,
\label{13.12}
\end{align}
we see that the decomposition theorem is recovered.

To underscore this result we note that (\ref{13.10}) can be inverted so as to write each gauge-invariant combination as a separate combination of components of $\delta W_{\mu\nu}$, viz.
\begin{eqnarray}
 \Omega^{2}\delta W_{00} &=&- \tfrac{2}{3} \tilde{\nabla}_{b}\tilde{\nabla}^{b}\tilde{\nabla}_{a}\tilde{\nabla}^{a}\alpha,
 \label{13.13}
\end{eqnarray}
\begin{eqnarray}
\tilde\nabla_a\tilde\nabla^a(\Omega^2 \delta W_{0i}) - \tilde\nabla_i \tilde\nabla^a(\Omega^2  \delta W_{0a})&=&- \tfrac{1}{2} \tilde{\nabla}_{b}\tilde{\nabla}^{b}\tilde{\nabla}_{a}\tilde{\nabla}^{a}(\overset{..}{B}_{i}-\overset{...}{E}_{i}) + \tfrac{1}{2} \tilde{\nabla}_{c}\tilde{\nabla}^{c}\tilde{\nabla}_{b}\tilde{\nabla}^{b}\tilde{\nabla}_{a}\tilde{\nabla}^{a}(B_{i} -\dot{E}_{i}),
\label{13.14}
\end{eqnarray}
\begin{eqnarray}
&&\tilde\nabla_a \tilde\nabla^a \big[ \tilde\nabla_b \tilde\nabla^b (\Omega^2 \delta W_{ij})- \tilde\nabla_i \tilde\nabla^l(\Omega^2  \delta W_{jl}) -  \tilde\nabla_j \tilde\nabla^l(\Omega^2  \delta W_{il})\big]+\tfrac{1}{2}\delta_{ij}\tilde\nabla_a \tilde\nabla^a\big[  \tilde\nabla^k \tilde\nabla^l(\Omega^2  \delta W_{kl})-\tilde\nabla_b \tilde\nabla^b(\Omega^2\delta^{ab}\delta W_{ab})\big]
\nonumber\\
&&+\tfrac{1}{2} \tilde\nabla_i\tilde\nabla_j \big[ \tilde\nabla^k \tilde\nabla^l(\Omega^2  \delta W_{kl}) + \tilde\nabla_a \tilde\nabla^a(\Omega^2 \delta^{ab}\delta W_{ab})\big] 
=\tilde\nabla_a \tilde\nabla^a \tilde\nabla_b \tilde\nabla^b\left[\tilde\nabla_c \tilde\nabla^c - \partial_0^2\right]^2E_{ij}.
\label{13.15}
\end{eqnarray}
On setting $\delta W_{\mu\nu}=0$, each of these relations can now be integrated separately, with the decomposition theorem then following. 

For completeness we also note that for  SVT3 fluctuations around the (\ref{9.2}) background $k\neq 0$ Robertson-Walker metric given in (\ref{9.1}) with $\tilde{\gamma}_{ij}dx^idx^j=dr^2/(1-kr^2)+r^2d\theta^2+r^2\sin^2\theta d\phi^2$ and with $\Omega$ actually now more generally being an arbitrary function of $\tau$ and $x^i$, the conformal gravity $\delta W_{\mu\nu}$ is given by 
\begin{eqnarray}
\delta W_{00}&=& - \tfrac{2}{3} \Omega^{-2} (\tilde\nabla_a\tilde\nabla^a + 3k)\tilde\nabla_b\tilde\nabla^b \alpha,
 \nonumber\\ 
\delta W_{0i}&=& -\tfrac{2}{3} \Omega^{-2}  \tilde\nabla_i (\tilde\nabla_a\tilde\nabla^a + 3k)\dot\alpha
+\tfrac12 \Omega^{-2}\left[ \tilde\nabla_a\tilde\nabla^a (\tilde\nabla_b \tilde\nabla^b-\partial_0^2)(B_i-\dot{E}_i) -2k(2k+\partial_0^2)(B_i-\dot{E}_i)\right],
  \nonumber\\ 
\delta W_{ij}&=& -\tfrac{1}{3} \Omega^{-2} \left[ \tilde{\gamma}_{ij} \tilde\nabla_a\tilde\nabla^a (\tilde\nabla_b \tilde\nabla^b +2k-\partial_0^2)\alpha - \tilde\nabla_i\tilde\nabla_j(\tilde\nabla_a\tilde\nabla^a - 3\partial_0^2)\alpha \right]
\nonumber\\
&& +\tfrac{1}{2} \Omega^{-2} \left[ \tilde\nabla_i ( \tilde\nabla_a\tilde\nabla^a -2k-\partial_0^2) \partial_0 (B_j-\dot{E}_j) 
+  \tilde\nabla_j ( \tilde\nabla_a\tilde\nabla^a -2k-\partial_0^2) \partial_0 (B_i-\dot{E}_i)\right]
\nonumber\\
&&+ \Omega^{-2}\left[ (\tilde\nabla_a\tilde\nabla^a-\partial_0)^2 E_{ij} - 4k (\tilde\nabla_a\tilde\nabla^a - k-2\partial_0^2)E_{ij} \right],
\label{13.16}
\end{eqnarray}
where again $\alpha = \phi + \psi + \dot B - \ddot E$. These equations can be inverted, and yield
\begin{align}
(\Omega^2\delta W_{00})= - \tfrac{2}{3}  (\tilde\nabla_a\tilde\nabla^a + 3k)\tilde\nabla_b\tilde\nabla^b \alpha,&
\nonumber\\
(\tilde\nabla_a\tilde\nabla^a-2k)(\Omega^2\delta W_{0i}) - \tilde\nabla_i \tilde\nabla^a (\Omega^2\delta W_{0a}) =
\tfrac{1}{2} (\tilde\nabla_a\tilde\nabla^a - 2k - \partial_0^2)(\tilde\nabla_b\tilde\nabla^b + 2k)(\tilde\nabla_c\tilde\nabla^c -2k)(B_i-\dot{E}_i),&
\nonumber\\
(\tilde\nabla_a\tilde\nabla^a-2k)(\tilde\nabla_b\tilde\nabla^b-3k)(\Omega^2\delta W_{ij})
+ \tfrac{1}{2} \tilde\nabla_i\tilde\nabla_j\big[ \tilde\nabla^a\tilde\nabla^b (\Omega^2\delta W_{ab}) + (\tilde\nabla_a\tilde\nabla^a +4k)(\tilde{\gamma}^{bc}(\Omega^2\delta W_{bc}))\big]&
\nonumber\\
+\tfrac{1}{2} \tilde{\gamma}_{ij} \big[ (\tilde\nabla_a\tilde\nabla^a-4k)\tilde\nabla^b\tilde\nabla^c (\Omega^2\delta W_{bc})-(\tilde\nabla_a\tilde\nabla^a\tilde\nabla_b\tilde\nabla^b -2k \tilde\nabla_a\tilde\nabla^a +4k^2)(\tilde{\gamma}^{bc}(\Omega^2\delta W_{bc}))\big]&
\nonumber\\
-(\tilde\nabla_a\tilde\nabla^a -3k)(\tilde\nabla_i\tilde\nabla^b (\Omega^2\delta W_{jb}) + \tilde\nabla_j \tilde\nabla^b (\Omega^2\delta W_{ib}))&
\nonumber\\
=(\tilde\nabla_a\tilde\nabla^a-2k)(\tilde\nabla_b\tilde\nabla^b-3k)\left[ (\tilde\nabla_a\tilde\nabla^a-\partial_0)^2 E_{ij} - 4k (\tilde\nabla_a\tilde\nabla^a - k-2\partial_0^2)E_{ij} \right].&
\label{13.17}
\end{align}
With this separation of the gauge-invariant combinations we again have the decomposition theorem.

\subsection{SVT4 Conformal Gravity Cosmological Fluctuations} 

For conformal gravity SVT4 fluctuations associated with the metric $g_{\mu\nu}+h_{\mu\nu}$ where the background metric $g_{\mu\nu}$ is of the conformal to flat form given in (\ref{13.7}), we recall that for completely arbitrary conformal factor $\Omega(x)$ the fluctuation $\delta W_{\mu\nu}$ is given by the remarkably simple expression  \cite{Amarasinghe2018}

\begin{eqnarray}
\delta W_{\mu\nu}&=&\frac{1}{2}\Omega^{-2}\bigg{(}\partial_{\sigma}\partial^{\sigma}\partial_{\tau}\partial^{\tau}[\Omega^{-2}K_{\mu\nu}]
-\partial_{\sigma}\partial^{\sigma}\partial_{\mu}\partial^{\alpha}[\Omega^{-2}K_{\alpha\nu}]
-\partial_{\sigma}\partial^{\sigma}\partial_{\nu}\partial^{\alpha}[\Omega^{-2}K_{\alpha\mu}]
\nonumber\\
&+&\frac{2}{3}\partial_{\mu}\partial_{\nu}\partial^{\alpha}\partial^{\beta}[\Omega^{-2}K_{\alpha\beta}]+\frac{1}{3}\eta_{\mu\nu}\partial_{\sigma}\partial^{\sigma}\partial^{\alpha}\partial^{\beta}[\Omega^{-2}K_{\alpha\beta}]\bigg{)},
\label{13.18}
\end{eqnarray} 
where all derivatives are four-dimensional derivatives with respect to a flat Minkowski metric, and where $K_{\mu\nu}$ is given by $K_{\mu\nu}=h_{\mu\nu}-(1/4)g_{\mu\nu}g^{\alpha\beta}h_{\alpha\beta}$. If we now make the SVT4 expansion
\begin{eqnarray}
h_{\mu\nu}=\Omega^2(x)\left[-2\eta_{\mu\nu}\chi+2\partial_{\mu}\partial_{\nu}F
+ \partial_{\mu}F_{\nu}+\partial_{\nu}F_{\mu}+2F_{\mu\nu}\right],
\label{13.19}
\end{eqnarray}
where the derivatives and the transverse and tracelessness  $\partial^{\mu}F_{\mu}=0$, $\partial^{\nu}F_{\mu\nu}=0$, $\eta^{\mu\nu}F_{\mu\nu}=0$ conditions are with respect to a flat Minkowski background, we find that (\ref{13.18}) reduces to
\begin{eqnarray}
\delta W_{\mu\nu}&=&\Omega^{-2}\partial_{\sigma}\partial^{\sigma}\partial_{\tau}\partial^{\tau}F_{\mu\nu}.
\label{13.20}
\end{eqnarray} 
This expression is remarkable not just in its simplicity but in the fact that all components of $F_{\mu\nu}$ are completely decoupled from each other, with (\ref{13.20}) being diagonal in the $\mu,\nu$ indices. Since (\ref{13.20}) only contains $F_{\mu\nu}$ with none of $\chi$, $F$ or $F_{\mu}$ appearing  in it, unlike in the Einstein gravity SVT4 case where one needs initial conditions to establish the decomposition theorem, in the conformal gravity SVT4 case the decomposition theorem is automatic.

\section{The Decomposition Theorem is for Gauge Invariant Quantities and not for Separate Scalar, Vector and Tensor Components}
\label{S14}

In the SVT4 study of fluctuations around a de Sitter background that we presented in Sec. \ref{S6} we had found that one of the gauge-invariant combinations was given by $\alpha=\dot{F}+\tau \chi +F_0$ (see (\ref{6.54})). In this combination $F$ and $\chi$ are scalars while $F_0$ is the fourth component of the vector $F_{\mu}$. In solving the fluctuation equations in this case we actually solved for the gauge-invariant combinations and not for the individual scalar, vector and tensor sectors. In the solution we found that $\alpha=0$.  Thus would entail only that  $\dot{F}+\tau \chi =-F_0$. However, decomposition with respect to scalars, vectors and tensors would in addition entail that $\dot{F}+\tau \chi=0$, and $F_0=0$, something that would not be warranted as it is not required by the fluctuation equations, while moreover not being a gauge-invariant decomposition of the 
gauge-invariant $\alpha$. Thus given this example we in general see that one should only look for a decomposition theorem for gauge-invariant combinations and not look for one for the separate scalar, vector and tensor sectors as gauge invariance can in general intertwine them. Since it might perhaps be thought that this is an artifact of using SVT4 we now present two examples in which it occurs in SVT3. One is fluctuations around an anti de Sitter background, and the other is fluctuations around a completely general conformal to flat background.

\subsection{Fluctuations Around an Anti de Sitter Background}
\label{S14a}
For an anti de Sitter background in four dimensions we have
\begin{eqnarray}
ds^2 &=& \Omega^2(z)\left[ dt^2 - dx^2-dy^2-dz^2\right]= -g_{\mu\nu}dx^{\mu} dx^{\nu},\quad
\Omega(z) = \frac{1}{Hz},
\nonumber\\
R_{\lambda\mu\nu\kappa} &=& -H^2(g_{\mu\nu}g_{\lambda\kappa} -g_{\lambda\nu}g_{\mu\kappa}),
\quad R_{\mu\kappa} =3H^2 g_{\mu\kappa},\quad R = 12H^2,
\nonumber\\
G_{\mu\nu} &=& -3H^2 g_{\mu\nu},\quad T_{\mu\nu} = 3H^2 g_{\mu\nu}.
\label{14.1}
\end{eqnarray}
We take the fluctuations to have the standard SVT3 form given in (\ref{1.42a}), viz. 
\begin{eqnarray}
ds^2 &=&- \Omega^2(z)\left( \eta_{\mu\nu}+ f_{\mu\nu}\right) dx^\mu dx^\nu,
\nonumber\\
f_{00} &=& -2 \phi,\quad f_{0i} = \tilde\nabla_i B + B_i,\quad \tilde{\nabla}^iB_i=0,
\nonumber\\
f_{ij} &=& -2 \psi \delta_{ij} + 2\tilde\nabla_i\tilde\nabla_j E + \tilde\nabla_i E_j
+ \tilde\nabla_i E_j + 2E_{ij},\quad \tilde{\nabla}^iE_i=0,\quad \tilde{\nabla}^iE_{ij}=0,\quad \delta^{ij}E_{ij}=0, 
\label{14.2}
\end{eqnarray}
where $\tilde{\nabla}_i$ and $\tilde{\nabla}^i=\delta^{ij}\tilde{\nabla}_j$ are defined with respect to a flat three-dimensional background $\eta_{ij}dx^idx^j=\delta_{ij}dx^idx^j$.

On defining
\begin{eqnarray}
\alpha &=& \phi +\psi+\dot B - \ddot E, \quad \delta = \phi -\psi + \dot B - \ddot E + \frac{2}{z}(\tilde\nabla_3 E + E_3),
\label{14.3}
\end{eqnarray}
following some algebra we find that the components of $\Delta_{\mu\nu}=\delta G_{\mu\nu}+\delta T_{\mu\nu}=\delta G_{\mu\nu}+ 3\Omega^2 H^2 f_{\mu\nu}$ are given by 
\begin{eqnarray}
g^{\mu\nu}\Delta_{\mu\nu}&=& -12 H^2 \alpha - 3 H^2 z^2 \overset{..}{\alpha} + 3 H^2 z^2 \overset{..}{\delta} + 12 H^2 \delta + H^2 z^2 \tilde\nabla^{2}{}\alpha - 3 H^2 z^2 \tilde\nabla^{2}{}\delta 
\nonumber \\ 
&& + 6 H^2 z \tilde{\nabla}_{3}\delta +6 H^2 z (\dot{B}_3-\ddot{E}_3)+24 H^2 E_{33},
  \nonumber\\ 
\delta^{ij}\Delta_{ij}&=& -9 z^{-2} \alpha - 3 \overset{..}{\alpha} + 3 \overset{..}{\delta} + 9 z^{-2} \delta - 2 \tilde\nabla^{2}{}\delta + z^{-1} \tilde{\nabla}_{3}\alpha + 5 z^{-1} \tilde{\nabla}_{3}\delta +6 z^{-1} (\dot{B}_3-\ddot{E}_3)+18 z^{-2} E_{33},
 \nonumber\\ 
\Delta_{00}&=& 3 z^{-2} \alpha - 3 z^{-2} \delta -  \tilde\nabla^{2}{}\alpha + \tilde\nabla^{2}{}\delta + z^{-1} \tilde{\nabla}_{3}\alpha -  z^{-1} \tilde{\nabla}_{3}\delta -6 z^{-2} E_{33},
 \nonumber\\ 
\Delta_{11}&=& -3 z^{-2} \alpha -  \overset{..}{\alpha} + \overset{..}{\delta} + 3 z^{-2} \delta -  \tilde\nabla^{2}{}\delta + \tilde{\nabla}_{1}\tilde{\nabla}_{1}\delta + z^{-1} \tilde{\nabla}_{3}\alpha + z^{-1} \tilde{\nabla}_{3}\delta +2 z^{-1} (\dot{B}_3-\ddot{E}_3) + \tilde{\nabla}_{1}(\dot{B}_1-\ddot{E}_1)
\nonumber \\ 
&&- \overset{..}{E}_{11} + 6 z^{-2} E_{33} + \tilde\nabla^{2}{}E_{11} + 4 z^{-1} \tilde{\nabla}_{1}E_{13} - 2 z^{-1} \tilde{\nabla}_{3}E_{11},
 \nonumber\\ 
\Delta_{22}&=& -3 z^{-2} \alpha -  \overset{..}{\alpha} + \overset{..}{\delta} + 3 z^{-2} \delta -  \tilde\nabla^{2}{}\delta + \tilde{\nabla}_{2}\tilde{\nabla}_{2}\delta + z^{-1} \tilde{\nabla}_{3}\alpha + z^{-1} \tilde{\nabla}_{3}\delta +2 z^{-1} (\dot{B}_3-\ddot{E}_3) + \tilde{\nabla}_{2}(\dot{B}_2-\ddot{E}_2) 
\nonumber \\ 
&& - \overset{..}{E}_{22}+ 6 z^{-2} E_{33} + \tilde\nabla^{2}{}E_{22} + 4 z^{-1} \tilde{\nabla}_{2}E_{23} - 2 z^{-1} \tilde{\nabla}_{3}E_{22},
 \nonumber\\ 
\Delta_{33}&=& -3 z^{-2} \alpha -  \overset{..}{\alpha} + \overset{..}{\delta} + 3 z^{-2} \delta -  \tilde\nabla^{2}{}\delta -  z^{-1} \tilde{\nabla}_{3}\alpha + 3 z^{-1} \tilde{\nabla}_{3}\delta + \tilde{\nabla}_{3}\tilde{\nabla}_{3}\delta +2 z^{-1} (\dot{B}_3-\ddot{E}_3) + \tilde{\nabla}_{3}(\dot{B}_3-\ddot{E}_3)
\nonumber \\ 
&&  -\overset{..}{E}_{33}+ 6 z^{-2} E_{33} + \tilde\nabla^{2}{}E_{33} + 2 z^{-1} \tilde{\nabla}_{3}E_{33},
 \nonumber\\ 
\Delta_{01}&=& - \tilde{\nabla}_{1}\dot{\alpha} + \tilde{\nabla}_{1}\dot{\delta}+\tfrac{1}{2} \tilde\nabla^{2}{}(B_1-\dot{E}_1) + z^{-1} \tilde{\nabla}_{1}(B_3-\dot{E}_3) -  z^{-1} \tilde{\nabla}_{3}(B_1-\dot{E}_1)+2 z^{-1} \dot{E}_{13},
 \nonumber\\ 
\Delta_{02}&=& - \tilde{\nabla}_{2}\dot{\alpha} + \tilde{\nabla}_{2}\dot{\delta}+\tfrac{1}{2} \tilde\nabla^{2}{}(B_2-\dot{E}_2) + z^{-1} \tilde{\nabla}_{2}(B_3-\dot{E}_3) -  z^{-1} \tilde{\nabla}_{3}(B_2-\dot{E}_2)+2 z^{-1} \dot{E}_{23},
 \nonumber\\ 
\Delta_{03}&=& - z^{-1} \dot{\alpha} + z^{-1} \dot{\delta} -  \tilde{\nabla}_{3}\dot{\alpha} + \tilde{\nabla}_{3}\dot{\delta}+\tfrac{1}{2} \tilde\nabla^{2}{}(B_3-\dot{E}_3)+2 z^{-1} \dot{E}_{33},
 \nonumber\\ 
\Delta_{12}&=& \tilde{\nabla}_{2}\tilde{\nabla}_{1}\delta +\tfrac{1}{2} \tilde{\nabla}_{1}(\dot{B}_2-\ddot{E}_2) + \tfrac{1}{2} \tilde{\nabla}_{2}(\dot{B}_1-\ddot{E}_1)- \overset{..}{E}_{12} + \tilde\nabla^{2}{}E_{12} + 2 z^{-1} \tilde{\nabla}_{1}E_{23} + 2 z^{-1} \tilde{\nabla}_{2}E_{13} - 2 z^{-1} \tilde{\nabla}_{3}E_{12},
 \nonumber\\ 
\Delta_{13}&=& - z^{-1} \tilde{\nabla}_{1}\alpha + z^{-1} \tilde{\nabla}_{1}\delta + \tilde{\nabla}_{3}\tilde{\nabla}_{1}\delta +\tfrac{1}{2} \tilde{\nabla}_{1}(\dot{B}_3-\ddot{E}_3) + \tfrac{1}{2} \tilde{\nabla}_{3}(\dot{B}_1-\ddot{E}_1)- \overset{..}{E}_{13} + \tilde\nabla^{2}{}E_{13} + 2 z^{-1} \tilde{\nabla}_{1}E_{33},
 \nonumber\\ 
\Delta_{23}&=& - z^{-1} \tilde{\nabla}_{2}\alpha + z^{-1} \tilde{\nabla}_{2}\delta + \tilde{\nabla}_{3}\tilde{\nabla}_{2}\delta +\tfrac{1}{2} \tilde{\nabla}_{2}(\dot{B}_3-\ddot{E}_3) + \tfrac{1}{2} \tilde{\nabla}_{3}(\dot{B}_2-\ddot{E}_2)- \overset{..}{E}_{23} + \tilde\nabla^{2}{}E_{23} + 2 z^{-1} \tilde{\nabla}_{2}E_{33},
\label{14.4} 
\end{eqnarray}
where $\tilde\nabla^{2} = \delta^{ab} \tilde\nabla_a\tilde\nabla_b$. With $\Delta_{\mu\nu}$ being gauge invariant we recognize $\alpha$, $\delta$, $B_i-\dot{E}_i$ and $E_{ij}$ as being gauge invariant. We thus see that one of the gauge-invariant combinations, viz. $\delta$, depends on both scalars and vectors. Since our only purpose here is in establishing that one of the gauge-invariant SVT3 combinations does depend on both scalars and vectors, we shall not seek to solve $\Delta_{\mu\nu}=0$ in this particular case. Though if we were to we would only find expressions for $\alpha$, $\delta$, $B_i-\dot{E}_i$ and $E_{ij}$, and not for the separate scalar and vector components of $\delta$.

\subsection{Fluctuations Around a General Conformal to Flat Background}
\label{S14b}

In \cite{Amarasinghe2018} it was shown that for the arbitrary conformal to flat SVT3 fluctuations given in (\ref{1.42a}), viz.
\begin{eqnarray}
ds^2 &=&\Omega^2({\bf x},t)\left[(1+2\phi) dt^2 -2(\partial_i B +B_i)dt dx^i - [(1-2\psi)\delta_{ij} +2\partial_i\partial_j E + \partial_i E_j + \partial_j E_i + 2E_{ij}]dx^i dx^j\right]
\label{14.5}
\end{eqnarray}
with general $\Omega({\bf x},t)$, the metric sector gauge-invariant combinations are
\begin{eqnarray}
\alpha &=& \phi +\psi+\dot B - \ddot E, \quad \eta=\psi -\Omega^{-1}\dot{\Omega}(B-\dot E)+\Omega^{-1}\tilde\nabla^i\Omega(E_i+\tilde\nabla_i E),\quad B_i-\dot E_i,\quad E_{ij}.
\label{14.6}
\end{eqnarray}
Of these invariant combinations three are independent of $\Omega$ altogether and have been encountered frequently throughout this study, while only one, viz. $\eta$, actually depends on $\Omega$ at all. (For specific choices of $\Omega$ the quantity $-\Omega\dot{\Omega}^{-1}\eta$ reduces to the previously introduced $\beta$ in the de Sitter (\ref{7.3}) and to $\gamma$ in the Robertson-Walker (\ref{8.7}) and (\ref{9.2}), while $\alpha-2\eta$ reduces to the anti de Sitter $\delta$ given in (\ref{14.3}).) The invariant $\alpha$ involves scalars alone, the invariant $B_i-E_i$ involves vectors alone, the invariant $E_{ij}$  involves tensors alone, and only the invariant $\eta$ actually involves more than just one of the scalar, vector and tensor sets of modes, with it specifically involving both scalars and vectors. While $\eta$ must always involve scalars, if $\Omega$ has a spatial dependence $\eta$ will also involve the vector $E_i$. A spatial dependence for $\Omega$ is  encountered in our study of anti de Sitter fluctuations as shown in (\ref{14.3}), and is also encountered in SVT3 fluctuations around a Robertson-Walker background with $k\neq 0$,  where the background Robertson-Walker metric as shown in (\ref{13.6}) for $k<0$ (and in \cite{Amarasinghe2018} for $k>0$) is written in a conformal to flat form, with the conformal factor expressly being a function of both time and space coordinates. Thus in such cases we must expect $\Delta_{\mu\nu}$ to depend on $\eta$ itself and not be separable in separate scalar and vector sectors. While this issue is met for $k\neq 0$ Robertson-Walker fluctuations when the background metric is written in the conformal to flat form given in (\ref{13.6}), we note that it is not in fact met for fluctuations around a background Robertson-Walker geometry with metric $ds^2=\Omega^2(\tau)[d\tau^2-dr^2/(1-kr^2)-r^2d\theta^2-r^2\sin^2\theta d\phi^2]$ as given in (\ref{9.1}), since with $\Omega$ only depending on $\tau$ in that case, the gauge-invariant $\gamma = - \dot\Omega^{-1}\Omega \psi + B - \dot E$ as given in (\ref{9.12}) does not involve the vector sector modes. While one can of course transform the background metric given in (\ref{13.6}) into the background metric given in (\ref{9.1}) by a coordinate transformation with fluctuations around the two metrics thus describing the same physics, the very structure of (\ref{14.6}) shows that one cannot simply separate in scalar, vector tensor components at will. Rather one must first separate in gauge-invariant combinations, and only if these combinations turn out not to intertwine any of the scalar, vector and tensor sectors could one then separate in each of the scalar, vector and tensor sectors. Moreover, while one can find a form for the background metric in which there is no such intertwining in the $k\neq 0$ Robertson-Walker case (for $k=0$ $\Omega$ only depends on $\tau$ and so there is no intertwining), this only occurs because of the specific purely $\tau$-dependent form that  the $k\neq 0$ $\Omega$ just happens to take.  For more complicated $\Omega$ there would be no coordinate transformation that would eliminate the intertwining, and so it is of interest to study the spatially dependent $\Omega$ situation in and of itself.

While we of course do not need to explicitly solve for fluctuations around a $k\neq 0$ Robertson-Walker metric when written in a conformal to flat form since in Secs. \ref{S9}, \ref{S10}, and \ref{S11} we already have solved for fluctuations around the same geometry when written in the general coordinate equivalent form given in (\ref{9.1}), it is nonetheless of interest to explore the structure of fluctuations around the conformal to flat form for a $k\neq 0$ Robertson-Walker geometry. In particular it is of interest to show that the $\eta$ invariant given in (\ref{14.6}) actually behaves quite differently from all the other invariants.   In (\ref{9.43a}) to (\ref{9.47a}) we had obtained some kinematic relations (i.e., relations that do not involve the evolution equations) that express the gauge-invariant combinations in terms of the  $f_{\mu\nu}$ components of the fluctuation metric. Inspection of these relations and of (\ref{9.12}) shows there is  only one, viz. that for the relevant $\eta$ in that case, that depends on $\Omega$. Now the relations given  in (\ref{9.43a}) to (\ref{9.47a}) were derived for fluctuations around the (\ref{9.1}) metric. If we now set $k=0$ in these relations so that the $\tilde{\nabla}$ derivative now refers to flat spacetime, we would anticipate that for fluctuations around (\ref{13.6}) the relations for the gauge-invariant combinations that do not involve $\Omega$ might be replaced by 

\begin{align}
\tilde{\nabla}^b\tilde{\nabla}_b\tilde{\nabla}^a\tilde{\nabla}_a\alpha&=-\frac{1}{2}\tilde{\nabla}^b\tilde{\nabla}_b\tilde{\nabla}^i\tilde{\nabla}_if_{00}
+\frac{1}{4}\tilde{\nabla}^a\tilde{\nabla}_a\left(-\tilde{\nabla}^b\tilde{\nabla}_bf+\tilde{\nabla}^m\tilde{\nabla}^nf_{mn}\right)
+\partial_0\tilde{\nabla}^b\tilde{\nabla}_b\tilde{\nabla}^if_{0i}
\nonumber\\
&-\frac{1}{4}\partial^2_0\left(3\tilde{\nabla}^m\tilde{\nabla}^nf_{mn}-\tilde{\nabla}^a\tilde{\nabla}_af\right),
\label{14.7}
\end{align}
\begin{align}
&\tilde{\nabla}^a\tilde{\nabla}_a\tilde{\nabla}^i\tilde{\nabla}_i(B_j-\dot{E_j})=\tilde{\nabla}^i\tilde{\nabla}_i (\tilde{\nabla}^a\tilde{\nabla}_af_{0j}-\tilde{\nabla}_j\tilde{\nabla}^af_{0a})
-\partial_0\tilde{\nabla}^a\tilde{\nabla}_a\tilde{\nabla}^if_{ij}
+\partial_0\tilde{\nabla}_j\tilde{\nabla}^a\tilde{\nabla}^bf_{ab},
\label{14.8}
\end{align}
\begin{align}
2\tilde{\nabla}^a\tilde{\nabla}_a\tilde{\nabla}^b\tilde{\nabla}_bE_{ij}
&=\tilde{\nabla}^a\tilde{\nabla}_a\tilde{\nabla}^b\tilde{\nabla}_bf_{ij}
+\tfrac{1}{2}\tilde{\nabla}_i\tilde{\nabla}_j\left[\tilde{\nabla}^a\tilde{\nabla}^bf_{ab}+\tilde{\nabla}^a\tilde{\nabla}_af\right]-\tilde{\nabla}^a\tilde{\nabla}_a(\tilde{\nabla}_i\tilde{\nabla}^bf_{jb}+\tilde{\nabla}_j\tilde{\nabla}^bf_{ib})
\nonumber\\
&+\tfrac{1}{2}\tilde{\gamma}_{ij}\left[\tilde{\nabla}^a\tilde{\nabla}_a\tilde{\nabla}^b\tilde{\nabla}^cf_{bc}
-\tilde{\nabla}_a\tilde{\nabla}^a\tilde{\nabla}_b\tilde{\nabla}^bf\right],
\label{14.9}
\end{align}
(where $f=\delta^{ab}f_{ab}$), with $\alpha$ still being given by (\ref{14.6}). Explicit calculation shows that this anticipation is actually borne out, with (\ref{14.7}), (\ref{14.8}) and (\ref{14.9}) being found to hold for the fluctuations given in (\ref{14.5}), no matter how arbitrary $\Omega$ might be. 

Now in our discussion of the fluctuations associated with the conformal gravity $\delta W_{\mu\nu}$ we had obtained the relations given in (\ref{13.10}) and their inversion as given in (\ref{13.13}), (\ref{13.14}) and (\ref{13.15}). We now note that these relations involve the same gauge-invariant combinations as the ones that appear in (\ref{14.7}), (\ref{14.8}) and (\ref{14.9}), viz. $\alpha$, $B_i-\dot{E_i}$ and $E_{ij}$, with $\eta$ not appearing. That $\eta$ could not appear in $\delta W_{\mu\nu}$ is because $W_{\mu\nu}$ is traceless so that on allowing for four coordinate transformations $\delta W_{\mu\nu}$ can only involve five quantities (a one-component $\alpha$, a two-component $B_i-\dot{E}_i$, and a two-component $E_{ij}$). In this sense then we should think of $\alpha$, $B_i-\dot{E_i}$ and $E_{ij}$ as a unit, with $\eta$ needing to be treated independently.

Since $W_{\mu\nu}$ is zero in a conformal to flat background, it is associated with a background $T_{\mu\nu}$ that is zero, with $\delta W_{\mu\nu}$ then being gauge invariant on its own as $\delta T^{\mu\nu}$ is then zero. Thus to determine a gauge-invariant relation that does involve $\eta$ we should look for a purely geometric gauge-invariant fluctuation relation that does not involve $\delta T_{\mu\nu}$. However, none is immediately available as we have already used up $\delta W_{\mu\nu}$, and in general a fluctuation such as $\delta G_{\mu\nu}$ would not be gauge invariant on its own. However, there is one situation in which not $\delta G_{\mu\nu}$ but  $\delta(g^{\mu\nu}G_{\mu\nu})$ is gauge invariant on its own, namely in the radiation era where $T_{\mu\nu}$, and thus $G_{\mu\nu}$, are traceless, with $\delta (g^{\mu\nu}T_{\mu\nu})$ then being zero, and with the quantity $\delta(g^{\mu\nu}G_{\mu\nu})$ then being gauge invariant on its own.

Thus in a radiation era conformal to flat $k\neq 0$ Robertson-Walker background case as given by (\ref{14.5}) we evaluate 
\begin{eqnarray}
g^{\mu\nu}G_{\mu\nu}=6\ddot{\Omega}\Omega^{-3}-6\Omega^{-3}\tilde{\nabla}_a\tilde{\nabla}^a\Omega=0,
\label{14.10}
\end{eqnarray}
and on setting $\ddot{\Omega}=\tilde{\nabla}_a\tilde{\nabla}^a\Omega$ obtain 
\begin{eqnarray}
\delta(g^{\mu\nu} G_{\mu\nu})&=& -6 \dot{\alpha} \dot{\Omega} \Omega^{-3} - 12 \dot{\eta} \dot{\Omega} \Omega^{-3} - 12 \overset{..}{\Omega} \alpha \Omega^{-3} - 6 \overset{..}{\eta} \Omega^{-2} - 2 \Omega^{-2} \tilde{\nabla}_{a}\tilde{\nabla}^{a}\alpha + 6 \Omega^{-2} \tilde{\nabla}_{a}\tilde{\nabla}^{a}\eta  - 6 \Omega^{-3} \tilde{\nabla}_{a}\Omega \tilde{\nabla}^{a}\alpha
\nonumber \\ 
&& + 12 \Omega^{-3} \tilde{\nabla}_{a}\Omega \tilde{\nabla}^{a}\eta -12 (B^{a}-\dot{E}^a) \Omega^{-3} \tilde{\nabla}_{a}\dot{\Omega} - 6 (\dot{B}^{a}-\ddot{E}^a) \Omega^{-3} \tilde{\nabla}_{a}\Omega +12 E^{ab} \Omega^{-3} \tilde{\nabla}_{b}\tilde{\nabla}_{a}\Omega,
\label{14.11}
\end{eqnarray}
where $\alpha$, $B_i-\dot{E}_i$ and $E_{ij}$ are as before, with $\eta$ now being given by the form given in (\ref{14.6}). We thus establish that in the conformal to flat case the appropriate $\eta$ is indeed given by the spatially-dependent $\eta=\psi -\Omega^{-1}\dot{\Omega}(B-\dot E)+\Omega^{-1}\tilde\nabla^i\Omega(E_i+\tilde\nabla_i E)$, just as required.

\subsection{Taking Advantage of the Gauge Freedom}
\label{14c}

In \cite{Amarasinghe2018} infinitesimal gauge transformations of the form 
\begin{eqnarray}
\bar{h}_{\mu\nu}=h_{\mu\nu}-\nabla _{\mu}\epsilon_{\nu}-\nabla _{\nu}\epsilon_{\mu}
\label{14.12}
\end{eqnarray}
acting on the conformal to flat (\ref{14.5}) with arbitrary $\Omega$ were considered. On introducing gauge parameters
\begin{eqnarray}
\epsilon_{\mu}=\Omega^2(x)f_{\mu},\quad f_{0}=-T,\quad f_i=L_i+\tilde{\nabla}_iL,\quad \delta^{ij}\tilde{\nabla}_jL_i=\tilde{\nabla}^iL_i=0,
\label{14.13}
\end{eqnarray}
the following transformation relations were found
\begin{eqnarray}
\bar{\phi}&=&\phi-\dot{T}-\Omega^{-1}[T\partial_0+(L_i+\tilde{\nabla}_iL)\delta^{ij}\partial_j]\Omega,\quad \bar{B}=B+T-\dot{L},\quad \bar{\psi}=\psi+\Omega^{-1}[T\partial_0+(L_i+\tilde{\nabla}_iL)\delta^{ij}\partial_j]\Omega,
\nonumber\\
\bar{E}&=&E-L,\quad \bar{B}_i=B_i-\dot{L}_i,\quad \bar{E}_i=E_i-L_i, \quad \bar{E}_{ij}=E_{ij},
\label{14.14}
\end{eqnarray}
with the elimination of the gauge parameters leading directly to the gauge-invariant combinations shown in (\ref{14.6}). We now specialize to a particular gauge, and pick the gauge parameters so that 
\begin{eqnarray}
L_i=E_i,\quad L=E, \quad B+T-\dot{L}=0.
\label{14.15}
\end{eqnarray}
With this choice (\ref{14.6}) simplifies to
\begin{eqnarray}
\alpha &=& \phi +\psi, \quad \eta=\psi ,\quad B_i,\quad E_{ij},
\label{14.16}
\end{eqnarray}
and now $\eta$ only depends on scalars. The combinations given in (\ref{14.16}) constitute the longitudinal or conformal-Newtonian gauge, a gauge that is often considered in cosmological perturbation theory (see e.g. \cite{Mukhanov1992, Bertschinger1996}). We thus see that using the gauge freedom one can find gauges in which there is no intertwining of scalars and vectors, so that for them a decomposition theorem in the separate scalar, vector and tensor sectors is achievable.

\section{Conclusions}
\label{S15}

In analyzing cosmological perturbations it is very convenient to use the ten degree of freedom scalar, vector, tensor basis for the fluctuations, where this basis is defined according to how these various components transform under three-dimensional rotations. As a basis, this basis is related to the ten-component $h_{\mu\nu}$ basis through integrals of the various $h_{\mu\nu}$ components, to thus be intrinsically non-local. Moreover, the existence of these various integrals requires good asymptotic convergence, and one can only set up the scalar, vector, tensor basis in the first place if one has such good asymptotic boundary conditions. Once one has the scalar, vector, tensor basis, for the cosmologically relevant de Sitter and Robertson-Walker background geometries one can form a set of six basis combinations (two scalars, a two-component transverse vector, and a two-component transverse, traceless rank two tensor) that are gauge invariant,  with the second-order derivative standard gravity cosmological fluctuation equations immediately being writable in terms of them, with gauge invariance then being manifest.

Without needing to specify any choice of gauge at all, one can solve the cosmological fluctuation equations directly, and isolate each of the relevant six gauge-invariant combinations by constructing higher-derivative equations, each one of which involves only one of the six gauge-invariant combinations. In the cosmological perturbation literature it is common to separate out these various combinations not at this higher-derivative level but at the level of the second-order equations themselves by assuming that in any given cosmological evolution equation the scalar, vector and tensor components that appear in it solve each such evolution equation separately, the decomposition theorem. In this paper we actually solve the cosmological evolution equations exactly, and are thus able to check if the decomposition theorem actually holds. We find that solely by imposing the same asymptotic boundary conditions at $r=\infty$ that are needed to set up the scalar, vector, tensor basis in the first place, one then indeed does get the decomposition theorem for fluctuations around a flat background, around a de Sitter background or around a spatially flat Robertson-Walker background with $k=0$. However, for fluctuations around a Robertson-Walker background with $k\neq 0$ one additionally has to require that fluctuations be well-behaved at $r=0$ in order to get the decomposition theorem to hold. The distinction between these two classes of solution is that in the first three (flat, de Sitter, $k=0$ Robertson-Walker) the background metric can be written as a conformal factor times a flat Minkowski metric, where the conformal factor depends on no more than the conformal time $\tau$. In the second class ($k\neq 0$ Robertson-Walker) the background metric can be written as a conformal factor times a flat Minkowski metric, where the conformal factor depends on both $r$ and $\tau$  (cf. (\ref{13.5})), or written in the coordinate equivalent form of a conformal factor times a static Robertson-Walker metric, where the conformal factor depends on the conformal time alone (cf. (\ref{9.1})). Thus if one writes a general cosmological background as a conformal factor times a flat Minkowski metric, then asymptotic boundary conditions alone will in general only secure the decomposition theorem if the conformal factor depends solely on the conformal time  \cite{footnote12}.

In conformal to flat background geometries in which the conformal factor depends on both time and space coordinates the requisite gauge-invariant combinations of metric fluctuations will contain one combination in which the scalar and vector components are intertwined. In such cases one does not in fact have a decomposition theorem for the separate scalar, vector and tensor fluctuation components, but instead one can only have a decomposition theorem for the gauge-invariant combinations themselves. However using the underlying gauge freedom in the theory one can find gauges in which the scalars and vectors are not intertwined. In such gauges one can have a decomposition theorem for the separate scalar, vector and tensor fluctuation components.

Given the lack of manifest covariance (though not of covariance itself) in defining a basis with respect to three-dimensional rotations, we introduced an alternate procedure for describing fluctuations, one in which the fluctuations are defined in a basis in which the components transform as scalars, vectors and tensors under four-dimensional general coordinate transformations. With this basis the cosmological fluctuation equations greatly simplify, and while one can again break up the fluctuation equations into separate sectors at the higher-derivative level, in general we find that even with boundary conditions we do not obtain a decomposition theorem in which the fluctuations separate at the level of the fluctuation equations themselves. Nonetheless, one still can obtain a decomposition theorem, but one instead needs initial conditions. However, if the cause of the perturbation is the introduction of a new source that is not a fluctuation of any source that is present in the background equations of motion, then since it is the new source that is causing the perturbation of the background in the first place, the decomposition theorem then follows for both the three and the four dimensional scalar, vector, tensor formalisms for all of the cosmologically relevant  backgrounds of interest to us in this paper. 

\appendix
\numberwithin{equation}{section}
\setcounter{equation}{0}
\section{Equivalence of the SVT Formalism to the Projection Operator Approach}
\label{SA}

\subsection{Vector Fields}

One of the key virtues of the  SVT3 and SVT4 formalisms is that both of them enable us to construct fluctuation equations that are gauge invariant. Thus  while a general $h_{\mu\nu}$ has ten components, because of the freedom to impose four coordinate transformations only six of the components are physical. Both the SVT3 and SVT4 formalisms then automatically provide fluctuation equations that only depend on six combinations of the terms in the SVT3 or SVT4 expansions of $h_{\mu\nu}$. Since the SVT3 and SVT4 components are related to the components of $h_{\mu\nu}$ via integral relations such as that given in (\ref{2.12}), viz. 
 \begin{eqnarray}
B=\int d^3yD^{(3)}(\mathbf{x}-\mathbf{y})\tilde{\nabla}_y^ih_{0i},\quad B_i=h_{0i}-\tilde{\nabla}_i\int d^3yD^{(3)}(\mathbf{x}-\mathbf{y})\tilde{\nabla}_y^ih_{0i},
\label{A.1a}
\end{eqnarray}
the SVT3 and SVT4 components are intrinsically non-local, with their very existence requiring that the associated integrals exist. Thus asymptotic boundary conditions are built into their very existence. Interestingly, as discussed in \cite{Mannheim2005} and \cite{Amarasinghe2018},  there is another way to implement gauge invariance using non-local operators, namely to use a projection operator approach.   We now show that this approach is equivalent to the SVT approach.

To discuss the application of the projection operator approach to rank two tensors such as $h_{\mu\nu}$ we first apply it to a four-dimensional gauge field $A_{\mu}$. Thus in analog to (\ref{A.1a}) we set
\begin{eqnarray}
A_{\mu}&=&A^T_{\mu}+\partial_{\mu}\int d^4x^{\prime}D(x-x^{\prime})\partial^{\alpha}A_{\alpha}=A^T_{\mu}+A^L_{\mu},
\nonumber\\
A^T_{\mu}&=&A_{\mu}-\partial_{\mu}\int d^4x^{\prime}D(x-x^{\prime})\partial^{\alpha}A_{\alpha},\quad A_{\mu}^L=\partial_{\mu}\int d^4x^{\prime}D(x-x^{\prime})\partial^{\alpha}A_{\alpha},
\label{A.2a}
\end{eqnarray}
where $\partial_{\mu}\partial^{\mu}D(x-x^{\prime})=\delta^4(x-x^{\prime})$, where $A^T_{\mu}$ obeys the transverse condition $\partial^{\mu}A^T_{\mu}=0$, and where $A_{\mu}^L$ is longitudinal. The utility of this expansion is that under  $A_{\mu}\rightarrow A_{\mu}+\partial_{\mu}\chi$ the transverse $A^T_{\mu}$ transforms as 
\begin{eqnarray}
A^T_{\mu}\rightarrow A_{\mu}+\partial_{\mu}\chi-\partial_{\mu}\int d^4x^{\prime}D(x-x^{\prime})\partial^{\alpha}A_{\alpha}
-\partial_{\mu}\int d^4x^{\prime}D(x-x^{\prime})\partial^{\alpha}\partial_{\alpha}\chi=A_{\mu}^T,
\label{A.3a}
\end{eqnarray}
assuming integration by parts. Thus with integration by parts the transverse $A^T_{\mu}$ is automatically gauge invariant. In addition we note that $A_{\mu}^T$ obeys 
\begin{eqnarray}
\partial_{\nu}\partial^{\nu}A^T_{\mu}=\partial_{\nu}\partial^{\nu}A_{\mu}-\partial_{\mu}\partial^{\nu}A_{\nu}=\partial^{\nu}F_{\nu\mu}.
\label{A.4a}
\end{eqnarray}
Thus just as with the use of the non-local SVT formalism for gravity, the use of the non-local $A_{\mu}^T$ enables us to write the Maxwell equations entirely in terms of gauge-invariant quantities. With $A_{\mu}^L$ being the derivative of a scalar function it is pure gauge, and thus cannot appear in the gauge-invariant Maxwell equations. Moreover, while there may be an integration by parts issue for $A_{\mu}^T$, there is none for $\partial_{\nu}\partial^{\nu}A^T_{\mu}$ as it is equal to the gauge-invariant quantity $\partial_{\nu}\partial^{\nu}A_{\mu}-\partial_{\mu}\partial^{\alpha}A_{\alpha}$, just as it must be since the Maxwell equations are gauge invariant.
In the SVT language, with (\ref{A.1a}) and (\ref{A.2a}) only involving scalars and vectors, we can think of (\ref{A.1a}) as an SV3 decomposition of the 3-component $h_{0i}$, and (\ref{A.2a}) as an SV4 decomposition of the 4-component $A_{\mu}$.

An alternate way of understanding these results is to introduce a projection operator
\begin{equation}
\Pi_{\mu\nu}=\eta_{\mu\nu}-\frac{\partial}{\partial x^{\mu}}\int
d^4x^{\prime}D(x-x^{\prime})\frac{\partial}{\partial x^{\prime \nu}},
\label{A.5a}
\end{equation}
as we can then rewrite $A_{\mu}^T$  as 
\begin{equation}
A_{\mu}^{T}=\Pi_{\mu\nu}A^{\nu}.
\label{A.6a}
\end{equation}
As introduced, $\Pi_{\mu\nu}$ obeys the projector algebra relations
\begin{align}
\Pi_{\mu\nu}\Pi^{\nu}_{\phantom{\sigma}\sigma}
&=\Pi_{\mu\sigma},
\nonumber \\
\Pi_{\mu\nu}A^{T \nu}&= A^{T}_{\mu}-\partial_{\mu}\int
d^4x^{\prime}D(x-x^{\prime})
\partial_{\nu}A^{T\nu}(x^{\prime})=A^{T}_{\mu},
\nonumber \\
\Pi_{\mu\nu}A^{L \nu}&=\partial_{\mu}\int
d^4x^{\prime}D(x-x^{\prime})\partial_{\nu}A^{\nu}(x^{\prime})
-\partial_{\mu}\int
d^4x^{\prime}D(x-x^{\prime})
\partial_{\nu}\partial^{\nu}\int
d^4x^{\prime\prime}D(x^{\prime}-x^{\prime\prime})
\partial_{\sigma}A^{\sigma}(x^{\prime\prime})=0.
\label{A.7a}
\end{align}
In the SVT4 language we set  $A_{\mu}=A_{\mu}^T+\partial_{\mu}A$, and can thus identify 
\begin{eqnarray}
A_{\mu}^T=  \Pi_{\mu\nu}A^{\nu},\quad A_{\mu}^L=\partial_{\mu}A=(\eta_{\mu\nu}-\Pi_{\mu\nu})A^{\nu}.
\label{A.8a}
\end{eqnarray}
For vector fields the SVT formalism is thus equivalent to the projector formalism.  Having now established this equivalence for vector fields, we turn now to tensor fields.

\subsection{Transverse and Longitudinal Projection Operators for Flat Spacetime Tensor Fields}

For tensor fields we introduce 4-dimensional flat  spacetime transverse and longitudinal projection operators \cite{Mannheim2005,Amarasinghe2018}: 
\begin{eqnarray}
T_{\mu\nu\sigma\tau}&=&\eta_{\mu\sigma}\eta_{\nu\tau}
-\partial_{\mu}\int d^4x^{\prime}D(x-x^{\prime})
\eta_{\nu\tau}\partial_{\sigma}
-\partial_{\nu}\int d^4x^{\prime}D(x-x^{\prime})
\eta_{\mu\sigma}\partial_{\tau}
\nonumber \\
&+&\partial_{\mu}\partial_{\nu}\int
d^4x^{\prime}D(x-x^{\prime})\partial_{\sigma}\int
d^4x^{\prime\prime}D(x^{\prime}-x^{\prime\prime})
\partial_{\tau},
\nonumber\\
L_{\mu\nu\sigma\tau}&=&\partial_{\mu}\int d^4x^{\prime}D(x-x^{\prime})
\eta_{\nu\tau}\partial_{\sigma}
+\partial_{\nu}\int d^4x^{\prime}D(x-x^{\prime})
\eta_{\mu\sigma}\partial_{\tau}
\nonumber \\
&-&\partial_{\mu}\partial_{\nu}\int
d^4x^{\prime}D(x-x^{\prime})\partial_{\sigma}\int
d^4x^{\prime\prime}D(x^{\prime}-x^{\prime\prime})
\partial_{\tau}.
\label{A.9a}
\end{eqnarray}
As constructed, these projectors obey a standard projector algebra
\begin{eqnarray}
&&T_{\mu\nu\sigma\tau}T^{\sigma\tau}_{\phantom{\sigma\tau}\alpha\beta}=
T_{\mu\nu\alpha\beta},\quad
L_{\mu\nu\sigma\tau}L^{\sigma\tau}_{\phantom{\sigma\tau}\alpha\beta}
=L_{\mu\nu\alpha\beta},
\nonumber \\
&&T_{\mu\nu\sigma\tau}L^{\sigma\tau}_{\phantom{\sigma\tau}\alpha\beta}=
0,\quad
L_{\mu\nu\sigma\tau}T^{\sigma\tau}_{\phantom{\sigma\tau}\alpha\beta}
=0,\quad L_{\mu\nu\sigma\tau}
+T_{\mu\nu\sigma\tau}
=\eta_{\mu\sigma}\eta_{\nu\tau}.
\label{A.10a}
\end{eqnarray}
In terms of these projectors we define transverse and longitudinal components $h^{T}_{\mu\nu}$ and $h^{L}_{\mu\nu}$ of $h_{\mu\nu}$ according to
\begin{eqnarray}
T_{\mu\nu\sigma\tau}h^{\sigma\tau}&=&h^{T}_{\mu\nu}=h_{\mu\nu}
-\partial_{\mu}\int
d^4x^{\prime}D(x-x^{\prime})\partial_{\sigma}
h^{\sigma}_{\phantom{\sigma}\nu}(x^{\prime}) 
-\partial_{\nu}\int d^4x^{\prime}D(x-x^{\prime})
\partial_{\kappa}h^{\kappa}_{\phantom{\kappa}\mu}(x^{\prime})
\nonumber \\
&&+\partial_{\mu}\partial_{\nu}\int
d^4x^{\prime}D(x-x^{\prime})\partial_{\sigma}\int
d^4x^{\prime\prime}D(x^{\prime}-x^{\prime\prime})
\partial_{\kappa}h^{\sigma\kappa}(x^{\prime\prime}),
\nonumber\\
L_{\mu\nu\sigma\tau}h^{\sigma\tau}&=&h^{L}_{\mu\nu}=\partial_{\mu}\int
d^4x^{\prime}D(x-x^{\prime})\partial_{\sigma}
h^{\sigma}_{\phantom{\sigma}\nu}(x^{\prime}) 
+\partial_{\nu}\int d^4x^{\prime}D(x-x^{\prime})
\partial_{\kappa}h^{\kappa}_{\phantom{\kappa}\mu}(x^{\prime})
\nonumber \\
&&-\partial_{\mu}\partial_{\nu}\int
d^4x^{\prime}D(x-x^{\prime})\partial_{\sigma}\int
d^4x^{\prime\prime}D(x^{\prime}-x^{\prime\prime})
\partial_{\kappa}h^{\sigma\kappa}(x^{\prime\prime}).
\label{A.11a}
\end{eqnarray}
Assuming integration by parts these components obey
\begin{eqnarray}
\partial_{\nu}h^{T\mu\nu}
=0,\quad 
\partial_{\nu}h^{L\mu\nu}=\partial_{\nu}h^{\mu\nu}.
\label{A.12a}
\end{eqnarray}
With $h^{T}_{\mu\nu}$ transforming as $h^{T}_{\mu\nu}\rightarrow h^{T}_{\mu\nu}$ under $h_{\mu\nu}\rightarrow h_{\mu\nu}-\partial_{\mu}\epsilon_{\nu}-\partial_{\nu}\epsilon_{\mu}$ as long as we can integrate by parts, we see that, as introduced, $h^{T}_{\mu\nu}$ is both transverse and gauge invariant. 

On evaluation we obtain 
\begin{align}
\frac{1}{2}[\partial_{\mu}\partial_{\nu}h^{T}
+\partial_{\alpha}\partial^{\alpha}h_{\mu\nu}^{T}]
-\frac{1}{2}\eta_{\mu\nu}\partial_{\sigma}\partial^{\sigma}
h^{T}
=\frac{1}{2}[\partial_{\mu}\partial_{\nu}h
-\partial_{\mu}\partial_{\lambda}h^{\lambda}_{\phantom{\lambda}\nu}
-\partial_{\nu}\partial_{\lambda}h^{\lambda}_{\phantom{\lambda}\mu}
+\partial_{\alpha}\partial^{\alpha}h_{\mu\nu}]
-\frac{1}{2}\eta_{\mu\nu}[\partial_{\alpha}\partial^{\alpha}h
-\partial_{\sigma}\partial_{\lambda}h^{\sigma\lambda}],
\label{A.13a}
\end{align}
where $h^{T}$ is given by
\begin{equation}
h^{T}=\eta^{\alpha\beta}h_{\alpha\beta}^{T}
=h -\partial_{\nu}\int
d^4x^{\prime}D(x-x^{\prime})\partial_{\sigma}
h^{\sigma\nu}(x^{\prime}),
\label{A.14a}
\end{equation}
with $h=\eta^{\alpha\beta}h_{\alpha\beta}$. On recognizing the right-hand side of  (\ref{A.13a}) as $\delta R_{\mu\nu}-\frac{1}{2}\eta_{\mu\nu}\delta R=\delta G_{\mu\nu}$,
we obtain 
\begin{eqnarray}
&&\delta G_{\mu\nu}=\tfrac{1}{2}[\partial_{\mu}\partial_{\nu}h^{T}
+\partial_{\alpha}\partial^{\alpha}h_{\mu\nu}^{T}]
-\frac{1}{2}\eta_{\mu\nu}\partial_{\sigma}\partial^{\sigma}
h^{T}.
\label{A.15a}
\end{eqnarray}
We thus write the perturbed Einstein tensor entirely in terms of the non-local, gauge invariant, six degree of freedom $h_{\mu\nu}^T$.

To make contact with the SVT4 expansion we insert
\begin{eqnarray}
h_{\mu\nu}=-2\eta_{\mu\nu}\chi+2\partial_{\mu}\partial_{\nu}F
+ \partial_{\mu}F_{\nu}+\partial_{\nu}F_{\mu}+2F_{\mu\nu}
\label{A.16a}
\end{eqnarray}
into $h_{\mu\nu}^T$,  to obtain
\begin{eqnarray}
h^T_{\mu\nu}=-2\eta_{\mu\nu}\chi+2F_{\mu\nu}+2\partial_{\mu}\partial_{\nu}\int d^4D(x-x^{\prime})\chi(x^{\prime}),\quad h^T=-6\chi.
\label{A.17a}
\end{eqnarray}
With $\delta G_{\mu\nu}$ being written in terms of the projected $h^T_{\mu\nu}$, we see that it is written in terms of the SVT4 $F_{\mu\nu}$ and $\chi$. However as written, $h_{\mu\nu}^T$ contains an integral term in (\ref{A.17a}). To eliminate it we extend transverse projection to transverse-traceless projection.

\subsection{Transverse-Traceless Projection Operators for Flat Spacetime Tensor Fields}

In \cite{Mannheim2005} and \cite{Amarasinghe2018} two further projectors were introduced
\begin{eqnarray}
Q_{\mu\nu\sigma\tau}&=&\frac{1}{3}\left[\eta_{\mu\nu}
-\partial_{\mu}\partial_{\nu}\int d^4x^{\prime}D(x-x^{\prime})\right]
\left[\eta_{\sigma\tau}-\partial^{\prime}_{\sigma}\int
d^4x^{\prime\prime}D(x^{\prime}-x^{\prime\prime})\partial^{\prime\prime}_{\tau}\right],
\nonumber\\
P_{\mu\nu\sigma\tau}&=&T_{\mu\nu\sigma\tau}-Q_{\mu\nu\sigma\tau}.
\label{A.18a}
\end{eqnarray}
They obey the projector algebra
\begin{eqnarray}
T_{\mu\nu\sigma\tau}Q^{\sigma\tau}_{\phantom{\sigma\tau}\alpha\beta}
&=&Q_{\mu\nu\alpha\beta},\quad
Q_{\mu\nu\sigma\tau}T^{\sigma\tau}_{\phantom{\sigma\tau}\alpha\beta}
=Q_{\mu\nu\alpha\beta},\quad
Q_{\mu\nu\sigma\tau}Q^{\sigma\tau}_{\phantom{\sigma\tau}\alpha\beta}
=Q_{\mu\nu\alpha\beta}, 
\nonumber\\
P_{\mu\nu\sigma\tau}Q^{\sigma\tau\alpha\beta}&=&0,\quad
Q_{\mu\nu\sigma\tau}P^{\sigma\tau\alpha\beta}=0,\quad
P_{\mu\nu\sigma\tau}P^{\sigma\tau}_{\phantom{\sigma\tau}\alpha\beta}
=P_{\mu\nu\alpha\beta}.
\label{A.19a}
\end{eqnarray}
The projector $P_{\mu\nu\sigma\tau}$ projects out the traceless piece of $h^T_{\mu\nu}$, while $Q_{\mu\nu\sigma\tau}$ projects out its complement, and they implement
\begin{eqnarray}
P_{\mu\nu}^{\phantom{\mu\nu}\sigma\tau}h^T_{\sigma\tau}=h^{T\theta}_{\mu\nu},\quad 
Q_{\mu\nu}^{\phantom{\mu\nu}\sigma\tau}h^T_{\sigma\tau}
=h^T_{\mu\nu}-h^{T\theta}_{\mu\nu},
\label{A.20a}
\end{eqnarray}
with $h^{T\theta}_{\mu\nu}$ being both traceless and transverse. With $Q_{\mu\nu}^{\phantom{\mu\nu}\sigma\tau}$ implementing $Q_{\mu\nu}^{\phantom{\mu\nu}\sigma\tau}h^L_{\sigma\tau}=0$, $P_{\mu\nu}^{\phantom{\mu\nu}\sigma\tau}$ implements $P_{\mu\nu}^{\phantom{\mu\nu}\sigma\tau}h^L_{\sigma\tau}=0$ as well, to thus implement 
\begin{eqnarray}
 P_{\mu\nu}^{\phantom{\mu\nu}\sigma\tau}h_{\sigma\tau}=h^{T\theta}_{\mu\nu}.
\label{A.21a}
\end{eqnarray}
$P_{\mu\nu\sigma\tau}$ is thus a traceless projector not just for the transverse $h_{\mu\nu}^T$ but for the full $h_{\mu\nu}$ as well. We can thus introduce its complementary projection operator $U_{\mu\nu\sigma\tau}=\eta_{\mu\sigma}\eta_{\nu\tau}-P_{\mu\nu\sigma\tau}$, as it obeys
\begin{eqnarray}
P_{\mu\nu\sigma\tau}U^{\sigma\tau\alpha\beta}&=&0,\quad
U_{\mu\nu\sigma\tau}P^{\sigma\tau\alpha\beta}=0,\quad
U_{\mu\nu\sigma\tau}U^{\sigma\tau}_{\phantom{\sigma\tau}\alpha\beta}
=U_{\mu\nu\alpha\beta},
\nonumber\\
U_{\mu\nu}^{\phantom{\mu\nu}\sigma\tau}h_{\sigma\tau}&=&h_{\mu\nu}-h^{T\theta}_{\mu\nu}=
h^{L\theta}_{\mu\nu}+\frac{1}{3}\eta_{\mu\nu}\eta^{\sigma\tau}h_{\sigma\tau}
-\frac{1}{3}\partial_{\mu}\partial_{\nu}\int d^4y D(x-y)\eta^{\sigma\tau}h_{\sigma\tau},
\label{A.22a}
\end{eqnarray}

Given (\ref{A.20a}) and (\ref{A.18a}) we obtain 
\begin{eqnarray}
h^{T\theta}_{\mu\nu}= h^{T}_{\mu\nu}-\frac{1}{3}\eta_{\mu\nu}\eta^{\sigma\kappa}h^{T}_{\sigma\kappa}
+\frac{1}{3}\partial_{\mu}\partial_{\nu}\int d^4y D(x-y)\eta^{\sigma\kappa}h^{T}_{\sigma\kappa},
\label{A.23a}
\end{eqnarray}
Inserting (\ref{A.17a}) into (\ref{A.23a}) yields
\begin{eqnarray}
h^{T\theta}_{\mu\nu}=2F_{\mu\nu},
\label{A.24a}
\end{eqnarray}
with $\chi$ dropping out. Finally, in terms of $h^{T\theta}_{\mu\nu}$ we can rewrite (\ref{A.15a}) as 
\begin{eqnarray}
&&\delta G_{\mu\nu}=
\tfrac{1}{2}\partial_{\alpha}\partial^{\alpha}h_{\mu\nu}^{T\theta}
-\tfrac{1}{3}\eta_{\mu\nu}\partial_{\sigma}\partial^{\sigma}h^{T}
+\tfrac{1}{3}\partial_{\mu}\partial_{\nu}h^{T}.
\label{A.25a}
\end{eqnarray}
Then with 
\begin{eqnarray}
F_{\mu\nu}=\tfrac{1}{2}h_{\mu\nu}^{T\theta}, \quad \chi=-\tfrac{1}{6}h^{T},
\label{A.26a}
\end{eqnarray}
we can rewrite (\ref{A.25a}) as 
\begin{eqnarray}
\delta G_{\mu\nu}&=&\partial_{\alpha}\partial^{\alpha}F_{\mu\nu}+2\eta_{\mu\nu}\partial_{\alpha}\partial^{\alpha}\chi-2\partial_{\mu}\partial_{\nu}\chi.
\label{A.27a}
\end{eqnarray}
We recognize (\ref{A.27a}) as the expression for $\delta G_{\mu\nu}$ as given in (\ref{3.10}) when $D=4$, and with $h^T_{\mu\nu}$ and thus $h^{T\theta}_{\mu\nu}$ and $h^T$ being gauge invariant, we confirm that given integration by parts $F_{\mu\nu}$ and $\chi$ are gauge invariant, just as noted in Sec. \ref{S3}. Thus with (\ref{A.26a})
we establish the equivalence of the  SVT4 decomposition and the projection operator technique.

As a further example of this equivalence we note that for conformal gravity fluctuations around a flat spacetime background (\ref{13.18}) takes the form
\begin{eqnarray}
\delta W_{\mu\nu}=\frac{1}{2}\bigg{(}\partial_{\sigma}\partial^{\sigma}\partial_{\tau}\partial^{\tau}K_{\mu\nu}
-\partial_{\sigma}\partial^{\sigma}\partial_{\mu}\partial^{\alpha}K_{\alpha\nu}
-\partial_{\sigma}\partial^{\sigma}\partial_{\nu}\partial^{\alpha}K_{\alpha\mu}
+\frac{2}{3}\partial_{\mu}\partial_{\nu}\partial^{\alpha}\partial^{\beta}K_{\alpha\beta}+\frac{1}{3}\eta_{\mu\nu}\partial_{\sigma}\partial^{\sigma}\partial^{\alpha}\partial^{\beta}K_{\alpha\beta}\bigg{)},
\label{A.28a}
\end{eqnarray} 
where all derivatives are four-dimensional derivatives with respect to a flat Minkowski metric, and where $K_{\mu\nu}$ is given by $K_{\mu\nu}=h_{\mu\nu}-(1/4)\eta_{\mu\nu}\eta^{\alpha\beta}h_{\alpha\beta}$. Inserting (\ref{A.11a}) and (\ref{A.23a}) into (\ref{A.28a}) yields
\begin{eqnarray}
\delta W_{\mu\nu}&=&\frac{1}{2}\partial_{\sigma}\partial^{\sigma}\partial_{\tau}\partial^{\tau}h^{T\theta}_{\mu\nu}.
\label{A.29a}
\end{eqnarray} 
With the insertion of (\ref{A.16a}) into (\ref{A.28a}) yielding 
\begin{eqnarray}
\delta W_{\mu\nu}&=&\partial_{\sigma}\partial^{\sigma}\partial_{\tau}\partial^{\tau}F_{\mu\nu},
\label{A.30a}
\end{eqnarray} 
(cf. (\ref{13.20}) with $\Omega=1$), we recover (\ref{A.24a}), and again confirm the equivalence of the  SVT4 decomposition and the projection operator technique.

\subsection{Transverse and Longitudinal Projection Operators for Curved Spacetime Tensor Fields}

For curved spacetime with background metric $g_{\mu\nu}$ it is convenient to  define a 2-index propagator
\begin{equation} 
[g^{\nu}_{\phantom{\nu}\beta}\nabla_{\tau}\nabla^{\tau}
+\nabla_{\beta}\nabla^{\nu}]D^{\beta}_{\phantom{\beta}\sigma}
(x,x^{\prime}) =g^{\nu}_{\phantom{\nu}\sigma}(-g)^{-1/2}\delta^4
(x-x^{\prime}).
\label{A.31a}
\end{equation}
In terms of it we introduce \cite{Mannheim2005}
 \begin{eqnarray} 
T_{\mu\nu\sigma\tau}&=&g_{\mu\sigma}g_{\nu\tau}- \nabla_{\mu}\int
d^4x^{\prime}(-g)^{1/2}
D_{\nu\sigma}(x,x^{\prime})
\nabla_{\tau} 
-\nabla_{\nu}\int
d^4x^{\prime}(-g)^{1/2}
D_{\mu\sigma}(x,x^{\prime})\nabla_{\tau},
\nonumber \\
L_{\mu\nu\sigma\tau}&=&\nabla_{\mu}\int
d^4x^{\prime}(-g)^{1/2}
D_{\nu\sigma}(x,x^{\prime})
\nabla_{\tau} 
+\nabla_{\nu}\int
d^4x^{\prime}(-g)^{1/2}
D_{\mu\sigma}(x,x^{\prime})
\nabla_{\tau}.
\label{A.32a}
\end{eqnarray}
These projection operators close on the projector algebra given in (\ref{A.10a}). As such, they effect
 $T_{\mu\nu\sigma\tau}h^{\sigma\tau}=
h^{T}_{\mu\nu}$ and $L_{\mu\nu\sigma\tau}h^{\sigma\tau}=
h^{L}_{\mu\nu}$, where
\begin{equation} 
h^{T}_{\mu\nu}=h_{\mu\nu}-\nabla_{\mu}\int
d^4x^{\prime}(-g)^{1/2}
D^{\nu}_{\phantom{\nu}\sigma}(x,x^{\prime})
\nabla_{\tau}h^{\sigma\tau}  
-\nabla_{\nu}\int
d^4x^{\prime}(-g)^{1/2}
D^{\mu}_{\phantom{\mu}\sigma}(x,x^{\prime})
\nabla_{\tau}h^{\sigma\tau},
\label{A.33a}
\end{equation}
\begin{equation} 
h^{L}_{\mu\nu}=\nabla_{\mu}\int
d^4x^{\prime}(-g)^{1/2}
D^{\nu}_{\phantom{\nu}\sigma}(x,x^{\prime})
\nabla_{\tau}h^{\sigma\tau} 
+\nabla_{\nu}\int
d^4x^{\prime}(-g)^{1/2}
D^{\mu}_{\phantom{\mu}\sigma}(x,x^{\prime})
\nabla_{\tau}h^{\sigma\tau}.
\label{A.34a}
\end{equation}

The utility of constructing these projected states is that under a gauge transformation $h_{\mu\nu}$ transforms into $h_{\mu\nu}-\nabla_{\mu}\epsilon_{\nu}-\nabla_{\nu}\epsilon_{\mu}$. However, we see that this is precisely the structure of $h^{L}_{\mu\nu}$. The longitudinal component of $h_{\mu\nu}$ can thus be removed by a gauge transformation, and the fluctuation Einstein equations can only depend on the 6-component $h^{T}_{\mu\nu}$. However, unlike the flat background case where one can write $\delta G_{\mu\nu}$ itself entirely in terms of $h^T_{\mu\nu}$, in the curved background case there must be a background $T_{\mu\nu}$, and thus it is only in the full $\delta G_{\mu\nu}+8\pi G \delta T_{\mu\nu}$ that the metric fluctuations can be described entirely by $h^T_{\mu\nu}$. If we introduce a quantity $\delta T^T_{\mu\nu}$ in which the dependence on $\epsilon_{\mu}$ has been excluded (i.e. under a gauge transformation $\delta T_{\mu\nu}\rightarrow \delta T^T_{\mu\nu}$ plus a function of $\epsilon_{\mu}$, and this function of $\epsilon_{\mu}$ cancels against an identical function of $\epsilon_{\mu}$ in $\delta G_{\mu\nu}$ \cite{footnote13}), then following the commuting of some derivatives,  the fluctuation equations take the form \cite{Mannheim2005}
\begin{eqnarray} 
\delta G_{\mu\nu}+8\pi G \delta T_{\mu\nu}
&=&\frac{1}{2}[\nabla_{\mu}\nabla_{\nu}h^{T}
+R^{\sigma}_{\phantom{\sigma}\mu}h_{\sigma\nu}^{T}
+R^{\sigma}_{\phantom{\sigma}\nu}h_{\sigma\mu}^{T}
-2R_{\mu\lambda\nu\sigma}h^{T\lambda\sigma}
+\nabla_{\alpha}\nabla^{\alpha}h_{\mu\nu}^{T}]
\nonumber \\
&-&\frac{1}{2}R^{\sigma}_{\phantom{\sigma}\sigma}h^{T}_{\mu\nu}
+\frac{1}{ 2}g_{\mu\nu}R_{\alpha\beta}h^{T\alpha\beta}
-\frac{1}{2}g_{\mu\nu}\nabla_{\alpha}\nabla^{\alpha}h^{T}
+8\pi G \delta T^T_{\mu\nu}=0.
\label{A.35a}
\end{eqnarray}
The SVT4 fluctuations around a de Sitter background as given in (\ref{6.16}) to (\ref{6.19}) and around a general Robertson-Walker background as given in (\ref{12.9}) are special cases of (\ref{A.35a}), with the only metric fluctuations that appear in (\ref{6.19}) and (\ref{12.9}) being $F_{\mu\nu}$ and $\chi$, viz. just the six degrees of freedom associated with $h^T_{\mu\nu}$.

\subsection{D-dimensional SVTD Transverse-Traceless Projection Operators for Curved Spacetime Tensor Fields}

Rather than generalize the general curved spacetime transverse and longitudinal projection technique to the general transverse-traceless case, we have instead  found it more convenient to generalize the SVTD discussion given in Secs. \ref{S3} and \ref{S6} to general curved spacetime background fluctuations. To this end we take $h_{\mu\nu}$ to be of the form:
\begin{eqnarray}
h_{\mu\nu} &=& 2F_{\mu\nu}+W_{\mu\nu}+S_{\mu\nu},
\label{A.36a}
\end{eqnarray}
where
\begin{align}
W_{\mu\nu} =\nabla_\mu W_\nu + \nabla_\nu W_\mu - \frac{2}{D}g_{\mu\nu}\nabla^\alpha W_\alpha,\quad S_{\mu\nu}=\frac{1}{D-1}\left( g_{\mu\nu}\nabla_\alpha \nabla^\alpha - \nabla_\mu\nabla_\nu\right)\int d^Dx^{\prime}[-g(x^{\prime})]^{1/2}D^{(D)}(x,x^{\prime}) h(x^{\prime}),
\label{A.37a}
\end{align}
with $D(x,x^{\prime})$ obeying
\begin{eqnarray}
\nabla_\alpha \nabla^\alpha D^{(D)}(x,x^{\prime}) =[-g(x)]^{-1/2}\delta^{(D)}(x-x^{\prime}).
\label{A.38a}
\end{eqnarray}
From (\ref{A.37a}) we obtain
\begin{eqnarray}
g^{\mu\nu}W_{\mu\nu}=0,\quad g^{\mu\nu}S_{\mu\nu}=h,
\label{A.39a}
\end{eqnarray}
\begin{eqnarray}
\nabla^\nu h_{\mu\nu} &=& \nabla^\nu W_{\mu\nu} + \nabla^\nu S_{\mu\nu}
\label{A.40a}
\end{eqnarray}
as the conditions that $F_{\mu\nu}$ be transverse and traceless. From (\ref{A.40a}) we obtain 
\begin{align}
\left[g_{\nu\alpha} \nabla_\beta \nabla^\beta +\nabla_\alpha \nabla_\nu - \frac{2}{D}\nabla_\nu\nabla_\alpha\right] W^\alpha
=\nabla^\alpha h_{\alpha\nu} - \frac{1}{D-1}\left(\nabla_\nu \nabla_\alpha\nabla^\alpha - \nabla_\alpha\nabla^\alpha \nabla_\nu\right)
\int d^Dx^{\prime}[-g(x^{\prime})]^{1/2} D^{(D)}(x,x^{\prime}) h(x^{\prime}),
\label{A.41a}
\end{align}
and by commuting derivatives can rewrite (\ref{A.41a}) as
\begin{align}
&\left[g_{\nu\alpha}\nabla_\beta\nabla^\beta + \left(\frac{D-2}{D}\right)\nabla_\nu \nabla_\alpha - R_{\nu\alpha}\right]W^\alpha
= \nabla^\alpha h_{\alpha\nu} - \frac{1}{D-1}R_{\nu\alpha}\nabla^\alpha \int d^Dx^{\prime}[-g(x^{\prime})]^{1/2}D^{(D)}(x,x^{\prime}) h(x^{\prime}).
\label{A.42a}
\end{align}

To solve for $W_{\mu}$ it is convenient to use the bitensor formalism in which we define $G_{\alpha}^{(D)\beta}(x,x^{\prime})=e^a_{\alpha}(x)e^{\beta}_a(x^{\prime})$ where the D-dimensional $e^a_{\alpha}(x)$ vierbeins obey $g_{\mu\nu}(x)=\eta_{ab}e^{a}_{\mu}(x)e^{b}_{\nu}(x)$, with $a$ and $b$ referring to a fixed D-dimensional basis. With this bitensor definition $e^a_{\alpha}(x)$ and $e^{\beta}_a(x^{\prime})$ are acting in separate spaces, but  at $x=x^{\prime}$ we obtain $G_{\alpha}^{(D)\beta}(x,x)=g_{\alpha}^{\phantom{\alpha}\beta}(x)$. On the introducing the propagator that satisfies 
\begin{eqnarray}
\left[g_{\nu\alpha}\nabla_\beta\nabla^\beta + \left(\frac{D-2}{D}\right)\nabla_\nu \nabla_\alpha - R_{\nu\alpha}\right]D_{(D)}^{\alpha\gamma}(x,x^{\prime}) &=& G_{\nu}^{(D)\gamma}(x,x^{\prime}) [-g(x^{\prime})]^{-1/2} \delta^{(D)}(x-x^{\prime}),
\label{A.43a}
\end{eqnarray}
we can solve for $W_{\mu}$ as
\begin{align}
W_{\mu}(x) = \int d^Dx^{\prime}[-g(x^{\prime})]^{1/2} D_{\mu}^{(D)\sigma}(x,x^{\prime})\left[ \nabla^{\rho}_{x^{\prime}} h_{\sigma\rho}(x^{\prime})-
\frac{1}{D-1}R_{\sigma\rho}(x^{\prime})\nabla^{\rho}_{x^{\prime}} \int d^Dx^{\prime\prime}[-g(x^{\prime\prime})]^{1/2} D^{(D)}(x^{\prime},x^{\prime\prime}) h(x^{\prime\prime})\right].
\label{A.44a}
\end{align}

Next we decompose $W_{\mu}$ into transverse and longitudinal components viz.
\begin{eqnarray}
W_{\mu} &=&W^T_{\mu}+W^L_{\mu}=F_{\mu}+\nabla_{\mu}H,\quad  \nabla^{\mu}F_{\mu}=0,\quad H=\int d^Dx^{\prime}[-g(x^{\prime})]^{1/2}D^{(D)}(x,x^{\prime})\nabla^\sigma W_\sigma(x^{\prime}),
\label{A.45a}
\end{eqnarray}
with $h_{\mu\nu}$ then taking the form
\begin{align}
h_{\mu\nu}&= 2F_{\mu\nu} + \nabla_\mu F_\nu + \nabla_\nu F_\mu + 2 \nabla_\mu\nabla_\nu H - \frac{2}{D}g_{\mu\nu}\nabla_\alpha \nabla^\alpha H 
\nonumber\\
&+\frac{1}{D-1}\left( g_{\mu\nu}\nabla_\alpha \nabla^\alpha - \nabla_\mu\nabla_\nu\right)\int d^Dx^{\prime}[-g(x^{\prime})]^{1/2} D^{(D)}(x,x^{\prime}) h(x^{\prime}).
\label{A.46a}
\end{align}
Upon further defining
\begin{align}
F &= H - \frac{1}{2(D-1)} \int d^Dx^{\prime}[-g(x^{\prime})]^{1/2} D^{(D)}(x,x^{\prime}) h(x^{\prime}),
\nonumber\\
\chi &= \frac{1}{D}\nabla_\alpha\nabla^\alpha H - \frac{1}{2(D-1)}\nabla_\alpha\nabla^\alpha\int d^Dx^{\prime}[-g(x^{\prime})]^{1/2} D^{(D)}(x,x^{\prime}) h(x^{\prime}),
\label{A.47a}
\end{align}
we may express $h_{\mu\nu}$ in the SVTD form:
\begin{eqnarray}
h_{\mu\nu} &=& -2g_{\mu\nu}\chi + 2\nabla_\mu\nabla_\nu F + \nabla_\mu F_\nu + \nabla_\nu F_\mu + 2F_{\mu\nu}
\label{A.48a},
\end{eqnarray}
where
\begin{align}
\chi &= \frac{1}{D}\nabla^\sigma W_{\sigma}  - \frac{1}{2(D-1)}h,\quad F_{\mu} = W_{\mu}^T=W_{\mu} -\nabla_\mu \int d^Dx^{\prime}[-g(x^{\prime})]^{1/2} D^{(D)}(x,x^{\prime})\nabla^{\sigma}W_\sigma(x^{\prime}),
\nonumber\\
F &= \int d^Dx^{\prime}[-g(x^{\prime})]^{1/2} D^{(D)}(x,x^{\prime}) \left(\nabla^\sigma W_{\sigma}(x^{\prime})  - \frac{1}{2(D-1)}h(x^{\prime})\right),
\nonumber\\
2F_{\mu\nu} &= h_{\mu\nu}+2g_{\mu\nu}\chi - 2\nabla_\mu\nabla_\nu F - \nabla_\mu F_\nu - \nabla_\nu F_{\mu}.
\label{A.49a}
\end{align}
We thus generalize the SVTD approach to the arbitrary D-dimensional curved spacetime background.


\begin{thebibliography}{99}
\bibitem{footnote1} While the flat spacetime limit of (\ref{1.21a}) is $\delta^{ij}\partial_i\partial_jD=0$, we note that (\ref{1.21a}) is not the covariant generalization of $\delta^{ij}\partial_i\partial_jD=0$ since (\ref{1.21a}) contains a term that depends on the spatial curvature of the background geometry, a term that vanishes in the flat spacetime limit. Covariantizing flat spacetime expressions would miss such terms.
\bibitem{footnote2} Since one is allowed to make gauge transformations with gauge functions that diverge asymptotically, a function that is bounded in a given gauge could become unbounded under a gauge transformation. However, for gauge-invariant functions (viz. those that appear in the fluctuation equations) this problem cannot occur. So if a gauge-invariant function is bounded in one gauge it is bounded in every gauge.
\bibitem{footnote3} What is happening here is that while every longitudinal vector can be written as the derivative of a scalar, not every derivative of a scalar is longitudinal, since it could be that $\partial_i\partial^i$ acting on the scalar is zero, with the derivative of the scalar then being transverse. Moreover, the $\partial_i\int d^3yD^{(3)}(\mathbf{x}-\mathbf{y})\partial^y_jA^j({\bf y})$ integral term that appears in (\ref{2.17}) could only be non-zero if $\partial_jA^j$ is not zero, i.e. if $A_i$ is not transverse. with $\partial_i\int d^3yD^{(3)}(\mathbf{x}-\mathbf{y})\partial^y_jA^j({\bf y})$ necessarily being longitudinal since $\partial^i\partial_i\int d^3yD^{(3)}(\mathbf{x}-\mathbf{y})\partial^y_jA^j({\bf y})=\partial_jA^j$ is non-zero if $A_i$ is not transverse. Equation (\ref{2.17}) is thus the unique way to decompose a bounded vector into its transverse and longitudinal components. In addition, we note that extracting the transverse component $V_i$ of a vector $A_i$ can also be done in momentum space, where one sets $V_i=A_i-k_ik^jA_j/k^2$ so that $k^iV_i=0$. With  $A_i=V_i+k_ik^jA_j/k^2$ being local in momentum space, in coordinate space one has the non-local (2.17). The SVT decomposition is thus intrinsically non-local.
\bibitem{footnote4} While for the moment we only need $\delta W_{\mu\nu}$ to be gauge invariant for fluctuations around flat spacetime, it turns out \cite{Amarasinghe2018} that it is also gauge invariant for fluctuations around any background that is conformal to flat, a point we return to below.
\bibitem{footnote5} With the antisymmetric Maxwell tensor $F_{\mu\nu}$ being invariant under a local electromagnetic gauge transformation and with the Maxwell action being of the form $\int d^4x(-g)^{1/2}F_{\mu\nu}F^{\mu\nu}$, we see that conformal gravity is its gravitational analog.
\bibitem{footnote6} In (\ref{3.10}) we recognize $\nabla_{\alpha}\nabla^{\alpha}F_{\mu\nu}$ as the transverse ($T$), traceless ($\theta$) piece $\delta G_{\mu\nu}^{T\theta}$ of $\delta G_{\mu\nu}$, and can thus write (\ref{4.7}) as $\delta W_{\mu\nu}=\nabla_{\alpha}\nabla^{\alpha}\delta G_{\mu\nu}^{T\theta}$, to thus relate $\delta W_{\mu\nu}$ and $\delta G_{\mu\nu}^{T\theta}$. Further discussion of transverse-traceless components of rank two tensors is given in the Appendix.
\bibitem{footnote7} While one could in principle consider any general coordinate invariant action involving arbitrarily high powers of Riemann-tensor-based quantities, fluctuations around flat spacetime being  gauge invariant for any of them, as we can see from the structure of the SVT3 fluctuations given in (\ref{2.6}), for fluctuations around flat spacetime we never need to go beyond fourth-order derivative actions in order to fully fix the relationship between the SVT3 and SVT4 bases.
\bibitem{footnote8} We can actually use (\ref{6.12}) for any background, since  as defined in (\ref{6.12}) $F_{\mu\nu}$ would automatically be traceless in any background. However, the resulting relation between $\nabla^{\nu}h_{\mu\nu}$ and $W_{\mu}$ would be very complicated because when $\nabla^{\nu}$ acts on the scalar integral  term, which we denote by $S$,  we have $\nabla^{\nu}\nabla_{\mu}\nabla_{\nu}S=\nabla_{\mu}\nabla^{\nu}\nabla_{\nu}S-R_{\mu\nu}\nabla^{\nu}S$, and in general $R_{\mu\nu}$ is not as simple as it is in a de Sitter geometry. In the Appendix we address this point by providing a treatment of SVTD fluctuations around a general curved spacetime background.
\bibitem{footnote9} In general, if some specific $\delta G_{\mu\nu}$ around some general background contains some non-gauge-invariant terms, then not only must $-8\pi G\delta T_{\mu\nu}$ compensate for them, from an inspection of $\delta G_{\mu\nu}$ alone we effectively know ahead of time what the generic form of $ \delta T_{\mu\nu}$ has to be. Or equivalently, from an inspection of $\delta T_{\mu\nu}$ alone we effectively know ahead of time what the generic form of $ \delta G_{\mu\nu}$ has to be. In addition we note that because $F$ and $F_{\mu}$ do not appear in the gauge-invariant (\ref{6.19}), no information can be obtained  by applying the covariant curl of the type introduced in (\ref{1.36a}).
\bibitem{footnote13} For fluctuations around a de Sitter background for instance, $T_{\mu\nu}=3H^2g_{\mu\nu}$, $\delta T_{\mu\nu}=3H^2(h^T_{\mu\nu}+h^L_{\mu\nu})$ with $\delta T^T_{\mu\nu}=3H^2h^T_{\mu\nu}$. Also we note that while an interplay between the transverse sectors of $\delta G_{\mu\nu}$ and $\delta T_{\mu\nu}$ is needed in order to enforce gauge invariance, no such interplay is needed for the Maxwell $\partial_{\alpha}\partial^{\alpha} A_{\mu}^T=J_{\mu}$ since $J_{\mu}$ is transverse, obeying $\partial_{\mu}J^{\mu}=0$ in flat space and $\nabla_{\mu}J^{\mu}=0$ in curved space.
\end{thebibliography}
\end{document}